\documentclass[prx,twocolumn,balancelastpage,superscriptaddress,floatfix,longbibliography,nofootinbib]{revtex4-2}
\pdfoutput=1

\usepackage{stylesheet}
 
\begin{document}

\notoc

\title{Continuous-variable quantum state designs: theory and applications}

\author{Joseph~T.~Iosue}
\email{jtiosue@umd.edu}
\affiliation{\JQI}
\affiliation{\QUICS}

\author{Kunal~Sharma}
\email{kunals@ibm.com}
\affiliation{\QUICS}
\affiliation{IBM~Quantum,~IBM~T.J.~Watson~Research~Center,~Yorktown~Heights,~NY~10598,~USA}

\author{Michael~J.~Gullans}
\affiliation{\QUICS}

\author{Victor~V.~Albert}
\affiliation{\QUICS}

\date{\today}

\begin{abstract}
We generalize the notion of quantum state designs to infinite-dimensional spaces.
We first prove that, under the definition of continuous-variable (CV) state~$t$-designs from~\href{http://doi.org/10.1007/s00220-014-1894-3}{Comm.~Math.~Phys.~\textbf{326}, 755 (2014)}, no state designs exist for~$t\geq2$. 
Similarly, we prove that no CV unitary $t$-designs exist for $t\geq 2$.
We propose an alternative definition for CV state designs, which we call rigged $t$-designs, and provide explicit constructions for~$t=2$. 
As an application of rigged designs, we develop a design-based shadow-tomography protocol for CV states. 
Using energy-constrained versions of rigged designs, we define an average fidelity for CV quantum channels and relate this fidelity to the CV entanglement fidelity.
As an additional result of independent interest, we establish a connection between torus $2$-designs and complete sets of mutually unbiased bases.
\end{abstract}

\maketitle

\section{Introduction \& summary}

It is useful in a wide variety of fields to be able to efficiently
calculate uniform averages of polynomial functions over points in
a space. Prominent examples include Gaussian quadrature rules \cite{Gauss1866} and spherical designs \cite{Delsarte1977,Hardin1996}, which reduce integrals of polynomials to weighted
sums of polynomial values at particular points. 
More generally, a
$t$-\textit{design} over a space is a set of points picked in
such a way that averaging any polynomial of degree $\leq t$ over
the design is equivalent to uniformly averaging the same polynomial over the space.
Gaussian quadrature rules and spherical designs are $t$-designs over
the hypercube and hypersphere, respectively, and closely related ideas can be formulated for simplices and tori  \cite{stroud_approximate_1971,cools_constructing_1997,hammer_numerical_1956,baladram_on_2018,kuperberg_numerical_2004,kuperberg_numerical_ec_2004,victoir} as well as general topological spaces \cite{seymour1984averaging}.

Designs also have a number of important applications in quantum theory.
A quantum state $t$-design is an ensemble of quantum states such that expectation values of homogeneous polynomials of degree $t$ or less in the amplitudes of quantum states are the same whether the averaging is performed uniformly over all states or over only the states in the design~\cite{hoggarTDesignsProjectiveSpaces1982,hoggarParametersTDesignsFPd1984,bannaiTightDesignsCompact1985,wootters1989optimal,renesSymmetricInformationallyComplete2004,klappenecker2005mutually, dankert2005efficient, scott_tight_2006,ambainisQuantumTdesignsTwise2007,roberts_chaos_2017,Kueng2015}. 
State, unitary, and spherical~\cite{dankertExactApproximateUnitary2009, ambainisQuantumTdesignsTwise2007} designs are important tools in tomography~\cite{renesSymmetricInformationallyComplete2004,scott_tight_2006,PhysRevLett.108.110503,aaronsonShadowTomographyQuantum2018,huangPredictingManyProperties2020,huangProvablyEfficientMachine2022,Acharya2021}, state distinction~\cite{ambainisQuantumTdesignsTwise2007,kueng2016distinguishing}, randomized benchmarking~\cite{emersonScalableNoiseEstimation2005,knillRandomizedBenchmarkingQuantum2008,dankertExactApproximateUnitary2009,scalablemagesan,cross2016scalable}, fidelity estimation~\cite{nielsenEntanglementFidelityQuantum1996,horodeckiGeneralTeleportationChannel1999,nielsenSimpleFormulaAverage2002,dankertExactApproximateUnitary2009,magesanGateFidelityFluctuations2011,luExperimentalEstimationAverage2015,Bravyi2021}, cryptography \cite{ambainis2004small,hayden2004randomizing}, sensing \cite{OAG16,kimmel2017phase}, fundamental physics \cite{roberts_chaos_2017,miInformationScramblingComputationally2021,sekino2008fast,hayden2007black}, and error correction \cite{conwaySpherePackingsLattices1999,Wu2010,Lacerda2016,Lacerda2017}.

Both the original formulation of designs and its quantum counterparts hold only for finite-dimensional spaces.
This means that none of the applications proven to work through the use of designs, e.g., quantum state fidelity relations \cite{emersonScalableNoiseEstimation2005,knillRandomizedBenchmarkingQuantum2008,dankertExactApproximateUnitary2009,scalablemagesan,cross2016scalable,horodeckiGeneralTeleportationChannel1999,nielsenSimpleFormulaAverage2002} and design-based tomographic protocols \cite{aaronsonShadowTomographyQuantum2018,huangPredictingManyProperties2020,huangProvablyEfficientMachine2022,Acharya2021}, carry over naturally to countably \textit{infinite}-dimensional spaces. 
Such spaces are important for quantum applications because they describe quantum systems whose natural degrees of freedom are continuous variables (CVs), e.g., electromagentic modes of optical or microwave cavities, or mechanical modes of harmonic oscillators.

Formulating a notion of designs would unlock important abilities for CV systems. We proceed to do so in this paper, summarizing both our formalism and several fleshed-out applications below.

\prg{Non-existence of CV designs}

A first attempt to define state $t$-designs for CV systems was made in Ref.~\cite{blume-kohout_curious_2014}. The authors showed that a particular set of CV states --- the Gaussian states \cite{Weedbrook2012} --- does not form a CV $2$-design. This is perhaps surprising since Gaussian unitaries are the infinite-dimensional analog of finite-dimensional Clifford unitaries, which themselves can form $2$-designs~\cite{webbCliffordGroupForms2016,zhuCliffordGroupFails2016,zhuMultiqubitCliffordGroups2017,graydonCliffordGroupsAre2021}. Similarly, Ref.~\cite{zhuang_scrambling_2019} defined the notion of CV unitary $t$-designs and argued that Gaussian unitaries do not form a 2-design. These results leave open the question of whether CV state (unitary) designs require non-Gaussian states~(unitaries). 

In this work, we answer this open question and prove that CV state and unitary $t$-designs do not exist for any $t\geq 2$. Our results hold for any separable, infinite-dimensional Hilbert space, not just the space $L^2(\mathbb{R})$ associated with CV quantum systems. 
Thus, even the inclusion of non-Gaussian states and unitaries does not help in defining $t$-designs over CV systems. 

Our proof relies on the connection between state designs and simplex designs. We first show that infinite-dimensional simplex $t$-designs do not exist for $t\geq 2$. 
Then, using the simple fact that the complex probability amplitudes of any pure quantum state can be parameterized by a simplex (for the moduli) and a torus (for the phases), we show by contradiction that infinite-dimensional state designs do not exist either. 

\prg{Rigged designs}
We show that removing the requirement for states to be normalizable yields a meaningful extension of the notion of designs. We define \textit{rigged $t$-designs} that utilize states in a rigged Hilbert space, the Hilbert space populated by, e.g., the non-normalizable eigenstates of the oscillator position and momentum operators. We construct several examples for rigged $2$-designs, thus proving that rigged state designs exist even though CV state $(t\geq 2)$-designs do not.

In particular, it is well-known that there is no notion of \textit{uniform} integration over $L^2(\bbR)$, and our proof that CV $t$-designs do not exist for $t\geq 2$ proves that there is no form of integration over $L^2(\bbR)$ that has even basic qualities that \textit{mimick} uniform integration. Rigged designs get around this shortcoming by expanding the integration space to the set of all non-normalizable states in a rigged Hilbert space---specifically, the space $S(\bbR)'$ of \textit{tempered distributions}. 
We construct a measure on $S(\bbR)'$ that mimicks the qualities of a uniform measure over infinite-dimensional quantum states, and we then construct designs on this space.

Our first rigged design consists of Fock states as well as the phase states, which form a well-known positive operator-valued measure (POVM) that is optimal for measuring the angle of rotation induced on a mode \cite{susskindQuantumMechanicalPhase1964,carruthers_phase_1968,helstromQuantumDetectionEstimation1969,holevoCovariantMeasurementsImprimitivity1984,Bergou1991,mathewsSimultaneousUncertaintiesCosine1974,shapiroQuantumPhaseMeasurement1991} (see \cite[Sec. 3.9]{Holevo2011} for an exposition). The other examples combine Fock states with the cosine and sine states (and rotated states thereof), close relatives of the phase states \cite{carruthers_phase_1968}. 
In all cases, an extra parameter is induced on the phase states via evolution by a ``Kerr'' Hamiltonian $\hat{n}^2$, with $\hat n$ the occupation number operator \cite{girvin2014circuit}.

\prg{Design-based shadows}

The ability to use rigged $t$-designs as POVMs lends itself to a natural extension of shadow tomography \cite{aaronsonShadowTomographyQuantum2018,huangPredictingManyProperties2020,huangProvablyEfficientMachine2022,Acharya2021} to CV systems. In finite-dimensional versions of such protocols, one generates a classical snapshot of an unknown quantum state by performing random measurements according to the states from a $2$- or $3$-design. Then the expectation values over several observables can be efficiently and accurately estimated using these classical snapshots \cite{huangPredictingManyProperties2020}. 

We propose a CV shadow tomography protocol based on the Kerred-phase-state and Fock-state rigged $2$-design. 
The advantage of our protocol is that it maintains the key feature of the original qubit shadow protocols; namely, the ability to efficiently measure many observables using only a set of ``shadow'' snapshots of a particular form.
This protocol can be generalized to an efficient multi-mode protocol using a recent result \cite{Gandhari2022}. 
We discuss how our rigged CV shadows can be used for CV entanglement verification.

Although our design-based shadow protocol is more experimentally taxing than, e.g., CV shadows based on conventional homodyne or photon parity measurements \cite{Gandhari2022}, it can be implemented by combining and improving previously demonstrated experimental techniques. 
In order to utilize our first (second, third) rigged two-design as a POVM in the lab, one needs to be able to evolve the system under a Kerr Hamiltonian and then apply the phase (cosine, sine) state POVM.
In addition, one needs to alternatively measure in the Fock-state basis.
All three aspects of this protocol --- CV phase measurements \cite{Wiseman95,Martin2020}, photon-number resolution (e.g., \cite{Schuster2007}), and engineered Kerr evolution \cite{Holland2015,Elliott2018,Zhang2022} --- have been realized in some form in microwave cavities coupled to superconducting qubits \cite{girvin2014circuit}.  
Providing an experimentally realizable implementation of our protocol that can achieve the same scaling as our predicted sample complexity is an interesting avenue for future work.

\prg{Approximate CV designs}

Another natural question to ask is whether \textit{approximate} CV state designs exist in $L^2(\mathbb{R})$. Or, can the notion of designs be defined over the CV states that satisfy some energy constraints? We provide a solution to this problem by regularizing the rigged CV designs.

To approximate our rigged designs with sets of normalized states, we use operators called \textit{regularizers}, which correspond to different cutoffs of the infinite-dimensional space. For example, a regularizer that projects onto a low-energy finite-dimensional subspace corresponds to a hard cutoff, i.e., a maximum-energy constraint. 
A regularizer that smoothly decays with increasing energy but has support on the full infinite-dimensional space corresponds to a soft cutoff, i.e., an average-energy constraint. 
By analogy to numerical quadrature rules on the real line, a sharp cutoff is akin to restricting the domain of integration to a compact interval, while a smooth cutoff is akin to endowing the line with a Gaussian measure.
Moreover, certain regularizers allow us to extend the notion of a frame potential~\cite{renesSymmetricInformationallyComplete2004,klappenecker2005mutually,scottOptimizingQuantumProcess2008} to infinite dimensions.

Regularizers (\textit{a.k.a.}\ cooling or damping operators) and related ideas have been employed in works on CV quantum error-correcting codes \cite{Gottesman2001,mol}\cite[Appx. B]{Menicucci2014}, uniform continuity for quantum entropies \cite{winter2016tight}, energy-constrained capacities \cite{sharma2018bounding} and distances \cite{winter2017energy,shirokov2018energy} of CV channels, and CV cryptographic protocols \cite{Culf}.

\prg{Average CV fidelity}

Armed with regularized-rigged designs, we extend the well-known notion of average fidelity (over all states) of a quantum channel from finite-dimensional \cite{emersonScalableNoiseEstimation2005,knillRandomizedBenchmarkingQuantum2008,dankertExactApproximateUnitary2009,scalablemagesan,cross2016scalable} to CV systems.
In previous such extensions, systems were limited to the setting where the average fidelity between operations is estimated over an ensemble of coherent states~\cite{braunstein2000criteria,furusawa1998unconditional,namiki2008fidelity,namiki2008fidelity,lvovsky2009continuous,chiribella2013optimal,bai2018test,wu2019efficient,farias2021certification,sharma2022optimal}. 
Other approaches to benchmarking CV operations rely on witnesses that are lower bounds to the true average fidelity over an ensemble of Gaussian states \cite{wu2019efficient,farias2021certification}, while energy-constrained diamond-distance based performance estimates require knowledge of the noise model in experimental approximations and are often computationally taxing \cite{winter2017energy,sharmaCharacterizingPerformanceContinuousvariable2020,becker2020convergence,becker2021energy,sharma2022optimal,lami2021quantum,mishra2022optimal}.

We provide two different definitions of the average fidelity of a CV quantum channel. These formulas can be directly employed to estimate the average fidelity between CV quantum gates and their experimental approximations \cite{sharmaCharacterizingPerformanceContinuousvariable2020}. 
Our formulation yields an experimental procedure to estimate the average fidelity of an arbitrary CV quantum gate without requiring the knowledge of the noise involved in experimental implementations. 
As a concrete example, we estimate the average fidelity between an ideal displacement operation and its experimental approximation \cite{sharmaCharacterizingPerformanceContinuousvariable2020}, suggesting that an average over coherent states only is not a good approximation to an average over all CV states.

\prg{Average-to-entanglement fidelity relation}
Another interesting open question in CV information theory is to establish a relation between the average channel fidelity and the entanglement fidelity, similar to the finite-dimensional setting \cite{horodeckiGeneralTeleportationChannel1999,nielsenSimpleFormulaAverage2002}. 
In this work, we solve this open problem and establish connections between average and entanglement fidelities for CV operations. 

We utilize the conventional notion of single-mode CV entanglement fidelity, namely, the fidelity over a two-mode squeezed vacuum state \cite{H12,AS17}. We then evaluate our average fidelity over states in the corresponding regularized-rigged design. Combining these two fidelity formulas, we establish a simple relation between the average gate fidelity and the entanglement fidelity for CV operations.

\prg{Relating designs to MUBs}
As an additional result of independent interest, we find a relationship between torus $2$-designs and complete sets of mutually unbiased bases \cite{durtMutuallyUnbiasedBases2010}, and we prove that the condition of mutually unbiasedness can be replaced by a torus $2$-design condition.

\prg{Outline}
The rest of the paper is meant to succinctly relay the results and is structured as follows. In \cref{sec:finite}, we introduce finite-dimensional designs. In \cref{sec:infinite}, we develop the notion of infinite-dimensional designs and prove that CV state and unitary $t$-designs do not exist for any $t\geq2$. In \cref{sec:rigged}, we then define rigged designs and provide explicit constructions for rigged 2-designs. In \cref{sec:regularized-rigged-designs}, we introduce regularized rigged designs. 
In \cref{sec:applications}, we study applications of rigged and regularized rigged designs. 
In particular, in \cref{sec:shadows}, we introduce the shadow tomography formalism to CV quantum states.
In \cref{sec:entanglement-verification}, we discuss how such rigged CV shadows can be used for CV entanglement verification.
Next, in \cref{sec:fidelity}, we define various notions of the average fidelity of a CV quantum channel using regularized rigged $2$-designs. We then prove a relationship between the CV entanglement and average fidelities. 
Finally, in \cref{sec:conclusion}, we conclude with a brief summary and discuss open questions.

\section{Finite dimensional designs}
\label{sec:finite}

In this section, we review relevant prior results on finite-dimensional state designs, making contact with designs on simplices and tori.

Quantum state designs reduce integrals of polynomials over all quantum states to averages over a discrete set.
Let $\bbC^d$ denote a $d$-dimensional Hilbert space with orthonormal basis $\set{\ket n}_{n=0}^{d-1}$. Due to their normalization and global-phase redundancy, quantum states in this space correspond to points in the complex-projective space $\bbC\bbP^{d-1}$~\cite{bengtsson_geometry_2008,nakahara_geometry_2018}.
A non-trivial complex-projective $t$-design is a set of states $X \subsetneq \bbC\bbP^{d-1}$, sampled according to some probability measure $\mu$, satisfying \cite{hoggarTDesignsProjectiveSpaces1982,hoggarParametersTDesignsFPd1984,bannaiTightDesignsCompact1985,wootters1989optimal,renesSymmetricInformationallyComplete2004,klappenecker2005mutually, dankert2005efficient, scott_tight_2006,ambainisQuantumTdesignsTwise2007,roberts_chaos_2017}
\begin{equation}
    \Expval_{\psi \in X} f(\psi) = \int_{\bbC\bbP^{d-1}} f(\psi) \dd{\psi}
\end{equation}
for any polynomial $f(\psi)$ of degree $t$ or less in the amplitudes of $\psi$ and degree $t$ or less in the conjugate amplitudes. The canonical measure $\dd{\psi}$ on the set of such quantum states, called the \textit{Fubini-Study measure}~\cite{bengtsson_geometry_2008,nakahara_geometry_2018}, is the unique unit-normalized volume measure 
that is invariant under the action of the unitary group $\U(d)$ (see \cref{ap:measure-theory} for more details).

The above conventional relation can be lifted into a relation between particular operators by using the fact that polynomials of degree up to \(t\) in state degrees of freedom can be expressed as expectation values of operators with \(t\) copies of the state.

Consider, for example, $t=2$ and an arbitrary polynomial $f(\psi)=\sum_{j,k,l,m=0}^{d-1}f_{jklm}\bar{\psi}_{j}\bar{\psi}_{k}\psi_{l}\psi_{m}$ in the  amplitudes $\psi_j \coloneqq \langle j|\psi\rangle$ and their conjugates $\bar \psi_j$, with complex coefficients $f_{jklm}$. This polynomial can equivalently be expressed as an expectation value of a bipartite operator $\hat f$ with respect to two copies of $|\psi\rangle\langle\psi|$,
\begin{salign}
    f(\psi)&=\sum_{j,k,l,m=0}^{d-1}f_{jklm}\langle l|\psi\rangle\langle m|\psi\rangle\langle\psi|j\rangle\langle\psi|k\rangle\\
    &=\Tr\left(\hat{f}~|\psi\rangle\langle\psi|^{\otimes2}\right)~,
\end{salign}
where $\hat{f}=\sum_{j,k,l,m=0}^{d-1}f_{jklm}|j\rangle|k\rangle\langle l|\langle m|$. 
Using this relation, we see that $X$ is a $t$-design if and only if
\begin{equation}\label{eq:cpdesign1}
    \Expval_{\psi\in X} \parentheses{\ket\psi\bra\psi}^{\otimes t} 
    = \int_{\bbC\bbP^{d-1}} \parentheses{\ket\psi\bra\psi}^{\otimes t} \dd{\psi}~.
\end{equation}

Next, we can use representation theory (see \cref{ap:cp-haar-integral} for details) to solve the integral on the right-hand side, yielding
\begin{equation}\label{eq:haar-integral}
    \int_{\bbC\bbP^{d-1}} (\ket\psi\bra\psi)^{\otimes t} \dd\psi = \frac{\Pi_t^{(d)}}{\Tr\Pi_t^{(d)}} ~,
\end{equation}
where $\Pi_t^{(d)}$ is the projector onto the permutation-invariant (\textit{a.k.a.}\ symmetric~\cite{harrow_church_2013}) subspace of $(\bbC^d)^{\otimes t}$, the $t$-fold tensor product of the original space, and $\Tr$ is the trace function.
When $t=1$, this integral reduces to a resolution of the identity.
For higher $t$, the resolution can only be of the symmetric subspace since the $t$-fold tensor product of any state is symmetric under all permutations (see \cref{ap:symmetric-projector} for details).

Combining the above manipulations yields the following ``operator-level'' definition of a complex-projective $t$-design,
\begin{equation}\label{def:cpdesign}
    \Expval_{\psi\in X} \parentheses{\ket\psi\bra\psi}^{\otimes t} 
    = \frac{\Pi_t^{(d)}}{\Tr\Pi_t^{(d)}}~.
\end{equation}

Designs can be obtained via the convenient parameterization of pure states in terms of a simplex and a torus. State amplitudes can be written as
\begin{equation}\label{eq:simplex-torus-par}
    \langle j|\psi\rangle = \sqrt{p_j}~\e^{\i\phi_j}~,
\end{equation}
where the probabilities $p_j$ add up to one due to normalization, and the phases $\phi_j$ are $2\pi$-periodic (with $\phi_0$ set to zero to remove global-phase redundancy).
By definition, the probability distribution defined by $p_j$ is a point on the $(d-1)$-simplex,
\begin{equation}
    \Delta^{d-1} \coloneqq \bigg\{(p_0, \dots, p_{d-1}) \in [0,1]^d ~~\bigg\vert~~ \sum_{j=0}^{d-1} p_j = 1 \bigg\},
\end{equation}
while the vector of phases parameterizes a $(d-1)$-torus $T^{d-1}$.
Hence, volume integration over all states is reduced to volume integration over the simplex and the torus \cite{bengtsson_geometry_2008,nakahara_geometry_2018} (see \cref{ap:simplex-torus-to-cp} for details).
This naturally makes contact with simplex and torus designs.

Simplex and torus designs are defined in similar fashion to complex-projective designs.
A set $X\subset \Delta^{m}$ of probability vectors is an $m$-simplex $t$-design if for all tuples $a = (a_1, \dots, a_t) \in \set{0,1,\dots, m}^t$,
\begin{equation}\label{eq:simplex-design}
    \Expval_{q\in X} \prod_{i=1}^t q_{a_i} = \int_{\Delta^m} \prod_{i=1}^t p_{a_i}\dd{p},
\end{equation}
where $\dd{p}$ is the standard measure on the simplex.
A set of angles $X \subset T^m$ is an $m$-torus $t$-design if for all tuples $a = (a_1, \dots, a_t) \in \set{1,2,\dots, m}^t$ and $b = (b_1, \dots, b_t) \in \set{1,2,\dots, m}^t$,
\begin{equation}\label{eq:torus-design}
     \Expval_{\theta \in X}\prod_{i=1}^t \e^{\i(\theta_{a_i} - \theta_{b_i})} = \int_{T^m} \prod_{i=1}^t \e^{\i(\phi_{a_i} - \phi_{b_i})} \dd{\phi},
\end{equation}
where $\dd{\phi}$ is the standard measure on the torus.
We discuss various constructions of simplex and torus designs in \cref{ap:simplex-designs} and \cref{ap:torus-designs}, respectively.

There is a bilateral connection between complex-projective designs and designs on the corresponding simplices and tori.
Denoting $\pi$ as the ``Born-rule'' map that produces the vector of probabilities $(p_n)_{n=0}^{d-1}$ from a state $|\psi\rangle$,
the set $\pi(X)$ is a simplex $t$-design for any complex-projective $t$-design $X$~\cite{kuperberg_numerical_2004,czartowski_isoentangled_2020} (see \cref{ap:cp-to-simplex} for details).
On the other hand, a combination of a simplex and a torus $t$-design of appropriate dimensions yields a complex-projective $t$-design~\cite{kuperberg_numerical_2004}.
We provide a proof of these latter connections and present various combinations that yield complex-projective $2$-designs for all $d$ in \cref{ap:simplex-torus-to-cp}.

Our simplex designs from \cref{eq:simplex-design} are more commonly referred to as simplex positive, interior (or boundary) cubature rules \cite{cools_constructing_1997,stroud_approximate_1971,hammer_numerical_1956,baladram_on_2018,kuperberg_numerical_2004,kuperberg_numerical_ec_2004}.
Our torus $t$-designs from \cref{eq:torus-design} closely resemble \emph{trigonometric cubature rules}~\cite{cools_constructing_1997}, but the two are not equivalent.
In \cref{ap:torus}, we show that torus designs are equivalent to a special case of torus cubature rules from Ref.~\cite{kuperberg_numerical_2004}.
We then establish a connection between torus $2$-designs and mutually unbiased bases (MUBs), which might be of independent interest. To the best of our knowledge, this connection has not been previously discussed.

\section{Continuous-variable designs}
\label{sec:infinite}

In this section, we develop the notion of continuous-variable (CV) designs and present our main results in \cref{thm:cv-design-nonexist} and Corollary~\ref{cor:nonexistence-unitary}.

Let $L^2(\bbR)$ denote an infinite-dimensional, separable Hilbert space of square-integrable functions on the real line, with a countable \textit{Fock-state} (\textit{a.k.a.}\ photon number-state or occupation number-state) basis $\set{\ket n \mid n \in \bbN_0}$, where $\bbN_0$ denotes the natural numbers including zero. 
We note that all separable Hilbert spaces are isomorphic to $L^2(\bbR)$. We call unit-norm vectors in $L^2(\bbR)$ CV quantum states.

The right-hand side of Eq. \eqref{eq:haar-integral} is straightforward to generalize to infinite dimensions.
Let $\Pi_t$ denote the projector onto the symmetric subspace of $t$ copies of $L^2(\bbR)$ (see \cref{ap:symmetric-projector}). For any tuples $a = (a_1,\dots, a_t) \in \bbN_0^t$ and $b = (b_1, \dots, b_t) \in \bbN_0^t$, 
\begin{equation}
    \Pi_t(a;b) \coloneqq \parentheses{\bigotimes_{i=1}^t \bra{a_i}} \Pi_t \parentheses{\bigotimes_{i=1}^t \ket{b_i}}
\end{equation}
denotes the matrix elements of $\Pi_t$.
The trace of this projector, $\Tr\Pi_t$, is infinite, but we can simply omit it from the equation.

The left-hand side of Eq.~\eqref{eq:haar-integral} is unfortunately impossible to generalize to infinite dimensions \cite{huntPrevalenceTranslationinvariantAlmost1992}.
Since $L^2(\bbR)$ is infinite dimensional, there is no finite Haar measure on its corresponding unitary group $\U(L^2(\bbR))$~\cite[Sec.~5]{grigorchukAmenabilityErgodicProperties2015}. 
Therefore, there is no natural unitarily invariant volume measure on the set of CV quantum states. 
However, if one \textit{could} define the unitarily invariant volume integration over all CV states, Schur's lemma \textit{would} imply that the resulting integration would be proportional to $\Pi_t$. 
Therefore, in principle, infinite-dimensional state designs can be defined using the definition of complex-projective designs in Eq.~\eqref{def:cpdesign}, but without the $\Tr\Pi_t$ term.

An infinite-dimensional design may be parameterized by points in a noncompact space with a non-normalizable measure. To accommodate this, we relax the assumption that the parameter space of a design is a probability space and instead assume it is a generic measure space --- a triple consisting of $X\subset L^2(\bbR)$, a collection $\Sigma$ of all reasonable subsets of $X$ called a $\sigma$-algebra, and a measure $\mu$ (see \cref{ap:CVdesign} for details). The only difference from a probability space is that $\mu(X)$, the measure on the entire space, no longer has to be finite.

Combining the above ideas, we define CV designs as abstract measure spaces that average to the unnormalized symmetric-subspace projector.

\begin{definition}
\label{def:cv-design}
Let $X \subset L^2(\bbR)$. The measure space $(X,\Sigma,\mu)$ is a \emph{continuous-variable $t$-design} if
\begin{equation}
    \int_X \parentheses{\ket\psi\bra\psi}^{\otimes t} \dd{\mu(\psi)} = \Pi_t,
\end{equation}
where we use the weak (Pettis) integral.
\footnote{The use of the weak integral in the definition of CV designs is well-motivated. For the purposes of designs, the weak (Pettis) integral is more natural than the strong (Bochner) integral because we are generally interested in averaged functions of $\psi$. Ultimately, we will prove that CV $t$-designs do not exist for $t\geq 2$, which immediately implies the result for the case of the strong integral as well.}
In other words, for all tuples $a,b \in \bbN_0^t$,
\begin{equation}
    \int_X \parentheses{\prod_{i=1}^t \bra{a_i} \ket\psi \bra\psi \ket{b_i}} \dd{\mu(\psi)} = \Pi_t(a; b)~.
\end{equation}
\end{definition}

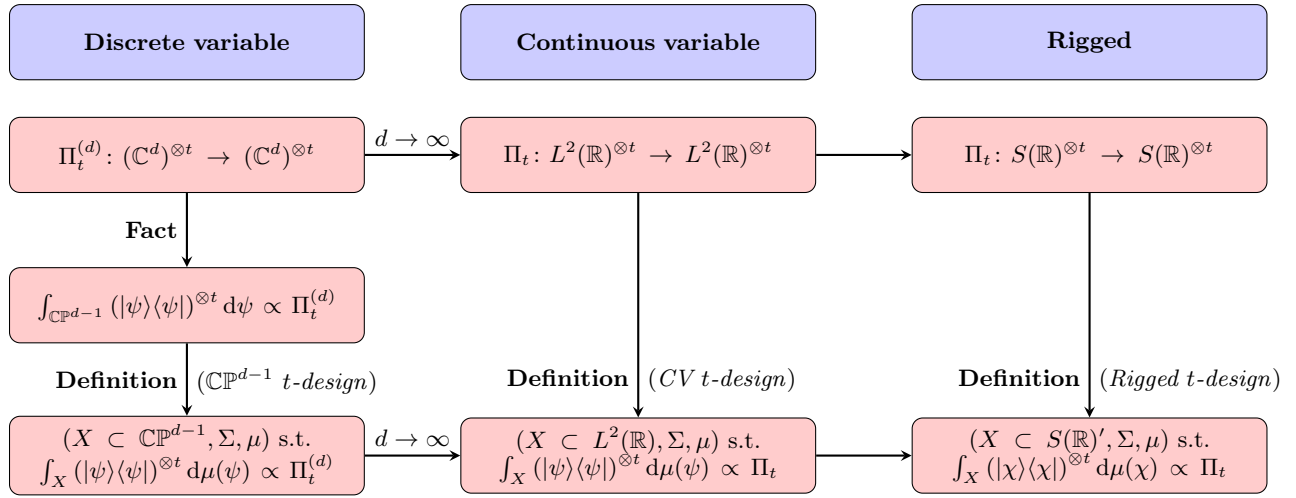
\begin{figure*}
    \centering

\tikzstyle{block} = [rectangle, rounded corners, minimum width=3cm, minimum height=1cm,text centered, text width=4.5cm, draw=black, fill=red!20]
\tikzstyle{title} = [rectangle, rounded corners, minimum width=3cm, minimum height=1cm,text centered, text width=4.5cm, draw=black, fill=blue!20]
\tikzstyle{arrow} = [thick,->,>=stealth]

\begin{tikzpicture}[node distance=2cm]

    \hypersetup{hidelinks}
    
    \node (cp-title) [title] {\bf Discrete variable};
    
    \node (cv-title) [title, right of=cp-title, xshift=4cm] {\bf Continuous variable};
    
    \node (cv-title) [title, right of=cv-title, xshift=4cm] {\bf Rigged};
    
    \node (finite-pi) [block, below of=cp-title, yshift=.5cm] {$\Pi_t^{(d)}\colon (\bbC^{d})^{\otimes t} \to (\bbC^{d})^{\otimes t}$};
    
    \node (cp-integral) [block, below of=finite-pi] {$\int_{\bbC\bbP^{d-1}} \parentheses{\ket\psi\bra\psi}^{\otimes t} \dd{\psi} \propto \Pi_t^{(d)}$};
    
    \node (cp-design) [block, below of=cp-integral] {$(X \subset \bbC\bbP^{d-1},\Sigma,\mu)$ s.t. $\int_X \parentheses{\ket{\psi}\bra{\psi}}^{\otimes t} \dd{\mu(\psi)} \propto \Pi_t^{(d)}$};

    \node (infinite-pi) [block, right of=finite-pi, xshift=4cm] {$\Pi_t\colon L^2(\bbR)^{\otimes t} \to L^2(\bbR)^{\otimes t}$};

    \node (cv-design) [block, right of=cp-design, xshift=4cm] {$(X \subset L^2(\bbR),\Sigma,\mu)$ s.t. $\int_X \parentheses{\ket{\psi}\bra{\psi}}^{\otimes t} \dd{\mu(\psi)} \propto \Pi_t$};

    \node (rigged-pi) [block, right of=infinite-pi, xshift=4cm] {$\Pi_t\colon S(\bbR)^{\otimes t} \to S(\bbR)^{\otimes t}$};
    
    \node (rigged-design) [block, right of=cv-design, xshift=4cm] {$(X \subset S(\bbR)',\Sigma,\mu)$ s.t. $\int_X \parentheses{\ket{\chi}\bra{\chi}}^{\otimes t} \dd{\mu(\chi)} \propto \Pi_t$};

    \draw [arrow] (finite-pi) -- node[anchor=east] {\bf \hyperref[eq:haar-integral]{Fact}} (cp-integral);
    
    \draw [arrow] (cp-integral) -- node[anchor=east] {\bf \hyperref[def:cpdesign]{Definition}} node[anchor=west, yshift=-.03cm] {\hyperref[def:cpdesign]{(\emph{$\bbC\bbP^{d-1}$ $t$-design})}} (cp-design);

    \draw [arrow] (infinite-pi) -- node[anchor=east, yshift=-.97cm] {\bf \hyperref[def:cv-design]{Definition}} node[anchor=west, yshift=-1cm] {\hyperref[def:cv-design]{(\emph{CV $t$-design})}} (cv-design);

    \draw [arrow] (rigged-pi) -- node[anchor=east, yshift=-.97cm] {\bf \hyperref[def:rigged-design]{Definition}} node[anchor=west, yshift=-1cm] {\hyperref[def:rigged-design]{(\emph{Rigged $t$-design})}} (rigged-design);

    \draw [arrow] (finite-pi) -- node[anchor=south] {$d\to\infty$} (infinite-pi);
    
    \draw [arrow] (infinite-pi) -- (rigged-pi);

    \draw [arrow] (cp-design) -- node[anchor=south] {$d \to \infty$} (cv-design);
    
    \draw [arrow] (cv-design) -- (rigged-design);

\end{tikzpicture}

    \caption{Sketch of definitions of finite-dimensional designs,  continuous-variable (CV) designs and rigged designs. The key point is the absence of the middle block in the middle and right columns. A generalization of the middle block to the continuous-variable case is ill-defined, as discussed in \cref{sec:infinite}. Therefore, to define CV designs, we simply skip the middle step, as discussed in \cref{def:cv-design}. An alternative characterization/definition of CV and rigged designs is described in \cref{ap:C-inf-designs}.
    }
    \label{fig:cv-design-def}
\end{figure*}

\cref{def:cv-design} is a formalized version of the definition of CV state $t$-designs given in Ref.~\cite{blume-kohout_curious_2014}. We note that \cref{def:cv-design} bypasses the issue of defining a volume measure on the set of CV quantum states. We do not perform any integration on the set of all states and instead \textit{require} a design to match the projector onto the symmetric subspace. This construction is illustrated in \cref{fig:cv-design-def}.

There is an alternative motivation for \cref{def:cv-design} that we describe in detail in \cref{ap:C-inf-designs}. It is based on the following observation in finite dimensions. Integration over the set $\bbC\bbP^{d-1}$ of $d$-dimensional quantum states is equivalent to integration over $\bbC^d$ with $d$ independent zero-mean, unit-variance complex Gaussian distributions. The integration is over each of the $d$ amplitudes of the quantum state with respect to the Gaussian measure, and the resulting state is then normalized. 
We can similarly put an infinite product of Gaussian measures on the space $\bbC^\infty$ and then define a CV $t$-design to be a measure space over $L^2(\bbR)$ that matches integration over $\bbC^\infty$ for polynomials of degree $t$ or less. We show in \cref{ap:C-inf-designs} that this definition is equivalent to \cref{def:cv-design}.

Since $\Pi_1 = \bbI$, where $\bbI$ denotes the infinite-dimensional identity operator, any orthonormal basis for $L^2(\bbR)$ or POVM is a CV $1$-design. For example, the photon-number basis $\ket n$ satisfies $\sum_{n\in\bbN_0}\ket n \bra n = \bbI$, which corresponds to a photon counting measurement. Coherent states $\{\ket\alpha\}$ also form a $1$-design as they satisfy~$\int_\bbC \ket\alpha\bra\alpha \frac{\Dd{2}{\alpha}}{\pi} = \bbI$, which corresponds to a heterodyne measurement. Finally, the eigenstates of $\cos(\phi)\hat x + \sin(\phi) \hat p$ form a $1$-design, which corresponds to a homodyne measurement.

In \cref{sec:finite}, we argued that a complex-projective design on $\bbC^d$ gives rise to a simplex design.
Similarly, in \cref{ap:nonexistence}, we prove that the existence of CV $t$-designs implies the existence of infinite-dimensional simplex $t$-designs. Here, we define a infinite-dimensional simplex design by starting with a finite-dimensional simplex integration over the unit-normalized Lebesgue measure and then removing the normalization requirement of the measure as we take the dimension to infinity.
 
By construction, a CV $1$-design induces an infinite dimensional simplex $1$-design by converting the amplitudes of a quantum state to probabilities via the Born rule. For example, the simplex design induced by the CV~$1$-design $\set{\ket n \mid n\in\bbN_0}$ is a set of probability distributions $\set{p^{(n)} \mid n \in \bbN_0}$. Here $p^{(n)} = (p_0^{(n)}, p_1^{(n)}, \dots)$ is a probability distribution over $\bbN_0$ defined as $p^{(n)}_i = \delta_{in}$.

As for $t>1$ designs, we prove that \textit{no set of CV states}, Gaussian or not, forms a CV $t$-design for any $t \geq 2$ (see \cref{ap:CVdesign} for proofs).

\begin{theorem}
\label{thm:cv-design-nonexist}
For any $t \geq 2$, continuous-variable state $t$-designs do not exist.
\end{theorem}

The non-existence of state $(t\geq 2)$-designs immediately implies non-existence of unitary $(t\geq 2)$-designs because their existence would imply the existence of state designs.

\begin{corollary}\label{cor:nonexistence-unitary}
For any $t \geq 2$, continuous-variable unitary $t$-designs do not exist.
\end{corollary}

To prove \cref{thm:cv-design-nonexist}, we show that infinite-dimensional simplex $t$-designs do not exist for $t \geq 2$, and then invoke the connection between state and simplex designs described in \cref{sec:finite}. The non-existence of infinite-dimensional simplex designs can be understood as follows. All simplex $(t\geq 2)$-designs require at least one point near the centroid of the simplex. The centroid of a finite-dimensional simplex $\Delta^{d-1}$ is the point $(1/d, \dots, 1/d)$. However, for the infinite-dimensional case, the centroid is no longer a valid point on the probability simplex. In the context of quantum states, this translates to the fact that uniform superpositions of all Fock states are not normalizable. 
We are therefore motivated to remove the requirement that elements of CV $t$-designs are normalized states. 

\section{Rigged designs: \texorpdfstring{\\}{} definition \& constructions}
\label{sec:rigged}

The non-existence of CV $t$-designs for $t>1$ stems from the requirement that elements of said designs, according to \cref{def:cv-design}, belong to $L^2(\mathbb{R})$ and thus should have finite norm. We are therefore motivated to develop a new notion of CV designs that allows for non-normalizable states.

To include non-normalizable states in a CV design, we need to consider a set larger than $L^2(\bbR)$.
We consider the space of \textit{tempered distributions}, denoted as $S(\bbR)'\supset L^2(\bbR)$, which contains infinitely squeezed position or momentum states as well as oscillator phase states \cite{susskindQuantumMechanicalPhase1964,carruthers_phase_1968,helstromQuantumDetectionEstimation1969,holevoCovariantMeasurementsImprimitivity1984,Bergou1991,Holevo2011}.
Despite being awkwardly called ``states'', these and other distributions may not be normalizable.

The use of distributions, whether for our purposes or for CV measurement protocols such as homodyne detection \cite{lvovsky2009continuous}, is only well-defined for those CV states for which inner products with tempered distributions are finite.
This class consists of those states which admit finite expectation values of all powers of the occupation number operator $\hat{n}=\sum_{n\in\bbN_0} n\ket n \bra n$, making up the \textit{Schwartz space} $S(\bbR) \subset L^2(\bbR)$ \cite{becnel_schwartz_2015}.
Together, the three spaces of interest make up the Gelfand triple
$
    S(\bbR) \subset L^2(\bbR) \subset S(\bbR)'~,
$
the standard \textit{rigged Hilbert space} for a quantum harmonic oscillator  \cite{rudin_functional_1991,gieres_mathematical_2000,madrid_role_2005}.

We modify \cref{def:cv-design} to include tempered distributions. The motivation for our modification is summarized in \cref{fig:cv-design-def}.
\begin{definition}
\label{def:rigged-design}
Let $X \subset S(\bbR)'$.
The measure space $(X, \Sigma, \mu)$ is called a \emph{rigged $t$-design} if
\begin{equation}\label{eq:rigged-design}
    \int_X \parentheses{\ket\chi\bra\chi}^{\otimes t'}\dd{\mu(\chi)} = \alpha_{t'}\Pi_{t'}
\end{equation}
for all positive integers $t' \leq t$, where $\alpha_{t'} \in (0,\infty)$, where we use the weak (Pettis) integral.
In other words, for all tuples $a,b \in \bbN_0^{t'}$,
\begin{equation}
    \int_X \parentheses{\prod_{i=1}^{t'} \bra{a_i} \ket\chi \bra\chi \ket{b_i}} \dd{\mu(\chi)} = \alpha_{t'}\Pi_{t'}(a; b)
\end{equation}
for $t\leq t'$.
\end{definition}

Analogously to what is discussed below \cref{def:cv-design}, there is an alternative motivation \cref{def:rigged-design} that we describe in \cref{ap:C-inf-designs}. Recall that we described an equivalent definition of CV designs to be measure spaces over $L^2(\bbR)$ that match integration over $\bbC^\infty$ with an infinite product of Gaussian measures. In \cref{ap:C-inf-designs}, we further show $\bbC^\infty \setminus S(\bbR)'$ has measure zero in $\bbC^\infty$, so that the integration over $\bbC^\infty$ is equivalent to integration over $S(\bbR)'$. It follows therefore that rigged $t$-designs exist for any $t\in \bbN$, since we can simply take the aforementioned measure space over $S(\bbR)'$ to be our design. 
This design is however not desirable since it involves infinite-dimensional integration. We thus look for more manageable measure spaces that form rigged designs.

Inclusion of distributions circumvents the no-go \cref{thm:cv-design-nonexist} and allows us to construct several examples of rigged $2$-designs.
Our first example consists of Fock states $\{|n\rangle\}_{n\in\bbN_0}$ and a family of distributions that we call \textit{Kerred phase states} $|\theta\rangle_{\varphi}$ --- tempered distributions defined informally as
\begin{equation}\label{eq:kerred-phase}
|\theta\rangle_{\varphi} \coloneqq \frac{1}{\sqrt{2\pi}} \sum_{n \in \bbN_0 } \exp\bargs{\i(\theta n + \varphi n^2)} \ket n,
\end{equation}
and formally as functionals mapping $|\psi\rangle\in S(\bbR)$ to
\begin{equation}
\psi(\theta,\varphi)
\coloneqq \prescript{}{\varphi}{\braket{\theta \vert \psi}}
= \frac{1}{\sqrt{2\pi}} \sum_{n \in \bbN_0 } \exp\bargs{-\i(\theta n + \varphi n^2)} \braket{n\vert \psi}.
\end{equation}
The Kerred phase ``states'' consist of oscillator phase states
\cite{susskindQuantumMechanicalPhase1964,carruthers_phase_1968,helstromQuantumDetectionEstimation1969,holevoCovariantMeasurementsImprimitivity1984,Bergou1991,mathewsSimultaneousUncertaintiesCosine1974,shapiroQuantumPhaseMeasurement1991,Holevo2011},
evolved up to some ``time'' $\theta$ under a Hamiltonian $\hat{n}^2$ associated with the optical Kerr effect.
In \cref{ap:rigged}, we prove that
\begin{equation}
    \label{eq:kerr-phase-rig-2-design}
    \frac{1}{2}\sum_{n\in\bbN_0} \parentheses{\ket n \bra n}^{\otimes t} + \frac{1}{2}\int_{-\pi}^\pi \dd{\theta}\int_{-\pi}^\pi \dd{\varphi} \parentheses{\prescript{}{\varphi}{\ket{\theta}}\!\bra{\theta}_\varphi}^{\otimes t} = \alpha_t\Pi_t
\end{equation}
for $t=1$ and $t=2$, where $\alpha_1 = \pi+1/2$ and $\alpha_2=1$.

To show that the above set is a design, we extend simplex and torus $2$-designs to the rigged regime (see \cref{ap:rigged} for details).
The integration over the two phases $\{\theta,\varphi\}$ corresponds to a torus $2$-design.
The Fock states $|n\rangle$ correspond to extremal points of a simple simplex $2$-design consisting of extremal points and the centroid in the finite-dimensional case, with the centroid vanishing in the infinite-dimensional case (as discussed in \cref{sec:infinite}).
By removing the normalization condition, we define an ``non-normalizable centroid'', which corresponds to a uniform superposition of Fock states $\ket{\theta=0}_{\varphi=0}$. Combining such a state with the aforementioned torus $2$-design gives the Kerred phase states.

Oscillator phase states are (left) eigenstates of the oscillator phase operator $Z=\sum_{n\in\bbN_0}|n+1\rangle\langle n|$ \cite{susskindQuantumMechanicalPhase1964,carruthers_phase_1968,helstromQuantumDetectionEstimation1969,holevoCovariantMeasurementsImprimitivity1984,Bergou1991,mathewsSimultaneousUncertaintiesCosine1974,shapiroQuantumPhaseMeasurement1991,Holevo2011}, an analogue of the oscillator raising operator but without the square-root factor.
Both the phase and raising operators do not admit right eigenstates, but $\pm$-superpositions of each operator with its adjoint yield (anti-)Hermitian operators that admit well-known distributions as eigenstates.
Superpositions of lowering and raising operators admit position and momentum states as eigenstates, respectively, while superpositions of the phase operator and its adjoint admit the \textit{cosine} and \textit{sine states}, respectively \cite{carruthers_phase_1968}.
In \cref{ap:rigged}, we show that these two sets of states, when evolved under the Kerr Hamiltonian and combined with Fock states, make up two more examples of rigged $2$-designs. 
More generally, since $Z$ is unitarily related to $Z \e^{\i\omega}$ via a Fock-space rotation $\e^{\i\omega\hat n}$, eigenstates of a linear combination of $Z\e^{\i\omega}$ and its conjugate should similarly yield a distinct set of designs for any $\omega$.

We do not provide constructions of useful rigged $3$-designs.
As shown with an example in \cref{ap:rigged}, not all simplex $2$-designs can be extended to infinite dimensions.
Thus, the difficulty in constructing a rigged $3$-design lies is finding a simplex $3$-design that is well-behaved enough to be extended to infinite dimensions.
We leave this exciting open question for future work.

\section{Regularized Rigged Designs}
\label{sec:regularized-rigged-designs}

Our rigged designs consist of non-normalizable states, but some applications require approximate versions of such designs that consist of physical quantum states.
One way to approximate is to simply truncate the Fock space, corresponding to a \textit{hard} or \textit{maximum-energy cutoff}.
This brings us back to finite dimensions, reducing rigged designs to ordinary quantum state designs.
Another way, possible only with our infinite-dimensional formulation, is to impose a \textit{soft} or \textit{average-energy cutoff} that maintains the ability for states to have infinite support in Fock space.
Both cutoffs can be encompassed in a general regularization formalism.

Let the \textit{regularizer} $R$ be a positive-semidefinite operator that yields a corresponding ``regularized projector''
\begin{equation}\label{eq:norm-proj}
    \Pi_t^{(R)} \coloneqq R^{\otimes t} \Pi_t R^{\otimes t}\quad\text{such that}\quad \Tr \Pi_t^{(R)}<\infty~.
\end{equation}
The two aforementioned energy cutoffs correspond, respectively, to regularizers
\begin{equation}\label{eq:cutoffs}
  R=\begin{cases}
P_{d}\coloneqq\sum_{n=0}^{d-1}|n\rangle\langle n| & \text{hard cutoff, }d\in\bbN_0\\
R_{\beta}\coloneqq \e^{-\beta\hat{n}} & \text{soft cutoff, }\beta>0
\end{cases}~,
\end{equation}
but our formalism allows for more general $R$.
We construct regularized designs by applying a regularizer to elements of a rigged design.

Suppose $(X, \Sigma, \mu)$ is a rigged $t$-design satisfying $\alpha_t = 1$.
Regularization by an appropriate regularizer, such as $R_\beta$ and $P_d$, converts $X$ into a set of normalized states
\begin{equation}
    Y \coloneqq \set{\ket\psi = R \ket\chi / \norm{R \ket\chi} \mid \ket\chi \in X},
\end{equation}
with corresponding $\sigma$-algebra $\Sigma_Y$ and measure
\begin{equation}\label{eq:rigged-measure}
    \dd{\nu\pargs{\ket\psi}} = \dd{\mu(\chi)} \cdot \norm{R \ket\chi}^{2t}/ \Tr\Pi_t^{(R)}~.
\end{equation}
These regularized designs average to $\Pi_t^{(R)}/\Tr\Pi_t^{(R)}$ instead of $\Pi_t$,
\begin{align}
    \int_Y \parentheses{\ket\psi\bra\psi}^{\otimes t}\dd{\nu(\psi)}
    &= \int_X \parentheses{R\ket\chi\bra\chi R}^{\otimes t} \frac{\dd{\mu(\chi)}}{\Tr\Pi_t^{(R)}}\nonumber\\
    &=\frac{\Pi_t^{(R)}}{\Tr\Pi_t^{(R)}} ~.\label{eq:r-reg-designs}
\end{align}

The use of normalized states allows us to promote $Y$ to a probability space.
By taking the trace of both sides of \cref{eq:r-reg-designs} and applying assumption (\ref{eq:norm-proj}),
we see that the measure $\nu$ is automatically normalized,
$
    1
    = \int_Y \braket{\psi\vert\psi}^t \dd{\nu(\psi)}
    = \nu(Y)
$.
This allows us to express $\int_Y (\cdot)\dd{\nu(\psi)}$ as a statistical expectation $\Expval_{\psi\in Y} (\cdot)$ of states in $Y$ sampled according to the distribution defined by $\nu$ (see \cref{ap:regularized-rigged} for details). This yields the definition below, with a related definition of regularized CV unitary designs provided in \cref{ap:cv-unitary-designs}.

\begin{definition}
\label{def:r-reg-designs}
Let $Y \subset L^2(\bbR)$.
The probability space $(Y, \Sigma_Y, \nu)$ is called an \emph{$R$-regularized rigged $t$-design} if
\begin{equation}\label{eq:reg-rigged-design}
    \Expval_{\psi\in Y} \parentheses{\ket\psi\bra\psi}^{\otimes t} = \frac{\Pi_t^{(R)}}{\Tr\Pi_t^{(R)}}~.
\end{equation}
\end{definition}

Analogous to the discussion below \cref{def:cv-design,def:rigged-design}, there is again an alternative motivation for \cref{def:r-reg-designs} that we detail in \cref{ap:C-inf-designs}. Recall that an infinite product of zero-mean, unit-variance Gaussian measures on $\bbC^\infty$ forms a rigged $t$-design. We show in \cref{ap:C-inf-designs} that if the variance of the $i^{\rm th}$ measure is instead $\lambda_i$ such that the diagonal operator $R_{ii} = \lambda_i$ is trace class ($\sum_i \lambda_i < \infty$), then the resulting measure space is a $\sqrt R$-regularized rigged $t$-design for any $t\in\bbN$. Importantly, with this measure, $\bbC^\infty \setminus L^2(\bbR)$ has measure zero in $\bbC^\infty$ so that the design is a measure space over $L^2(\bbR)$ as desired.

We now consider regularizing the Fock-state and Kerred phase-state design \eqref{eq:kerr-phase-rig-2-design} with the soft-energy cutoff $R = R_\beta = \e^{-\beta \hat n}$ (\ref{eq:cutoffs}).
Denote the regularized Kerred phase states (\textit{a.k.a.}\ phase coherent states \cite{dodonov_nonclassical_2002}) as
\begin{equation}
\ket{\tilde\theta}_\varphi \coloneqq \frac{R_\beta \ket\theta_\varphi}{\lVert R_\beta \ket\theta_\varphi\rVert} = \sqrt{1-\e^{-2\beta}}\sum_{n=0}^\infty \e^{-\beta n + \i \theta n + \i \varphi n^2}\ket n~,
\end{equation}
such that $\lVert \ket{\tilde\theta}_\varphi \rVert = 1$.
Then it follows that
\begin{align}\label{eq:reg-rigged}
    \sum_{n\in\bbN_0} w_n \ket n \bra n^{\otimes 2} + f_\beta
    \int
    \prescript{}{\varphi}{\ket{\tilde\theta}\bra{ \tilde\theta}}_\varphi^{\otimes 2}  \dd\theta \dd\varphi =\frac{\Pi_2^{(R_\beta)}}{\Tr \Pi_2^{(R_\beta)}},
\end{align}
where the limit of integration for both $\theta$ and $\varphi$ is $[-\pi,\pi]$, $f_\beta \coloneqq \cosh\beta/(\e^{\beta}(2\pi)^2)$, and $w_n = \frac{4  \sinh^2\beta \cosh\beta}{\e^{\beta(4n+3)}} $.

Given a fixed average-energy constraint $E$, it is natural to define an energy-constrained state design consisting of states $\{\psi\}$, such that each state in the design satisfies $\Tr(\hat{n}\psi)\leq E$.
Our regularized-rigged design does not satisfy this condition explicitly as it contains Fock states $|n\rangle$ with $n>E$, as shown in \cref{eq:reg-rigged}.
However, the contribution of large $n$ terms is suppressed by the $w_n$ coefficient in \cref{eq:reg-rigged}, which decays exponentially with $n$.
Thus, our regularized-rigged designs are good approximations to energy-constrained state designs.
It is an interesting open question to further develop the framework for energy-constrained state designs; we make some headway in this direction by formulating constrained integration in \cref{ap:constrained-integration}, albeit for the finite-dimensional case.

As another example, we show in \cref{ap:dispalced-fock} that displaced Fock states form regularized $2$-designs for which the regularizer is the maximum-energy cutoff from \cref{eq:cutoffs}, granted that we are allowed to use negative weights in the combination.

An important feature not inherited from the finite-dimensional case is that, in general, an $R$-regularized rigged $t$-design is not an $R$-regularized rigged $(t-1)$-design.
For example, if $Y$ is an $R$-regularized rigged $2$-design,
then
\begin{equation}\label{eq:r-reg-designs-test}
    \Expval_{\psi\in Y} \ket\psi\bra\psi = \frac{\Pi_1^{(R)}}{2\Tr\Pi_2^{(R)}} \parentheses{(\Tr R^2)\bbI + R^2} \neq \frac{\Pi_1^{(R)}}{\Tr\Pi_1^{(R)}},
\end{equation}
violating \cref{eq:reg-rigged-design} for $t=1$.
Similarly, if $Y$ is an $R$-regularized rigged $3$-design, then
\begin{equation}\label{eq:r-reg-designs-test2}
\begin{aligned}
    \Expval_{\psi\in Y} (\ket\psi\bra\psi)^{\otimes 2} = &\frac{\Pi_2^{(R)}}{3\Tr \Pi_3^{(R)}} \\
    \times \bigg((\Tr R^2) &\bbI  \otimes \bbI + \bbI \otimes R^2 + R^2 \otimes \bbI\bigg)
\end{aligned}
\end{equation}
instead of $\Pi_2^{(R)} / \Tr\Pi_2^{(R)}$.

Notice that, as $R$ gets closer to the identity in \cref{eq:r-reg-designs-test,eq:r-reg-designs-test2}, $\Tr R^2$ dominates the remaining terms. This behavior holds for general $t$. As described further in \cref{ap:regularized-rigged}, if $Y$ is an $R$-regularized rigged $t$-design, then it is \textit{almost} an $R$-regularized rigged $(t-1)$-design in the sense that
\begin{equation}
    \Expval_{\psi\in Y} (\ket\psi\bra\psi)^{\otimes (t-1)} \approx \frac{\Pi_{t-1}^{(R)}}{\Tr \Pi_{t-1}^{(R)}}\parentheses{1+\bigO{1/\Tr R^2}}.
\end{equation}

We conclude this section by generalizing the \textit{frame potential} from finite dimensions \cite{renesSymmetricInformationallyComplete2004,klappenecker2005mutually,scottOptimizingQuantumProcess2008} to regularized rigged $t$-designs.
For a positive definite (and therefore invertible) regularizer $R$, we define the frame potential of an ensemble $\calG$ over unit vectors in $L^2(\bbR)$ to be
\begin{equation}\label{eq:frame-potential}
    V_t^{(R)}(\calG) \coloneqq \Expval_{\psi,\phi \in \calG} \abs{\bra\psi R^{-1} \ket \phi}^{2t}.
\end{equation}
In \cref{ap:frame-potential}, we prove the following proposition regarding the frame potential.

\begin{proposition}\label{prop:frame-potential}
    Let $R$ be positive definite.
    For any ensemble $\calG$,
    \begin{equation}
        V_t^{(R)}(\calG) \geq \frac{1}{\Tr \Pi_t^{(R)}},
    \end{equation}
    with equality if and only if $\calG$ is an $R$-regularized rigged $t$-design.
\end{proposition}

Note the presence of the $R^{-1}$ in \cref{eq:frame-potential}. We will see something similar in \cref{sec:fidelity}, where we find that finite-dimensional formulas nicely generalize to infinite-dimensions by introducing factors of $R^{-1}$ to $R$-regularized rigged designs.

\section{Applications of rigged designs}
\label{sec:applications}

In \cref{sec:shadows}, we develop a shadow tomography protocol for CV systems based on rigged CV designs.
In \cref{sec:entanglement-verification}, we show how such CV rigged shadows can be used for entanglement verification.
In \cref{sec:fidelity}, we develop the notion of the average fidelity of a CV quantum channel by using regularized-rigged $2$-designs, relate this fidelity to the CV entanglement fidelity, and compare various fidelities for the case of the pure loss channel.

\subsection{Design-based CV shadows}
\label{sec:shadows}

The main idea behind finite-dimensional shadow tomography protocols is to perform random measurements of an unknown state to create classical snapshots through which many properties of the same unknown state can be efficiently predicted \cite{aaronsonShadowTomographyQuantum2018,huangPredictingManyProperties2020,huangProvablyEfficientMachine2022,Acharya2021}.
One can perform $\bigO{\log M}$ random measurements of an unknown state $\rho$ to accurately predict the expectation values of $M$ different observables with high probability.
Each measurement for one such protocol yields a \textit{shadow} of the form $3|e \rangle\langle e|-I$ on each qubit of the system, where $e$ is an eigenstate of one of the qubit Pauli matrices, and $I$ is the two-by-two identity.
The number of measurements needed is independent of the dimension of the Hilbert space, a property that can be proven using designs \cite{huangPredictingManyProperties2020}.

Shadow tomography can be framed in terms of informationally-complete positive operator-valued measures (POVMs), which include quantum state $(t\geq 2)$-designs \cite{acharyaInformationallyCompletePOVMbased2021}.
The concept of POVMs extends to infinite dimensions in such a powerful way that POVM elements can even be tempered distributions [\citealp{Holevo2011}; \citealp[Appx.~A]{Culf}].
Such POVMs are widely used.
For example, homodyne measurements correspond to measurements in the position-state POVM or its rotated counterparts~\cite{Raymer_Lvovsky}, while measuring in the phase-state POVM is optimal for determining the angle induced by a phase-space rotation~\cite[Sec. 3.9]{Holevo2011}.

Utilizing rigged designs as infinite-dimensional POVMs, we develop a CV shadow tomography protocol (see \cref{ap:shadows} for more details).
Here, our goal is to determine $\angles{\calO_j} \coloneqq \Tr(\rho \calO_j)$ for a collection of $M$ single-mode observables $\calO_1,\dots,\calO_M$, where $\rho$ is an unknown infinite-dimensional state which we can access on a quantum device.
We first describe a protocol utilizing a rigged $3$-design, and then describe a protocol utilizing a rigged $2$-design such as the one constructed in \cref{eq:kerr-phase-rig-2-design}. The former case is slightly more general and easier to describe, but we have not yet constructed useful rigged $3$-designs. We leave this question for future work.

\subsubsection{CV shadows with rigged three-designs}

Let $(X,\Sigma,\mu)$ denote a rigged $3$-design, which implies that $\int_X (\ket\chi\bra\chi)^{\otimes t}\dd{\mu(\chi)} = \alpha_t \Pi_t$ for $t\in\set{1,2,3}$ and $\alpha_t \in (0,\infty)$.
Without loss of generality, let $\alpha_1=1$, rescaling the measure $\mu$ if necessary.
Then, it follows that the design resolves the identity,
\begin{equation}
  \int_X \ket\chi\bra\chi \dd{\mu(\chi)} = \bbI~,
\end{equation}
and therefore, $\nu\colon A \mapsto \int_A \ket\chi\bra\chi \dd{\mu(\chi)}$ is a POVM.

Recall that a POVM maps subsets, which correspond to collections of measurement outcomes, to bounded, nonnegative self-adjoint operators (see \cref{ap:measure-theory} for a measure theory review and \cref{ap:shadows} for a short review on POVMs).
Sampling from such a POVM results in sampling measurement outcomes from the probability measure $\mu'\colon A \mapsto \Tr(\rho \nu(A))$.
We denote the measurement outcome corresponding to  $\chi$ as $c(\chi)$ that we then store on a classical computer.

Suppose that we measure $N$ times from $\mu'$, resulting in outputs $\set{c(\chi_1),\dots,c(\chi_N)}$. Each of these outputs corresponds to a CV shadow
\begin{equation}
  \hat\rho_i \coloneqq \frac{2}{\alpha_2}\ket{\chi_i}\bra{\chi_i} - \bbI~.
\end{equation}
Note that $\ket{\chi_i}$ is not generally a physical quantum state but instead a tempered distribution.
Fortunately, this is unimportant, since we are simply storing a description of $\ket{\chi_i}$ on a classical computer.

Using the classical snapshot and the classical description of observables $\calO_j$, one can compute
\begin{equation}\label{eq:shadow-obs}
    \hat o_j \coloneqq \frac{1}{N}\sum_{i=1}^N \Tr(\hat\rho_i \calO_j)~.
\end{equation}
On average, this yields the right answer: by the rigged $2$-design property of $X$, $\Expval[\hat o_j] = \angles{\calO_j}$, where the expectation value is taken over measurement outcomes.
Moreover, convergence to the right answer depends only on the features of $\calO_j$: using the rigged $3$-design property of $X$, we find that $\variance(\hat o_j) = \bigO{\frac{(\Tr \abs{\calO_j})^2}{N}}$ in the large-$N$ limit (see \cref{ap:shadows} for details). 
We perform the aforementioned procedure $K$ times, resulting in a collection $C_j = \{\hat o_j^{(1)}, \dots, \hat o_j^{(K)}\}$. Following Ref.~\cite[Thm.~1]{huangPredictingManyProperties2020}, for each $j$, the median of $C_j$ is within $\varepsilon$ of $\angles{\calO_j}$ with probability at least $1-\delta$ provided that
\begin{subequations}
\begin{align}
    &N = \bigO{\frac{1}{\varepsilon^2}(\max_j \Tr\abs{\calO_j})^2},\\
    &K = \bigO{\log(M/\delta)}.
\end{align}
\end{subequations}
In other words, using a shadow tomography procedure with a rigged $3$-design, we can accurately determine the expectation values of $M$ observables using only $\sim \log M$ measurements, provided that each observable $\calO_j$ is reasonably well-behaved; that is, provided that $\max_j \Tr \abs{\calO_j}$ is not too large.

\subsubsection{CV shadows with rigged two-designs}

If we had only used a rigged $2$-design in the above protocol, we would still have that $\Expval[\hat o_j] = \angles{\calO_j}$.
For certain observables $\calO_j$, we can show that a rigged $2$-design is sufficient to give reasonable bounds on the variance by following an analogous result in finite dimensions from Ref.~\cite{acharyaInformationallyCompletePOVMbased2021}.

As before, suppose we have a collection of $N$ shadows $\hat \rho_1,\dots,\hat\rho_N$ sampled from the POVM defined by the rigged $2$-design, yielding estimates $\hat o_j$ (\ref{eq:shadow-obs}).
We pick observables that satisfy
\begin{equation}
  c < \Tr(\hat \rho_i \calO_j) < d
\end{equation}
for some $c < d \in \bbR$ almost surely for every shadow $\hat\rho_i$.
Then, to achieve a success probability of at least $1-\delta$ and maximum additive error $\varepsilon$, we need only
\begin{equation}
    N \geq \log\pargs{\frac{2M}{\delta}} \frac{(d-c)^2}{2\varepsilon^2}
\end{equation}
shadows to determine $\angles{\calO_j}$ for each~$1\leq j \leq M$.

For concreteness, we consider a simple example of the rigged $2$-design shadow protocol.
Let each observable $\calO_j$ be of the form $\calO_j = \ket{a_j}\bra{b_j} + \ket{b_j}\bra{a_j}$ for $a_j,b_j\in\bbN_0$.
We use the rigged $2$-design from \cref{eq:kerr-phase-rig-2-design} consisting of Fock states and Kerred phase states. 
The explicit sampling step for this procedure is worked out in \cref{subsec:worked-examples}.
Using the explicit form of $\ket{\theta}_\varphi$, it follows that for any possible shadow $\hat\rho_i$ coming from this design, $|\Tr(\hat\rho_i \calO_j)| < 1/5$. Therefore, to determine the $M$ observables $\{\calO_j\}$ to a maximum additive error $\varepsilon$ with success probability at least $1-\delta$, we need only
\begin{equation}
    N \geq \log\pargs{\frac{2M}{\delta}}\frac{2}{25\varepsilon^2}
\end{equation}
measurements.

\subsection{Entanglement verification}
\label{sec:entanglement-verification}

In finite dimensions, classical shadows of a quantum state allows for the checking of many entanglement witnesses on that state \cite{huangPredictingManyProperties2020}.
Indeed, the same result holds for design-based CV shadows.

From Ref.~\cite[Thm.~2.2]{houConstructingEntanglementWitnesses2010}, for infinite-dimensional states $\rho$, $\rho$ is entangled if and only if there exists a finite-rank operator $A$ and a real number $\alpha$ such that $\alpha + \Tr(\rho A) < 0$ and $\alpha + \Tr(\sigma A) \geq 0$ for all separable states $\sigma$.
Since $A$ is finite rank, the expectation value of $A$ with respect to a rigged shadow is finite even though the rigged shadow is not a normalizable quantum state. Hence, the use of rigged shadows (obtained from very few measurements of $\rho$) allows one to test many candidate witnesses $A$ in order to determine if $\rho$ is entangled.

\subsection{Fidelities of CV quantum channels}
\label{sec:fidelity}

We develop the notions of the average fidelity of a continuous-variable (CV) quantum channel as well as their relationship to the CV entanglement fidelity.
Such notions require approximate (i.e., regularized) versions of our rigged designs.
We work out the case of a general positive semi-definite regularizer $R$, but note that the reader should keep in mind the two physically relevant hard- and soft-energy cases (\ref{eq:cutoffs}), corresponding to $R = P_d \coloneqq \sum_{n=0}^{d-1}\ket n \bra n$ and $R = R_\beta\coloneqq \e^{-\beta \hat n}$, respectively.
Finally, we benchmark the performance of a displacement operation by evaluating various fidelities for the case of the loss channel in \cref{sec:loss-channel}.

\subsubsection{Average fidelity of CV quantum channels}

In a $d$-dimensional Hilbert space, quantum states belong to a compact space $\bbC\bbP^{d-1}$. Therefore, one can define quantities that are averaged over all quantum states.
In particular, for a quantum channel $\calD$, the \textit{average channel fidelity} is defined as \cite{nielsenEntanglementFidelityQuantum1996,horodeckiGeneralTeleportationChannel1999,nielsenSimpleFormulaAverage2002,dankertExactApproximateUnitary2009,magesanGateFidelityFluctuations2011,luExperimentalEstimationAverage2015}
\begin{equation}\label{eq:average-channel-fidelity}
    \overline F(\calD) \coloneqq \int_{\bbC\bbP^{d-1}} \bra\psi \calD(\ket {\psi } \bra{ \psi })\ket\psi \dd{\psi},
\end{equation}
quantifying how close $\calD$ is to an identity channel on average.
Due to non-existence of a standard measure on infinite-dimensional space, as discussed in \cref{sec:infinite}, this formula cannot be extended to CV systems.

Since there are exactly two copies of $\ket {\psi } \bra{ \psi }$ in the integrand for the average fidelity, the integral over all states can be substituted with an average over any state 2-design $X$ using Eqs. (\ref{eq:haar-integral}-\ref{def:cpdesign}),
\begin{equation}\label{eq:average-channel-designs}
  \overline F(\calD)=\Expval_{\psi\in X} \bra\psi \calD(\ket {\psi } \bra{ \psi })\ket\psi~.
\end{equation}
The design provides a more manageable sample of states that is useful for estimating the average fidelity of quantum operations
\cite{emersonScalableNoiseEstimation2005,knillRandomizedBenchmarkingQuantum2008,
dankertExactApproximateUnitary2009,scalablemagesan,cross2016scalable}.
This formula \textit{can} be extended to infinite dimensions using normalized (i.e., regularized) versions of our rigged designs from \cref{sec:regularized-rigged-designs}.

Let $Y$ denote a regularized-rigged $2$-design with a general positive semi-definite regularizer $R$. 
There is more than one way to generalize the average fidelity from the finite-dimensional case, and we consider two average-fidelity quantities defined for a CV channel~$\calD$,
\begin{subequations}\label{eq:avg-fid}
\begin{align}\label{eq:avg-fid1}
    \overline F_1^{(R)}(\calD) &\coloneqq N_R \Expval_{\psi\in Y} \bra\psi R^{+} \calD(\psi) R^{+}\ket\psi ,\\
    \overline F_2^{(R)}(\calD) &\coloneqq \Expval_{\psi\in Y} \bra\psi \calD(\psi) \ket\psi~,\label{eq:avg-fid2}
\end{align}
\end{subequations}
where we use the short-hand notation $\calD(\psi) = \calD(|\psi\rangle\langle\psi|)$, and where the constant $N_R = \frac{\Tr R^4 + (\Tr R^2)^2}{\Tr R^2 + (\Tr R)^2}$. 
The second quantity faithfully uses two copies of the normalized state projections $|\psi\rangle\langle\psi|$ sampled from the design, while the first can revert one copy back to its non-normalizable version using the Moore-Penrose inverse $R^{+}$ of the regularizer.

As a sanity check, let us employ a hard-energy cutoff and plug in the regularizer $R = P_d = \sum_{n=0}^{d-1} |n\rangle\langle n|$ from \cref{eq:cutoffs} into \cref{eq:avg-fid}.
This essentially recovers the finite-dimensional case.
Since the Moore-Penrose inverse of a projector is itself, the two average-fidelity quantities are equal for this case.
Moreover, if $\calD$ is trace-preserving for states within the subspace defined by $P_d$, then $\overline F_1^{(P_d)}(\calD) = \overline F_2^{(P_d)}(\calD) = \overline F(\calD)$, recovering the finite-dimensional design-based average fidelity from \cref{eq:average-channel-designs}.

In a setting relevant to CV states enjoying infinite support, such as coherent or squeezed states, one should consider a regularizer with no zero eigenvalues.
We prove in \cref{ap:average-fidelity} that an $R$-regularized rigged $2$-design is informationally-complete for states on the entire Fock space whenever $R$ is invertible. 
Therefore, choosing $R=P_d$ may not be a good approximation of average fidelity over all CV states.

\subsubsection{Average-to-entanglement fidelity relation}\label{sec:entanglement-fid}

In the finite-dimensional case, the \textit{entanglement fidelity} for a quantum channel $\calD$ on $\bbC^d$ is \cite{nielsenEntanglementFidelityQuantum1996,horodeckiGeneralTeleportationChannel1999,nielsenSimpleFormulaAverage2002,dankertExactApproximateUnitary2009}
\begin{equation}
    F_e(\calD) \coloneqq \bra\phi (\calI \otimes \calD)(\phi) \ket\phi~,
\end{equation}
where $\ket\phi \coloneqq \frac{1}{\sqrt{d}}\sum_{n=0}^{d-1} \ket n \otimes \ket n$ denotes a maximally entangled state, and $\calI$ is the identity channel. 
This fidelity quantifies how well entanglement with a reference system is preserved by~$\calD$.
We refer the reader to Ref.~\cite[Ap.~A]{albertPerformanceStructureSinglemode2018} for a nice review of the utility of the entanglement fidelity.
The entanglement fidelity is related to the average fidelity by the following simple formula \cite{horodeckiGeneralTeleportationChannel1999,nielsenSimpleFormulaAverage2002}
\begin{align}
    \label{eq:fd-relation}
    \overline F(\calD) = \frac{d F_e(\calD)+1}{d+1}~.
\end{align}
We can similarly relate our average-fidelity relations (\ref{eq:avg-fid}) to a CV version of entanglement fidelity.

Maximally entangled states become non-normalizable as $d\to\infty$, meaning that CV versions of such states also have to be regularized in order to define an analogous fidelity. We require that $R$ be diagonal in the Fock-state basis and define the regularized state
\begin{equation}\label{eq:tmsv}
    \ket{\phi_R} \coloneqq \frac{1}{\sqrt{\Tr R}} (R^{1/4} \otimes R^{1/4}) \sum_{n\in \bbN_0} \ket n \otimes \ket n~,
\end{equation}
a purification of the \textit{regularizer state} 
\begin{equation}\label{eq:thermal-state}
    \rho_R \coloneqq \Tr_2(\ket {\phi_R } \bra{ \phi_R }) = R/\Tr R~.
\end{equation}
The \textit{$R$-regularized CV entanglement fidelity} of a channel $\calD$ is then
\begin{equation}\label{eq:ent-fid}
    F_e^{(R)}(\calD) \coloneqq \bra{\phi_R}(\calI \otimes \calD)(\phi_R) \ket{\phi_R}.
\end{equation}

In \cref{ap:average-fidelity}, we show that both CV average-fidelity quantities from \cref{eq:avg-fid} are related to the CV entanglement fidelity (\ref{eq:ent-fid}) as
\begin{subequations}\label{eq:fids-ent-fid}
\begin{align}
    \overline F_1^{(R)}(\calD) &= \frac{d_R F_e^{(R)}(\calD) + \Tr[\calD(\rho_{R^2})RR^+]}{d_R + 1}, \label{eq:fid-ent-fid}\\
    \overline F_2^{(\sqrt R)}(\calD) &= \frac{d_R F_e^{(R)}(\calD) + d_R \Tr[\calD(\rho_R) \rho_R]}{d_R+1}.\label{eq:fid-ent-fid2}
\end{align}
\end{subequations}
Since we are assuming $R$ is diagonal, $RR^+$ is simply a projector onto the subspace for which $R$ has support.
For invertible $R$, this subspace is the whole space so that $RR^+ = \bbI$, and therefore \cref{eq:fid-ent-fid} yields a CV generalization of the finite-dimensional average-to-entanglement fidelity relation \eqref{eq:fd-relation}:
\begin{equation}
    \label{eq:fid-ent-fid-inv}
    \overline F_1^{(R)}(\calD) = \frac{d_R F_e^{(R)}(\calD) + 1}{d_R + 1},
\end{equation}
where the \textit{effective dimension} dictated by the regularizer is the inverse purity of the regularizer state (\ref{eq:thermal-state}),
\begin{equation}\label{eq:effective-dim}
    d_R \coloneqq 1/\Tr \rho_R^2 = (\Tr R)^2 / \Tr R^2~.
\end{equation}
This effective dimension in the infinite-dimensional case plays the role of, and reduces to, the actual dimension in the finite-dimensional case.

The above general formulation reduces to a more physically relevant one when the soft-energy cutoff $R = R_\beta = \e^{-\beta \hat n}$ (\ref{eq:cutoffs}) is used as the regularizer.
The state $\ket{\phi_R}$ (\ref{eq:tmsv}) becomes a Gaussian two-mode squeezed vacuum state (\textit{a.k.a.}\ thermofield double) with squeezing parameter
$
    r = \log[\frac{1+\e^{-\beta/2}}{\sqrt{1-\e^{-\beta}}}]
$ \cite{AS17},
while the regularizer state (\ref{eq:thermal-state}) becomes a thermal state whose ``inverse temperature'' $\beta>0$ sets the energy scale of states involved in the regularization.
The effective dimension (\ref{eq:effective-dim}) becomes
\begin{equation}\label{eq:effective-dim-energy}
d_R = 2\Tr(\rho_R \hat n)+1~,
\end{equation}
directly related to the mean energy of the thermal state.
Similar energy-dependent factors also replace dimensions in studies of uniform continuity for quantum entropies \cite{winter2016tight} as well as bounds on energy-constrained capacities of Gaussian channels \cite{sharma2018bounding}.
Since $\ket{\phi_R}$ is a Gaussian state, the corresponding CV entanglement fidelity may be extractable via reasonable experimental protocols \cite{bai2018test,wu2019efficient}. 

As for more general $R$, we emphasize that \cref{eq:fids-ent-fid,eq:fid-ent-fid-inv} only hold as is when $R$ is diagonal in the $\set{\ket n}$ basis even though \cref{eq:avg-fid} is well-defined even when $R$ is not diagonal. Of course, one has the freedom to arbitrarily choose the basis with respect to which the CV entanglement fidelity is defined, so $R$ being diagonal is not a substantial restriction.

Recall in \cref{eq:frame-potential,prop:frame-potential}, we saw that introducing factors of $R^{-1}$ into a definition of frame potential resulted in finite-dimensional formulae nicely generalizing to infinite-dimensions. We again see this effect present in \cref{eq:fid-ent-fid-inv}. The definition of $\overline F_1^{(R)}$ utilizes factors of $R^{-1}$ while $\overline F_2^{(R)}$ does not. As a consequence, the finite-dimensional relation \eqref{eq:fd-relation} involving $\overline F$ and $F_e$ very closely matches the infinite-dimensional relation \eqref{eq:fid-ent-fid-inv} involving $\overline F_1^{(R)}$ and $F_e^{(R)}$, whereas the relation involving $\overline F_2^{(R)}$ and $F_e^{(R)}$ \eqref{eq:fid-ent-fid2} contains a factor not present in the finite-dimensional case.

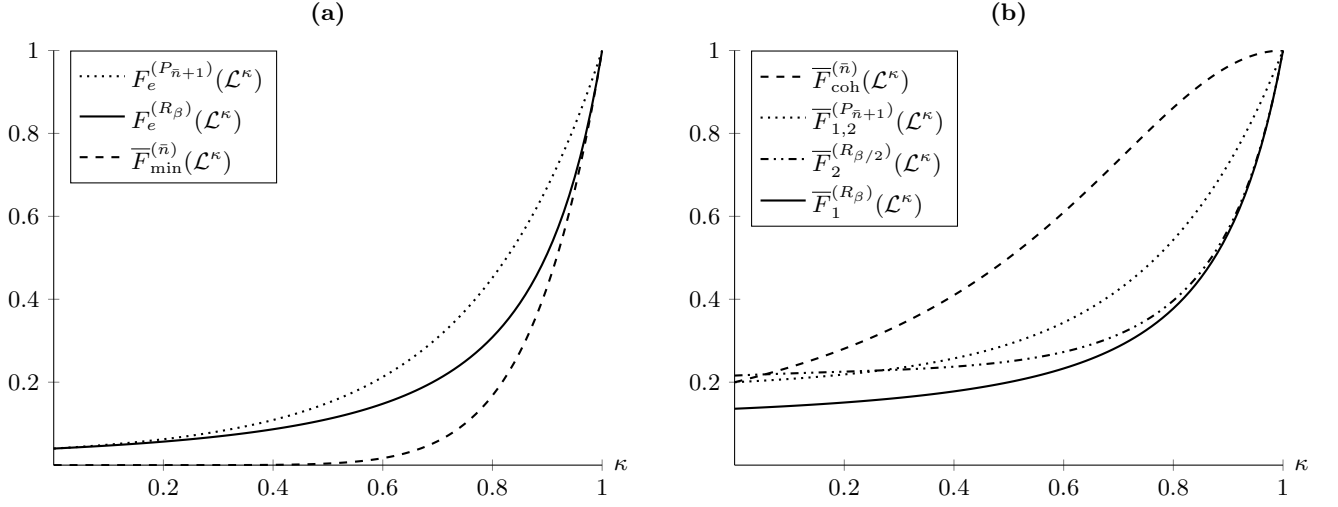
\begin{figure*}
    \centering
    \begin{minipage}{.495\textwidth}
    \begin{tikzpicture}
    \begin{axis}[
            width=\textwidth,
            height=.8\textwidth,
            xlabel={$\kappa$},
            title={\textbf{(a)}},
            legend cell align={left},
            legend style={
                at={(0.03,1.0)},anchor=north west,
            },
            axis y line=middle,
            axis x line=middle,
            x axis line style=-,
            y axis line style=-,
            x label style={xshift=13,yshift=-5}
        ]
        
        \addplot[thick, dotted] table [x=kappa, y=Fecut]{\relativepath loss.dat};
        \addlegendentry{$F_e^{(P_{\bar n + 1})}(\calL^\kappa)$}
        
        \addplot[thick, solid] table [x=kappa, y=Fe]{\relativepath loss.dat};
        \addlegendentry{$F_e^{(R_\beta)}(\calL^\kappa)$}
        
        \addplot[thick, dashed] table [x=kappa, y=Fmin]{\relativepath loss.dat};
        \addlegendentry{$\overline F_{\rm min}^{(\bar n)}(\calL^\kappa)$}

    \end{axis}
    \end{tikzpicture}
    \end{minipage}
    \hfill
    \begin{minipage}{.495\textwidth}
    \begin{tikzpicture}
    \begin{axis}[
            width=\textwidth,
            height=.8\textwidth,
            xlabel={$\kappa$},
            ymin=0, ymax=1,
            title={\textbf{(b)}},
            legend cell align={left},
            legend style={
                at={(0.03,1.0)},anchor=north west,
            },
            axis y line=middle,
            axis x line=middle,
            x axis line style=-,
            y axis line style=-,
            x label style={xshift=13,yshift=-5}
        ]
        
        \addplot[thick, dashed] table [x=kappa, y=Fcoh]{\relativepath loss.dat};
        \addlegendentry{$\overline F_{\rm coh}^{(\bar n)}(\calL^\kappa)$}

        \addplot[thick, dotted] table [x=kappa, y=F12]{\relativepath loss.dat};
        \addlegendentry{$\overline F_{1,2}^{(P_{\bar n + 1})}(\calL^\kappa)$}
        
        \addplot[thick, dash dot dot] table [x=kappa, y=F2]{\relativepath loss.dat};
        \addlegendentry{$\overline F_2^{(R_{\beta/2})}(\calL^\kappa)$}
        
        \addplot[thick, solid] table [x=kappa, y=F1]{\relativepath loss.dat};
        \addlegendentry{$\overline F_1^{(R_\beta)}(\calL^\kappa)$}

    \end{axis}
    \end{tikzpicture}
    \end{minipage}

    \caption{
    Various fidelity benchmarks for the pure-loss channel $\mathcal{L}^{\kappa}$ plotted vs the channel's transmissitivity $\kappa$, with the energy-constrained parameter $\bar{n} = 4$, and all other fidelity parameters being functions of $\bar n$ according to \cref{eq:pars}.
    {\bf(a)}
    Comparison of fidelities that utilize a reference mode: the CV entanglement fidelity $F_e$ \eqref{eq:ent-fid} with soft- and hard-energy constraints \eqref{eq:cutoffs} as well as the minimum energy-constrained entanglement fidelity $F_{\rm min}$ \eqref{eq:ec-channel-fid}.
    {\bf(b)}
    Comparison of our three average-fidelity quantities --- the soft-energy constrained average fidelities $\overline{F}_1^{(R_\beta)}$ \eqref{eq:fid-ent-fid} and $\overline{F}_2^{(R_{\beta/2})}$ \eqref{eq:fid-ent-fid2} as well as the hard-energy constrained case $\overline{F}_{1,2}^{(P_d)}$ --- with the fidelity $\overline{F}^{\bar n}_{\rm coh}$ \eqref{eq:coherent-state-fid} calculated by averaging over an ensemble of coherent states.
    The qualitatively different behavior of the coherent-state fidelity suggests that it may not be a good approximation to averages over CV states.
    }
    \label{fig:loss}
\end{figure*}

\subsubsection{Fidelity benchmarks for displacement operations}
\label{sec:loss-channel}

We compare the fidelity quantities introduced in this section to known quantities for the case of the pure loss channel, $\calD = \calL^{\kappa}$ \cite{ivan_operator-sum_2011}, with transmissitivity $\kappa\in[0,1]$.
This case is relevant to benchmarking the performance of displacement operations that are implemented via a non-ideal two-mode beam-splitter, with the transmissivity characterizing the degree of nonideality \cite{sharmaCharacterizingPerformanceContinuousvariable2020,paris1996displacement}.
All quantities described below are computed analytically in \cref{ap:loss-channel}.

In order to put all quantities on as equal of a footing, we set them to be a function of a fixed energy scale $\bar n$ using the following convention (with other choices possible).
For the soft- and hard-energy regularizers, $R_\beta = e^{-\beta \hat n}$ and $P_d = \sum_{n=0}^{d-1} |n\rangle\langle n|$ (\ref{eq:cutoffs}), respectively, we set
\begin{equation}\label{eq:pars}
    \beta=\log(1+1/\bar{n})\quad\quad\text{and}\quad\quad d=\lfloor\bar{n}\rfloor+1~.
\end{equation}
The soft-energy cutoff then corresponds to an average energy of $\bar n$ for the regularizer thermal state (\ref{eq:thermal-state}) and  an effective dimension $d_R = 2\bar n + 1$ (\ref{eq:effective-dim-energy}).

Our first comparison is between all fidelities that utilize a reference mode. This comparison is between the CV entanglement fidelity (\ref{eq:ent-fid}), with either soft- or hard-energy regularization, and its minimum energy-constrained version~\cite{nair2018, sharmaCharacterizingPerformanceContinuousvariable2020}
\begin{equation}\label{eq:ec-channel-fid}
    \overline F_{\rm min}^{(\bar n)}(\calL^\kappa) \coloneqq \min_{\substack{\psi_{EA}: \Tr(\hat{n}_A \psi_A) \leq \bar n}} \bra\psi \calL^\kappa(\psi_{EA})\ket\psi_{EA}~,
\end{equation}
consisting of an optimization of the CV entanglement fidelity over all input states whose average energy on the mode acted on by the channel is bounded by $\bar n$.

The three reference-mode fidelities $\{F_e^{(P_{d})},\allowbreak F_e^{(R_{\beta})},\allowbreak F_{\rm min}^{(\bar n)}\}$ are plotted for $\bar n = 4$ and all transmissivities $\kappa \in [0,1]$ in \cref{fig:loss}(a).
All quantities decrease in similar fashion with decreasing transmissitivy, with the soft-energy fidelity following the scaling of the minimum case slightly better than the hard-energy fidelity near unity transmissitivity.
Due to the parameterization picked in Eq. (\ref{eq:pars}), the entanglement fidelities for the two energy constraints are equal for zero transmissivity, $F_e^{(P_{d})} = F_e^{(R_{\beta})} = 1/(\bar n+1)^2$ at $\kappa = 0$.

Our second comparison is between fidelities that do not utilize a reference mode. 
This set includes both of our CV average fidelities from \eqref{eq:avg-fid}, each with either a soft- or a hard-energy constraint.
These are related to the entanglement fidelity of a CV channel via \cref{eq:fid-ent-fid,eq:fid-ent-fid2}, respectively.
Since the pure-loss channel is trace preserving on the subspace defined by $P_d$, two of these four fidelities are equal in the case of the hard-energy constraint, $\overline F_1^{(P_d)} = \overline F_2^{(P_d)} \eqqcolon \overline F_{1,2}^{(P_d)}$.
This comparison also includes the average fidelity of the pure-loss channel over an ensemble of coherent states,
\begin{equation}\label{eq:coherent-state-fid}
    \overline F_{\rm coh}^{(\bar n)}(\calL^\kappa) \coloneqq \int_\bbC p(\alpha) \bra\alpha \calL^\kappa(\alpha) \ket\alpha \Dd{2}\alpha,
\end{equation}
where $\ket\alpha$ denotes the coherent state specified by $\alpha\in\bbC$.
We choose the density function to be $p(\alpha) = \frac{1}{\pi\bar n}\e^{-\abs{\alpha}^2 / \bar n}$ to ensure that the average occupation number of the ensemble of coherent states is $\int_\bbC p(\alpha)\abs{\alpha}^2 \Dd{2}{\alpha} = \bar n$.

The four average-fidelity quantities $\{\overline F_{1,2}^{(P_d)},\allowbreak \overline{F}_1^{(R_\beta)},\allowbreak \overline{F}_2^{(R_{\beta/2})},\allowbreak \overline F_{\rm coh}^{(\bar n)}\}$ are plotted for $\bar n = 4$ and all transmissivities $\kappa$ in \cref{fig:loss}(b).
Note that the average fidelity over an ensemble of coherent states does not qualitatively match the other fidelities. In particular, the concavity of $\overline F_{\rm coh}$ near unity transmissivity is different from the other fidelity quantities. This may be related to the fact that an ensemble of coherent states only forms a CV $1$-design, whereas the other fidelities are defined with respect to various notions of $2$-designs.
This result suggests that the coherent-state average may not be a useful approximation for an average over all CV states.

\section{Conclusion}
\label{sec:conclusion}

In this work, we study quantum state designs in finite and infinite dimensions. 
In finite dimensions, we review a method for constructing complex-projective designs using simplex and torus designs. 
In particular, we establish a relationship between torus designs and complete sets of mutually unbiased bases. 

We then prove a no-go theorem implying that a naïve extension of the definition of state designs to infinite dimensions fails.
Similarly, we prove that CV unitary $t$-designs do not exist for any $t\geq 2$. 
Prior to our work, it was proven~\cite{blume-kohout_curious_2014} (argued~\cite{zhuang_scrambling_2019}) that Gaussian resources are not sufficient to form CV state (unitary) designs. 
Our no-go theorem establishes a stronger result implying that even non-Gaussian resources are not sufficient to form CV designs.

The lack of CV designs is due to a restriction to using only normalizable states.
We successfully extend the notion of state designs to infinite dimensions by proposing a new definition of CV state designs using non-normalizable states. 
These non-normalizable states belong to a rigged Hilbert space, and we provide various constructions of such rigged $2$-designs consisting of Fock states and oscillator phase states \cite{susskindQuantumMechanicalPhase1964,carruthers_phase_1968,helstromQuantumDetectionEstimation1969,holevoCovariantMeasurementsImprimitivity1984,Bergou1991,Holevo2011} subject to Kerr-Hamiltonian evolution.

As an application of rigged designs, we extended the formalism of shadow tomography \cite{aaronsonShadowTomographyQuantum2018,huangPredictingManyProperties2020,huangProvablyEfficientMachine2022,Acharya2021} to CV systems. 
We show that our rigged $2$-designs and, if useful ones exist, rigged $3$-designs can yield efficient shadow-based protocols.
It is an interesting direction to experimentally implement our design-based CV shadow tomography protocol based on rigged $2$-designs and compare it with other protocols based on homodyne or heterodyne measurements~\cite{Raymer_Lvovsky}, which can also be formulated within a shadow-like framework (albeit without the use of designs)~\cite{Gandhari2022}. 
The POVMs defined by the rigged $2$-designs that we constructed are highly non-Gaussian. It is an exciting open theoretical and experimental direction to develop techniques to measure from such POVMs.

We construct approximate CV designs by regularizing the elements of rigged designs. These regularized-rigged designs consist of physical quantum states and therefore can be used to define information-theoretic quantities, such as fidelities, for CV quantum channels. In particular, we define various notions of the average fidelity of a CV channel. We then establish a relation between the average fidelity and the entanglement fidelity of a CV channel. Our result is a natural generalization of finite-dimensional formulas \cite{nielsen_quantum_2010}, where the dimension is replaced by the effective dimension that depends on the mean energy of the input state to the channel. It is an interesting open question to develop efficient experimental methods to prepare states belonging to regularized-rigged designs introduced in our work.
On the theory side, it may be interesting to determine a relationship between the energy-constrained diamond distance \cite{shirokov2018energy,winter2017energy} and the average fidelity introduced in our work.

As discussed in \cref{sec:loss-channel}, an important application of regularized-rigged designs is to estimate the average fidelity between an ideal unitary and its experimental approximations. We emphasize again that our results are applicable directly when analytical expressions of an ideal unitary gate and its experimental approximation are known. Instead of estimating the average fidelity over a subset of states such as coherent states, one can calculate a good approximation of the average fidelity over \textit{all} states using our regularized-rigged designs.

We construct rigged and regularized-rigged CV state $2$-designs, leaving the interesting question of constructing useful CV state $t$-designs for $t\geq 3$ to future work.
Another interesting direction is to develop the notion of energy-constrained CV state designs, where each state in the design satisfies a fixed energy constraint. 
Our regularized-rigged state designs are good approximations of energy-constrained CV state designs. 

Our rigged designs are defined on the Hilbert space $L^2(\bbR)$ of a single mode, but can formally be mapped into any other countably infinite Hilbert space because all such spaces are isomorphic.
A mapping like this from the single-mode space to the space $L^2(\bbR^n)$ of multiple modes is likely to be physically obscure.
An interesting future topic would be to develop designs for other spaces, such as multiple modes, rotors and rigid bodies \cite{mol}, using states natural to those spaces.
For example, we anticipate that designs similar to our Kerred phase-state designs can be formulated for the space of the planar rotor, $L^2(\text{U}(1))$ \cite[Sec. IV.B]{mol}. Similarly, cross-Kerr interactions \cite{girvin2014circuit} may provide a recipe for rigged designs for multiple modes.

We also prove that CV unitary $t$-designs do not exist for any $t\geq 2$. A natural research question is whether, similar to rigged CV state designs, there exists a reasonable notion of CV operator designs. We introduce one such notion in this work, leaving the interesting and important question of how to construct such designs to future work.

Finally, another interesting avenue to explore is that of designs for function spaces. In \cref{ap:C-inf-designs}, we showed how our rigged designs can be interpreted as designs over infinite-dimensional function spaces. Can this theory be further generalized to other functional integrals, such as e.g.~path integrals? In particular, in field theories, one is typically interested in correlators (i.e.~polynomials in the fields) of various degrees; a $t$-design is therefore a space of fields that match all correlators up to degree $t$. Can designs be defined and used in this context? Refs.~\cite{cameronSimpsonRuleNumerical1951,gelfandIntegrationFunctionalSpaces1960,brushFunctionalIntegralsStatistical1961,konheimNumericalEvaluationWiener1967}, which contain a small number of cubature rules for Weiner integrals, may be a useful place to start.

\begin{acknowledgments}
We thank Steve Flammia, Daniel Gottesman, Jonas Helsen, Greg Kuperberg, Richard Kueng, Yi-Kai Liu, Alireza Seif, Niklas Galke, and T.~C.~Mooney for helpful discussions.
MJG and VVA thank Srilekha Gandhari, Thomas Gerrits and Jake Taylor for discussions and collaborations on \cite{Gandhari2022}.
JTI thanks the Joint Quantum Institute at the University of Maryland for support through a JQI fellowship.
JTI acknowledges funding by the DoE ASCR Accelerated Research in Quantum Computing program (award No.~DE-SC0020312), DoE QSA, NSF QLCI (award No.~OMA-2120757), NSF PFCQC program, the DoE ASCR Quantum Testbed Pathfinder program (award No.~DE-SC0019040), AFOSR, ARO MURI, AFOSR MURI, and DARPA SAVaNT ADVENT.  
KS acknowledges support from the Department of Defence.
MJG acknowledges support from NIST grant 70NANB21H055\_0. VVA acknowledges financial support from NSF QLCI grant OMA-2120757, and thanks Olga Albert and Ryhor Kandratsenia for providing daycare support throughout this work.
Contributions to this work by NIST, an agency of the US government, are not subject to US copyright.
Any mention of commercial products does not indicate endorsement by NIST.
\end{acknowledgments}

\toc

\clearpage
\onecolumngrid\appendix
\renewcommand{\tocname}{Appendices}

\section{Pointers to Appendices}

In \cref{ap:review,ap:finite-designs,ap:CVdesign,ap:applications,ap:torus}, we provide proofs of our main results and summarize relevant background material on continuous-variable (CV) information theory. \cref{ap:review} covers relevant definitions from measure theory and properties of projectors onto the symmetric subspace of a separable Hilbert space. \cref{ap:simplex-designs,ap:torus-designs} review finite-dimensional simplex and torus designs. \cref{ap:cp-haar-integral,ap:simplex-torus-to-cp,ap:cp-to-simplex} review complex-projective designs and their relationship to simplex and torus designs. Using simplex and torus designs, we develop a design formalism for constrained complex-projective integration in \cref{ap:constrained-integration}. 
To the best of our knowledge, the formalism developed in \cref{ap:constrained-integration} is novel. 
Similarly, to the best of our knowledge, \cref{def:cv-design,thm:torus-two-design,prop:minimal-torus-2design} from \cref{ap:torus-designs} are new, though we prove in \cref{ap:torus} that \cref{def:cv-design} is equivalent to a previous definition given in Ref.~\cite{kuperberg_numerical_2004}.
A relationship between simplex, torus, and complex-projective designs was described Ref.~\cite{kuperberg_numerical_2004}. 
We further extend on this relationship in \cref{ap:cp-haar-integral,ap:simplex-torus-to-cp,ap:cp-to-simplex}.

Readers who are familiar with finite-dimensional complex-projective designs may wish to begin directly from \cref{ap:CVdesign}.
\cref{ap:CVdesign,ap:applications} discuss the main results of this paper. In \cref{ap:nonexistence,ap:nonexistence-unitary}, we prove that CV state and unitary $t$-designs do not exist for $t>1$. In \cref{ap:rigged,ap:regularized-rigged}, we define and construct rigged and regularized rigged designs, which are generalizations of CV state designs.
In \cref{ap:C-inf-designs}, we discuss an alternative characterization of CV, rigged, and reguarized rigged designs based on integration over infinite-dimensional Gaussian measures.
In \cref{ap:cv-unitary-designs}, using regularized rigged designs, we propose a new definition of an approximate CV unitary $t$-design. 
In \cref{ap:shadows}, we develop the formalism for CV shadows based on rigged designs. We then define the average fidelity of a CV channel based on regularized rigged designs in \cref{ap:average-fidelity}. Finally, in \cref{ap:torus}, we establish a relationship between torus $2$-designs and complete sets of mutually unbiased bases that, to the best of our knowledge, had not been previously established.

\tableofcontents

\section{Preliminaries}
\label{ap:review}

In this section, we summarize some definitions and prior results relevant for the rest of the appendix. We point readers to \cite{H12,AS17} and \cite{royden_real_2010} for background on continuous-variable information theory and measure theory, respectively.

\medskip

Throughout this manuscript, $\bbN$ and $\bbN_0$ denote the sets of positive and non-negative integers, respectively. A $t$-fold Cartesian product $\bbN_0 \times \dots \times \bbN_0$ will be denoted by $\bbN_0^t$. $\bbZ_d$ will be the integers modulo $d$, $\bbZ_d = \set{0,\dots,d-1}$. 
 
\medskip

\noindent\textbf{States}: We will consider continuous-variable states (normalized vectors) in the separable infinite dimensional Hilbert space $\mathcal{H} = L^2(\mathbb{R})$. Separable Hilbert spaces, by definition, have a \textit{Schauder} or \textit{Hilbert Space basis}; any vector in a separable Hilbert space can be written as $\sum_{n=0}^\infty \alpha_n \ket{v_n}$ for some Schauder basis $\set{v_n}$ which is always guaranteed to exist \cite[ch.~17.1]{blanchard_mathematical_2015}. For concreteness, when discussing an explicit basis, we will use the standard Fock basis on $L^2(\bbR)$, denoted by $\set{\ket n \mid n\in \bbN_0}$.  In the position representation, a Fock state $\ket n$ is $\psi_n(x) = \braket{x \vert n} = \frac{\pi^{-1/4}}{\sqrt{n! 2^n}} \e^{-x^2/2} H_n(x)$, where $H_n$ is the $n^{\rm th}$~Hermite polynomial. We will also consider $d$ dimensional qudit states, where the Hilbert space is $\bbC^d$. We will fix an orthonormal basis of $\bbC^d$ and denote it as $\set{\ket{0}, \dots, \ket{d-1}}$. Qudit states belong to complex-projective space $\bbC\bbP^{d-1}$, which is described more in \cref{ap:measure-theory}.

\bigskip 

\noindent \textbf{m-torus and m-simplex}: The $m$-torus is denoted by $T^m \cong [0,2\pi)^m = (\bbR / 2\pi \bbZ)^m$. The unit-normalized Lebesgue measure on $T^m$ is given by $\dd{\phi} \coloneqq \frac{1}{(2\pi)^m}\dd{\phi_1} \dots \dd{\phi_m}$. Moreover, the $m$-simplex is defined as
\begin{equation}
    \Delta^m = \set{(p_0, \dots, p_m) \in [0,1]^{m+1} \vert \sum_{i=0}^m p_i = 1}.
\end{equation}

Any integration over $\Delta^m$ can be defined using the 
unit-normalized Lebesgue measure on  $\Delta^m$ as follows:
\begin{equation}
    \int_{\Delta^m} f(p) \dd{p} = m!\int_{[0,1]^{m+1}} f(p_0, \dots, p_m) \delta(1-p_0 - \dots - p_m) \dd{p_0}\dots \dd{p_m},
\end{equation}
where $\delta$ is the Dirac delta function and $f(p)$ is any function over $p$.

\subsection{Measure theory}
\label{ap:measure-theory}

In this section, we summarize definitions and key theorems from measure theory. We point readers to \cite{royden_real_2010} for more details. For a concise introduction to basic concepts in measure theory, we recommend video lectures in \cite{brighter-side-of-mathematics}, which serves as much of the inspiration for our summary below.

For a finite set $X$, the most natural way to assign a measure (i.e. ``size'' or ``volume'') to subsets of $X$ is by cardinality. However, for many applications, this method breaks down for infinitely large sets. Intuitively speaking, measure theory is a way to generalize the notion of determining the size of a subset to infinitely large sets. To begin, fix a possibly infinite set $X$. We will denote the power set of $X$ by $\calP(X)$.

To assign generalized ``volumes'' to subsets of $X$, we are looking for a map $\mu\colon \Sigma \to [0,\infty]$, where $\Sigma \subseteq \calP(X)$ is some collection of subsets of $X$. For a subset $A \subseteq X$, let $A \in \Sigma$. We assign the volume, or measure, of $A$ in $X$ to be $\mu(A)$. Notice that the codomain of $\mu$ is the positive extended real line $[0,\infty]$, which we define to be $[0,\infty) \cup \{\infty \}$. This notation signifies $[0,\infty)$ as the standard nonnegative part of $\bbR$, and $\{\infty \}$ as the set containing the symbol $\infty$. In other words, we include $\infty$ in the codomain of the measure $\mu$. For all $r \in [0,\infty]$, the symbol $\infty$ is defined by the following three rules:
\begin{equation}
    r+\infty \coloneqq \infty~, \qquad r \cdot \infty \coloneqq \begin{cases} 0&\text{if }r = 0\\\infty&\text{otherwise}~,\end{cases} \qquad \infty - \infty \text{ undefined}~.
\end{equation}
The domain $\Sigma$ of $\mu$ is the collection of all measurable subsets of $X$, where a measure is assigned to each element of $\Sigma$ by $\mu$. In particular, the collection of measurable sets should satisfy:
\begin{enumerate}
    \item $\emptyset, X \in \Sigma$, i.e., a volume can be assigned to the empty set and the whole set $X$.
    \item If $A \in \Sigma$, then the complement of $A$, $A^c = X \setminus A$, should also be in $\Sigma$, i.e., if $A$ is measurable, the complement of $A$ should also be measurable.
    \item If a countable collection of sets $A_i$ are in $\Sigma$, then their union $\bigcup_i A_i$ should also be in $\Sigma$.
\end{enumerate}
A set $\Sigma \subseteq \calP(X)$ satisfying these aforementioned properties is called a \emph{$\sigma$-algebra}.

Given a set $X$ and a $\sigma$-algebra $\Sigma$ on $X$, one can then formally define a measure $\mu \colon \Sigma \to [0,\infty]$. $\mu$ should generalize the properties of volume, and therefore must satisfy the following two conditions:
\begin{enumerate}
    \item $\mu(\emptyset) = 0$, i.e., the empty set has zero volume.  
    \item For any countable collection of pairwise disjoint sets $A_j \in \Sigma$, $\mu(\bigcup_j A_j) = \sum_{j} \mu(A_j)$, i.e., the volume of a region is the sum of the volumes of its constituents. 
\end{enumerate}
The triplet $(X, \Sigma, \mu)$ is called a \emph{measure space}. A measure $\mu$ on $X$ is called \emph{$\sigma$-finite} if $X$ is the union of at most countably many subsets of finite measure. In other words, if their exist a countable collection $A_1$, $A_2$, $\dots \in \Sigma$ such that $\bigcup_j A_j = X$ and each $A_j$ satisfies $\mu(A_j) < \infty$, then $\mu$ is $\sigma$-finite. For example, consider $X=\bbR$, and $A_j = (j-1.1, j+0.1) \cup (-j-0.1, -j+1.1)$. The length of each $A_j$ is $2.4$ which is finite, and the countable union $\cup_j A_j = \bbR$. Hence, $\bbR$ with the measure $\mu((a,b)) = b-a$ is $\sigma$-finite.

One can show that for many cases of interest, not all subsets can be measurable (i.e.~$\Sigma \neq \calP(X)$) if the measure is desired to satisfy certain properties. For example, in the case of $X = \bbR$,  we desire the measure to have the properties $\mu(r+A) = \mu(A)$ for all $r\in\bbR$ and $A \subseteq X$, and $\mu([a,b]) = b-a$. One can prove that such a $\mu\colon \calP(X) \to [0,\infty]$ cannot exist. Hence, in general, one must restrict the $\sigma$-algebra $\Sigma$ to not be the entire power set. The most important $\sigma$-algebra on $\bbR$ is the Borel $\sigma$-algebra, which is the smallest $\sigma$-algebra that contains all open sets in $\bbR$ equipped with the standard topology. The most important measure on $\bbR$ is the Lebesgue measure, which satisfies the two properties above. Given the product of two $\sigma$-finite measure spaces, one can define a unique product measure space. Using this construction, one can construct the Lebesgue measure on $\bbR^2$, and this can be reinterpreted as a Lebesgue measure on $\bbC$.

A crucial feature of measure spaces is the concept of \emph{$\mu$-almost-everywhere}, often abbreviated $\mu$-a.e., or just a.e.~if the measure is clear. A property is said to hold $\mu$-a.e.~if it is true everywhere except on a subset that is contained inside a subset of measure zero. For example, the rationals $\bbQ$ are contained within a measurable subset of measure zero in the reals $\bbR$ with respect to the Lebesgue measure. In fact, $\bbQ$ is itself measurable. Therefore, the property that ``$r \in \bbR$ is irrational'' holds a.e.~with respect to the Lebesgue measure. One important property that will show up often is $f \leq g$ a.e.~for two measurable functions $f, g$. This means that the set $\set{x \in X \mid g(x) > f(x)}$ is contained within a measurable set with measure zero.

Between two measure spaces $(X_1, \Sigma_1, \mu_1)$ and $(X_2, \Sigma_2, \mu_2)$, a map $f \colon X_1 \to X_2$ is called \emph{measurable} if the preimage of measurable sets is measurable, meaning that $f^{-1}(A_2) \in \Sigma_1$ for all $A_2 \in \Sigma_2$. For $X_2 = \bbR$ with the standard Lebesgue measure, if a map $f\colon X \to \bbR$ is measurable then for every $r\in \bbR$ the preimage $f^{-1}(\{r\}) = \set{x \in X_1 \mid f(x) = r}$ is measurable \cite[p.~359-360]{royden_real_2010}. Intuitively, in order to integrate over a function, we must be able to determine the measure of the domain for which that function takes a certain value. For example, in the case of a bump function $f\colon \bbR \to \bbR$ where $f(r) = c$ whenever $r \in A$ and zero otherwise, the integral of $f$ is defined as $c \mu_1(A) = c \mu_1(f^{-1}(\{ c\}))$. Therefore, we must require that $f^{-1}(\{ c \})$ be measurable.

We will now briefly describe the intuition for Lebesgue integration. For a measurable set $A \in \Sigma$, the indicator function $\ind_A(x)$ is defined to be $1$ if $x \in A$ and $0$ otherwise. A \emph{simple function} is any function of the form $f_{\rm sim} =\sum_{i=1}^m \alpha_i \ind_{A_i}$ for each $\alpha_i \in \bbR$ and $A_i \in \Sigma$. The Lebesgue integral of $f_{\rm sim}$ is defined as $\int_X f_{\rm sim} \dd{\mu} \coloneqq \sum_{i=1}^m \alpha_i \mu(A_i)$. Let $\calS$ denote the set of all simple functions. For a non-simple, nonnegative function $f$, the Lebesgue integral of $f$ is defined as a supremum over all simple functions
\begin{equation}
    \int_X f \dd{\mu} \coloneqq \sup_{\substack{f_{\rm sim} \in \calS \\f_{\rm sim}\leq f \text{ a.e.}}} \int_X f_{\rm sim} \dd{\mu}.
\end{equation}
Finally, for a general measurable function $f$, the Lebesgue integral of $f$ is defined in terms of the integral of nonnegative functions by $\int_X f \dd{\mu} = \int_X \max(0,f)\dd{\mu} -  \int_X \max(0,-f)\dd{\mu}$, and is hence defined only if both $\max(0,f)$ and $\max(0,-f)$ are integrable since $\infty - \infty$ is undefined.

A measurable function $f$ is said to be \emph{integrable} if $\int_X \abs{f} \dd{\mu}$ is finite. One basic fact about Lebesgue integration is that if $f \leq g$ almost everywhere, then $\int_X f \dd{\mu} \leq \int_X g \dd{\mu}$. Also, the integral is linear, so that $\int_X(f+g)\dd{\mu} = \int_X f \dd{\mu} + \int_X g \dd{\mu}$. Oftentimes, we will include an integration parameter for clarity. We define the notation 
\begin{align}
\int_X f(x)\dd{\mu(x)} \coloneqq \int_X f \dd{\mu}~.
\end{align}

The space $L^t(X, \Sigma, \mu)$ is the set measurable functions (identified if they agree almost everywhere) $f\colon X \to \bbR$ for which $\int_X \abs{f}^t \dd{\mu} < \infty$. Define the $t$-norm to be $\norm{f}_t \coloneqq \parentheses{\int_X \abs{f}^t \dd{\mu}}^{1/t}$. With respect to the $t$-norm, $L^t(X, \Sigma,\mu)$ is a Banach space. $L^2(X,\Sigma,\mu)$ is a Hilbert space with respect to the inner product $\langle f, g \rangle = \int_X f g \dd{\mu}$ in the real case, and similarly in the complex case but with $g \to \bar g$. A bounded sequence in $L^t(X, \Sigma,\mu)$ is a sequence of measurable maps $(f_i)_{i\in \bbN}$ for which $\norm{f_i}_t < M$ for some finite number $M \in \bbR$. When the $\sigma$-algebra and measure are clear from context, we denote $L^t(X, \Sigma, \mu)$ as $L^t(X)$. For example, when $L^2(\bbR)$ is written, the $\sigma$-algebra and measure are assumed to be the standard Borel $\sigma$-algebra and Lebesgue measure on $\bbR$.

\bigskip
\noindent\textbf{Theorems and lemmas}
\bigskip

We now state and discuss various theorems that are used in our proofs. First, we review the Lebesgue Dominated Convergence Theorem, which provides a condition under which a limit can be brought inside of an integral.

\begin{theorem}[Lebesgue Dominated Convergence Theorem {\cite[Chapter~18.3]{royden_real_2010}}]
\label{thm:dom-conv}
Let $(X,\Sigma,\mu)$ be a measure space and $(f_n)_{n\in \bbN}$ a sequence of measurable functions on $X$ for which $f_n \to f$ pointwise almost everywhere on~$X$ and the function~$f$ is measurable. Assume there is a nonnegative function $g$ that is integrable over $X$ and dominates the sequence $(f_n)_{n\in \bbN}$ on $X$ in the sense that $\abs{f_n} \leq g$ almost everywhere on $X$ for all $n$. Then $f$ is integrable over $X$ and $\lim_{n\to\infty}\int_X f_n \dd{\mu} = \int_X f \dd{\mu}$.
\end{theorem}

As a simple example of the Lebesgue Dominated Convergence Theorem, consider the sequence of functions $f_n(x) = \e^{-n x^2}$ on $\bbR$. $(f_n)_{n\in\bbN}$ converges pointwise to the zero function, because for every fixed $x$, $\lim_{n\to\infty}f_n(x) = 0$. Every $f_n$ is bounded above by $g(x) = \e^{-x^2}$ for all $x$, and the integral of $g$ over $\bbR$ is finite. Hence, $\lim_{n\to\infty} \int_\bbR f_n \dd{\mu} = \int_\bbR \lim_{n\to\infty}f_n \dd{\mu} = 0$. If we instead just compute the integral, we find that $\int_\bbR f_n \dd{\mu} = \sqrt{\pi/n}$, which indeed goes to zero as $n\to\infty$.

\medskip 

Next, we state the Riesz Weak Compactness Theorem, which forms the backbone of our proof that continuous-variable state $t$-designs do not exist for $t \geq 2$.

\begin{theorem}[Riesz Weak Compactness Theorem {\cite[Chapter~19.5]{royden_real_2010}}]
\label{thm:rwct}
Let $(X,\Sigma,\mu)$ be a $\sigma$-finite measure space. Let $1 < t < \infty$ and $t'$ such that $1/t+1/t' = 1$. If $(p_i)_{i=0}^\infty$ is a bounded sequence in $L^t(X) = L^t(X,\Sigma,\mu)$, then there exist a subsequence $(p_{i_k})_{k=0}^\infty$ of $(p_i)_{i=0}^\infty$ and a function $q \in L^t(X)$ for which
\begin{equation}
    \forall h \in L^{t'}(X)\colon~\lim_{k\to\infty}\int_X p_{i_k} h \dd{\mu} = \int_X q h \dd{\mu}.
\end{equation}
\end{theorem}

Indeed, this theorem will be the main ingredient in our proof of the non-existence of continuous-variable $t$-designs. Notice that this theorem does not hold for $t=1$, but rather $t>1$. This will ultimately be the reason why our proof of non-existence of continuous-variable state $t$-designs only holds for $t \geq 2$. This is a nice sanity check, since \cref{ex:oscillator-1-design} shows explicit examples of continuous-variable state $1$-designs. The proof of the Riesz Weak Compactness Theorem uses that $L^t(X)$ is a reflexive Banach space for all $1<t<\infty$. For each such $t$, $L^{t'}(X)$ is naturally isomorphic to the dual space of $L^t(X)$. However, $L^1(X)$ is not the dual of $L^\infty(X)$. Since the Riesz Weak Compactness Theorem is so important for this work, we present a simple example to help understand the theorem.

\begin{example}
  Suppose $(f_n)_{n\in\bbN}$ is a sequence of functions $f_n\colon [0,1] \to \bbR$ defined by $f_n = \sqrt{n}\ind_{[0,1/n]} \in L^2([0,1])$, where the indicator function \(\ind_{[a,b]}(x)\) on an interval \([a,b]\) is $1$ for any \(x\) in the interval, and $0$ elsewhere. The norm of $f_n$ is $\norm{f_n}_2^2 = \int_{[0,1]} \abs{f_n}^2 \dd{\mu} = n \mu([0,1/n]) = 1$. Hence, $(f_n)_{n\in\bbN}$ is a bounded sequence in $L^2([0,1])$. Therefore, there is a subsequence $(f_{n_k})_{k\in \bbN}$ and a function $q\in L^2([0,1])$ for which
  \begin{equation}
      \forall h \in L^2([0,1])\colon ~~ \lim_{k\to\infty}\int_{[0,1]} f_{n_k} h \dd{\mu} = \int_{[0,1]} q h \dd{\mu}.
  \end{equation}
  Consider the constant function $h = 1\in L^2([0,1])$. We can explicitly compute the left-hand side to be $\lim_{k\to\infty} \int_{[0,1]} \sqrt{n_k}\ind_{[0,1/n_k]} = \lim_{k\to\infty} \sqrt{n_k} / n_k = 0$. Similarly, one can consider functions $h_{a,b} = \ind_{[a,b]}$ for any $a \leq b \in [0,1]$ and compute the left-hand side to be zero for all choice of $a$ and $b$. Therefore, it must be that $\int_{[0,1]} q \ind_{[a,b]} \dd{\mu} = 0$ for all $a$ and $b$, meaning that $q$ must be zero almost everywhere. 
  
  One says that the subsequence $(f_{n_k})$ converges \emph{weakly} to $0$ in $L^2([0,1])$. This is to be contrasted with strong convergence. If $(f_{n_k})$ were to converge strongly to $0$ in $L^2([0,1])$, then $\lim_{k\to\infty} \int_{[0,1]} \abs{f_{n_k} - 0}^2 \dd{\mu} = 0$, which is clearly not the case. It is not hard to see that the full sequence $(f_n)$ also converges weakly to $0$.
  
  Consider instead the sequence $(f_n)_{n\in\bbN}$ defined by $f_n = (-1)^n$. The sequence is bounded in $L^2([0,1])$. Therefore, there is a subsequence $(f_{n_k})_{k\in \bbN}$ that converges weakly to a function $q\in L^2([0,1])$. This example shows why the Riesz Weak Compactness Theorem only proves that a \emph{subsequence} weakly converges to $q$, as opposed to the whole sequence. In this case, one can take the subsequence of even $n$ so that $f_n =1$ or odd $n$ so that $f_n = -1$. These subsequences then converge weakly (and strongly) to $1$ and $-1$ respectively, but the full sequence $(f_n)_{n\in\bbN}$ does not converge weakly to anything in $L^2([0,1])$.
\end{example}

Notice that the Riesz Weak Compactness Theorem requires a $\sigma$-finite measure space. Thus, in order to use the theorem, we will need to be able to ensure that our measure space is $\sigma$-finite. The following lemma will allow us to do this.

\begin{lemma}[{\cite[Ch.~18.2,~Prop.~9]{royden_real_2010}}] 
\label{lem:sigma-finite}
Let $(X,\Sigma,\mu)$ be a measure space and $f\colon X \to [0,\infty]$ a nonnegative integrable function on $X$. Then $f$ is finite almost everywhere and the set $\set{x \in X \mid f(x) > 0}$ is $\sigma$-finite.
\end{lemma}

\bigskip 
\noindent \textbf{Haar measure}
\bigskip 

Suppose $(G,\cdot)$ is a compact Hausdorff topological group. Let $\Sigma$ be the Borel $\sigma$-algebra on $G$; that is, the smallest $\sigma$-algebra that contains all open sets of $G$. A measure $\mu$ is called \emph{left-invariant} if $\mu(A) = \mu(gA)$ for all $g \in G$ and $A \in \Sigma$. The Haar measure on $G$ is the unique left-invariant measure satisfying $\mu(G) = 1$. The finite dimensional unitary group $G = \U(d)$ is compact, as therefore can be equipped with the measure $\mu_{\rm Haar}$ satisfying
\begin{equation}
    \int_{\U(d)} \dd{\mu_{\rm Haar}}(U) = 1, \qquad
    \int_{\U(d)} f(U) \dd{\mu_{\rm Haar}}(U) = \int_{\U(d)} f(VU) \dd{\mu_{\rm Haar}}(U) = \int_{\U(d)} f(UV) \dd{\mu_{\rm Haar}}(U)
\end{equation}
for any $V \in \U(d)$. 

The Haar measure on $\U(d)$ induces a unitarily invariant measure on complex-projective space $\bbC\bbP^{d-1}$. The construction is summarized as follows \cite{sternbergGroupTheoryPhysics2003,bengtsson_geometry_2008,nakahara_geometry_2018}. The unitary group is defined on $\bbC^d$ with respect to an inner product. The unit sphere $S^{2d-1} \subset \bbC^d$ can be viewed as an embedding into $\bbC^d$, and consists of all unit normalized vectors in $\bbC^d$. The inner product on $\bbC^d$ remains defined on $S^{2d-1}$. The set of all quantum states in $\bbC^d$ is $\bbC\bbP^{d-1} \coloneqq S^{2d-1} / \U(1)$. Modding out by $\U(1)$ represents the irrelevance of a global phase factor. In particular, $S^{2d-1}$ can  be viewed as a fiber-bundle, with $\bbC\bbP^{d-1}$ the base space and $\U(1)$ the fiber on top of each point in $\bbC\bbP^{d-1}$. The bundle projection $\pi\colon S^{2d-1} \to \bbC\bbP^{d-1}$ induces a map between the tangent spaces $\pi_\ast\colon T_p S^{2d-1} \to T_{\pi(p)}\bbC\bbP^{d-1}$ (the pushforward map). With some care, $\pi_\ast$ can be used to construct a Hermitian metric on $T\bbC\bbP^{d-1}$ via the inner product on $S^{2d-1}$. The real part of such a metric defines a Riemannian metric $g$, which can then be used to define a volume form, called the Fubini-Study volume form, on $\bbC\bbP^{d-1}$ in the usual way. By construction, the Fubini-Study volume form is unitarily invariant with respect to the definition of unitary via the inner product on $S^{2d-1}$. In particular, $\pi$ is a Riemannian submersion and the resulting metric is the unique unitarily invariant metric up to scaling.
In this way, we have defined a unitarily invariant measure on $\bbC\bbP^{d-1}$. Integrals with respect to this measure are denoted $\int_{\bbC\bbP^{d-1}} f(\psi) \dd{\psi}$. By unitary invariance, $\int_{\bbC\bbP^{d-1}} f(U\psi)\dd{\psi} = \int_{\bbC\bbP^{d-1}} f(\psi)\dd{\psi}$ for all $U \in \U(d)$.

A (nonfinite) Haar measure can also be defined on non-compact groups provided that they are locally compact. We will not discuss this fact much here, other than to say that the unitary group on $L^2(\bbR)$, $\U(L^2(\bbR))$, does not have a Haar measure since it is not locally compact \cite[sec.~5]{grigorchukAmenabilityErgodicProperties2015}. Hence, there is no natural way to integrate over all unitaries acting on the space of continuous-variable quantum states.

\subsection{Projector onto symmetric subspace}
\label{ap:symmetric-projector}

In this section, we summarize the analytical form of projectors onto the symmetric subspace of a separable Hilbert space $\mathcal{H}$, either finite- or infinite dimensional. Let $S_t$ denote a group of permutations of $t$ elements. For any $\sigma \in S_t$, let  $W_{\sigma}\colon \mathcal{H}^{\otimes t} \to \mathcal{H}^{\otimes t}$ denote a unitary operator that transforms the Fock basis as follows: 
\begin{equation}
    \label{eq:wsigma}
    W_{\sigma} \ket{n_1} \otimes \dots \otimes \ket{n_t} = \ket{n_{\sigma^{-1}(1)}} \otimes \dots \otimes \ket{n_{\sigma^{-1}(t)}}~.
\end{equation}

Let $\Pi_t\colon \calH^{\otimes t} \to \calH^{\otimes t}$ denote the projector onto the symmetric subspace of $\mathcal{H}^{\otimes t}$, i.e., the subspace isomorphic to the quotient space $\mathcal{H}^{\otimes t} / \{v-W_\sigma v \mid v \in \mathcal{H}^{\otimes t}, \sigma \in S_t\}$. Using $W_{\sigma}$, $\Pi_t$ can be defined as follows.

\begin{claim}
For each $t \in \bbN$, let $\Pi_t$ denote a projector onto the symmetric subspace of $\mathcal{H}^{\otimes t}$. Then
\begin{equation}
    \label{eq:symm-proj-perms}
    \Pi_t = \frac{1}{\abs{S_t}} \sum_{\sigma \in S_t} W_\sigma~.
\end{equation}
\end{claim}

For completeness, we outline an algebraic proof of $\Pi_t$. It can also be proven using group-theoretic tools, as is shown in e.g.~\cite[sec. 2]{roberts_chaos_2017}, and using linear algebra, as shown in \cite[prop.~1]{harrow_church_2013}. For a complete discussion on the symmetric projector, we refer to Harrow's ``The Church of the Symmetric Subspace'' \cite{harrow_church_2013}.
\begin{proof}
We denote the set of permutations of a vector $v\in \calH^{\otimes t}$ as $P(v) \coloneqq \set{W_\sigma v \mid \sigma\in S_t}$, where $W_{\sigma}$ is given by \cref{eq:wsigma}. Let $\calB = \set{\ket{i} \mid i \in \set{1,\dots,\dim \calH}}$ be an orthonormal basis of $\calH$, and $\calB^{\otimes t}$ the corresponding orthonormal basis of $\calH^{\otimes t}$. 

As an example, suppose $t=4$ and $v = \ket{1}\otimes \ket{2} \otimes \ket{1} \otimes \ket {1} \in \calB^{\otimes t}$. We will use the notation $v!$ to mean $v! = 3! \cdot 1!$, since $\ket 1$ occurs three times and $\ket 2$ one time. Similarly, suppose $t=5$ and $v = \ket{5}^{\otimes 3}\otimes \ket{1}^{\otimes 2}$. Then $v! = 3!\cdot 2!$. One can then verify that
\begin{equation}
\calB' = \set{\frac{1}{\sqrt{v! \abs{S_t}}} \sum_{\sigma \in S_t} W_\sigma v \mid v \in \calB^{\otimes t}}~,
\end{equation}
is an orthonormal basis of the symmetric subspace of $\calH^{\otimes t}$. For any $v \in \calB^{\otimes t}$ and $u \in \calB'$, the following holds:
\begin{equation}
\braket{v\vert u} = \sqrt{\frac{v!}{\abs{S_t}}}\delta_{u \in {\rm span}(P(v))} = \begin{cases}
 \sqrt{\frac{v!}{\abs{S_t}}}&\text{if } u \in {\rm span}(P(v))\\
 0&\text{otherwise.}
\end{cases}
\end{equation}

Using one dimensional subspaces spanned by the basis vectors in $\mathcal{B}^{\prime}$, the projector onto the symmetric subspace of $\mathcal{H}^{\otimes t}$ can be represented as
\begin{equation}
    \Pi_t = \sum_{u \in \calB'} \vert u\rangle \langle u\vert~.
\end{equation} 
Next, we determine the matrix elements of $\Pi_t$ in the basis $\calB$. Let $v,w \in \cal B$. Then
\begin{salign}
    \bra{v}\Pi_t \ket{w} &= \sum_{u \in \calB'} \braket{v\vert u} \braket{u\vert w}\\
    &= \frac{\sqrt{v!w!}}{\abs{S_t}}\sum_{u \in \calB'} \delta_{u \in {\rm span}(P(v))}\delta_{u \in {\rm span}(P(w))}\\
    &= \frac{v!}{\abs{S_t}}\delta_{v \in P(w)}.
\end{salign}
Finally, consider the matrix elements of $\frac{1}{\abs{S_t}}\sum_{\sigma \in S_t} W_\sigma$:
\begin{salign}
    \bra{v} \parentheses{\frac{1}{\abs{S_t}}\sum_{\sigma \in S_t} W_\sigma} \ket{w} &= \frac{1}{\abs{S_t}} \sum_{\sigma \in S_t}\bra v W_\sigma \ket w\\
    &= \frac{v!}{\abs{S_t}} \delta_{v \in P(w)},
\end{salign}
which proves the claim.
\end{proof}

We now define some more notation for the matrix elements of the symmetric projector.

\begin{definition}
\label{def:pi-lambda}
For any tuples $a, b \in \bbN_0^t$, define
\begin{salign}
     \Pi_t(a;b) &\equiv \Pi_t(a_1,\dots,a_t; b_1,\dots, b_t)\\
     &\coloneqq \parentheses{\bigotimes_{i=1}^t \bra{a_i}} \Pi_t \parentheses{\bigotimes_{i=1}^t \ket{b_i}},
\end{salign}
and define 
\begin{equation}
    \Lambda_t(a) \equiv \Lambda_t(a_1,\dots,a_t) \coloneqq \Pi_t(a;a).
\end{equation}
\end{definition}

For this appendix, we will need some properties of $\Lambda$. Clearly, $\Lambda_t(a) > 0$ for all tuples $a \in \bbN_0^t$. Similarly, $\Lambda_t(a) = \Lambda_t(\sigma(a))$ for any $\sigma \in S_t$. Additionally, $\Lambda_t(a_1,\dots,a_t) = \Lambda_t(a_1+1,\dots,a_t+1)$. Finally, for a tuple $b \in \bbN_0^{t-1}$ and a number $i \in \bbN_0$, denote the direct sum as the tuple $b \oplus (i) \coloneqq \parentheses{b_1, \dots, b_{t-1}, i}$.  For any fixed $b \in \bbN_0^{t-1}$, there exists an $N\in \bbN$ such that for all $m,m' > N\colon$ $\Lambda_t(b \oplus (m)) = \Lambda_t(b \oplus (m'))$. This means that, for example
\begin{equation}
    \label{eq:lambdat_limit}
    \lim_{m\to\infty} \Lambda_t(0,\dots,0,m) > 0.
\end{equation}

To get a handle on these definitions and properties, consider the example for $t=2$.

\begin{example}
\label{ex:pi2}
In the case of $t=2$,
\begin{equation}
    \Pi_2 = \frac{1}{2}\parentheses{ W_{(1)(2)} + W_{(12)}} = \frac{1}{2}\parentheses{\bbI + S},
\end{equation}
where $S$ is the SWAP operator. We used cyclic notation for permutations, so that $(1)(2)$ is the identity permutation and $(12)$ is the other permutation in $S_2$. One can also find $\Pi_2$ by summing over projectors onto an orthonormal set of symmetric states, as
\begin{equation}
\begin{aligned}
    \Pi_2 &= \ket{0}\ket{0}\bra{0}\bra{0} \\
    &\quad + \parentheses{\frac{\ket 0 \ket 1+\ket 1 \ket 0}{\sqrt{2}}}\parentheses{\frac{\bra 0 \bra 1+\bra 1 \bra 0}{\sqrt{2}}}\\
    &\quad + \ket{1}\ket{1}\bra{1}\bra{1} \\
    &\quad +\parentheses{\frac{\ket 0 \ket 2+\ket 2 \ket 0}{\sqrt{2}}}\parentheses{\frac{\bra 0 \bra 2+\bra 2 \bra 0}{\sqrt{2}}} + \dots.
\end{aligned}
\end{equation}
Therefore, 
\begin{equation}
\Pi_2(a_1,a_2;b_1,b_2) = \frac{1}{2}\parentheses{\delta_{a_1b_1}\delta_{a_2b_2} + \delta_{a_1b_2}\delta_{a_2b_1}},
\end{equation}
and $\Lambda_2(a_1,a_2) = \frac{1}{2}\parentheses{1+\delta_{a_1a_2}}$.
\end{example}

\section{Finite dimensional designs}
\label{ap:finite-designs}

\subsection{Simplex designs}
\label{ap:simplex-designs}

A simplex $t$-design, more commonly referred to as a \emph{(positive, interior) simplex cubature rule} in the literature \cite{cools_constructing_1997,stroud_approximate_1971,hammer_numerical_1956,baladram_on_2018,kuperberg_numerical_2004,kuperberg_numerical_ec_2004}, is a set of points on the simplex and a weight function that exactly integrates polynomials of degree $t$ or less.

\begin{definition}[Simplex design]
Let $P \subset \Delta^{m}$ be a finite set, and $u\colon P \to \bbR_{>0}$ be a weight function on $P$. Let $\dd{p}$ denote the standard unit normalized Lebesgue measure on the simplex. The pair $(P, u)$ is called an $m$ dimensional \emph{simplex $t$-design} if, for all tuples $a = (a_1, \dots, a_t) \in \set{0,1,\dots, m}^t$,
\begin{equation}
    \sum_{q \in P} w(q) \prod_{i=1}^t q_{a_i} = \int_{\Delta^m} \prod_{i=1}^t p_{a_i}\dd{p}.
\end{equation}
The pair $(P, u)$ defines a probability ensemble, and we will therefore define $\Expval_{q\in P} g(q) \coloneqq \sum_{q \in P} w(q)g(q)$ for any function $g$.
\end{definition}

Since the coordinates of a point on the simplex sum to $1$, by summing over one of the $a_i$ on both sides, we find that a simplex $t$-design is automatically a simplex $(t-1)$-design. The measure $\dd{p}$ is proportional to $\delta(1-p_0 - \dots - p_n)\dd{p_0}\dots \dd{p_m}$. For $\beta \in \bbN_0^{m+1}$ and $p \in \Delta^m$, define $p^\beta \coloneqq \prod_{i=0}^m p_i^{\beta_i}$. For example, if $m=2$ and $\beta = (0,2,1)$, then $p^{\beta} = p_1^2 p_2$. Then one can compute the moments from the Dirichlet distribution \cite{stroud_approximate_1971}
\begin{equation}
    \label{eq:simplex-t-moment}
    \int_{\Delta^m}p^\beta \dd{p} = \frac{m!}{(m+\beta_0 + \dots + \beta_m)!} \prod_{i=0}^m \beta_i!.
\end{equation}

We will list various simplex $t$-designs. We will use the notation
\begin{equation}
f^{(i)} = (\underbrace{0,\dots,0}_{i}, 1, \underbrace{0,\dots,0}_{m-i}),
\end{equation}
so that $f_j^{(i)} = \delta_{ij}$. In this way, a point $p \in \Delta^m$ is written as $p = \sum_{i=0}^m p_i f^{(i)}$. Denote the centroid of the simplex by $c = \frac{1}{m+1}\sum_{i=0}^m f^{(i)}$.

\begin{theorem}[Extremal points of the unit $m$-simplex form a one-design]
    Let $P$ be the set $P = \set{f^{(i)} \mid i \in \set{0, \dots, m}}$, and $u$ the constant map $w(f^{(i)}) = 1/(m+1)$. The pair $(P, u)$ is a simplex one-design.
\end{theorem}
\begin{proof}
We must prove that $\frac{1}{m+1}\sum_{i =0}^m g(f^{(i)}) = n! \int_{\Delta^m} g(p) \dd{p}$ for any linear polynomial $g(p) = p_j$. The left-hand side is then $\frac{1}{m+1}\sum_{i=0}^m \delta_{ij} = 1/(m+1)$, and the right-hand side is $\frac{1}{m+1}$ by \cref{eq:simplex-t-moment}.
\end{proof}

\begin{theorem}[Centroid of the unit $m$-simplex forms a one-design]
    Let $P$ be the set $P = \set{c}$ and $u$ the map $u(c) = 1$. Then the pair $(P, u)$ is a simplex one-design.
\end{theorem}
\begin{proof}
Clearly $g(c) = 1/(m+1)$ for any linear polynomial $g(p) = p_j$.
\end{proof}

\begin{theorem}[Extremal points plus the centroid of the unit $m$-simplex form a two-design]
    \label{thm:simplex-ext-centroid-design}
    Let $P$ be the set $P = \{c \} \cup \set{f^{(i)} \mid i \in \set{0, \dots, m}}$, and $u$ the map defined by $u(c) = \frac{m+1}{m+2}$ and $u(f^{(i)}) = \frac{1}{(m+1)(m+2)}$. Then the pair $(P, u)$ is a simplex two-design.
\end{theorem}
\begin{proof}
It suffices to prove that
\begin{equation}
    \frac{m+1}{m+2}g(c) + \frac{1}{(m+1)(m+2)} \sum_{i=0}^m g(f^{(i)}) = m! \int_{\Delta^m} g(p) \dd{p}
\end{equation}
for any quadratic polynomial $g(p) = p_j p_k$. By \cref{eq:simplex-t-moment}, the right-hand side equals $\frac{1+\delta_{jk}}{(m+1)(m+2)}$. The left-hand side is
\begin{salign}
    \frac{m+1}{m+2}g(c) + \frac{1}{(m+1)(m+2)} \sum_{i=0}^m g(f^{(i)}) &= \frac{1}{(m+1)(m+2)} + \frac{1}{(m+1)(m+2)} \sum_{i=0}^m \delta_{ij}\delta_{ik}\\
    &= \frac{1+\delta_{jk}}{(m+1)(m+2)},
\end{salign}
as desired.
\end{proof}

\begin{theorem}[Simplex two-design {\cite[thm.~2]{hammer_numerical_1956}}, {\cite[cor.~4.1]{baladram_on_2018}}]
    \label{thm:symmetrical-simplex-two-design}
    Let $r = 1/\sqrt{m+2}$. Let $v^{(i)} = r f^{(i)} + (1-r) c$. Let $P$ be the set $P = \set{v^{(i)} \mid i \in \set{0,\dots,m}}$, and $u$ the constant map $u(v^{(i)}) = 1/(m+1)$. Then the pair $(P, u)$ is a simplex two-design.
\end{theorem}

The simplex $2$-design in \cref{thm:symmetrical-simplex-two-design} utilizes $m+1$ points in $\Delta^m$, which is in fact the best that can be done \cite[tab.~1]{cools_constructing_1997}.

\subsection{Torus designs}
\label{ap:torus-designs}

We define torus designs analogously to simplex designs. We let $T = [0,2\pi)$.

\begin{definition}
\label{def:torus-design}
Let $S \subset T^m$ be a finite set, and $v \colon S \to \bbR_{> 0}$ be a weight function on $S$. Let $\dd{\phi}$ denote the standard unit normalized Lebesgue measure on the torus. The pair $(S, v)$ is called an $m$ dimensional \emph{torus $t$-design} if, for all tuples $a = (a_1, \dots, a_t) \in \set{1,2,\dots, m}^t$ and $b = (b_1, \dots, b_t) \in \set{1,2,\dots, m}^t$,
\begin{equation}
    \sum_{\theta \in S}v(\theta) \prod_{i=1}^t \e^{\i(\theta_{a_i} - \theta_{b_i})} = \int_{T^m} \prod_{i=1}^t \e^{\i(\phi_{a_i} - \phi_{b_i})} \dd{\phi}.
\end{equation}
The pair $(S, v)$ defines a probability ensemble, and we will therefore define $\Expval_{\theta \in S} g(\e^{\i \theta_1}, \dots, \e^{\i \theta_m}) \coloneqq \sum_{\theta \in S} v(\theta) g(\e^{\i \theta_1}, \dots, \e^{\i \theta_m})$ for any function $g$.
\end{definition}

\medskip

It follows from the definition that a torus $t$-design is always a torus $(t-1)$-design. For example, suppose $t = 2$, and let $a = (1,j)$ and $b = (1, k)$ for any $j$ and $k$. Then it is clear that the $2$-design $(S, v)$ also satisfies the $1$-design condition. By definition, a torus $t$-design must match integration on polynomials $g(s) = g(s_1,\dots, s_m)$ that are degree $t$ in $s$ and degree $t$ in degree $\bar s$. One could generalize the definition to match integration on polynomials that are degree $t$ in $s$ and degree $t'$ in $\bar s$. We call the corresponding sets $(t,t')$ torus designs. In this way, a torus $t$-design is a shorthand notation for a $(t,t)$-design.

The definition of a torus $t$-design closely resembles the definition of a \emph{trigonometric cubature rule} \cite{cools_constructing_1997}, however they are not equivalent.
To the best of our knowledge, the notion of general torus cubature was first proposed in Ref.~\cite{kuperberg_numerical_2004}, where it was formulated as a generalization of trigonometric cubature rules in terms of algebraic tori. Our definition of a $T^m$ $t$-design corresponds to the definition in Ref.~\cite{kuperberg_numerical_2004} of an order $t$ cubature rule on the maximal torus $T(\PSU(m+1)) \cong T^m$ with an algebraic structure given by a faithful orbit of its linear action by conjugation on the vector space of $(m+1) \times (m+1)$ complex matrices. Here, $\PSU(m+1)$ is the projective special unitary group, which is the special unitary group $\SU(m+1)$ modulo its center. In \cref{ap:torus}, we show the equivalence of the two definitions, as well as comment on the relationship to standard trigonometric cubature and to complete sets of mutually unbiased bases.

We now construct various torus designs.

\begin{theorem}[$1$-design on the $m$-torus]
    Let $S$ be the set
    \begin{equation}
        S = \set{\parentheses{0, 2\pi q/m, 2\pi 2q/m, \dots, 2\pi (n-1)q/m} \mid q \in \bbZ_m},
    \end{equation}
    and $v$ the constant map $v(\phi) = 1/m$. Then the pair $(S, v)$ is an $m$-torus $1$-design.
\end{theorem}
\begin{proof}
It is sufficient to check for $g(s) = s_a \bar s_b$.
\begin{equation}
    \frac{1}{m}\sum_{q\in \bbZ_m} \e^{2\pi\i a q / m}\e^{-2\pi\i b q / m}
    = \frac{1}{m}\sum_{q\in \bbZ_m} \e^{2\pi\i q(a-b)/m}\\
    = \delta_{ab}.
\end{equation}
Meanwhile,
\begin{equation}
    \int_{T^m} \e^{\i \phi_a}\e^{-\i \phi_b} \dd{\phi}
    = \frac{1}{(2\pi)^2}\int_{T^2} \e^{\i(\phi_a - \phi_b)} \dd{\phi_a}\dd{\phi_b}\\
    = \delta_{ab}. \qedhere
\end{equation}
\end{proof}

\begin{theorem}[$t$-design on the $m$-torus (concatenation of $t$-designs on each factor of $S^1$)]
\label{thm:torus-t-design}
    Let $S$ be the set
    \begin{equation}
        S = \set{\parentheses{2\pi d_1/(t+1), 2\pi d_2/(t+1), \dots, 2\pi d_m/(t+1)} \mid d \in \bbZ_{t+1}^m},
    \end{equation}
    and $v$ the constant map $v(\phi) = (t+1)^{-m}$. Then the pair $(S, v)$ is an $m$-torus $t$-design.
\end{theorem}
\begin{proof}
It is sufficient to check for $g(s) = s_{a_1}\dots s_{a_t} \bar s_{b_1}\dots s_{b_t}$.
\begin{equation}
    \frac{1}{(t+1)^m} \sum_{d \in \bbZ_{t+1}^m} \exp\bargs{\frac{2\pi\i}{t+1} (d_{a_1} + \dots + d_{a_t})} \exp\bargs{-\frac{2\pi\i}{t+1} (d_{b_1} + \dots + d_{b_t})}
    = \begin{cases}
      1&\text{if } a \text{ is a permutation of } b\\
      0&\text{otherwise.}
    \end{cases}
\end{equation}
Meanwhile,
\begin{equation}
    \int_{T^m} \e^{\i (\phi_{a_1} + \dots + \phi_{a_t})} \e^{-\i (\phi_{b_1} + \dots + \phi_{b_t})} \dd{\phi}
    = \begin{cases}
      1&\text{if } a \text{ is a permutation of } b\\
      0&\text{otherwise.}
    \end{cases}~. \qedhere
\end{equation}
\end{proof}

\begin{theorem}[Efficient $2$-design on the $m$-torus]
    \label{thm:torus-two-design}
    Define $p$ to be the smallest prime number strictly larger than $\max(2, m)$ (by the prime number theorem, $p \in \bigO{m + \log m}$). Let $S$ be the set
    \begin{equation}
        S = \set{\parentheses{0,2\pi (q_1+q_2)/p,2\pi (2q_1+4q_2)/p, \dots, 2\pi ((m-1)q_1+(m-1)^2 q_2)/p  } \mid  q_1 \in \bbZ_{p}, q_2 \in \bbZ_{p}}
    \end{equation}
    and $v$ the constant map $v(\phi) = 1/p^2$. Then the pair $(S, v)$ is an $m$-torus $2$-design.
\end{theorem}
\begin{proof}
    It suffices to prove that
    \begin{equation}
        \int_{T^m} \e^{\i(\phi_i + \phi_j - \phi_k - \phi_l)} \dd{\phi} = \frac{1}{p^2}\sum_{\theta \in S}\e^{\i(\theta_i + \theta_j - \theta_k- \theta_l) }.
    \end{equation}
    The right-hand side is
    \begin{salign}
        \frac{1}{p^2}\sum_{\theta \in S}\e^{\i(\theta_i + \theta_j - \theta_k- \theta_l) } &= \frac{1}{p^2}\sum_{q_1,q_1 \in \bbZ_p}\e^{\frac{2\pi \i}{p} q_1(i + j - k- l) }\e^{\frac{2\pi \i}{p} q_2(i^2 + j^2 - k^2- l^2) }\\
        &= \delta_{i+j,k+l}\delta_{i^2 + j^2, k^2+ l^2}\\
        &= \begin{cases}
        1&\text{if } i=l \land j=k \text{ or } i=k \land j=l\\
        0&\text{otherwise},
        \end{cases}
    \end{salign}
    where we used \cref{lem:diophantine} in the last line.
    The left-hand side is
    \begin{salign}
        \int_{T^m} \e^{\i(\phi_i + \phi_j - \phi_k - \phi_l)} \dd{\phi} &= \frac{1}{(2\pi)^m}\int_{[0,2\pi]^m} \e^{\i(\phi_i + \phi_j - \phi_k - \phi_l)} \dd{\phi_1}\dots\dd{\phi_m}\\
        &= \begin{cases}
        1&\text{if } i=l \land j=k \text{ or } i=k \land j=l\\
        0&\text{otherwise},
        \end{cases}
    \end{salign}
    which is equal to the right-hand side.
\end{proof}

The torus $2$-design in \cref{thm:torus-two-design} utilizes what Ref.~\cite{van_dam_mutually_2011} calls the ``$ax^2 + bx$ construction'' that is utilized in constructions of complete sets of mutually unbiased bases. From Ref.~\cite{klappenecker2005mutually}, it is known that such sets form complex-projective $2$-designs. Hence, we can now understand the ``$ax^2 + bx$ construction'' as a torus $2$-design. The ``$ax^2 + bx$'' construction utilizes the following Diophantine system.

\begin{lemma}
\label{lem:diophantine}
Let $\bbF_p$ be the finite field with $p$ elements for an odd prime $p$. Let $F$ be either $\bbF_p$ or $\bbZ$, and let addition, multiplication, and equality be with respect to $F$ (e.g. for $F = \bbF_p$, $a = b$ is the same as $a \equiv b \pmod{p}$). The Diophantine system of equations
\begin{equation}
    a+b = c+d, \quad a^2 + b^2 = c^2 + d^2
\end{equation}
is solved only by solutions of the form
\begin{equation}
    (a=c)\land (b=d), \qquad \text{or}~~ (a=d)\land(b=c).
\end{equation}
\end{lemma}
\begin{proof}
  Plugging the first equation into the second equation, we find that $(c+d-a)^2 + b^2 = c^2 + d^2$. Simplifying yields $ab=cd$.  If $b = 0$, then either $c$ or $d$ must equal zero, so that the solution is of the desired form. If $b\neq 0$, then
  \begin{salign}
      &ab + b^2 = bc + bd\\
      &cd + b^2 = bc + bd\\
      &(d-b)(c-b) = 0.
  \end{salign}
  Therefore, either $b=c$ or $b=d$. Along with $a+b=c+d$, this proves the claim.
\end{proof}

\cref{thm:torus-two-design} can be generalized to the case where we allow $p$ to be any positive integer power of a prime, because \cref{lem:diophantine} can be generalized to the case of any Galois (finite) field.

An $m$-torus $1$-design trivially requires at least $m$ elements. To conclude this subsection, we show that an $m$-torus $2$-design requires at least $m(m-1)+1$ elements.

\begin{proposition}\label{prop:minimal-torus-2design}
Let $(S, v)$ be a $T^m$ $2$-design. Then $\abs{S} > m(m-1)$.
\end{proposition}
\begin{proof}
The torus $2$-design condition can be expressed as follows. Let
\begin{equation}
    \Gamma = \set{(0,\dots, 0), (1,-1,0,\dots,0), (1,0,-1,0,\dots,0), \dots, (-1,1,0,\dots,0),\dots, (0,\dots,0,-1,1)},
\end{equation}
so that $\abs{\Gamma} = m(m-1)+1$. Let each $\phi\in S$ label a basis element of $V\coloneqq \bbC^{\abs{S}}$ so that $\set{\ket\phi \mid \phi \in S}$ is an orthonormal basis of $V$. Then for $k\in\Gamma$, define $\ket k = \sum_{\phi\in S} \sqrt{v(\phi)} \e^{\i k \cdot \phi} \ket\phi$. The $2$-design condition is summed up by $\braket{k\vert k'} = \delta_{kk'}$. Hence, $\set{\ket k \mid k \in \Gamma}$ must be orthonormal in $V$, meaning that $\abs{\Gamma} \leq \dim V = \abs{S}$.
\end{proof}

\subsection{Complex-projective Haar integral}
\label{ap:cp-haar-integral}

For integration over the set of $d$ dimensional qudit states $\bbC\bbP^{d-1}$, one finds \cite{roberts_chaos_2017,scott_tight_2006}
\begin{equation}
    \int_{\bbC\bbP^{d-1}} \parentheses{\ket{\psi} \bra{\psi}}^{\otimes t} \dd{\psi} = \frac{\Pi_t^{(d)}}{\Tr \Pi_t^{(d)}},
\end{equation}
where $\dd{\psi}$ denotes the unitarily invariant Fubini-Study volume form on the complex-projective space $\bbC\bbP^{d-1}$, and $\Pi_t^{(d)}\colon (\bbC^d)^{\otimes t} \to (\bbC^d)^{\otimes t}$ is the projector onto the symmetric subspace of $(\bbC^d)^{\otimes t}$ defined in \cref{def:pi-lambda} \cite[sec.~4.5,~4.7,~7.6]{bengtsson_geometry_2008} \cite[ex.~8.8]{nakahara_geometry_2018}. We will begin by showing this equality.

Each $\ket{\psi}$ lives in the finite dimensional Hilbert space $\calH = \bbC^d$. For any integer $t$, the tensor product $\calH^{\otimes t}$ splits up into a direct sum of the symmetric and other subspaces of $\calH^{\otimes t}$. For example, in the case of $t=2$, $\calH^{\otimes 2} \cong \calH^{\rm sym}_2 \oplus \calH_2^{\rm asym}$. Consider the representation of the group of unitaries acting on $\calH$, $\rho\colon \U(\calH) \to \U(\calH^{\otimes t})$, defined by $U \mapsto U^{\otimes t}$. The subspace $\calH_t^{\rm sym}$ is invariant under $\rho$. One can see this by noting that for any unitary $U$, $\rho(U)\calH_t^{\rm sym} = \calH_t^{\rm sym}$ since $U^{\otimes t}$ acts symmetrically on the tensor product  factors. Therefore, the representation $\rho$ can be decomposed into a direct sum of irreducible representations, one of which being $\calH_t^{\rm sym}$.

We can now invoke Schur's lemma, which states that if a nonzero operator $M$ on an irrep space commutes with every element of that irrep, then $M$ is proportional to the identity on that irrep space. In our case, $M = \int_{\bbC\bbP^{d-1}}\parentheses{\ket{\psi}\bra{\psi}}\dd{\psi}$. The irrep space of interest is $\calH_t^{\rm sym}$. The elements of the irrep are unitaries $U^{\otimes t}$. Due to the unitary invariance of the Fubini-Study metric, one finds that $M$ commutes with all unitaries of the form $U^{\otimes t}$. Therefore, by Schur's lemma, $M$ must be proportional to the identity on $\calH_t^{\rm sym}$, which is precisely $\Pi_t^{(d)}$. Finally, the Fubini-Study volume measure is normalized such that the volume of $\bbC\bbP^{d-1}$ is unity. Hence, $\Tr M = 1$, meaning that the proportionality constant must be $1/\Tr \Pi_t^{(d)}$.

Next, we discuss integration over $\bbC\bbP^{m}$ with respect to the Fubini-Study volume form where $m=d-1$, and show that it can be expressed as integration over a flat simplex and a flat torus. For a formal treatment of this fact, see \cite[sec.~4.5, 4.7, 7.6]{bengtsson_geometry_2008} \cite[ex.~8.8]{nakahara_geometry_2018}. One first constructs the Fubini-Study volume form (see \cref{ap:measure-theory}). Then, one constructs a coordinate transformation mapping the simplex cross the torus to a coordinate patch of $\bbC\bbP^m$. Pulling back the volume form along this coordinate transformation yields the volume form on the simplex cross the torus.

Here, we will instead give an informal treatment.
Define $p_0 = 1-\sum_{i=1}^m p_i$ and $\phi_0 = 0$. Then the $p_i$ and $\phi_i$ parameterize a quantum state $\sqrt{p_0}\ket 0 + \sum_{i=1}^m \sqrt{p_i}\e^{\i \phi_i}\ket i$. To define a valid state, the $p$ are elements of the probability simplex $\Delta^m = \set{p=(p_0, \dots, p_m) \in [0,1]^{m+1} \mid \sum_{i=0}^m p_i = 1}$ and the $\phi$ are elements of the torus $T^m = [0,2\pi)^m$. We'll denote the Lebesgue measure on $\Delta^m$ by $\dd{p} = \prod_{i} \dd{p_i}$, and on $T^m$ by $\dd{\phi} = \prod_{i=1}^m \dd{\phi_i}$. One can easily perform the integration over the simplex and torus to find that $\frac{m!}{(2\pi)^m} \dd{p}\dd{\phi}$ is a normalized volume measure such that ${\rm vol}(\Delta^m \times T^m) = 1$. Consider a quantum state in $\bbC^{m+1}$ parameterized by $\alpha_i \in \bbC$ as $\ket{\psi} = \sum_{n=0}^{m} \alpha_n \ket n$, the natural measure is $\Dd{2}{\alpha_0}\dots \Dd{2}{\alpha_m}$. Applying the polar coordinate transformation $\alpha_n = \sqrt{p_n} \e^{\i \phi_n}$ and keeping track of Jacobian factors, the measure becomes proportional to $\dd{p}\dd{\phi}$.

In conclusion, we have determined that
\begin{equation}
    \label{eq:cp-simplex-torus}
    \int_{\bbC\bbP^m} \parentheses{\ket{\psi}\bra{\psi}}^{\otimes t} \dd{\psi} = \frac{m!}{(2\pi)^m}\int_{\Delta^m \times T^m} \parentheses{\ket{p,\phi}\bra{p,\phi}}^{\otimes t} \dd{p}\dd{\phi},
\end{equation}
where
\begin{equation}
\ket{p,\phi} = \sqrt{p_0} \ket{0} + \sum_{j=1}^m \sqrt{p_j}\e^{\i \phi_j} \ket{j}.
\end{equation}

\subsection{Complex-projective designs from simplex and torus designs}
\label{ap:simplex-torus-to-cp}

For finite $d$, an ensemble $\calE$ over $\bbC\bbP^{d-1}$ is a complex-projective $t$-design if
\begin{equation}
    \EX{\ket{\psi}\in\calE}{\parentheses{\ket{\psi}\bra{\psi}}^{\otimes t}} = \int_{\bbC\bbP^{d-1}}\parentheses{\ket{\psi}\bra{\psi}}^{\otimes t} \dd{\psi}.
\end{equation}
Again let $m=d-1$. The characterization of the integral over $\bbC\bbP^m$ given in \cref{eq:cp-simplex-torus} motivates the construction of complex-projective designs via constructions of simplex and torus designs. Such a construction was also noted in \cite[thm.~4.1]{kuperberg_numerical_2004}. In particular, \cref{eq:cp-simplex-torus} consists of a product of integrals of the form $m!\int_{\Delta^m} \sqrt{\prod_{i=1}^{2t} p_{j_i}} \dd{p}$ and $\frac{1}{(2\pi)^m}\int_{T^m} \exp\bargs{\i\sum_{i=1}^{t} (\phi_{j_i} - \phi_{j_{t+i}}) } \dd{\phi}$. The latter integral can be evaluated by a $t$-design on the torus, and is equal to $1$ whenever $\sum_{i=1}^t (\phi_{j_i} - \phi_{j_{t+i}}) = 0$ regardless of $\phi$, and zero otherwise. In other words, it is only nonzero when the $j_i$'s are paired. But when the $j_i$'s are paired, the term $\sqrt{\sum_{i=1}^{2t} p_{j_i}}$ becomes a monomial of degree $t$ in $p$. Hence, the resulting integral can be evaluated with a simplex $t$-design. We summarize with the following theorem.

\begin{theorem}
\label{thm:cp-from-simplex-torus}
Let $P$ be a $t$-design on the $m$-simplex $\Delta^m$, meaning that $P$ is an ensemble over $\Delta^m$ such that
\begin{equation}
    \EX{q \in P}{g(q)} = \int_{\Delta^m} g(p) \dd{p}
\end{equation}
for any polynomial $g(p) = g(p_0,\dots,p_m)$ of degree less than or equal to $t$. Similarly, let $S$ be a $t$-design on the $m$-torus $T^m$, meaning that $S$ is an ensemble over $T^m$ such that
\begin{equation}
    \EX{\theta \in S}{g(\e^{\i \theta_1}, \dots, \e^{\i \theta_m})} = \int_{T^m} g(\e^{\i \phi_1}, \dots, \e^{\i \phi_m}) \dd{\phi}
\end{equation}
for any polynomial $g(s) = g(s_1,\dots, s_m)$ of degree $t$ in $s$ and degree $t$ in $\bar s$. Then $D = P \times S$ is a $t$-design on $\bbC\bbP^m$, meaning that
\begin{equation}
    \EX{(p,\phi) \in D}{\parentheses{\ket{p,\phi}\bra{p,\phi}}^{\otimes t}} = \int_{\bbC\bbP^m} \parentheses{\ket\psi\bra\psi}^{\otimes t} \dd{\psi},
\end{equation}
with $\ket{p,\phi} \coloneqq \sum_{j=0}^{m} \sqrt{p_j} \e^{\i\phi_j} \ket j$.
\end{theorem}

We can state this in terms of weight functions as follows. Let $(P, u)$ be a $\Delta^{d-1}$ $t$-design, and $(S, v)$ be a $T^d$ $t$-design. Define $D \coloneqq \set{\ket{p,\phi} \mid p \in P, \phi \in S}$, and
\begin{equation}
    \label{eq:simplex-torus-cp-w}
    w(\ket{p,\phi}) \coloneqq  u(p)\sum_{\substack{\phi' \in S \text{ st}\\ \ket{p,\phi} = \ket{p,\phi'}}} v(\phi').
\end{equation}
Then $(D, w)$ is a $\bbC\bbP^{d-1}$ $t$-design.
Morally, $w(\ket{p,\phi})$ is essentially $u(p)v(\phi)$. However, the map $(p,\phi) \mapsto \ket{p,\phi}$ is not bijective; specifically, if $p$ is on the boundary $\partial \Delta^{d-1}$, then for any $\phi$ there are many $\phi'$ satisfying $\ket{p,\phi} = \ket{p,\phi'}$. Therefore, the definition of $w$ must be modified accordingly, as is done in \cref{eq:simplex-torus-cp-w}.

\medskip

We will now construct explicit complex-projective designs by concatenating simplex and torus designs given in \cref{ap:simplex-designs,ap:torus-designs}. For this subsection, we will use the following notation for complex-projective $t$-designs. Fix a set $D \subset \mathbb{CP}^m$ of points in $\mathbb{CP}^m$, and let $w\colon D \to \bbR_{>0}$ be a weight function. The pair $(D, w)$ is a complex-projective $t$-design if
\begin{equation}
    \sum_{\ket\xi \in D} w(\ket\xi) \parentheses{\ket\xi\bra\xi}^{\otimes t} = \int_{\bbC\bbP^m} \parentheses{\ket\psi\bra\psi}^{\otimes t} \dd{\psi}.
\end{equation}

\medskip

\textbf{Construction 1:} Combining the simplex $2$-design from \cref{thm:simplex-ext-centroid-design} and the torus $2$-design from \cref{thm:torus-two-design}, we find that for any $m \in \bbN_0$, the pair $(D, w)$ is a complex-projective $2$-design, where $p$ is the smallest prime number strictly larger than $\max(2,m)$, $D$ is the set
\begin{equation}
    \label{eq:mub}
    D = \set{\ket i \mid i \in \set{0,\dots, m}} \cup \set{\ket{q_1,q_2} \mid q_1, q_2 \in \bbZ_p},
\end{equation}
and $w\colon D \to \bbR_{>0}$ is the map defined by $w(\ket i) = \frac{1}{(m+1)(m+2)}$ and $w(\ket{q_1, q_2}) = \frac{m+1}{(m+2)p^2}$. By the prime number theorem, $p \in \bigO{m + \log m}$. Here $\ket{q_1, q_2} \coloneqq \frac{1}{\sqrt{m+1}}\sum_{j=0}^m \e^{2\pi \i (q_1 j + q_2 j^2) / p} \ket{j}$. When $m+1$ is prime, this reduces to the well-known complete set of mutually unbiased bases given in \cite{wootters_optimal_1989} (indeed this can be generalized to whenever $d=m+1$ is a prime power). For prime $d$, this complex-projective design is uniformly weighted. However, for non-prime $d$, the weights are not uniform.

\medskip

\textbf{Construction 2:} We can construct a uniformly weighted complex-projective $2$-design for all $m$ that uses $p^2 (m+1)$ points by combining the simplex $2$-design from \cref{thm:symmetrical-simplex-two-design} and the torus $2$-design from \cref{thm:torus-two-design}. Define $r = 1/\sqrt{m+2}$ and the state
\begin{equation}
    \ket{\ell,q_1, q_2} \coloneqq \sqrt{\frac{1+rm}{m+1} } \e^{2\pi\i (q_1 \ell + q_2 \ell^2) / p} \ket \ell + \sqrt{\frac{1-r}{m+1}}\sum_{j\neq \ell} \e^{2\pi\i (q_1 j + q_2 j^2) / p} \ket{j},
\end{equation}
the set
\begin{equation}
    \label{eq:unweighted-cp-2-design}
    D = \set{\ket{\ell,q_1, q_2} \mid \ell \in \set{0,\dots, m}, q_1, q_2 \in \bbZ_p},
\end{equation}
and the constant map $w(\ket{\ell,q_1,q_2}) = (n+1)^{-1}p^{-2}$. Then the pair $(D, w)$ is a complex-projective $2$-design.

\bigskip

One can also construct a complex-projective $\geq 3$-design for all $n$ by combining the simplex designs given in \cite{kuperberg_numerical_2004,kuperberg_numerical_ec_2004} with the torus design given in \cref{thm:torus-t-design}.

We note that if one relaxes the requirement that the weights be nonnegative, then one can construct signed complex-projective designs by using signed simplex and torus designs. For example, simple and explicit simplex signed $t$-designs are given for all odd $t$ in \cite[thm.~4]{grundmann_invariant_1978}. We leave this for future work.

\subsection{Simplex designs from complex-projective designs}
\label{ap:cp-to-simplex}

In this subsection, we will discuss the opposite direction to \cref{thm:cp-from-simplex-torus}; namely, that complex-projective $t$-designs give rise to simplex $t$-designs via the projection $\pi\colon \bbC\bbP^m \to \Delta^m$ defined by $\ket\psi \mapsto (\abs{\braket{0\vert \psi}}^2, \dots, \abs{\braket{m\vert \psi}}^2)$. Such a construction was also pointed out in \cite{kuperberg_numerical_2004,czartowski_isoentangled_2020}. This will be the first step in our proof of the nonnexistence of continuous-variable $(t\geq 2)$-designs. We will show that a continuous-variable design gives rise to an infinite dimensional analogue of a simplex design via a lemma analogous to \cref{lem:simplex-design-from-cp}, and then show that such infinite dimensional simplex designs do not exist for $t\geq 2$. Hence, it is useful to discuss the finite dimensional case first.

\begin{lemma}
\label{lem:simplex-design-from-cp}
Let $D$ be a $t$-design on $\bbC\bbP^m$. Then $\pi(D)$ is a $t$-design on $\Delta^m$.
\end{lemma}
\begin{proof}
Since $D$ is a design on $\bbC\bbP^m$, which satisfies by definition
\begin{salign}
    \EX{\ket\xi \in D}{\parentheses{\ket\xi\bra\xi}^{\otimes t}} &= \int_{\bbC\bbP^m} \parentheses{\ket\psi\bra\psi}^{\otimes t}\dd{\psi}\\
    &= \int_{\Delta^m \times T^m} \parentheses{\ket{p,\phi}\bra{p,\phi}}^{\otimes t}\dd{p}\dd{\phi},
\end{salign}
where the last line comes from \cref{eq:cp-simplex-torus}. Sandwiching this equation by $\bra{a_1}\dots\bra{a_t}$ and $\ket{a_1}\dots\ket{a_t}$, we find
\begin{equation}
    \EX{\ket\xi \in D}{\abs{\braket{a_1\vert\xi}}^2\dots \abs{\braket{a_t\vert\xi}}^2} = \int_{\Delta^m \times T^m} p_{a_1} \dots p_{a_t} \dd{p}\dd{\phi},
\end{equation}
and hence
\begin{equation}
    \EX{q \in \pi(D)}{q_{a_1}\dots q_{a_t}} = \int_{\Delta^n} p_{a_1} \dots p_{a_t} \dd{p}.
\end{equation}
Therefore, the ensemble $\pi(D)$ matches the integral over $\Delta^n$ for degree $t$ monomials, and thus by linearity matches for all polynomials of degree $t$ or less.
\end{proof}

In terms of weight functions, we can write this as follows. Let $\pi^{-1}$ denote the preimage of $\pi$. If $(D, w)$ is a $\bbC\bbP^m$ $t$-design, then $(\pi(D), u)$ is a $\Delta^m$ $t$-design, where
\begin{equation}
\begin{aligned}
    u \colon &\pi(D) \to \bbR_{>0}\\
    &p \mapsto \sum_{\psi \in \pi^{-1}(p)}w(\psi).
\end{aligned}
\end{equation}
\cref{lem:simplex-design-from-cp} tells us that
\begin{equation}
  \begin{aligned}
    \int_{\Delta^n} p_{a_1} \dots p_{a_t} \dd{p} &= \frac{1}{\Tr \Pi_t^{(d)}} \bra{a_1}\dots\bra{a_t} \Pi_t^{(d)} \ket{a_1}\dots\ket{a_t}\\
    &= \frac{1}{\Tr \Pi_t^{(d)}} \Lambda_t^{(d)}(a_1, \dots, a_t),
  \end{aligned}
\end{equation}
where recall that $\Lambda_t^{(d)}$ and $\Pi_t^{(d)}$ are defined in \cref{def:pi-lambda}. One can then define an infinite dimensional simplex design analogously to how we define continuous-variable designs in \cref{def:oscillator-t-design}. In particular, to get something well-defined in the infinite limit, we remove the $\Tr\Pi_t$ normalization, and we replace the $\Expval$ by an integral over an arbitrary measure space.

\begin{definition}[Infinite dimensional simplex $t$-design]
\label{def:inf-simplex-design}
Let $(X, \Sigma, \mu)$ be a measure space, and fix an integer $t\in \bbN$. Let $p=(p_i)_{i \in \bbN_0}$ be a sequence of measurable maps $p_i \colon X \to [0,1]$. If
\begin{equation}
    \sum_{i\in \bbN_0} p_i(x) = 1 ~~\mu\text{-a.e.~in }X,
\end{equation}
and
\begin{equation}
    \forall a \in \bbN_0^t\colon~ \int_X \prod_{j=1}^t p_{a_j}(x) \dd{\mu(x)} = \Lambda_t(a),
\end{equation}
then $[(X, \Sigma, \mu), p]$ is an \emph{infinite dimensional simplex $t$-design}.
\end{definition}

\medskip

In this definition, $\Lambda_t$ is defined in terms of $\Pi_t \colon \calH^{\otimes t} \to \calH^{\otimes t}$ given in \cref{def:pi-lambda}, and $\calH$ is a infinite dimensional separable Hilbert space, e.g.~$L^2(\bbR)$. In the next subsection, we will show in \cref{lem:oscillator-t} that without loss of generality, the measure space for an infinite dimensional simplex-design design can be taken to be $\sigma$-finite. Then in \cref{lem:no-oscillator-t}, we will show that no infinite simplex $t$-designs exist for any $t \geq 2$.

\begin{example}[Infinite dimensional simplex $1$-design]
    When $t = 1$, we have that $\Lambda_1(a) = 1$ for any $a \in \bbN_0$. We have many infinite dimensional simplex $1$-designs. For example, let $X = \bbN_0$, $\Sigma = \calP(X)$, and $\mu$ be the standard counting measure $\mu(A) = \abs{A}$. Finally, for $x \in X$, let $p_a(x) = \delta_{ax}$. Then
    \begin{equation}
        \int_X p_a(x) \dd{\mu(x)} = \sum_{x \in \bbN_0} \delta_{ax} = 1 = \Lambda_1(a),
    \end{equation}
    as desired.
\end{example}

\subsection{Constrained complex-projective integration}
\label{ap:constrained-integration}

We now briefly describe one consequence of the formalism developed so far. This subsection is essentially unrelated to the rest of the paper, but interesting nonetheless.
We will sketch the consequence with an example using the number operator, though we note that it can be generalized.

Define the number operator $\hat n$ by $\hat n \ket n = n \ket n$. Consider the constraint on $\ket\psi\in\bbC\bbP^{d-1}$ that $\bra\psi \hat n \ket \psi = \calN$ for some constant $\calN$. Since the constraint is diagonal in the chosen basis, it only acts on the simplex part of $\bbC\bbP^{d-1}$. In particular, while integration over $\bbC\bbP^{d-1}$ involves integration over the simplex $\Delta^{d-1}$, integration over $\bbC\bbP^{d-1}$ with the constraint that $\bra\psi \hat n \ket \psi = \calN$ involves integration over the simplex $\tilde\Delta^{d-2}$, where
\begin{equation}
    \tilde\Delta^{d-2} \coloneqq \set{p \in \Delta^{d-1} \mid \sum_{n=0}^{d-1} n p_n = \mathcal N}.
\end{equation}
Recall from the Krein–Milman theorem that any compact convex subset of Euclidean space is the convex hull of its extremal points.
The simplex $\Delta^{d-1}$ is the convex hull of its $d$ extremal points $(1,0,\dots, 0)$, $(0,1,\dots, 0)$, $\dots$, $(0,0,\dots, 1)$. The simplex $\tilde\Delta^{d-2}$ is also the convex hull of its $d-1$ extremal points, but its extremal points are more complicated and depend on $\calN$. In particular, we let $b^{(i)}$ denote the $i^{\rm th}$ extremal point of $\tilde \Delta^{d-2}$, so that $b^{(i)} = (b_0^{(i)}, \dots, b_{d-1}^{(i)})$. Then it is easy to check that the extremal points are
\begin{equation}
    b^{(i)}_j = \begin{cases}
    \parentheses{1-\frac{\calN}{i+1}}\delta_{j0} + \frac{\calN}{i+1}\delta_{j,i+1} &\text{if } i + 1 \geq \calN\\
    \parentheses{1-\frac{\calN - i -1}{d - i - 2}}\delta_{j,i+1} + \frac{\calN - i -1}{d - i - 2} \delta_{j,d-1} & \text{if } i + 1 \leq \calN,
    \end{cases}
\end{equation}
where $i \in \set{0,\dots, d-2}$ and $j \in \set{0,\dots, d-1}$.

It then follows, analogously to \cref{thm:cp-from-simplex-torus}, that a $t$-design on the constrained $\bbC\bbP^{d-1}$ space can be constructed from $t$-designs on $\tilde\Delta^{d-2}$ and $T^d$. Furthermore, the simplex $\tilde\Delta^{d-2}$ can be parameterized via baryocentric coordinates in terms of the standard simplex $\Delta^{d-2}$. In particular, a point in $\Delta^{d-2}$ defines a particular convex combination of the extremal points of $\tilde\Delta^{d-2}$, which gives a point in $\tilde\Delta^{d-2}$. Therefore, one can compute the integral $\int_{\bbC\bbP^{d-1}} \parentheses{\ket\psi\bra\psi}^{\otimes t} \delta(\calN - \bra\psi \hat n \ket\psi) \dd{\psi}$ up to proportionality by using simplex and torus designs. We note that such a construction does not work if the $\delta(\calN - \bra\psi \hat n \ket\psi)$ constraint is replaced with $\Theta(\calN - \bra\psi \hat n \ket\psi)$, where $\Theta$ is the Heaviside step function. This is for a slightly subtle reason. The $\delta$ constraint results in a measure on $\tilde\Delta^{d-2}$ that is, up to proportionality, the standard Lebesgue measure. On the other hand, the $\Theta$ constraint results in a more complicated measure, and indeed this measure mixes the contributions of the torus and the simplex in the integral. As such, the resulting integral is no longer over a simple product of a simplex and torus, but rather over a more complicated combination of the two.

The $\delta$ constraint that fixed $\bra\psi\hat n\ket\psi = \calN$ is interesting nonetheless. By using any of the simplex $1$- and $2$-designs from \cref{ap:simplex-designs} and any of the torus $1$- and $2$-designs from \cref{ap:torus-designs}, we can compute the following integrals, up to proportionality, in terms of the extremal points $b^{(i)}$:
\begin{align}
    &\int_{\bbC\bbP^{d-1}} \ket\psi\bra\psi \delta(\calN - \bra\psi\hat n \ket \psi) \dd{\psi} \propto \frac{1}{d-1}\sum_{k=0}^{d-1}\ket k \bra k \sum_{j=0}^{d-2} b_k^{(j)},  \\
    \begin{split}
        &\int_{\bbC\bbP^{d-1}} (\ket\psi\bra\psi)^{\otimes 2} \delta(\calN - \bra\psi\hat n \ket \psi) \dd{\psi} \propto
        \frac{1}{d(d-1)} \sum_{i,j =1}^{d-2} \parentheses{1+\delta_{ij}} \bigg[\\
        &\qquad\qquad \sum_{k_1,k_2 =1}^{d-1} b^{(i)}_{k_1}b^{(j)}_{k_2} \parentheses{\ket{k_1}\ket{k_2}\bra{k_1}\bra{k_2}+\ket{k_1}\ket{k_2}\bra{k_2}\bra{k_1}} + \sum_{k =1}^{d-1} b^{(i)}_k b^{(j)}_k \ket{k}\ket{k}\bra{k}\bra{k}\bigg].
    \end{split}
\end{align}
If we, for example, fix $\calN = 1$, then the result is
\begin{equation}
    \int_{\bbC\bbP^{d-1}} \ket\psi\bra\psi \delta(1 - \bra\psi\hat n \ket \psi) \dd{\psi} \propto
    \parentheses{1-\frac{H_{d-1}}{d-1}}\ket 0 \bra 0 + \frac{1}{d-1}\sum_{k=1}^{d-1} \frac{1}{k} \ket k \bra k,
\end{equation}
where $H_{d-1} = \sum_{k=1}^{d-1}1/k$ is the $(d-1)^{\rm th}$ harmonic number.

\section{Continuous-variable designs} \label{ap:CVdesign}

In extending the definition of complex-projective designs to the infinite dimensional case of continuous-variables, one encounters the issue that $\Tr \Pi_t$ is not finite. Hence, in accordance with the definition of continuous-variable designs given in \cite{blume-kohout_curious_2014}, we remove the trace in the denominator of $\Pi_t / \Tr\Pi_t$ and replace the equality with a proportionality. By a simple rescaling of the ensemble, the proportionality constant can be made arbitrary. Thus, we can in fact keep the equality. We are therefore tempted to define a continuous-variable $t$-design as an ensemble $\calE$ satisfying
\begin{equation}
    \text{``}~\EX{\calE}{\parentheses{\ket\psi\bra\psi}^{\otimes t}} = \Pi_t.~\text{''}
\end{equation}
However, since $\Tr \Pi_t$ is infinite, it follows that the ensemble $\calE$ must not be compact, making $\Expval_{\calE}$ ill-defined. We therefore replace the expectation value with an integral over an arbitrary measure space $(X, \Sigma, \mu)$. Here $X$ is a set, $\Sigma$ a $\sigma$-algebra on $X$, and $\mu\colon \Sigma \to \bbR_{\geq 0} \cup \{\infty\}$ is a measure on $X$. Finally, we arrive at the precise definition of a continuous-variable $t$-design on $L^2(\bbR)$.

\begin{definition}[Continuous-variable state $t$-design]
\label{def:aesthetic-cv-design}
Let $X \subset L^2(\bbR)$, $(X, \Sigma, \mu)$ be a measure space, and fix a positive integer $t \in \bbN$. Let $\Pi_t\colon L^2(\bbR)^{\otimes t}\to L^2(\bbR)^{\otimes t}$ be as in \cref{def:pi-lambda}. If
\begin{equation}
    \int_X \parentheses{\ket\psi\bra\psi}^{\otimes t} \dd{\mu(\psi)} = \Pi_t,
\end{equation}
where we use the weak (Pettis) integral,
then $(X, \Sigma, \mu)$ is a \emph{continuous-variable state $t$-design}. 
Hence, we require a design to satisfy
\begin{equation}
    \int_X \parentheses{\prod_{i=1}^t \braket{a_i \vert \psi} \braket{\psi \vert b_i}} \dd{\mu(\psi)} = \Pi_t(a_1,\dots,a_t; b_1,\dots, b_t)
\end{equation}
for all tuples $a,b \in \bbN_0^t$.
\end{definition}

The motivation for this definition of continuous-variable state designs is summarized in \cref{fig:cv-design-def}. An alternative characterization of continuous-variable state designs is given in \cref{ap:C-inf-designs}. If one is familiar with weighted complex-projective designs, as defined in e.g.~Ref.~\cite{scott_tight_2006}, then one can imagine that the measure $\mu$ is a Lebesgue-Stieltjes measure coming from a weight function. 
For the purposes of designs, the weak (Pettis) integral is more natural than the strong (Bochner) integral because we are generally interested in averaged functions of $\psi$. Ultimately, we will prove that continuous-variable $t$-designs do not exist for $t\geq 2$, which immediately implies the result for the case of the strong integral as well.

By parameterizing states in $L^2(\bbR)$ with polar coordinates, one can arrive at an equivalent definition of continuous-variable $t$-designs.

\begin{definition}[Continuous-variable state $t$-design]
\label{def:oscillator-t-design}
Let $X$ be an arbitrary set, $(X,\Sigma,\mu)$ be a measure space, and fix an integer $t \in \bbN$. Let $p = (p_i)_{i \in \bbN_0}$ and $\phi = (\phi_i)_{i \in \bbN_0}$ be sequences of measurable maps $p_i\colon X \to [0,1]$ and $\phi_i\colon X \to \bbR$ satisfying $\sum_{i \in \bbN_0} p_i(x) = 1$ for almost all $x \in X$. Define the state $\ket{p(x),\phi(x)} \in L^2(\bbR)$ by 
\begin{equation}
    \ket{p(x),\phi(x)} \coloneqq \sum_{n \in \bbN_0} \sqrt{p_n(x)}\e^{\i\phi_n(x)} \ket{n}.
\end{equation}
Let $\Pi_t\colon L^2(\bbR)^{\otimes t}\to L^2(\bbR)^{\otimes t}$ be as in \cref{def:pi-lambda}. If
\begin{equation}
    \int_X \parentheses{\ket{p(x),\phi(x)}\bra{p(x),\phi(x)}}^{\otimes t} \dd{\mu(x)} = \Pi_t,
\end{equation}
where we use the weak (Pettis) integral,
then $\brackets{(X,\Sigma,\mu), p, \phi}$ is a \emph{continuous-variable state $t$-design}. 
Hence, we require a design to satisfy
\begin{equation}
    \label{eq:oscillator-t-design}
    \int_X \parentheses{\prod_{i=1}^t \braket{a_i \vert p(x),\phi(x)} \braket{p(x),\phi(x) \vert b_i}} \dd{\mu(x)} = \Pi_t(a_1,\dots,a_t; b_1,\dots, b_t)
\end{equation}
for all tuples $a,b \in \bbN_0^t$.
\end{definition}

\cref{def:oscillator-t-design} will be a more operationally useful definition for our purposes, but we emphasize that \cref{def:aesthetic-cv-design,def:oscillator-t-design} are equivalent definitions, where the latter is simply a different parameterization of the former.

\begin{proposition}
\label{prop:equiv-def}
\cref{def:aesthetic-cv-design,def:oscillator-t-design} are equivalent definitions.
\end{proposition}
\begin{proof}
For any $\ket\psi$ coming from the first definition, we get the sequences $p$ and $\phi$ for the second definition as $p_n = \abs{\braket{n \vert \psi}}^2$ and $\phi_n = \arg \braket{n \vert \psi}$. One can then normalize each $p_i$ by $p_i(x) \to p_i(x) / \sum_i p_i(x)$, and then absorb a factor of $\parentheses{\sum_i p_i(x)}^t$ into the measure.

Conversely for any measure space and sequences $p$ and $\phi$ coming from the second definition, we get the measure space $(X \subset L^2(\bbR), \Sigma, \mu)$ for the first definition since the parameterization defines states in $L^2(\bbR)$.
\end{proof}

We include \cref{def:aesthetic-cv-design} since it closer matches the standard definition of a weighted complex-projective design. In light of \cref{prop:equiv-def}, henceforth we will use \cref{def:oscillator-t-design}. To become better acquainted with this definition, consider the following example of a continuous-variable state $1$-design.

\begin{example}[Continuous-variable state $1$-designs]
  \label{ex:oscillator-1-design}
  Consider the measure space where $X = \bbN_0$, $\Sigma$ is the power-set $\calP(X)$, and $\mu$ is the standard counting measure on $\bbN_0$. Let $p_n\colon x\mapsto \delta_{nx}$, and $\phi_n\colon x\mapsto 0$. Then
  \begin{equation}
      \int_X \ket{p(x), \phi(x)}\bra{p(x),\phi(x)} \dd{\mu(x)} = \sum_{x\in\bbN_0} \ket{n=x}\bra{n=x} = \Pi_1,
  \end{equation}
  where note that $\Pi_1 = \bbI$. Hence, this is an example of a continuous-variable state $1$-design.
  
  Similarly, consider $\bbR_{> 0}$ with the standard Borel $\sigma$-algebra and Lebesgue measure. Consider also $[0,2\pi)$ with the normalized Lebesgue measure. Let $(X,\Sigma,\mu)$ be the unique product measure space for $X = \bbR_{> 0}\times [0,2\pi)$. For an element $x \in X$, notate $x = (r,\theta)$, for $r \in \bbR_{> 0}$ and $\theta \in [0,2\pi)$. Let $p_n\colon (r,\theta) \mapsto e^{-r}r^n/n!$ and $\phi_n\colon (r,\theta)\mapsto \theta n$. Then
  \begin{salign}
    \int_{X}&\ket{p(x),\phi(x)}\bra{p(x),\phi(x)}\dd{\mu(x)}\nonumber\\&=\frac{1}{2\pi}\sum_{n,m\in\bbN_{0}}\ket{n}\bra{m}\int_{0}^{\infty}\dd{r}\int_{0}^{2\pi}\dd{\theta}~\e^{\i\theta(n-m)}\e^{-r}\frac{r^{n/2+m/2}}{\sqrt{n!m!}}\\&=\sum_{n\in\bbN_{0}}\ket{n}\bra{n}\int_{0}^{\infty}\dd{r}~\e^{-r}\frac{r^{n}}{n!}\\&=\sum_{n\in\bbN_{0}}\ket{n}\bra{n}=\Pi_{1},
\end{salign}
  giving another example of a $1$-design. This $1$-design is more commonly written as
  \begin{equation}
      \int \ket\alpha\bra\alpha \frac{\Dd{2}{\alpha}}{\pi} = \bbI,
  \end{equation}
  where $\Dd{2}{\alpha} = \dd{\Re\alpha}\dd{\Im\alpha}$ and $\ket\alpha$ is a coherent state. Namely, coherent states form an overcomplete frame.
\end{example}

We emphasize that this definition completely sidesteps the issue that one cannot define a finite Haar measure on $\U(L^2(\bbR))$ since it is not a compact group, and indeed not even a locally compact group \cite[sec.~5]{grigorchukAmenabilityErgodicProperties2015}. See \cref{fig:cv-design-def} for a visualization. The issue is sidestepped by never considering the integral over all states in $L^2(\bbR)$. Instead, we note that the integral over all states for a finite dimensional Hilbert space gives a finite dimensional $\Pi_t$, and we extend the definition of a design to the infinite dimensional case by extending $\Pi_t$ to the infinite dimensional space $L^2(\bbR)$. This is exactly the approach that was taken in \cite{blume-kohout_curious_2014,zhuang_scrambling_2019}. Alternatively, in \cref{ap:C-inf-designs}, we \textit{do} explicitly consider integration over the infinite dimensional space $\bbC^\infty \supset L^2(\bbR)$.

\subsection{Non-existence of continuous-variable state designs}
\label{ap:nonexistence}

It has been shown that no set of Gaussian states can form a continuous-variable $2$-design \cite{blume-kohout_curious_2014}. We extend this result to show that there do not exist continuous-variable $t$-designs for any $t > 1$. We emphasize that our proof in fact works for any separable Hilbert space $\calH$, not just $L^2(\bbR)$, since it only assumes the existence of a countable orthonormal basis.

\begin{theorem}
\label{thm:nonexistence}
No continuous-variable state or unitary $t$-designs exist for any integer $t \geq 2$.
\end{theorem}

\cref{thm:nonexistence} is an immediate consequence of \cref{lem:oscillator-t} and \cref{lem:no-oscillator-t} below. \cref{fig:proof-outline} provides an overview of the proof. We recommend reading this section first keeping in mind the specific example of $t=2$, where the explicit form of $\Pi_2$ and $\Lambda_2$ are given in \cref{ex:pi2}. After understanding this case, the extension to arbitrary $t \geq 2$ is straightforward.

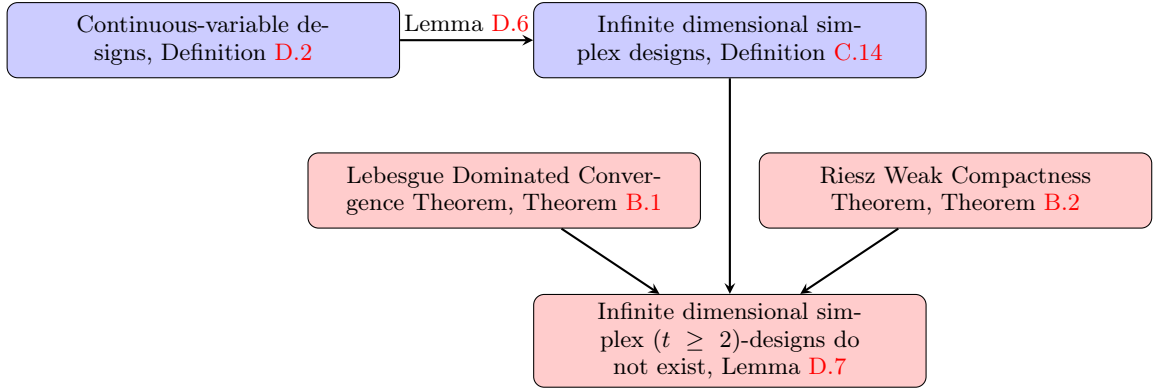
\begin{figure}
    \centering

    \tikzstyle{block} = [rectangle, rounded corners, minimum width=3cm, minimum height=1cm,text centered, text width=5cm, draw=black, fill=red!20]
    \tikzstyle{title} = [rectangle, rounded corners, minimum width=3cm, minimum height=1cm,text centered, text width=5cm, draw=black, fill=blue!20]
    \tikzstyle{arrow} = [thick,->,>=stealth]
    
    \begin{tikzpicture}[node distance=0cm]

        \node (simplex) [title] {Infinite dimensional simplex designs, \cref{def:inf-simplex-design}};
        
        \node (ldct) [block, below of=simplex, xshift=-3cm, yshift=-2cm] {Lebesgue Dominated Convergence Theorem, \cref{thm:dom-conv}};
        
        \node (rwct) [block, below of=simplex, xshift=3cm, yshift=-2cm] {Riesz Weak Compactness Theorem, \cref{thm:rwct}};
        
        \node (simplex-nonexist) [block, below of=simplex, yshift=-4cm] {Infinite dimensional simplex $(t\geq 2)$-designs do not exist, \cref{lem:no-oscillator-t}};
        
        \node (cv) [title, left of=simplex, xshift=-7cm] {Continuous-variable designs, \cref{def:oscillator-t-design}};

        \draw [arrow] (cv) -- node[anchor=south] {\cref{lem:oscillator-t}} (simplex);
        
        \draw [arrow] (simplex) -- (simplex-nonexist);
        
        \draw [arrow] (ldct) -- (simplex-nonexist);
        
        \draw [arrow] (rwct) -- (simplex-nonexist);

    \end{tikzpicture}

    \caption{An outline of the proof of the non-existence of continuous-variable $t$-designs for $t \geq 2$.}
    \label{fig:proof-outline}
\end{figure}

To begin, we will show that existence of continuous-variable state $t$-designs implies existence of simplex \(t\)-designs. 

\begin{lemma}
\label{lem:oscillator-t}
If a continuous-variable $t$-design exists, then there exists a $\sigma$-finite measure space $(X, \Sigma, \mu)$  and a sequence $p = (p_i)_{i=0}^\infty$ of measurable maps $p_i\colon X \to [0,1]$ satisfying
\begin{equation}
    \label{eq:cond1-t}
    \sum_{i=0}^\infty p_i(x) = 1 ~~\mu\text{-a.e. in }X,
\end{equation}
and
\begin{equation}
    \label{eq:cond2-t}
    \forall a \in \bbN_0^t \colon \int_X \prod_{i=1}^t p_{a_i}(x) \dd{\mu(x)} = \Lambda_t(a).
\end{equation}
\end{lemma}
\begin{proof}
Suppose a continuous-variable state $t$-design exists. Then \cref{eq:oscillator-t-design} holds for all tuples $a,b \in \bbN_0^t$, and $p$ satisfies \cref{eq:cond1-t} by \cref{def:oscillator-t-design}. Indeed, \cref{eq:cond1-t} is simply the requirement that the quantum states be normalized. Plugging in $a=b$ and $\Lambda_t(a) = \Pi_t(a;a)$ by definition, we get
\begin{salign}
    \Lambda_t(a) &= \int_X \parentheses{\prod_{i=1}^t \braket{a_i\vert p(x),\phi(x)} \braket{p(x),\phi(x) \vert a_i}} \dd{\mu(x)} \\
    &= \int_X \prod_{i=1}^t \abs{\braket{a_i \vert p(x),\phi(x)}}^2 \dd{\mu(x)} \\
    &= \int_X \prod_{i=1}^t p_{a_i}(x) \dd{\mu(x)}.
\end{salign}
Therefore, the measure space and sequence $p$ satisfy \cref{eq:cond1-t} and \cref{eq:cond2-t}. The only remaining thing to show is that $X$ can be $\sigma$-finite.

Consider the function $f = p_i^t$ whose codomain is clearly $[0,1]$. By \cref{eq:cond2-t}, $0 < \int_X f \dd{\mu} < \infty$. Hence, by \cref{lem:sigma-finite}, the preimage $f^{-1}((0,1]) = p_i^{-1}((0,1])$ is a $\sigma$-finite set. Since a countable union of $\sigma$-finite sets is $\sigma$-finite, it must be that $Y \coloneqq \bigcup_{i=0}^\infty p_i^{-1}((0,1])$ is $\sigma$-finite (also recall that any $\sigma$-finite set is measurable). The set $X \setminus Y$ is equal to $\bigcap_{i=0}^\infty p_i^{-1}(\set{0})$. \cref{eq:cond1-t} is required to hold almost everywhere in $X$. This means that the set of points for which it does not hold is contained within a measure zero subset of $X$. Clearly, \cref{eq:cond1-t} does not hold when $x \in X\setminus Y$. Therefore, $X \setminus Y$ is contained within a measure zero subset of $X$. Thus, if \cref{eq:cond2-t} holds on $X$, then it also holds on $Y$, and of course the same is true for \cref{eq:cond1-t}. 

Hence we have determined if \cref{eq:cond1-t,eq:cond2-t} are satisfied by the measure space $(X, \Sigma, \mu)$, then they are also satisfied by the measure space $(Y, \Sigma\rvert_Y, \mu\rvert_Y)$, where $\rvert_Y$ denotes the restriction to the subset $Y \subseteq X$. To see that $(Y, \Sigma\rvert_Y, \mu\rvert_Y)$ is a valid measure space, recall that we have already shown that $Y \in \Sigma$. Then one can straightforwardly check that $\Sigma\rvert_Y$ is a $\sigma$-algebra of $Y$ and $\mu\rvert_Y$ is a valid measure with respect to $\Sigma\rvert_Y$, so that the restriction of $(X,\Sigma,\mu)$ to $Y$ is a measure space (see e.g.~\cite[Ch.~17.1,~exercise 6]{royden_real_2010}). We have also shown that $(Y, \Sigma\rvert_Y, \mu\rvert_Y)$ is $\sigma$-finite.
In summary, we have shown that if a continuous-variable $t$-design exists, then there exists a measure space satisfying \cref{eq:cond1-t,eq:cond2-t}. We then showed that the existence of this measure space implies the existence of a $\sigma$-finite measure space satisfying \cref{eq:cond1-t,eq:cond2-t}, hence completing the proof.
\end{proof}

As we commented in \cref{lem:simplex-design-from-cp}, a complex-projective design gives rise to a simplex design. At a high level, \cref{lem:oscillator-t} is extending this fact to the continuous-variable regime. Similar to how we extended the definition of a complex-projective design to infinite dimensions, the analogous extension of a simplex design to infinite dimensions is the conditions in \cref{eq:cond1-t,eq:cond2-t}, as in \cref{def:inf-simplex-design}. The extra bit about $X$ being $\sigma$-finite is just a technical point needed so that \cref{thm:rwct} can be used in the next lemma.

Given \cref{lem:oscillator-t}, we immediately see that if no $\sigma$-finite measure space and sequence $p$ can satisfy \cref{eq:cond1-t,eq:cond2-t}, then no continuous-variable $t$-designs can exist. This is what we show in the following lemma.

\begin{lemma}
\label{lem:no-oscillator-t}
No $(X, \Sigma, \mu)$ and $(p_i)_{i=0}^\infty$ exist satisfying the conditions of \cref{lem:oscillator-t} for any $t \in \bbN_{\geq 2}$.
\end{lemma}
\begin{proof}
Assume by way of contradiction that such $(X,\Sigma,\mu)$ and $(p_i)$ exist. Because of \cref{eq:cond1-t}, it must be that for almost all $x \in X$, $\lim_{i\to\infty}p_i(x) = 0$. Since the sequence $(p_i)$ converges, it must be the case that every subsequence $(p_{i_k})_{k=0}^\infty$ of $(p_i)$ also converges to the same point; $\lim_{k\to\infty} p_{i_k}(x) = 0$ for almost all $x$. For any tuple $j \in \bbN_0^t$, define $g(x) = \prod_{l=1}^t p_{j_l}(x)$, which is in $L^1(X) = L^1(X,\Sigma,\mu)$ (i.e. $\int_X g \dd{\mu} < \infty$) by \cref{eq:cond2-t}. Consider the sequence $(f_{i_k})_{k=0}^\infty$ where $f_{i_k}(x) = p_{i_k}(x) g(x)$ for any $j \in \bbN_0^t$. Then $f_{i_k}$ converges pointwise to the zero function $f(x) = 0$ almost everywhere as $k\to\infty$, and $f$ is obviously measurable. Since the codomain of $p_{i_k}$ is $[0,1]$, it follows that $f_{i_k} \leq g$ for all $i_k$. Therefore, $(f_{i_k})$ is a sequence in $L^1(X)$ and is dominated by a nonnegative integrable $g$. Hence, we can apply the Dominated Convergence \cref{thm:dom-conv} to swap the limit and the integral and find that $\lim_{k\to\infty} \norm{f_{i_k}-f}_1 = 0$, giving
\begin{equation}
    \label{eq:f-t}
    \forall j \in \bbN_0^t\colon~\lim_{k\to\infty}\int_X p_{i_k}(x) \prod_{l=1}^t p_{j_l}(x) \dd{\mu(x)} = 0.
\end{equation}

Next we consider the sequence $(p_i)_{i=0}^\infty$, which is a bounded sequence in $L^t(X)$ since $\int_X p_i^t \dd{\mu} < \infty$ by \cref{eq:cond2-t}. Therefore, we can apply the Riesz Weak Compactness Theorem from \cref{thm:rwct} to find a subseqence $(p_{i_k})_{k=0}^\infty$ and a function $q$ for which for all $h \in L^{t'}(X)$,
\begin{equation}
    \label{eq:q-t}
    \lim_{k\to\infty} \int_X p_{i_k}(x)h(x) \dd{\mu(x)} = \int_X q(x)h(x) \dd{\mu(x)},
\end{equation}
where $t' = t/(t-1)$. Now we must prove that $q$ is zero almost everywhere.

First we show that $q$ must be nonnegative almost everywhere. Heuristically, this is because \(q\) is being substituted for a limit of probabilities, which themselves are always nonnegative. More technically, let $\ind_A$ be the indicator function, so that $\ind_A(x)$ is $1$ if $x \in A$, and zero otherwise. Since $X$ is $\sigma$-finite, there exists a sequence $(A_j)$ where $A_j$ is measurable $A_j \in \Sigma$, $\mu(A_j) < \infty$, and $X = \bigcup_{j=0}^\infty A_j$. Since $A_j$ has finite measure, $\ind_{A_j} \in L^{t'}(X)$. Plugging $h=\ind_{A_j}$ into \cref{eq:q-t}, we find that $\int_X q(x) \ind_{A_j}(x) \dd{\mu(x)} \geq 0$ for all $j$. Therefore, $\int_{A}q \dd{\mu} \geq 0$ for every $A \in \Sigma$ of finite measure, and we can build up $X$ from such $A$'s. This tells us that $q \geq 0$ almost everywhere.

Next we show that $q$ must be the zero function almost everywhere. For some $j \in \bbN_0^t$, we plug $h(x) = \prod_{l=1}^t p_{j_l}(x) \in L^{t'}(X)$ into \cref{eq:q-t}. Using \cref{eq:f-t} for the left hand side of \cref{eq:q-t}, this tells us that
\begin{equation}
    \forall j \in \bbN_0^t\colon~ \int_X q(x)\prod_{l=1}^t p_{j_l}(x)\dd{\mu(x)} = 0.
\end{equation}
Along with the fact that $q$ must be nonnegative almost everywhere, this implies that $q(x)$ must be zero almost everywhere whenever $p_j(x) \neq 0$ for any $j$. As such, $q$ must be zero almost everywhere on the set $\bigcup_{j=0}^\infty p_j^{-1}((0,1])$. But we showed in the proof of \cref{lem:oscillator-t} that $X \setminus \bigcup_{j=0}^\infty p_j^{-1}((0,1])$ is contained within a measure zero subset of $X$.

We have shown that $q$ is the zero function almost everywhere on $X$. Hence, \cref{eq:q-t} becomes
\begin{equation}
    \label{eq:h-t}
    \forall h \in L^{t'}(X)\colon~ \lim_{k\to\infty} \int_X p_{i_k}(x) h(x) \dd{\mu(x)} = 0.
\end{equation}
Plugging $h = p_0^{t-1}$ (which is in $L^{t'}(X)$ by \cref{eq:cond2-t}) into \cref{eq:h-t}, we arrive at $\lim_{k\to\infty} \int_X p_0^{t-1} p_{i_k}\dd{\mu} = 0$. But \cref{eq:cond2-t} tells us that $\lim_{k\to\infty} \int_X p_0^{t-1} p_{i_k}\dd{\mu} = \lim_{k\to\infty} \Lambda_t(0,\dots,0,i_k)$, which, from the definition of $\Lambda_t$ in terms of $\Pi_t$, is strictly positive as shown in \cref{eq:lambdat_limit}. We've reached a contradiction, hence completing the proof.
\end{proof}

The non-existence of state designs statement of \cref{thm:nonexistence} follows as an immediate corollary of \cref{lem:oscillator-t,lem:no-oscillator-t}. The non-existence of unitary designs follows straightforwardly from the non-existence of state designs, as we explain in the next subsection. Furthermore, \cref{thm:nonexistence} still holds even in the case when one allows $(X,\Sigma,\mu)$ to be a \textit{signed} measure space by a simple appeal to the Hahn Decomposition Theorem \cite{benedettoIntegrationModernAnalysis2009}. Indeed, using this theorem, one simply splits the signed measure space into two nonnegative measure spaces and then proceeds with the proof of \cref{thm:nonexistence}.

\subsection{Non-existence of continuous-variable unitary designs}
\label{ap:nonexistence-unitary}

\cref{thm:nonexistence} extends the results from \cite{blume-kohout_curious_2014}, where it is shown that the set of Gaussian states does not form a state $2$-design, and the results from \cite{zhuang_scrambling_2019}, where it is shown that the set of Gaussian unitaries does not form a unitary $2$-design. The non-existence of continuous-variable state $t$-designs for $t > 1$ immediately implies the non-existence of continuous-variable unitary $t$-designs for $t>1$, since any unitary design gives rise to a state design by \emph{twirling} a fiducial state. To be clear, we consider the definition of a continuous-variable unitary $2$-design given in \cite{zhuang_scrambling_2019}. Namely, a unitary $2$-design is any ensemble $\calE$ of unitaries satisfying
\begin{equation}
    \EX{\calE}{(U \otimes U) A (U \otimes U)^\dagger} \propto \frac{1}{2}\parentheses{\bbI \Tr\bargs{A} + S \Tr\bargs{S A}}
\end{equation}
for any operator trace-class operator $A$, where $S$ is the SWAP operator that swaps the elements of the tensor product space. Since this should hold for any $A$, we can substitute $A = \parentheses{\ket{\phi}\bra{\phi}}^{\otimes 2}$ for any fiducial state $\ket \phi$ (e.g. the zero Fock state $\ket 0$). We can then define a new ensemble over states $\calE' = \set{U \ket{\phi} \mid U \in \cal E}$. The result is
\begin{equation}
    \EX{\calE'}{\parentheses{\ket{\psi}\bra{\psi}}^{\otimes 2}} \propto \frac{1}{2}\parentheses{\bbI + S} = \Pi_2,
\end{equation}
which precisely matches the definition of a continuous-variable state $2$-design given in \cite{blume-kohout_curious_2014}, and the definition we use in \cref{def:oscillator-t-design}. Hence, by contraposition, if a state design does not exist, then a unitary design necessarily does not exist. 

This result holds generally for the definition of continuous-variable unitary $t$-designs given in \cite[footnote~89]{zhuang_scrambling_2019}. Specifically, a unitary $t$-design is an ensemble $\calE$ satisfying
\begin{equation}
    \EX{\calE}{U^{\otimes t} A U^{\dag~\otimes t}} \propto \frac{1}{\abs{S_t}}\sum_{\sigma \in S_t} W_\sigma \Tr(W_{\sigma}^{-1}A)
\end{equation}
for any trace-class operator $A$. Substituting $A = \parentheses{\ket\phi\bra\phi}^{\otimes t}$ for some fiducial state $\ket\phi$. Then define a new ensemble over states $\calE' = \set{U\ket\phi \mid U \in \calE}$. The result is then
\begin{equation}
    \EX{\calE'}{\parentheses{\ket\psi\bra\psi}^{\otimes t}} \propto \frac{1}{\abs{S_t}}\sum_{\sigma \in S_t} W_\sigma = \Pi_t,
\end{equation}
meaning that $\calE'$ is a continuous-variable state $t$-design according to \cref{def:oscillator-t-design}. By the nonnexistence of continuous-variable $(t\geq 2)$-designs then, such a unitary design does not exist for $t \geq 2$.

\subsection{Rigged continuous-variable state designs}
\label{ap:rigged}

The result of \cref{thm:nonexistence} is that no continuous-variable $(t\geq 2)$-designs exist. The main hindrance to the construction of continuous-variable state $(t\geq 2)$-designs is the requirement that the states be normalized. In particular, the proof did not rely on exactly what the states were normalized to, only that $\lim_{i\to\infty} p_i = 0$. Hence, the requirement that the states belong to $L^2(\bbR)$ inhibits the existence of continuous-variable designs. This motivates the approach taken in this section, where we construct \emph{rigged continuous-variable state designs} by relaxing the normalization condition, thus allowing unphysical states such as the infinite superposition state $\sum_{n\in \bbN_0} \ket{n}$. Specifically, the we will use elements of the standard rigged Hilbert space on top of $L^2(\bbR)$ to reconstruct $\Pi_t$. These elements are called \emph{tempered distributions} (see below); some familiar tempered distributions are the position eigenstates $\ket x$ and the momentum eigenstates $\ket p$. In the next section, we will reintroduce normalization via a soft energy cutoff, hence making the rigged continuous-variable designs a type of approximate continuous-variable design.

When we remove the normalization condition on the states, many of the finite dimensional complex-projective designs still do not naturally extend to the continuous-variable regime. For example, consider the complex-projective design given in \cref{eq:unweighted-cp-2-design}. We begin by making the states $\ket{m,q_1,q_2}$ non-normalizable by multiplying through by $\sqrt{n+1}$, and then take the $n\to\infty$ limit. In this limit, $r = 1/\sqrt{n+2} \to 0$, and hence $\sqrt{n+1}\ket{m,q_1,q_2} = \sqrt{n+1}\ket{m',q_1,q_2}$ for all $m$ and $m'$. One can straightforwardly check that these states do not reconstruct $\Pi_2$. In particular, these states only form a $1$-design as the underlying simplex design is only a $1$-design.

However, some finite dimensional complex-projective designs do extend to the continuous-variable regime when normalization is removed. For example, consider the following $\bbC\bbP^{d-1}$ $2$-design, when $d$ is prime, given in \cref{eq:mub}, which also happens to be a maximal set of mutually unbiased bases \cite{wootters_optimal_1989,klappenecker2005mutually}. Define the state $\ket{q_1}_{q_2} \equiv \prescript{}{q_2}{\ket{q_1}} \coloneqq \frac{1}{\sqrt{d}}\sum_{n = 0}^{d-1} \exp\bargs{\frac{2\pi \i}{d} \parentheses{q_1 n + q_2 n^2}} \ket{n}$. One can straightforwardly show that for each $q_2$, $\set{\ket{q_1}_{q_2} \mid q_1 \in \set{0, \dots, d-1}}$ is an orthonormal basis, and that
\begin{equation}
    \Pi_2 = \frac{1}{2} \sum_{n=0}^{d-1}\parentheses{\ket n \bra n}^{\otimes 2} + \frac{1}{2}\sum_{q_1, q_2 =0}^{d-1} \parentheses{\prescript{}{q_2}{\ket{q_1}}\!\bra{q_1}_{q_2}}^{\otimes 2} .
\end{equation}
The phases involved in this design utilize the so-called ``$ax^2 + bx$ construction'' described in \cite{van_dam_mutually_2011}. In the language of our paper, the ``$ax^2 + bx$ construction'' is alluding to a particular torus $2$-design construction, namely given in \cref{thm:torus-two-design}.

This design cannot be extended to a continuous-variable design because the states $\ket{q_1}_{q_2}$ are unphysical when $d$ is infinite. If we relax the normalization condition, however, we can reconstruct the infinite dimensional symmetric projector with an analogous design. 

\begin{theorem}
\label{thm:kerred-phase-design}
Define the non-normalizable state
\begin{equation}
\ket{\theta}_{\varphi} \equiv \prescript{}{\varphi}{\ket\theta} \coloneqq \frac{1}{\sqrt{2\pi}}\sum_{n\in \bbN_0} \exp\bargs{\i\parentheses{\theta n + \varphi n^2}} \ket{n}.
\end{equation}
Then
\begin{equation}
    \label{eq:nonnormalizable-theta-design}
    \Pi_2 = \frac{1}{2} \sum_{n\in \bbN_0} \parentheses{\ket{n}\bra{n}}^{\otimes 2} + \frac{1}{2} \int_{-\pi}^{\pi}\dd{\varphi}\int_{-\pi}^{\pi}\dd{\theta}\parentheses{\prescript{}{\varphi}{\ket{\theta}}\!\bra{\theta}_\varphi}^{\otimes 2}.
\end{equation}
\end{theorem}
\begin{proof}
\begin{salign}
    \int_{-\pi}^\pi \dd\varphi \int_{-\pi}^\pi \dd{\theta} &\braket{a\vert \theta}_\varphi \braket{b\vert \theta}_\varphi \prescript{}{\varphi}{\braket{\theta \vert c}}\prescript{}{\varphi}{\braket{\theta \vert d}}\\
    &= \frac{1}{(2\pi)^2}\int_{-\pi}^\pi \dd\varphi \int_{-\pi}^\pi \dd{\theta} \e^{\i \theta (a+b-c-d)} \e^{\i \varphi (a^2+b^2-c^2-d^2)}\\
    &= \begin{cases}
    1&\text{if } a+b=c+d \text{ and } a^2+b^2 = c^2 + d^2\\
    0&\text{otherwise}
    \end{cases}\\
    &= \begin{cases}
    1&\text{if } a=c\neq b = d \text{ or } a=d \neq b = c \text{ or } a=b=c=d\\
    0&\text{otherwise}
    \end{cases} \qquad \text{from \cref{lem:diophantine}}\\
    &= \delta_{ac}\delta_{bd} + \delta_{ad}\delta_{bc} - \delta_{ab}\delta_{ac}\delta_{ad}\\
    &= 2\Pi_2(a,b;c,d) - \delta_{ab}\delta_{ac}\delta_{ad}\\
    &= 2\Pi_2(a,b;c,d) - \sum_{n\in\bbN_0} \braket{a\vert n}\braket{b\vert n}\braket{n \vert c}\braket{n\vert d}. 
\end{salign}
\end{proof}

Strictly speaking, \cref{eq:nonnormalizable-theta-design} should say $\Pi_2\big\rvert_{S(\bbR)}$, since the $\bra{\theta}_\varphi$ are only defined on Schwartz space $S(\bbR) \subset L^2(\bbR)$. However, its action can be uniquely extended to all of $L^2(\bbR)$. This will be formalized later on in this section.

We can find analogous results using $\cos$ and $\sin$ states from \cite{carruthers_phase_1968}. We note that looking at the $\ket{\theta}_\varphi$ non-normalizable states was motivated by lifting the finite dimensional complex-projective design to the continuous-variable case. These states are Kerred phase states, which form a continuous projection-valued measure (PVM) \cite{carruthers_phase_1968}. From these states, we are then motivated to define the $\ket{\cos\theta}_\varphi$ and $\ket{\sin\theta}_\varphi$ states, which are defined in \cite{carruthers_phase_1968}, since they are similar to the Kerred phase states, but nicer in many ways. In particular, for each $\varphi$, they form an generalized orthogonal basis.

\begin{theorem}
\label{thm:cos-design}
Define the non-normalizable state 
\begin{equation}
\ket{\cos\theta}_\varphi \equiv \prescript{}{\varphi}{\ket{\cos\theta}} \coloneqq \sqrt{\frac{2}{\pi}}\sum_{n \in \bbN_0} \e^{\i \varphi n^2}\sin((n+1)\theta) \ket n.
\end{equation}
Then
\begin{equation}
    \Pi_2 = \frac{1}{4} \sum_{n\in \bbN_0} \parentheses{\ket{n}\bra{n}}^{\otimes 2} + \frac{1}{4} \int_{-\pi}^{\pi}\dd{\varphi} \int_0^{\pi}\dd{\theta} \parentheses{\prescript{}{\varphi}{\ket{\cos\theta}}\!\bra{\cos\theta}_\varphi}^{\otimes 2}
\end{equation}
\end{theorem}
\begin{proof}
\begin{salign}
    \int_{-\pi}^\pi \dd{\varphi}\int_0^\pi \dd{\theta} & \braket{a\vert \cos\theta}_\varphi \braket{b\vert \cos\theta}_\varphi \prescript{}{\varphi}{\braket{\cos\theta \vert c}}\prescript{}{\varphi}{\braket{\cos\theta \vert d}}\\
    &= \frac{4}{\pi^2}\int_{-\pi}^\pi \dd{\varphi}\int_0^\pi \dd{\theta} ~\e^{\i \varphi (a^2 + b^2 - c^2 - d^2)} \sin((a+1)\theta)\sin((b+1)\theta)\sin((c+1)\theta)\sin((d+1)\theta)\\
    &= \frac{8}{\pi} \delta_{a^2+b^2, c^2+d^2} \int_0^\pi \dd{\theta} \sin((a+1)\theta)\sin((b+1)\theta)\sin((c+1)\theta)\sin((d+1)\theta)\\
    \begin{split}
    &= \frac{1}{\pi}\delta_{a^2+b^2, c^2+d^2} \int_0^\pi \dd{\theta} \big[ \cos (\theta(a +b +c +d +4 ))-\cos (\theta(-a  +b  +c  +d +2 ))\\
    &\qquad -\cos (\theta(a -b +c +d +2 )) -\cos (\theta(a +b -c +d +2) )+\cos (\theta(a -b -c +d) )\\
    &\qquad -\cos (\theta (a +b +c -d +2) )+\cos (\theta(a -b +c -d) )+\cos (\theta(a +b -c -d ) ) \big]
    \end{split}\\
    \begin{split}
    &= \delta_{a^2+b^2, c^2+d^2}\big[\sinc (a +b +c +d +4 )-\sinc (-a  +b  +c  +d +2 ) -\sinc (a -b +c +d +2 )\\
    &\qquad -\sinc (a +b -c +d +2)+\sinc (a -b -c +d) - \sinc (a +b +c -d +2) \\
    &\qquad +\sinc (a -b +c -d )+\sinc(a +b -c -d )  \big]
    \end{split}\\
    \begin{split}
    &= \delta_{a^2+b^2, c^2+d^2}\big[-\delta_{a,b  +c  +d +2} -\delta_{b,a +c +d +2}\\
    &\qquad -\delta_{c,a +b +d +2} - \delta_{d,a +b +c +2} +\delta_{a+d, b+c} +\delta_{a +c, b+d}+\delta_{a +b, c +d}  \big],
    \end{split}
\end{salign}
where we used that $\int_0^\pi \cos(x\theta) \dd{\theta} = \pi \sinc x$, and $\sinc x = \frac{\sin(\pi x)}{\pi x}$ when $x \neq 0$, and $1$ when $x=0$. One can easily verify that there are no integer solutions to the Diophantine system $a^2+b^2=c^2+d^2$ and $a = b+c+d+2$. Thus,
\begin{equation}
    \int_{-\pi}^\pi \dd{\varphi}\int_0^\pi \dd{\theta} \braket{a\vert \cos\theta}_\varphi \braket{b\vert \cos\theta}_\varphi \prescript{}{\varphi}{\braket{\cos\theta \vert c}}\prescript{}{\varphi}{\braket{\cos\theta \vert d}} = \delta_{a^2+b^2, c^2+d^2}\parentheses{\delta_{a+d, b+c} +\delta_{a +c, b+d}+\delta_{a +b, c +d}}.
\end{equation}
We now focus on the three terms individually. The third term is solved in \cref{lem:diophantine} a $\delta_{a^2+b^2, c^2+d^2}\delta_{a +b, c +d} = \delta_{ac}\delta_{bd} + \delta_{ad}\delta_{bc} - \delta_{ab}\delta_{ac}\delta_{ad}$. The first term is only nonzero when $a,b,c,d$ solve
\begin{equation}
    a^2+b^2 = c^2 + d^2, \quad \text{and } ~~a+d=b+c.
\end{equation}
Plugging $a = b+c-d$ in, we find $(b+c-d)^2 + b^2 = c^2 + d^2$, or equivalently $(b+c)(b-d) = 0$. Therefore, the first term is only nonzero when $b=d$ and $a=c$. A similar analysis holds for the second term, where we find that it only nonzero when $b=c$ and $a=d$. Hence, we find that
\begin{salign}
    \int_{-\pi}^\pi \dd{\varphi}\int_0^\pi \dd{\theta} &\braket{a\vert \cos\theta}_\varphi \braket{b\vert \cos\theta}_\varphi \prescript{}{\varphi}{\braket{\cos\theta \vert c}}\prescript{}{\varphi}{\braket{\cos\theta \vert d}}\\
    &=  \delta_{ac}\delta_{bd} + \delta_{ad}\delta_{bc} + (\delta_{ac}\delta_{bd} + \delta_{ad}\delta_{bc} - \delta_{ab}\delta_{ac}\delta_{ad})\\
    &= 2(\delta_{ac}\delta_{bd} + \delta_{ad}\delta_{bc}) - \delta_{ab}\delta_{ac}\delta_{ad}\\
    &= 4\Pi_2(a,b;c,d) - \sum_{n\in\bbN_0} \braket{a\vert n}\braket{b\vert n}\braket{n \vert c}\braket{n\vert d},
\end{salign}
proving the result.
\end{proof}

\begin{theorem}
\label{thm:sin-design}
Define the non-normalizable state 
\begin{equation}
\ket{\sin\theta}_\varphi \equiv \prescript{}{\varphi}{\ket{\sin\theta}} \coloneqq \frac{1}{\sqrt{2\pi}}\sum_{n \in \bbN_0} \e^{\i \varphi n^2} \parentheses{\e^{\i (n+1)\theta} - \e^{-\i (n+1)(\theta-\pi)}} \ket n.
\end{equation}
Then
\begin{equation}
    \Pi_2 = \frac{1}{4} \sum_{n\in \bbN_0} \parentheses{\ket{n}\bra{n}}^{\otimes 2} + \frac{1}{4} \int_{-\pi}^{\pi}\dd{\varphi} \int_{-\pi/2}^{\pi/2}\dd{\theta} \parentheses{\prescript{}{\varphi}{\ket{\sin\theta}}\!\bra{\sin\theta}_\varphi}^{\otimes 2}
\end{equation}
\end{theorem}
\begin{proof}
\begin{salign}
    \int_{-\pi}^\pi \dd{\varphi}\int_{-\pi/2}^{\pi/2} \dd{\theta} & \braket{a\vert \sin\theta}_\varphi \braket{b\vert \sin\theta}_\varphi \prescript{}{\varphi}{\braket{\sin\theta \vert c}}\prescript{}{\varphi}{\braket{\sin\theta \vert d}}\\
    \begin{split}
    &= \frac{1}{(2\pi)^2}\int_{-\pi}^\pi \dd{\varphi}\int_{-\pi/2}^{\pi/2} \dd{\theta} ~\e^{\i \varphi (a^2 + b^2 - c^2 - d^2)} \parentheses{\e^{\i(a+1)\theta} - \e^{-\i(a+1)(\theta-\pi)}}\parentheses{\e^{\i(b+1)\theta} - \e^{-\i(b+1)(\theta-\pi)}}\\
    &\qquad \times \parentheses{\e^{-\i(c+1)\theta} - \e^{\i(c+1)(\theta-\pi)}}\parentheses{\e^{-\i(d+1)\theta} - \e^{\i(d+1)(\theta-\pi)}}
    \end{split}\\
    \begin{split}
    &= \frac{1}{2} \delta_{a^2+b^2, c^2+d^2} \big[\left((-1)^{a+b-c-d}+1\right) \sinc\left( (a+b-c-d)/2\right) \\
    &\qquad + (-1)^{b-c} \left((-1)^{a-b+c-d}+1\right) \sinc\left((a-b+c-d)/2\right)\\
       &\qquad +  (-1)^{a-c} \left((-1)^{a-b-c+d}+1\right) \sinc\left((a-b-c+d)/2\right)\\
       &\qquad + \text{terms that are always zero when } a^2+b^2=c^2+d^2
    \big]
    \end{split}\\
    &= \delta_{a^2+b^2, c^2+d^2} \parentheses{\delta_{a+b,c+d} + (-1)^{b-c}\delta_{a+c,b+d} + (-1)^{a-c}\delta_{a+d,b+c}}.
\end{salign}
From here we continue exactly as in \cref{thm:cos-design}, and find that 
\begin{equation}
    \int_{-\pi}^\pi \dd{\varphi}\int_{-\pi/2}^{\pi/2} \dd{\theta} \braket{a\vert \sin\theta}_\varphi \braket{b\vert \sin\theta}_\varphi \prescript{}{\varphi}{\braket{\sin\theta \vert c}}\prescript{}{\varphi}{\braket{\sin\theta \vert d}} = 4\Pi_2(a,b;c,d) - \sum_{n\in\bbN_0} \braket{a\vert n}\braket{b\vert n}\braket{n \vert c}\braket{n\vert d}
\end{equation}
completing the proof.
\end{proof}

We now state some facts about the non-normalizable states used above, the proof of which can be found in \cite[sec.~6]{carruthers_phase_1968}. Our definition of these states differs from the definition in \cite{carruthers_phase_1968} in that we have added an additional phase factor $\e^{\i \varphi \hat n^2}$, but this does not affect any of the following facts. For the following, we will use $\delta$ to denote the Dirac delta function on the interval $[-\pi,\pi]$, namely $\delta(\theta) \equiv \delta_{[-\pi,\pi]}(\theta) = \frac{1}{2\pi}\sum_{j \in \bbZ} \e^{\i \theta j}$.
\begin{salign}
    &\prescript{}{\varphi}{\braket{\cos\theta \vert \cos\theta'}}_\varphi = \delta(\theta-\theta')\\
    &\prescript{}{\varphi}{\braket{\sin\theta \vert \sin\theta'}}_\varphi = \delta(\theta-\theta')\\
    &\int_0^\pi \prescript{}{\varphi}{\ket{\cos\theta}}\!\bra{\cos\theta}_\varphi \dd{\theta} = \bbI\\
    &\int_0^\pi \prescript{}{\varphi}{\ket{\sin\theta}}\!\bra{\sin\theta}_\varphi \dd{\theta} = \bbI\\
    &\int_0^\pi \prescript{}{\varphi}{\ket{\theta}}\!\bra{\theta}_\varphi \dd{\theta} = \bbI.
\end{salign}
From these, we see that the $\cos$ and $\sin$ states form a generalized orthogonal basis for each $\varphi$, and the $\cos$, $\sin$, and $\theta$ states form a continuous PVM for each $\varphi$. More generally, one can consider $\gamma$-rotated sine/cosine states
\begin{equation}
    \ket{\theta}_{\varphi,\gamma} \coloneqq \frac{1}{\sqrt{8}} \sum_{n=0}^\infty \e^{\i\varphi n^2} \parentheses{\e^{\i (n+1)\theta} - \e^{-\i(n+1)(\theta-\gamma)}} \ket n,
\end{equation}
where $\ket{\theta}_{\varphi,0} = \ket{\cos\theta}_\varphi$ and $\ket{\theta}_{\varphi,\pi} = \ket{\sin\theta}_\varphi$.
Similar to above, for any fixed $\gamma$, summing over Fock states and integrating the $\ket{\theta}_{\varphi,\gamma}$ states over $\theta$ and $\varphi$ yields a rigged $2$-design.

Next, we restrict our attention to the $\ket{\cos\theta}_\varphi$, since a similar analysis holds for $\ket{\sin\theta}_\varphi$ and $\ket{\theta}_\varphi$. Let $\hat a$ and $\hat a^\dag$ be the standard annihilation and creation operators so that $\hat n = \hat a^\dag \hat a$.
The elements of $\{\ket{\cos\theta}_\varphi \mid \theta \in (0,\pi)\}$ are generalized eigenvectors -- or more precisely tempered distributions -- of the operator $\widehat{\cos(\theta)}_\varphi \coloneqq \frac{1}{2}\hat a^\dag (\hat n + 1)^{-1/2} \e^{\i \varphi (2\hat{n}+1)} + \text{h.c.}$, where h.c.~denotes the Hermitian conjugate of the first term. This can be concisely expressed in terms of the Susskind-Glogower phase operator \cite{susskindQuantumMechanicalPhase1964,Bergou1991} $\widehat{\e^{\i\theta}} \coloneqq \sum_{n\in\bbN_0} \ket{n+1}\bra{n}$, yielding $\widehat{\cos(\theta)}_\varphi = \frac{1}{2}\widehat{\e^{\i\theta }}\e^{\i\varphi (2\hat{n}+1)} + \text{h.c.}$.

The standard position state $\bra{x}$ is to be understood as a distribution in that it is a continuous linear functional on a subset of $L^2(\bbR)$ that will be described below. It is defined via the relation $\braket{x\vert \psi} \coloneqq \psi(x) = \int_\bbR \psi(x')\delta(x'-x)\dd{x'}$. Similarly, the momentum states $\bra{p}$ is understood as a distribution defined by $\braket{p \vert \psi} \coloneqq \frac{1}{\sqrt{2\pi}}\int_\bbR \e^{-\i p x} \psi(x) \dd{x}$. 

The $\prescript{}{\varphi}{\bra{\cos\theta}}$ can be understood analogously; namely, $\prescript{}{\varphi}{\bra{\cos\theta}}$ is defined by 
\begin{equation}
    \prescript{}{\varphi}{\braket{\cos\theta \vert \psi}} \coloneqq \sqrt{\frac{2}{\pi}}\sum_{n\in\bbN_0} \sin((n+1)\theta)\e^{-\i \varphi n^2} \braket{n \vert \psi},
\end{equation}
where $\braket{n\vert\psi}$ is the standard inner product $\int_\bbR \bar\psi_n(x) \psi(x) \dd{x}$ with $\psi_n(x) = \braket{x \vert n}$ the Fock state wavefunctions. 

We now formalize this intuitive understanding of the $\bra{\cos\theta}_\varphi$ states as distributions. References for this discussion are \cite{rudin_functional_1991} for a formal treatment, and \cite{gieres_mathematical_2000,madrid_role_2005} for a broad overview. We have found that the combination of Fock states with $\ket{\cos\theta}_\varphi$, $\ket{\sin\theta}_\varphi$, or $\ket{\theta}_\varphi$ distributions is enough to reconstruct $\Pi_2$. We call such designs \emph{rigged} designs, since the latter states live in the rigged Hilbert space on top of $L^2(\bbR)$.

The standard rigged Hilbert space of the harmonic oscillator is the Gelfand triple $S(\bbR) \subset L^2(\bbR) \subset S(\bbR)'$. $S(\bbR)$ is called \emph{Schwartz space}, and as a topological vector space it has a continuous dual space. $S(\bbR)'$ is called the space of \emph{tempered distributions}, and is the continuous dual space of $S(\bbR)$. For physical quantum states, we often desire that they have finite position, momentum, and energy moments. The first part of the Gelfand triple is the set of all such states, which for the harmonic oscillator is hence
\begin{equation}
S(\bbR) = \bigcap_{\alpha,\beta \in \bbN_0} \calD(\hat x^\alpha \hat p^\beta),
\end{equation}
where $\calD(M)$ denotes the maximal domain of the operator $M$. Since the domains of $\hat x$ and $\hat p$ are dense in $L^2(\bbR)$, it follows that $S(\bbR)$ is dense in $L^2(\bbR)$. 

$S(\bbR)$ is a Fr\'echet space, meaning that it is a topological vector space with a topology induced by a countable family of seminorms $\norm{\cdot}_{\alpha,\beta}$ defined by
\begin{equation}
    \norm{f}_{\alpha,\beta}\coloneqq\sup_{x\in\bbR}\abs{x^{\alpha}~\frac{\dd^{\beta}f}{\dd x^{\beta}}}~.
\end{equation}
An equivalent condition for a function $f$ to belong to $S(\bbR)$ is that $\norm{f}_{\alpha,\beta} < \infty$ for all $\alpha, \beta \in \bbN_0$. The topology induced by the seminorms is equivalent to the topology induced by the metric \cite[pg.~29]{rudin_functional_1991}
\begin{equation}
    d(f,g) = \sum_{\alpha,\beta \in \bbN_0} 2^{-\alpha-\beta} \frac{\norm{f-g}_{\alpha,\beta}}{1+\norm{f-g}_{\alpha,\beta}}.
\end{equation}
Equipped with the metric, we can check continuity of a map $T\colon S(\bbR) \to \bbC$ in the usual way. $T$ is continuous if for all sequences $(f_n)_{n\in \bbN}$, $f_n \xrightarrow{n\to\infty} f$ implies $T(f_n)\xrightarrow{n\to\infty} T(f)$. Strictly speaking, this is the definition of sequentially continuous, but continuity and sequential continuity are equivalent on metric spaces. Here $f_n \xrightarrow{n\to\infty} f$ means that $\forall \epsilon > 0, \exists N \in \bbN$ such that $\forall n \geq N\colon$ $d(f_n,f) < \epsilon$, and $T(f_n) \xrightarrow{n\to\infty} T(f)$ means similarly but with the metric on $\bbC$.

We are now interested in characterizing the continuous dual of $S(\bbR)$, denoted by $S(\bbR)'$, which is a subset of the algebraic dual. Hence, we restrict our attention to linear maps $T\colon S(\bbR) \to \bbC$. When $T$ is linear, $T(f) - T(g) = T(f-g)$. We also notice that $d(f,g) = 0$ if and only if $\norm{f-g}_{\alpha,\beta} = 0$ for all $\alpha, \beta$. Therefore, the condition that $T$ be a tempered distribution, meaning that $T \in S(\bbR)'$, is that it is linear and satisfies
\begin{equation}
    \label{eq:tempered-dist-continuity}
    \parentheses{\forall \alpha,\beta \in \bbN_0 \colon \lim_{m\to\infty} \norm{f_m}_{\alpha,\beta} = 0} \implies \parentheses{\lim_{m\to\infty} \abs{T(f_m)} = 0}
\end{equation}
for any sequence of functions $(f_m)_{m\in \bbN} \subset S(\bbR)$.

As described above, a rigged Hilbert space is a triplet $S(\bbR) \subset L^2(\bbR) \subset S(\bbR)'$. $S(\bbR)'$ is the ``bra space'' of tempered distributions. One can analogously construct the ``ket space'' of \emph{anti}linear continuous functionals on $S(\bbR)$. This space is often denoted as $S(\bbR)^\times$.

We will now revisit the $\bra{\cos\theta}_\varphi$, $\bra{\sin\theta}_\varphi$, and $\bra{\theta}_\varphi$ states and show that they each belong to $S(\bbR)'$, while their ket counterparts belong to $S(\bbR)^\times$. As before, we will restrict our attention to the $\bra{\cos\theta}_\varphi$ states, as the others are analogous. By construction, $\bra{\cos\theta}_\varphi$ is clearly linear. We now show that it is continuous. We use \cref{eq:tempered-dist-continuity}, and compute
\begin{salign}
    \lim_{m\to\infty}\abs{\prescript{}{\varphi}{\braket{\cos\theta\vert f_{m}}}}&=\lim_{m\to\infty}\abs{\sqrt{\frac{2}{\pi}}\sum_{n\in\bbN_{0}}\e^{-\i\varphi n^{2}}\sin((n+1)\theta)\braket{n\vert f_{m}}}\\&\leq\lim_{m\to\infty}\sum_{n\in\bbN_{0}}\abs{\braket{n\vert f_{m}}}\\&=\lim_{m\to\infty}\sum_{n\in\bbN_{0}}\frac{1}{\sqrt{n!}}\abs{\bra0\hat{a}^{n}\ket{f_{m}}}\\&=\lim_{m\to\infty}\sum_{n\in\bbN_{0}}\frac{1}{\sqrt{n!}}\abs{\int_{\bbR}\psi_{0}(x)\left(x+\frac{\mathrm{d}}{\mathrm{d}x}\right)^{n}f_{m}(x)\dd{x}}\\&\leq\lim_{m\to\infty}\sum_{n\in\bbN_{0}}\frac{1}{\sqrt{n!}}\left(\int_{\bbR}\abs{\psi_{0}(x)}\dd{x}\right)\left(\sup_{x}\abs{\left(x+\frac{\mathrm{d}}{\mathrm{d}x}\right)^{n}f_{m}(x)}\right)\\&\propto\lim_{m\to\infty}\sup_{x}\abs{\left(x+\frac{\mathrm{d}}{\mathrm{d}x}\right)^{n}f_{m}(x)}\\&=0.
\end{salign}
The last line comes by assumption from \cref{eq:tempered-dist-continuity}. The second to last line comes from the facts that $\sum_n 1/\sqrt{n!} < \infty$ and that $\int_\bbR \abs{\psi_0(x)} \dd{x} < \infty$, where \(\psi_0(x)\) is the position representation of the lowest Fock state as described in \cref{ap:review}. Hence, $\bra{\cos\theta}_\varphi$ is a tempered distribution.

From \cref{thm:nonexistence}, we know that CV 2-designs do not exist. However, \cref{thm:kerred-phase-design,thm:cos-design,thm:sin-design} show that rigged CV 2-designs do indeed exist, where we define a rigged CV design analogously to a standard CV design with the additional feature that tempered distributions are allowed.

\subsection{Regularized rigged state designs -- making rigged state designs physical}
\label{ap:regularized-rigged}

Suppose we have a construction of $\Pi_t$ in terms of unphysical states, so that $\Pi_t = \int_X \parentheses{\ket\chi\bra\chi}^{\otimes t} \dd{\mu}$, where $X$ is some measure space with measure $\mu$. We use $\chi$ to denote possibly non-normalizable states, and $\psi$ to denote properly normalized states. Define a Hermitian operator $R$ which we'll call the regularizer. For example, $R$ could be $\e^{-\beta \hat n}$, where $\hat n$ is the number operator diagonal in the Fock basis $\hat n \ket{i} = i \ket{i}$. Then,
\begin{salign}
    \Pi_t^{(R)} &\coloneqq R^{\otimes t} \Pi_t R^{\otimes t} \\
    &= \int_X \parentheses{R\ket\chi\bra\chi R}^{\otimes t} \dd{\mu}.
\end{salign}
As long as the amplitudes of each $\ket\chi$ do not grow too fast (indeed their growth is constrained by the condition that $\ket\chi$ be a tempered distribution; see below), the states $R\ket\chi$ will be normalizable. Define the normalized state corresponding to the tempered distribution $\chi$ as $\ket{\psi} \coloneqq R\ket\chi / \norm{R\ket\chi}$. Then
\begin{equation}
    \Pi_t^{(R)} = \int_X \parentheses{\ket{\psi}\bra{\psi}}^{\otimes t} \norm{R\ket\chi}^{2t}\dd{\mu}.
\end{equation}
One can then define a new measure $\nu$ which is $\mu$ weighted by the positive factor $\norm{R\ket\chi}^{2t} / \Tr \Pi_t^{(R)}$ (one can imagine using a Lebesgue-Stieltjes measure construction), thus giving
\begin{equation}
    \frac{\Pi_t^{(R)}}{\Tr \Pi_t^{(R)}} = \int_X \parentheses{\ket{\psi}\bra{\psi}}^{\otimes t} \dd{\nu}.
\end{equation}

The first thing to note is that by taking the trace of both sides one finds that $\nu(X) = 1$. Hence the measure space defined by $X$ and $\nu$ is a proper probability space. Next, suppose that $R=\e^{-\beta \hat n}$. The parameter $\beta$ is an inverse energy. $1/\beta$ fixes an energy scale of the states involved in the design. As $1/\beta \to \infty$, the energy of the states becomes infinite, and $\Pi_t^{(R)}$ looks more and more like $\Pi_t = \Pi_t^{(\bbI)}$. When $\beta$ is exactly zero, the equation becomes uninteresting since $\Pi_t / \Tr\Pi_t$ is just the zero operator.

Nevertheless, $\beta$ is a parameter that one can tune that enforces a soft energy cutoff. The smaller one tunes $\beta$, the more the ensemble resembles a continuous-variable state $t$-design. The soft energy cutoff $\e^{-\beta \hat n}$ was chosen to ensure physicality of the resulting states. In particular, $\e^{-\beta \hat n}$ will always take a tempered distribution to a state in $L^2(\bbR)$, whereas, for example, a soft-cutoff of the form $(\hat n + 1)^{-b}$ for some $b > 0$ will not always achieve this. We therefore use $\e^{-\beta \hat n}$ to make any rigged design into an approximate design composed of physical states. This is formalized in the following proposition.

\medskip

\begin{proposition}
If $\ket\chi$ is a tempered distribution, then $\e^{-\beta \hat n}\ket\chi$ is a state in $L^2(\bbR)$ for any $\beta > 0$.
\end{proposition}
\begin{proof}[Proof sketch]
From \cite[Thm.~3]{simonDistributionsTheirHermite1971}, any tempered distribution can be expressed as $\ket\chi = \sum_{n \in \bbN_0} a_n \ket n$.
We first calculate the norm of $\e^{-\beta \hat n}\ket\chi$;
\begin{equation}
    \bra\chi \e^{-2\beta \hat n} \ket\chi
    = \sum_{n\in\bbN_0} \abs{a_n}^2 \e^{-2\beta n}.
\end{equation}
We therefore find that $\e^{-\beta \hat n}\ket\chi \in L^2(\bbR)$ as long as $\abs{a_n}$ grows with $n$ asymptotically slower than exponential. Hence, to prove the proposition, we need to show that if $a_n$ grows exponentially or faster in $n$, then $\ket\chi$ is not a tempered distribution. This is proven in \cite[Thm.~3]{simonDistributionsTheirHermite1971}. For completeness, we show it here as well. We will use \cref{eq:tempered-dist-continuity} to show this.

Fix some sequence $(f_m)_{m\in \bbN}$ of states $f_m\in S(\bbR)$ satisfying 
\begin{equation}
    \forall \alpha,\beta \in \bbN_0\colon \lim_{m\to\infty} \norm{f_m}_{\alpha,\beta} = 0.
\end{equation}
Specifically, let $\ket{f_m} = \e^{-m} \sum_{n\in\bbN_0}\e^{-\varepsilon n}\ket n$ for some arbitrarily small $\varepsilon > 0$. Then, assuming the best case where $a_n$ grows exponentially as $a_n = \e^{\i \theta_n} \e^{\gamma n}$ for some $\gamma > 0$,
\begin{salign}
    \lim_{m\to\infty} \abs{\braket{\chi\vert f_m}}
    &= \lim_{m\to\infty} \abs{\sum_{n\in\bbN_0} \bar a_n \e^{-m-\varepsilon n}}\\
    &= \lim_{m\to\infty} \e^{-m} \abs{\sum_{n\in\bbN_0} \e^{-\i\theta_n} \e^{n(\gamma - \varepsilon)}}.
\end{salign}
Since $\varepsilon$ can be arbitrarily small, we can always choose it so that $\gamma - \varepsilon > 0$, and therefore the sum diverges no matter the choices of the phases $\theta_n$. Hence,
\begin{equation}
    \lim_{m\to\infty} \abs{\braket{\chi\vert f_m}} \neq 0,
\end{equation}
proving, by \cref{eq:tempered-dist-continuity}, that $\ket\psi$ is not a tempered distribution.
\end{proof}

\medskip

This proposition justifies our choice $\e^{-\beta \hat n}$ as the soft energy cutoff, since a cutoff such as $(1+\hat n)^{-b}$ does not satisfy the proposition for any $b$. However, there do exist rigged designs for which $(1+\hat n)^{-b}$ \emph{is} sufficient. For example, the $\ket{\theta}_\varphi$, $\ket{\cos\theta}_\varphi$, and $\ket{\sin\theta}_\varphi$ are all tempered distributions that generate rigged 2-designs, and $(1+\hat n)^{-2}\ket{\theta}_\varphi, (1+\hat n)^{-2}\ket{\cos\theta}_\varphi, (1+\hat n)^{-2}\ket{\sin\theta}_\varphi \in L^2(\bbR)$. Hence, one may suggest that for these rigged designs, one should use $(1+\hat n)^{-b}$ as a soft energy cutoff in place of $\e^{-\beta \hat n}$. However, one desirable property of physical quantum states is that all position, momentum, and energy moments are finite. In other words, one may desire that the states belong to $S(\bbR) \subset L^2(\bbR)$. One can straightforwardly show that, for example, $(1+\hat n)^{-b} \ket{\theta}_\varphi \notin S(\bbR)$ for any $b$, whereas $\e^{-\beta \hat n}\ket{\theta}_\varphi \in S(\bbR)$. This is another justification for the use of $\e^{-\beta \hat n}$.

\medskip

\begin{example}
Consider, for example, the rigged design given \cref{thm:kerred-phase-design}. Sandwiching the design with $R$ results in the normalized states $\sqrt{1-\e^{-2\beta}}\sum_{n\in\bbN_0} \e^{-\beta n + \i \theta n + \i \varphi n^2}\ket n$. Each of these states has energy $\coth(\beta)/2-1/2$. The design also still consists of the original Fock states $\ket n$, but the weight in front of each Fock state decays exponentially with $n$ as $\sim \e^{-\beta n}$. Thus, despite the fact that the design uses arbitrarily high energy states (i.e. $\ket n$ for all natural numbers $n$), the weight factor in front of these high energy states is exponentially small in the energy. Therefore, the design effectively uses states finitely upper bounded in energy, where the bound is tuned by $\beta$. We refer to \cite[sec.~5]{dodonov_nonclassical_2002} for a review of these states, which are related to so-called phase coherent states.
\end{example}

\medskip

We consider now an $R$-regularized rigged $t$-design $\calG$, which satisfies 
\begin{equation}
    \Expval_{\psi\in\calG}(\ket\psi\bra\psi)^{\otimes t} = \frac{ \Pi_t^{(R)} }{ \Tr \Pi_t^{(R)} }~.\label{eq:t-design-appx}   
\end{equation}
By tracing out e.g.~the last factor, we find $\Expval_{\psi\in\calG}(\ket\psi\bra\psi)^{\otimes (t-1)}\propto \Tr_t \Pi_t^{(R)}$. Recall from \cref{ap:symmetric-projector} that $\Pi_t = \frac{1}{t!}\sum_{\sigma \in S_t} W_\sigma$. Consider a permutation $\sigma \in S_t$ that leaves the last factor fixed. Let $\pi \in S_{t-1}$ be the permutation with the same cyclic decomposition as $\sigma$. For example, when $t=3$ and $\sigma = (12)(3)$ is the permutation swapping $1$ and $2$ and leaving $3$ fixed, then we set $\pi = (12)$. We see that for such a $\sigma$, $\Tr_t (R^{\otimes t} W_\sigma R^{\otimes t}) = (\Tr R^2) R^{\otimes (t-1)}W_\pi R^{\otimes (t-1)}$. Hence, the sum over all such permutations results in $(\Tr R^2)\sum_{\pi\in S_{t-1}} R^{\otimes (t-1)}W_\pi R^{\otimes (t-1)} = (\Tr R^2)\Pi_{t-1}^{(R)}$. For all other permutations $\tau$ that do not leave the $t^{\rm th}$ factor fixed, $\Tr_t W_\tau$ does not pick up a factor of $(\Tr R^2)$. We have hence found that $\Expval_{\psi\in\calG}(\ket\psi\bra\psi)^{\otimes (t-1)} \sim (\Tr R^2)\Pi_{t-1}^{(R)} + (\text{terms without }(\Tr R^2))$. Assuming that the regularizer $R$ is close to the identity so that $(\Tr R^2)$ is large and applying the above arguments to both the numerator and denominator of \cref{eq:t-design-appx}, we have thus found that an $R$-regularized rigged $t$-design $\calG$ satisfies
\begin{equation}
    \Expval_{\psi\in\calG}(\ket\psi\bra\psi)^{\otimes (t-1)} = \frac{\Pi_{t-1}^{(R)}}{\Tr \Pi_{t-1}^{(R)}} \parentheses{1+ \bigO{1/\Tr R^2}}.
\end{equation}
It is in this sense that an $R$-regularized rigged $t$-design is \textit{almost} an $R$-regularized rigged $(t-1)$-design up to factors of $1/\Tr R^2$.

In the special case when $R = P_d = \sum_{n=0}^{d-1}\ket n \bra n$, a $P_d$-regularized rigged $t$-design $\calG$ is simply a $\bbC\bbP^{d-1}$ $t$-design, and hence it is also \textit{exactly} a $(t-1)$-design. However, when $R$ is an invertible operator, the result is only a $(t-1)$-design up to terms of order $1/\Tr R^2$.

\subsubsection{Frame potential}
\label{ap:frame-potential}

In this section, we will generalize the well-known \textit{frame potential} from finite-dimensional state designs \cite{klappenecker2005mutually} to regularized rigged designs. Specifically, for a positive definite regularizer $R$, we define the frame potential of an ensemble $\calG$ (i.e.~a probability space over unit vectors in $L^2(\bbR)$) to be
\begin{equation}
    V_t^{(R)}(\calG) \coloneqq \Expval_{\psi,\phi \in \calG} \abs{\bra\psi R^{-1} \ket \phi}^{2t}.
\end{equation}
We prove the following proposition regarding $R$-regularized rigged $t$-designs and the frame potential.

\medskip
 
\begin{proposition}
    Let $R$ be positive definite.
    For any ensemble $\calG$,
    \begin{equation}
        V_t^{(R)}(\calG) \geq \frac{1}{\Tr \Pi_t^{(R)}},
    \end{equation}
    with equality if and only if $\calG$ is an $R$-regularized rigged $t$-design.
\end{proposition}
\begin{proof}
This proof is a modification of that of Ref.~\cite[Eq.~(3)]{klappenecker2005mutually}.
Let $E \coloneqq \Expval_{\psi\in\calG} (\ket\psi\bra\psi)^{\otimes t}$ and $\xi \coloneqq (R^{-1})^{\otimes t} E - \Pi_t^{(\sqrt{R})} / \Tr \Pi_t^{(R)}$. By recalling the definitions of regularized-rigged designs and of the symmetric projector \eqref{eq:symm-proj-perms}, we see that $\calG$ is an $R$-regularized rigged $t$-design if and only if $\xi = 0$, or equivalently, $\Tr \xi^2 = 0$.
We find that
\begin{salign}
    0
    &\leq \Tr \xi^2\\
    &= \Tr\bargs{(R^{-1})^{\otimes t} E (R^{-1})^{\otimes t} E} + \frac{\Tr \bargs{(\Pi_t^{\sqrt{R}})^2}}{\parentheses{\Tr \Pi_t^{(R)}}^2} - \frac{2}{\Tr\Pi_t^{(R)}} \Tr \bargs{E \Pi_t^{(\sqrt R)}(R^{-1})^{\otimes t}}\\
    &= V_t^{(R)}(\calG) + \frac{\Tr \Pi_t^{(R)}}{(\Tr \Pi_t^{(R)})^2} - \frac{2}{\Tr\Pi_t^{(R)}} \Tr \bargs{E \Pi_t}\\
    &= V_t^{(R)}(\calG) + \frac{1}{\Tr \Pi_t^{(R)}} - \frac{2}{\Tr\Pi_t^{(R)}}\\
    &= V_t^{(R)}(\calG) - \frac{1}{\Tr \Pi_t^{(R)}},
\end{salign}
with equality if and only if $\calG$ is an $R$-regularized rigged $t$-design. In the second to last line, we used that $E \Pi_t = E$ and that $\Tr E = 1$.
\end{proof}

\medskip

If $R$ is instead only positive \textit{semi}-definite and not invertible, then we can modify the definition of the frame potential to utilize the Moore-Penrose inverse $R^+$ in place of the inverse $R^{-1}$. The proposition then still holds as is, with the addition of the assumption that $RR^+ \calG = \calG$, where recall $RR^+$ is a projector onto the support of $R$.

Notice the presence of the $R^{-1}$ in the definition of the frame potential. We will also see such a presence in \cref{ap:average-fidelity} when generalizing fidelity quantities to infinite-dimensional spaces.

\subsection{Alternative characterization of continuous-variable, rigged, and regularized rigged designs}
\label{ap:C-inf-designs}

To generate a random state $\ket\psi\in\bbC\bbP^{d-1}$, one can equivalently choose $d$ amplitudes $\set{\alpha_i \in \bbC \mid i \in 0,\dots, d-1}$, where each $\alpha_i$ is drawn independently from the unit variance normal distribution $\calN(0, 1)$. The state $\frac{\sum_i \alpha_i \ket i}{\norm{\sum_i \alpha_i \ket i}}$ is then a random state drawn from $\bbC\bbP^{d-1}$.

Motivated by this and by Ref.~\cite[Sec.~4.1]{lunardiInfiniteDimAnalysis}, we consider integration on the Frech\'et space $\bbC^\infty = \prod_{i\in\bbN_0} \bbC$ with the product topology. Define $\delta_j\colon \bbC^\infty \to \bbC$ to be the projections $\delta_j(x) = x_j$. Let $\Sigma$ be the smallest $\sigma$-algebra on $\bbC^\infty$ such that $\delta_j$ is measurable for every $j$. Note that this corresponds to the Borel $\sigma$-algebra; that is, the product topology and the $\sigma$-algebra are both generated by sets of the form $A = \prod_{i\in\bbN_0} A_i$, where each $A_i$ is an open subset of $\bbC$ and only finitely many $A_i$ are proper.

Let $\calN(0, \lambda_i)$ be the Gaussian measure on $\bbC$ with mean $0$ and variance $\lambda_i$. Define the measure $\mu\colon \Sigma \to [0,\infty]$ by $\mu\coloneqq \bigotimes_{j\in\bbN_0} \calN(0,\lambda_j)$, where each $\lambda_j \in (0,\infty)$. The construction for such a measure is as follows. For $A = \prod_{i\in\bbN_0} A_i$ where all but finitely many $A_i$ satisfy $A_i=\bbC$, define $\mu(A) = \prod_{i\in\bbN_0}\calN(0,\lambda_i)(A_i)$. For every $i$ for which $A_i=\bbC$, $\calN(0,\lambda_i)(A_i) = 1$. Hence, $\mu$ is well-defined on such sets $A$ since the product is finite. From its definition on such sets $A$, $\mu$ can be uniquely extended to all of $\Sigma$ \cite[Thm.~10.6.1]{cohnMeasureTheory2013}.

Let $\set{\ket n \mid n \in \bbN_0}$ be a basis for $L^2(\bbR)$. For $z\in \bbC^\infty$, let $\ket z \coloneqq \sum_{n\in\bbN_0}z_n \ket n$. Any tempered distribution can be expressed as $\ket z$ for some $z\in\bbC^\infty$ satisfying certain conditions \cite[Thm.~3]{simonDistributionsTheirHermite1971}. We therefore define the following subsets of $\bbC^\infty$:
\begin{salign}
    S &\coloneqq \set{z \in \bbC^\infty \mid \ket z \in S(\bbR)} \\
    \ell^2_\bbC(\bbN_0) &\coloneqq \set{z \in \bbC^\infty \mid \ket z \in L^2(\bbR)} \\
    S' &\coloneqq \set{z \in \bbC^\infty \mid \ket z \in S(\bbR)'}.
\end{salign}

\begin{lemma}
    Suppose that $\lambda_i=1$ for all $i\in\bbN_0$. Then $\mu(S') = 1$ and therefore $\mu(\bbC^\infty \setminus S') = 0$.
\end{lemma}
\begin{proof}
    Note that
    \begin{equation}
        \int_{\bbC^\infty} \norm{\frac{1}{\hat n+1}\ket z}^2 \dd\mu(z)
        = \sum_{n\in\bbN_0}\frac{1}{(n+1)^2} \int_{\bbC^\infty} \abs{z_n}^2 \dd\mu(z)
        = \sum_{n\in\bbN_0}\frac{1}{(n+1)^2} < \infty.
    \end{equation}
    Therefore, $\norm{\frac{1}{\hat n+1}\ket z}^2 < \infty$ for almost all $z$.
    For any $z \in \bbC^\infty$, if $\norm{\frac{1}{\hat n+1}\ket z} < \infty$, then $\ket z \in S(\bbR)'$ \cite[Thm.~3]{simonDistributionsTheirHermite1971}. Hence $\ket z \in S(\bbR)'$ $\mu$-a.e.
\end{proof}

Through an analogous calculation with the integrand being $\norm{\ket z}^2$, one finds that if $\sum_{i\in\bbN_0}\lambda_i < \infty$, then $\mu(\ell^2_\bbC(\bbN_0)) = 1$ \cite[Rem~4.1.2]{lunardiInfiniteDimAnalysis}. Define $R$ to be the diagonal matrix with diagonal entries $\lambda_i$. If $R$ is the identity, then $\mu(S') = 1$, while if $R$ is trace-class, then $\mu(\ell^2_\bbC(\bbN_0)) = 1$.

It follows that if $R$ is the identity, integrals over the measure space $(\bbC^\infty, \Sigma, \mu)$ are equal to integrals over the restricted measure space $(S', \Sigma\rvert_{S'}, \mu\rvert_{S'})$. Similarly, if $R$ is trace-class, integrals over the measure space $(\bbC^\infty, \Sigma, \mu)$ are equal to integrals over the restricted measure space $(\ell^2_\bbC(\bbN_0), \Sigma\rvert_{S'}, \mu\rvert_{S'})$.

Next, we show that when $R$ is the identity, $(S', \Sigma\rvert_{S'}, \mu\rvert_{S'})$ is a rigged $t$-design for any $t\in \bbN$, and when $R$ is trace-class, $(\ell^2_\bbC(\bbN_0), \Sigma\rvert_{\ell^2_\bbC(\bbN_0)}, \mu\rvert_{\ell^2_\bbC(\bbN_0)})$ is an $R$-regularized rigged $t$-design for any $t\in \bbN$.
\footnote{More accurately, the image of $(S', \Sigma\rvert_{S'}, \mu\rvert_{S'})$ under the map $z\mapsto \ket z$ -- which is a measure space over $S(\bbR)'$ -- is a rigged $t$-design, and the image of $(\ell^2_\bbC(\bbN_0), \Sigma\rvert_{\ell^2_\bbC(\bbN_0)}, \mu\rvert_{\ell^2_\bbC(\bbN_0)})$ -- which is a measure space over $L^2(\bbR)$ -- is a regularized rigged $t$-design.}

Given the construction of our measure space over $\bbC^\infty$, integrals over polynomials in $z$ reduce to simple finite-dimensional Gaussian integration. For the purposes of designs, we are only interested in such polynomials. Consider
\begin{equation}
    \int_{\bbC^\infty} \prod_{i=1}^t \bra{a_i}\ket z \bra z \ket{b_i} \dd \mu(z)
    = \int_{\bbC^\infty} \prod_{i=1}^t z_{a_i}\bar z_{b_i} \dd \mu(z).
\end{equation}
for $a,b\in \bbN_0^t$. Since the integrand depends only on at most $2t$ elements of $z$, we can use Fubini's theorem so that the integral reduces to an integral over $\bbC^{2t}$ with the measure $\bigotimes \calN(0,\lambda_i)$. Then, one can easily check by induction (or just by using standard properties of Gaussian integrals) that the integral equals $\Pi_t^{(\sqrt{R})}(a;b)$, and therefore
\begin{equation}
    \int_{S'} \prod_{i=1}^t \ket z \bra z \dd \mu(z)
    = \Pi_t^{(\sqrt{R})}
\end{equation}
in the weak sense. When $R$ is the identity, $(S', \Sigma\rvert_{S'}, \mu\rvert_{S'})$ is a rigged $t$-design (for all $t\in \bbN_0$), and when $R$ is trace-class, $(\ell^2_\bbC(\bbN_0), \Sigma\rvert_{\ell^2_\bbC(\bbN_0)}, \mu\rvert_{\ell^2_\bbC(\bbN_0)})$ is a $\sqrt{R}$-regularized rigged $t$-design (for all $t \in \bbN_0)$.

\subsection{Displaced Fock states as negative-weight approximate designs}
\label{ap:dispalced-fock}

The projection onto the two-body symmetric subspace is (see \cref{ap:symmetric-projector})
\begin{equation}
\Pi_{2}=\frac{1}{2}\left(\bbI+\e^{\i\frac{\pi}{2}\left(a^{\dagger}-b^{\dagger}\right)\left(a-b\right)}\right)\,,
\end{equation}
where the second operator in the parentheses is the SWAP operator, and
$a$ ($b$) represents the lowering operator for the first (second)
mode. To simplify calculations, we apply the beam-splitter operation
\begin{equation}
U=\exp\bargs{\frac{\pi}{4}\left(a^{\dagger}b-ab^{\dagger}\right) }\,,\quad\quad\text{acting as}\quad\quad U^{\dagger}\begin{pmatrix}a\\
b
\end{pmatrix}U=\frac{1}{\sqrt{2}}\begin{pmatrix}a+b\\
b-a
\end{pmatrix}\,,
\end{equation}
which is equivalent to partitioning the two-mode Hilbert
space into a tensor product of a center-of-mass $L^{2}(\mathbb{R})$
factor whose corresponding coordinate is symmetric under SWAP and
an anti-symmetric factor whose coordinate is anti-symmetric~\cite[Sec.~III]{blume-kohout_curious_2014}.
In the Fock-space picture, this results in
\begin{equation}
U\Pi_{2}U^{\dagger}=\bbI\otimes\frac{1+\e^{\i\pi b^{\dagger}b}}{2}=\sum_{n\in\mathbb{N}_{0}}|n\rangle\langle n|\otimes\sum_{p\in\mathbb{N}_{0}}|2p\rangle\langle2p|\,,
\end{equation}
which now projects onto the entire symmetric factor and the even Fock-state
subspace of the anti-symmetric factor.

We now determine what happens if one sums up two copies of all displaced
versions of a particular Fock state $|\ell\rangle$. Using the fact
that SWAP acts on displacements as $UD_{\alpha}^{\otimes2}U^{\dagger}=D_{\alpha}\otimes \bbI$
and the fact that displacements form a unitary $1$-design, we have
\begin{subequations}
\begin{align}
U\left(\int\frac{\mathrm d^{2}\alpha}{\pi}D_{\alpha}^{\otimes2}|\ell\ell\rangle\langle\ell\ell|D_{-\alpha}^{\otimes2}\right)U^{\dagger} & =\int\frac{\mathrm d^{2}\alpha}{\pi}\left(D_{\alpha}\otimes \bbI\right)U|\ell\ell\rangle\langle\ell\ell|U^{\dagger}\left(D_{\alpha}^{\dagger}\otimes \bbI\right)\\
 & =\Tr_{1}\pargs{U|\ell\ell\rangle\langle\ell\ell|U^{\dagger}}\,,
\end{align}
\end{subequations}
where $\Tr_{1}$ is the partial trace over the first factor. 

We next write out $U$ as a direct sum of irreducible representations
of $\SU(2)$, with each representation acting on a sector of fixed
total occupation number. Irreducible representations of $\SU(2)$ are
known exactly in terms of the Wigner-$D$ matrices~\cite{VMH}, and
the matrix elements we will need are
\begin{equation}
c_{n}^{(\ell)}=\left|\langle2\ell-n,n|U|\ell,\ell\rangle\right|^{2}=\left|D_{2n-\ell,0}^{\ell}\left(0,-\frac{\pi}{2},0\right)\right|^{2}=\frac{\left(2\ell-2n\right)!\left(2n\right)!}{4^{\ell}\left[n!\left(\ell-n\right)!\right]^{2}}\,.
\end{equation}
Plugging this in yields
\begin{equation}
U\left(\int\frac{\mathrm d^{2}\alpha}{\pi}D_{\alpha}^{\otimes2}|\ell\ell\rangle\langle\ell\ell|D_{-\alpha}^{\otimes2}\right)U^{\dagger}=\bbI\otimes\sum_{n=0}^{\ell}c_{n}^{(\ell)}|2n\rangle\langle2n|\,.\label{eq:disp-fock}
\end{equation}
When $\ell=0$, we have $c_{0}^{(0)}=1$, corroborating the result
from~\cite[Sec. III]{blume-kohout_curious_2014}. For general $\ell$,
this result yields nonzero coefficients $c_{n}^{(\ell)}$ for all
Fock states $\leq2\ell$ in the anti-symmetric factor. 

We now linearly combine instances of \cref{eq:disp-fock} with
$\ell$ from zero to some $\ell_{\text{max}}$ and compensate the
$c_{n}^{(\ell)}$ using weights $b_{\ell}$ in front of each Fock
state. This yields
\begin{subequations}
\begin{align}
U\left(\sum_{\ell=0}^{\ell_{\text{max}}}b_{\ell}\int\frac{\mathrm d^{2}\alpha}{\pi}D_{\alpha}|\ell\ell\rangle\langle\ell\ell|D_{\alpha}^{\dagger}\right)U^{\dagger} & =\bbI\otimes\sum_{n=0}^{\ell_{\text{max}}}|2n\rangle\langle2n|\,,\label{eq:fock-design-1}\\
b_{\ell} & =\frac{1-\sum_{p=\ell+1}^{\ell_{\text{max}}}b_{p}c_{\ell}^{(L)}}{c_{\ell}^{(\ell)}}\,,
\end{align}
\end{subequations}
which yields $\Pi_{2}$ up to the Fock state $2\ell$ in the anti-symmetric
factor. However, some of the $b_{\ell}$'s are negative, meaning that
the right-hand side of Eq.~(\ref{eq:fock-design-1}) cannot be treated
as an expectation value of operators sampled according to a probability
distribution. The ensemble can be formulated in terms of a measure
space with a signed measure, and there may be schemes to sample from
such an ensemble \cite{temmeErrorMitigationShortDepth2017}.
Thus, displaced Fock states form a hard-energy regularized $2$-design
with a signed measure.

One may be tempted to take \(\ell_{max}\) to infinity. In this case, the coefficient \(b_{\ell_{max}}\to\infty\), showing that this regularized design does not yield a CV design and corroborating the no-go \cref{thm:nonexistence} (recall we extended \cref{thm:nonexistence} to the case of signed measure spaces at the end of \cref{ap:nonexistence}).

\subsection{Approximate continuous-variable unitary designs}
\label{ap:cv-unitary-designs}

In finite dimensions, a unitary design reconstructs the superoperator
\begin{equation}
    \mathcal{P}_t = \frac{1}{\abs{S_t}} \sum_{\sigma\in S_t} \Vert W_\sigma \rangle\langle W_\sigma \Vert.
\end{equation}
In this way, when acting on a fiducial state $\rho = \ket\phi\bra\phi$, one finds
\begin{salign}
    \mathcal{P}_t \Vert \rho^{\otimes t}\rangle &= \frac{1}{\abs{S_t}} \sum_{\sigma\in S_t}\Vert W_\sigma \rangle\langle W_\sigma \Vert \rho^{\otimes t}\rangle\\
    &= \frac{1}{\abs{S_t}} \sum_{\sigma\in S_t}\Vert W_\sigma \rangle \Tr\brackets{W_\sigma^{-1} \rho^{\otimes t}}\\
    &= \frac{1}{\abs{S_t}}\sum_{\sigma\in S_t} \Vert W_\sigma \rangle \\
    &= \Pi_t.
\end{salign}

From above, we have states in $L^2(\bbR)$ that construct the normalized symmetric projector $\Pi_t^{(R)} = R^{\otimes t}\Pi_t R^{\otimes t}$. In a similar way, let's normalize the superoperator $\mathcal{P}_t$. Define
\begin{equation}
    \mathcal{P}_t^{(R)} \coloneqq \frac{1}{\abs{S_t}}\sum_{\sigma\in S_t} \Vert R^{\otimes t}W_\sigma \rangle\langle R^{\otimes t} W_\sigma \Vert.
\end{equation}
Then, when acting on a fiducial state $\rho = \ket\phi\bra\phi$, one finds
\begin{salign}
    \mathcal{P}_t^{(R)} \Vert \rho^{\otimes t}\rangle &= \frac{1}{\abs{S_t}}\sum_{\sigma \in S_t} \Vert R^{\otimes t}W_\sigma \rangle\langle RW_\sigma \Vert \rho^{\otimes t}\rangle\\
    &= \frac{1}{\abs{S_t}}\sum_{\sigma \in S_t} \Vert R^{\otimes t} W_\sigma \rangle \mathrm{Tr}[R^{\otimes t}W_\sigma^{-1} \rho^{\otimes t}]\\
    &= \mathrm{Tr}[(R\rho)^{\otimes t}] \Pi_t^{(R)} \\
    &\propto \Pi_t^{(R)}.
\end{salign}

With this, we now define an approximate continuous-variable unitary $t$-design to be a collection of unitaries $U_i \colon L^2(\bbR) \to L^2(\bbR)$ that satisfy
\begin{equation}
    \sum_i \parentheses{\Vert U_i \rangle\langle U_i \Vert}^{\otimes t} = \mathcal{P}_t^{(R)}.
\end{equation}
As with rigged state designs, the parameterization $i$ of the unitaries, represented here heuristically as a sum, may constitute a measure space.
We leave determination of existence of such designs to future work.

\section{Applications of rigged and regularized rigged designs}
\label{ap:applications}

\subsection{Continuous-variable shadows}
\label{ap:shadows}

In this subsection, we use rigged designs to construct infinite-dimensional classical shadows of a quantum state $\rho$. With these shadows, one can for example efficiently compute the expectation value of many observables. Ref.~\cite{acharyaInformationallyCompletePOVMbased2021} phrased shadow tomography from Ref.~\cite{huangPredictingManyProperties2020} in terms of informationally-complete POVMs. We will generalize their discussion to infinite dimensions.

Specifically, suppose that the measure space $(X, \Sigma, \mu)$ is a rigged $3$-design. In other words,
\begin{equation}
\int_X (\ket\chi\bra\chi)^{\otimes t}\dd{\mu(\chi)} = \alpha_t \Pi_t
\end{equation}
for each $t\in \set{1,2,3}$, where $\alpha_1,\alpha_2,\alpha_3 \in (0,\infty)$ are some numbers. We assume without loss of generality that $\alpha_1 = 1$ (if not, just rescale the measure). Recall that we will use $\ket\chi$ to denote tempered distributions and $\ket\psi$ to denote physical quantum states. 

Let $\nu\colon\Sigma \to P(\calH)$, where $P(\calH)$ denotes the set of nonnegative operators on an underlying separable, infinite dimensional Hilbert space $\calH$, and define 
\begin{equation}
\nu(A) \coloneqq \int_A \ket\chi\bra\chi \dd{\mu(\chi)}~.
\end{equation}
This map is a positive operator-valued measure (POVM) because it satisfies the axioms
\begin{enumerate}
    \item $\nu(X) = \bbI$,
    \item $\nu(\emptyset) = 0$, and
    \item $\nu(\bigcup_i A_i) = \sum_i \nu(A_i)$ for countable collections of disjoint $A_i\in\Sigma$.
\end{enumerate}
The first axiom is satisfied since $X$ is a rigged $1$-design with $\alpha_1=1$. The second axiom is trivially satisfied. The third axiom follows from the $\sigma$-additivity of the measure $\mu$.

We can therefore measure a state $\rho$ with respect to the POVM $\nu$. As usual, associated to the POVM is a standard probability measure $\mu'$ defined by $\mu'(A) = \Tr[\rho \nu(A)]$. When measuring the state $\rho$ with the POVM $\nu$, we sample outcomes labeled by $\chi\in X$ from the probability measure $\mu'$. Indeed, we have the freedom to label the outcomes however we choose. In particular, suppose that to each tempered distribution (i.e.~non-normalizable, and therefore unphysical, quantum state) $\ket\chi\in X$, we associate a physical state $\ket{\psi_\chi}\in\calH$ of unit norm. Then the measurement channel representing the POVM $\nu$ is $\rho \mapsto \int_X \ket{\psi_\chi}\bra{\psi_\chi} \dd{\mu'(\chi)}$.

In the realm of shadow tomography, however, we have even more freedom than this. Once we measure from the POVM, we store a shadow on a classical computer and never need to physically prepare the shadow. Therefore, we do not \textit{need} to associate physical states $\ket{\psi_\chi}$ to the measurement outcomes of the POVM; we are free to associate the unphysical tempered distributions $\ket\chi$ to the measurement outcome corresponding to $\chi$. The resulting map representing the measurement process is then
\begin{equation}
    \calM(\rho) = \int_X \ket\chi\bra\chi \dd{\mu'(\chi)}.
\end{equation}
Since $\mu'$ is a probability measure, we will define the notation $\Expval_{\chi\in X'}(\cdot) \coloneqq \int_X (\cdot) \dd{\mu'(\chi)}$. Hence,
\begin{equation}
    \calM(\rho) = \Expval_{\chi\in X'}\ket\chi\bra\chi.
\end{equation}
$\calM$ is not a physical quantum channel; indeed, $\Tr\calM(\rho)$ is not finite. However, $\calM$ represents the process of measuring $\rho$ with respect to the physical POVM $\nu$ and storing the result classically. This part of the formalism, namely associating the infinite-trace operator $\ket\chi\bra\chi$ to the measurement outcome $\chi$, is the only part that differs from the finite dimensional case. In the finite dimensional case, the designs contain only physical states $\ket\psi$, and a physical density matrix $\ket\psi\bra\psi$ is associated to the measurement outcome $\psi$. Ultimately, since this part of the procedure is being done classically, this difference is inconsequential, and we continue exactly as we would in the finite dimensional case.

Using the fact that $X$ is a rigged $2$-design, we can evaluate 
\begin{salign}
    \calM(\rho)
    &= \int_X \ket\chi\bra\chi \dd{\mu'(\chi)}\\
    &= \int_X \ket\chi\bra\chi \Tr[\rho \ket\chi\bra\chi] \dd{\mu(\chi)}\\
    &= \Tr_1\bargs{(\rho\otimes \bbI) \int_X (\ket\chi\bra\chi)^{\otimes 2} \dd{\mu(\chi)}}\\
    &= \alpha_2 \Tr_1\bargs{(\rho\otimes \bbI) \Pi_2}\\
    &= \frac{\alpha_2}{2}\Tr_1\bargs{(\rho\otimes \bbI) (\bbI\otimes \bbI+S)}\\
    &= \frac{\alpha_2}{2} (\bbI + \rho),
\end{salign}
where $S$ is the SWAP operator. Hence, $\rho = \EX{\chi\in X'}{\frac{2}{\alpha_2}\ket\chi\bra\chi-\bbI}$, and therefore for any observable $\calO$,
\begin{equation}
    \angles{\calO} \coloneqq \Tr(\rho\calO) = \Expval_{\chi\in X'}\Tr\bargs{\parentheses{\frac{2}{\alpha_2}\ket\chi\bra\chi-\bbI}\calO}.
\end{equation}
Suppose that we make $N$ measurements. The output of the $i^{\rm th}$ measurement is a label $\chi_i$. We store the classical shadow $\hat\rho_i \coloneqq \frac{2}{\alpha_2}\ket{\chi_i}\bra{\chi_i}-\bbI$ on a classical computer. Therefore, after $N$ measurements, we have a classical collection $\set{\hat\rho_1,\dots,\hat\rho_N}$. Given sufficient information about our design and the observable, one can classically compute $\Tr(\hat \rho_i \calO)$. Define
\begin{equation}
    \hat o \coloneqq \frac{1}{N}\sum_{i=1}^N \Tr(\hat\rho_i \calO).
\end{equation}
By construction, $\Expval[\hat o] = \angles{\calO}$, where the expectation is taken over possible measurement outcomes. By Chebychev's inequality, $\Pr{}{\abs{\hat o - \Expval[\hat o]}\geq \epsilon} \leq \variance(\hat o)/\epsilon^2$, where
\begin{salign}
    \variance(\hat o)
    &= \sum_{i=1}^N \variance\pargs{\frac{1}{N}\Tr[\hat\rho_i \calO]}\\
    &= \frac{1}{N^2}\sum_{i=1}^N \variance\pargs{\Tr[\hat\rho_i \calO]}\\
    &= \frac{1}{N^2}\sum_{i=1}^N \variance_{\chi\in X'}\pargs{\Tr\bargs{\frac{2}{\alpha_2}\ket\chi\bra\chi \calO}-\Tr\calO}\\
    &= \frac{1}{N}\variance_{\chi\in X'}\pargs{\Tr\bargs{\frac{2}{\alpha_2}\ket\chi\bra\chi \calO}-\Tr\calO}\\
    &= \frac{1}{N}\Expval_{\chi\in X'}\parentheses{\Tr\bargs{\frac{2}{\alpha_2}\ket\chi\bra\chi \calO}-\Tr\calO}^2 - \frac{1}{N} \angles{\calO}^2\\
    &= \frac{1}{N}\Expval_{\chi\in X'}\parentheses{\Tr\bargs{\frac{2}{\alpha_2}\ket\chi\bra\chi \calO}^2+(\Tr\calO)^2-2\Tr\bargs{\frac{2}{\alpha_2}\ket\chi\bra\chi \calO}(\Tr\calO)} - \frac{1}{N} \angles{\calO}^2\\
    \begin{split}
    &= \frac{4}{N\alpha_2^2}\int_X\Tr[\ket\chi\bra\chi \calO]^2 \Tr(\rho\ket\chi\bra\chi)\dd{\mu(\chi)}\\
    &\qquad +\frac{1}{N}(\Tr\calO)^2 \int_X\Tr(\rho\ket\chi\bra\chi)\dd{\mu(\chi)}\\
    &\qquad -\frac{4}{N\alpha_2}(\Tr\calO)\int_X\Tr[\ket\chi\bra\chi \calO] \Tr(\rho\ket\chi\bra\chi)\dd{\mu(\chi)}\\
    &\qquad - \frac{1}{N} \angles{\calO}^2. 
    \end{split}
\end{salign}
Using that $X$ is a rigged $1$-, $2$-, and $3$-design, we find
\begin{salign}
    \variance(\hat o)
    &= \frac{4\alpha_3}{N\alpha_2^2}\Tr\bargs{(\rho\otimes\calO\otimes\calO)\Pi_3}
    +\frac{1}{N}(\Tr\calO)^2
    -\frac{4}{N\alpha_2}(\Tr\calO)\Tr\bargs{(\rho\otimes \calO)\Pi_2}
    - \frac{1}{N} \angles{\calO}^2\\
    &= \frac{2\alpha_3}{3N\alpha_2^2}\brackets{(\Tr\calO)^2 + 2 (\Tr\calO)\angles{\calO} + \Tr\calO^2 + 2\angles{\calO^2}}\\
    &\qquad +\frac{1}{N}(\Tr\calO)^2 -\frac{2}{N\alpha_2}(\Tr\calO)(\Tr\calO + \angles{\calO}) - \frac{1}{N} \angles{\calO}^2\nonumber\\
    &\in \bigO{\frac{(\Tr \abs{\calO})^2}{N}}.
\end{salign}
It then follows that
\begin{equation}
    \Pr{}{\abs{\hat o - \Expval[\hat o]} \geq \epsilon} \in \bigO{\frac{(\Tr \abs{\calO})^2}{N\epsilon^2}}.
\end{equation}
Consider computing the expectation value of $M$ observables $\calO_1,\dots,\calO_M$ using the same $N$ shadows $\hat\rho_i$ as above, and let $\hat o_i$ be the same as $\hat o$ from above but corresponding to $\calO_i$. Then, applying the union bound, we find
\begin{equation}
    \Pr{}{\max_i \abs{\hat o_i - \Expval[\hat o_i]} \geq \epsilon} \in \bigO{\frac{M\max_i (\Tr \abs{\calO_i})^2}{N\epsilon^2}}.
\end{equation}
Hence, to achieve a failure probability of at most $\delta$, we need $N \in \bigO{\frac{M \max_i(\Tr\abs{\calO_i})^2}{\delta \epsilon^2}}$.

\subsubsection{Using median-of-means}

We can do much better than this by using the median-of-means estimator as described in \cite[Thm.~1]{huangPredictingManyProperties2020} where we compute the median of $K$ sample means and each mean is taken with $N$ samples. Indeed, their theorem applies immediately, and we instead find that
\begin{equation}
    N \in \bigO{\frac{1}{\epsilon^2}\max_i (\Tr \abs{\calO_i})^2} \qquad \text{and} \qquad K \in \bigO{\log(M/\delta)}
\end{equation}
suffices to estimate each $\angles{\calO_i}$ to maximum additive error $\epsilon$ with success probability at least $1-\delta$. Thus, the total number of samples from $\rho$ needed to accurately predict $\angles{\calO_1}, \dots, \angles{\calO_M}$ scales as $\log M$.

Unfortunately, we have not yet found a useful rigged $3$-design (a rigged $3$-design is described in \cref{ap:C-inf-designs}, but it involves infinite-dimensional integration). The $3$-design condition was used to compute the variance $\variance(\hat o)$. 

One may wonder how well a rigged $2$-design works for shadow tomography. Since the variance calculation requires three copies of $\ket\chi\bra\chi$, the variance depends on the specific rigged $2$-design that is used. Here we will compute the variance with respect to the rigged $2$-design that uses the Kerred phase states; namely,
\begin{equation}
    \label{eq:shadows-rigged-2-design}
    \frac{1}{2\pi+1}\sum_{n\in\bbN_0}(\ket n \bra n)^{\otimes t} + \frac{2\pi}{2\pi+1}\int_{[0,2\pi]^2} (\prescript{}{\varphi}{\ket\theta}\!\bra\theta_\varphi)^{\otimes t} \frac{\dd{\theta}\dd{\varphi}}{2\pi} = \alpha_t\Pi_t
\end{equation}
for $t\in \set{1,2}$, where $\alpha_1=1$ and $\alpha_2 = 1/(\pi+1/2)$. The only term in the variance that is different is
\begin{salign}
    \int_X\Tr[\ket\chi\bra\chi \calO]^2 & \Tr(\rho\ket\chi\bra\chi)\dd{\mu(\chi)}\nonumber\\
    &= \frac{1}{2\pi+1}\sum_{n\in\bbN_0} (\bra n \calO \ket n)^2 \bra n \rho \ket n + \frac{1}{2\pi+1}\int_{[0,2\pi]^2} (\prescript{}{\varphi}{\bra\theta}\calO \ket\theta_\varphi)^2 \prescript{}{\varphi}{\bra\theta}\rho \ket\theta_\varphi \dd{\theta}\dd{\varphi}\\
    \begin{split}
    &= \frac{1}{2\pi+1}\sum_{n\in\bbN_0} \calO_{n,n}^2 \rho_{n,n} + \frac{1}{(2\pi+1)(2\pi)^3}\sum_{n_1,n_2,n_3\in\bbN_0}\sum_{m_1,m_2,m_3\in\bbN_0}\calO_{n_1,m_1} \calO_{n_2,m_2}\rho_{n_3,m_3}\\
    &\qquad \times \int_{[0,2\pi]^2} \e^{\i\theta(n_1+n_2+n_3-m_1-m_2-m_3)} \e^{\i\varphi(n_1^2+n_2^2+n_3^2-m_1^2-m_2^2-m_3^2)}  \dd{\theta}\dd{\varphi}
    \end{split}\\
    \begin{split}
    &= \frac{1}{2\pi+1}\sum_{n\in\bbN_0} \calO_{n,n}^2 \rho_{n,n} + \frac{1}{(2\pi+1)2\pi}\sum_{n_1,n_2,n_3\in\bbN_0}\sum_{m_1,m_2,m_3\in\bbN_0}\calO_{n_1,m_1} \calO_{n_2,m_2}\rho_{n_3,m_3}\\
    &\qquad \times \delta_{n_1+n_2+n_3,m_1+m_2+m_3}\delta_{n_1^2+n_2^2+n_3^2,m_1^2+m_2^2+m_3^2}.
    \end{split}
\end{salign}
Unfortunately, there is no obvious closed form simplification. We can however investigate specific cases. 
For example, consider the case when $\calO$ is diagonal in the Fock basis $\set{\ket n}$. Then this term simply becomes $\frac{1}{2} \angles{\calO^2} + \frac{1}{4\pi} (\Tr\calO)^2$. Hence, if we have a collection of $M$ observables $\calO_1,\dots,\calO_M$ that are each diagonal in the Fock state basis, then one needs only $\sim \log(M)\max_i (\Tr \abs{\calO_i})^2$ measurements of $\rho$ from the POVM defined by the rigged $2$-design to estimate $\angles{\calO_1}, \dots, \angles{\calO_M}$.

Perhaps a more interesting case is when $\calO = \ket{a}\bra{b} + \ket{b}\bra{a}$ for positive integers $a$ and $b$. Assume that $b>a$ and define $\Delta \coloneqq b - a > 0$. In this case, the term above becomes 
\begin{salign}
    \begin{split}
    &= \frac{1}{(2\pi+1)2\pi} \sum_{n_1,n_2,m_1,m_2\in \set{a,b}}\sum_{n_3,m_3\in\bbN_0} \calO_{n_1,m_1} \calO_{n_2,m_2}\rho_{n_3,m_3}\\
    &\qquad \times \delta_{n_1+n_2+n_3,m_1+m_2+m_3}\delta_{n_1^2+n_2^2+n_3^2,m_1^2+m_2^2+m_3^2}
    \end{split}\\
    \begin{split}
    &= \frac{1}{(2\pi+1)2\pi} \sum_{n_3,m_3\in\bbN_0}\big[\calO_{a,b}^2 \rho_{n_3,m_3}\delta_{2a+n_3,2b+m_3}\delta_{2a^2+n_3^2,2b^2+m_3^2} \\
    &\qquad + \calO_{b,a}^2 \rho_{n_3,m_3}\delta_{2b+n_3,2a+m_3}\delta_{2b^2+n_3^2,2a^2+m_3^2}\\
    &\qquad + 2\calO_{a,b}\calO_{b,a} \rho_{n_3,m_3}\delta_{n_3,m_3} \big] 
    \end{split}\\
    \begin{split}
    &= \frac{1}{(2\pi+1)2\pi} \big[ \sum_{n_3\in\bbN_0}\rho_{n_3,2(a-b)+n_3}\delta_{2a^2+n_3^2,2b^2+(2a+n_3-2b)^2} \\
    &\qquad + \sum_{n_3\in\bbN_0}\rho_{n_3,2(b-a)+n_3}\delta_{2b^2+n_3^2,2a^2+(2b+n_3-2a)^2}\\
    &\qquad + 2 \Tr \rho \big]
    \end{split}\\
    \begin{split}
    &= \frac{1}{\pi(2\pi+1)} + \frac{1}{(2\pi+1)2\pi} \big[ \sum_{n_3\geq 2\Delta}\rho_{n_3,n_3-2\Delta}\delta_{2a^2+n_3^2,2b^2+(n_3-2\Delta)^2} \\
    &\qquad + \sum_{n_3\in\bbN_0}\rho_{n_3,n_3+2\Delta}\delta_{2b^2+n_3^2,2a^2+(n_3+2\Delta)^2}\big]
    \end{split}\\
    &= \frac{1}{\pi(2\pi+1)} + \frac{1}{(2\pi+1)2\pi} \delta_{3b\geq a}\delta_{3a\geq b} \brackets{ 
    \rho_{(3b-a)/2,(3a-b)/2}  + \rho_{(3a-b)/2,(3b-a)/2} }\\
    &= \frac{1}{\pi(2\pi+1)}\parentheses{1+ \delta_{3b\geq a}\delta_{3a\geq b} \Re(\rho_{(3b-a)/2,(3a-b)/2}) }\\
    &\leq \frac{2}{\pi(2\pi+1)}.
\end{salign}
Hence, if we have a collection of $M$ observables of the form $\calO_i = \ket{a_i}\bra{b_i} + \ket{b_i}\bra{a_i}$, then we can accurately determine each $\angles{\calO_i}$ with only $\sim \log M$ measurements of $\rho$ using the rigged $2$-design.

\subsubsection{Using Hoeffding's inequality}

Again motivated by Ref.~\cite{acharyaInformationallyCompletePOVMbased2021}, we consider using Hoeffding's inequality and only using the $2$-design property. Hence, this section applies to rigged $2$-designs, of which we have constructed several. Specifically, suppose that we again consider estimating $\angles{\calO_j}$ with $N$ shadows by $\hat o_j = \frac{1}{N}\sum_{i=1}^N \Tr(\hat\rho_i \calO_j)$. If $-\infty < c < \Tr(\hat\rho_i \calO_j) < d < \infty$ almost surely for each shadow $\hat\rho_i$, then Hoeffding's inequality immediately implies that
\begin{equation}
    \Pr{}{\abs{\hat o_j - \Expval[\hat o_j]} \geq \epsilon} \leq 2\exp\bargs{-\frac{2N\epsilon^2}{(d-c)^2}}.
\end{equation}
Then, applying the union bound,
\begin{equation}
    \Pr{}{\max_j \abs{\hat o_j - \Expval[\hat o_j]} \geq \epsilon} \leq 2M\exp\bargs{-\frac{2N\epsilon^2}{(d-c)^2}}.
\end{equation}
Therefore, to achieve a failure probability of at most $\delta$, we need
\begin{equation}
    \label{eq:hoeffding-num-shadows}
    N\geq \log\pargs{\frac{2M}{\delta}}\frac{(d-c)^2}{2\epsilon^2}
\end{equation}
to compute the $M$ observables to additive accuracy $\epsilon$.

For instance, we consider the example from above where the observables are $\calO_j = \ket{a_j}\bra{b_j} + \ket{b_j}\bra{a_j}$ and we perform the shadows procedure with the rigged $2$-design given in \cref{eq:shadows-rigged-2-design}. One easily finds that $-2/\pi(\pi+1/2) \leq \Tr(\hat\rho_i \calO_j) \leq 2/\pi(\pi+1/2)$. Hence, we can determine the expectation value of the $M$ observables with error $\epsilon$ and failure probability at most $\delta$ with only $N\geq \log\pargs{\frac{2M}{\delta}}\frac{8}{\pi^2(\pi+1/2)^2\epsilon^2}$ measurements.

\subsubsection{Worked example}\label{subsec:worked-examples}

We now work through a simple, explicit example of using shadow tomography with the rigged $2$-design in \cref{eq:shadows-rigged-2-design} to determine the expectation value of $M$ observables with $\log M$ measurements. We let each observable be $\calO_j = \ket{a_j}\bra{b_j} + \ket{b_j}\bra{a_j} + \ket{c_j}\bra{c_j}$ for arbitrary nonnegative integers $a_j,b_j,c_j$. Suppose that we have access to a blackbox quantum device that prepares $\rho$, but we know nothing else about it.

\textbf{Generate shadows.} 
The first step is to describe a procedure to generate a classical shadow. Recall that single-qubit ``local Clifford'' shadows \cite{huangPredictingManyProperties2020} consist of choosing randomly between measuring in three different POVMs --- the three Bloch-sphere axes --- each yielding a binary outcome. In our case, for a single mode, we choose between measuring in either the discrete Fock-space POVM or a continuum of phase-state POVMs which differ by how much they have evolved under the Kerr Hamiltonian (quantified by \(\varphi\)). Each POVM has an infinite number of outcomes: the Fock-state POVM admits a countable infinity of outcomes indexed by Fock-state occupation number \(n\), while the phase-state POVMs have a compact continuous set of outcomes indexed by phase-state index \(\theta\).

From \cref{eq:shadows-rigged-2-design}, we generate a shadow as follows. First we draw a random number $x$ between $0$ and $1$. If $x\leq 1/(2\pi+1)$, then we measure $\rho$ in the Fock state basis $\set{\ket n \mid n\in\bbN_0}$. The result will be an integer $n \in \bbN_0$ and the classical shadow is then a classical label representing the operator $\hat\rho^{(n)} \coloneqq (2\pi+1)\ket n \bra n - \bbI$. If, on the other hand, $x > 1/(2\pi+1)$, then we draw a random number $\varphi$ between $0$ and $2\pi$ and measure $\rho$ with the continuous POVM defined by the operators $\{\prescript{}{\varphi}{\ket\theta}\bra\theta_\varphi \mid \theta\in[0,2\pi)\}$ and the measure $\dd{\theta}$. The output of such a measurement is an angle $\theta \in [0,2\pi)$ and the classical shadow is then a classical label representing the operator $\hat\rho^{(\theta,\varphi)} \coloneqq (2\pi+1) \prescript{}{\varphi}{\ket\theta}\bra\theta_\varphi - \bbI$.

\textbf{Classically compute expectation values w.r.t.~shadows.} 
For the shadow $\hat\rho^{(n)}$, we easily see that
\begin{equation}
    \Tr[\hat\rho^{(n)}\calO_j] = \Tr\bargs{((2\pi+1)\ket n \bra n - \bbI)\calO_j} = (2\pi+1)\delta_{n,c_j}-1.
\end{equation}
For the shadow $\hat\rho^{(\theta,\varphi)}$, we compute
\begin{salign}
    \Tr[\hat\rho^{(\theta,\varphi)}\calO_j]
    &= \Tr\bargs{((2\pi+1)\prescript{}{\varphi}{\ket\theta}\!\bra\theta_\varphi - \bbI)\calO_j}\\
    &= \frac{2\pi+1}{2\pi}\sum_{n,m\in\bbN_0}\e^{\i\theta(n-m)+\i\varphi(n^2-m^2)} \bra m \calO_j \ket n  - 1\\
    &= \frac{2\pi+1}{2\pi}\parentheses{
    \e^{\i\theta(a_j-b_j)+\i\varphi(a_j^2-b_j^2)}+\e^{\i\theta(b_j-a_j)+\i\varphi(b_j^2-a_j^2)} + 1
    } - 1\\
    &= (2+1/\pi)\cos(\theta(a_j-b_j)+\varphi(a_j^2-b_j^2)) + \frac{1}{2\pi}.
\end{salign}

\textbf{Choose the number of shadows to generate.}
We see that for every possible shadow $\hat\rho$ and every observable $\calO_j$, $-2-1/\pi + 1/2\pi \leq \Tr[\hat\rho\calO_j] \leq 2\pi$. Therefore, from \cref{eq:hoeffding-num-shadows}, we set 
\begin{equation}
    N = \ceil{\log\pargs{\frac{2M}{\delta}} \frac{(2\pi+2+1/\pi-1/2\pi)^2}{2\epsilon^2}} \approx \frac{36}{\epsilon^2}\log\pargs{\frac{2M}{\delta}}.
\end{equation}

\textbf{Estimate expectation values w.r.t.~state.} With all this in place, we can now classically compute each $\angles{\calO_j}$ to a maximum additive error of $\epsilon$ with success probability at least $1-\delta$. First, generate $N$ shadows with the procedure described above. Then, with those $N$ shadows, classically compute the mean expectation value of each observable $\calO_j$ over the $N$ shadows using the expressions derived above for $\Tr[\hat\rho^{(n)}\calO_j]$ and $\Tr[\hat\rho^{(\theta,\varphi)}\calO_j]$. With probability at least $1-\delta$, all of these $M$ means will be within $\epsilon$ of the true expectation values with respect to $\rho$.

\subsection{Fidelity calculations}
\label{ap:average-fidelity}

In this subsection, we derive the calculations shown in \cref{sec:fidelity}. Throughout this subsection, we let $\calE$ denote an $R$-regularized rigged $2$-design, meaning that $\calE$ is an ensemble over unit-normalized quantum states satisfying $\Expval_{\psi\in\calE} \parentheses{\ket\psi\bra\psi}^{\otimes 2} = \frac{\Pi_2^{(R)}}{\Tr \Pi_2^{(R)}}$. We assume that $R$ is positive semi-definite. Recall then that $\Pi_2^{(R)} = (R\otimes R) \Pi_2 (R \otimes R)$, and $\Pi_2 = \frac{1}{2}(\bbI + S)$. Therefore, $2\Tr\Pi_2^{(R)} = (\Tr R^2)^2 + \Tr R^4$. From this characterization, one easily computes that
\begin{salign}
    \Expval_{\psi\in\calE} \bra\psi A \ket\psi \ket\psi\bra\psi
    &= \Tr_1\bargs{ (A \otimes \bbI)\Expval_{\psi\in\calE} \ket\psi\bra\psi \otimes  \ket\psi\bra\psi}\\
    &= \frac{1}{2\Tr\Pi_2^{(R)}} \Tr_1\bargs{ (A \otimes \bbI) \parentheses{R^2 \otimes R^2 + (R^2 \otimes R^2) S } }\\
    &= \frac{1}{2\Tr\Pi_2^{(R)}} \brackets{ R^2 \Tr(R AR) + R^2 A R^2}\\
    &= \frac{R^2 \Tr(R AR) + R^2 A R^2}{(\Tr R^2)^2 + \Tr R^4}.\label{eq:expval-A}
\end{salign}
Furthermore,
\begin{salign}
    \Expval_{\psi\in\calE} \bra\psi A \ket\psi \bra\psi B \ket\psi
    &= \Tr\bargs{B \Expval_{\psi\in\calE} \bra\psi A \ket\psi \ket\psi\bra\psi}\\
    &= \Tr \frac{B R^2 \Tr(R AR) + B R^2 A R^2}{(\Tr R^2)^2 + \Tr R^4}\\
    &= \frac{\Tr(R B R) \Tr(R AR) + \Tr(R B R^2 A R)}{(\Tr R^2)^2 + \Tr R^4}.\label{eq:expval-A-B}
\end{salign}

We now study definitions of fidelity. We now assume that $R$ is diagonal in the $\hat n$ basis. We define a continuous-variable version of a maximally-entangled state as \cite{AS17}
\begin{equation}
    \ket{\phi_R} \coloneqq \frac{1}{\sqrt{\Tr R}} (R^{1/4} \otimes R^{1/4}) \sum_{n=0}^\infty \ket n \otimes \ket n.
\end{equation}
When $R = \e^{-\beta \hat n}$, $\ket{\phi_R}$ is a two-mode squeezed vacuum state; when $R = P_d$, $\ket{\phi_R}$ is a finite dimensional maximally entangled state. Define its reduced state on one mode by 
\begin{equation}
    \rho_R \coloneqq \Tr_2 \ket{\phi_R}\bra{\phi_R} = R / \Tr R.
\end{equation}

Let $\calD$ be a quantum channel with Kraus operators $K$ so that $\calD(\rho) = \sum_K K \rho K^\dag$. In analogy with the finite dimensional case, define the entanglement fidelity as
\begin{salign}
    F_e^{(R)}(\calD) 
    &\coloneqq \bra{\phi_R}(\calI \otimes \calD)(\phi_R) \ket{\phi_R}\\
    \begin{split}
    &= \frac{1}{(\Tr R)^2}\sum_K \sum_{n,m,j,k} (\bra n \otimes \bra n R^{1/2}) (\bbI \otimes K) (\ket m \otimes R^{1/2}\ket m) \\
    &\qquad\qquad \times (\bra j \otimes \bra j R^{1/2}) (\bbI \otimes K^\dag) (\ket k \otimes R^{1/2}\ket k)
    \end{split}\\
    &= \frac{1}{(\Tr R)^2} \sum_K \sum_{n,j} \bra n R^{1/2}K R^{1/2}\ket n \bra j R^{1/2}K^\dag R^{1/2}\ket j\\
    &= \sum_K \abs{\Tr(\rho_R K)}^2.\label{eq:ent-fid-kraus-operators}
\end{salign}
Furthermore, in analogy with the finite dimensional case, we define two ``average fidelity'' quantities,
\begin{salign}
    \overline F_1^{(R)}(\calD) &\coloneqq \frac{\Tr R^4 + (\Tr R^2)^2}{\Tr R^2 + (\Tr R)^2} \Expval_{\psi\in \calE} \bra\psi R^{+} \calD(\psi) R^{+}\ket\psi ,\\
    \overline F_2^{(R)}(\calD) &\coloneqq \Expval_{\psi\in \calE} \bra\psi \calD(\psi) \ket\psi.
\end{salign}
We again emphasize that these definitions are independent of which $R$-regularized rigged $2$-design $\calE$ is used since they involve only two copies of $\ket\psi$ and two copies of $\bra\psi$. Notice that when $R = P_d$, since the Moore-Penrose inverse $R^{+}$ of a projector is itself, we find that $\overline F_1^{(P_d)} = \overline F_2^{(P_d)}$.

By \cref{eq:expval-A-B}, we immediately find that
\begin{salign}
    \overline F_1^{(R)}(\calD)
    &= \frac{\Tr R^4 + (\Tr R^2)^2}{\Tr R^2 + (\Tr R)^2} \sum_K \Expval_{\psi\in \calE} \bra\psi R^{+} K \ket\psi\bra\psi K^\dag R^{+}\ket\psi\\
    &= \frac{\Tr R^4 + (\Tr R^2)^2}{\Tr R^2 + (\Tr R)^2} \sum_K \frac{\Tr(R K^\dag R^+ R) \Tr(R R^+ K R) + \Tr(R K^\dag R^+ R^2 R^+ K R)}{(\Tr R^2)^2 + \Tr R^4}\\
    &=  \frac{\sum_K \frac{d_R}{(\Tr R)^2}\Tr(R K^\dag R^+ R) \Tr(R R^+ K R) + \Tr(R^+ R^2 R^+ \calD(\rho_{R^2}))}{d_R + 1},
\end{salign}
where we define an effective dimension $d_R \coloneqq (\Tr R)^2 / \Tr R^2$. Since we are assuming $R$ to be diagonal, $RR^+ = R^+ R$. Furthermore, by definition of the Moore-Penrose inverse, $RR^+ R = R$ and $R^+ R R^+ = R^+$. Therefore,
\begin{salign}
    \overline F_1^{(R)}(\calD)
    &=  \frac{d_R \sum_K\abs{\Tr(\rho_R K)}^2 + \Tr(R R^+ \calD(\rho_{R^2}))}{d_R + 1}\\
    &=  \frac{d_R F_e^{(R)}(\calD) + \Tr(R R^+ \calD(\rho_{R^2}))}{d_R + 1}.\label{eq:f2-fe}
\end{salign}
We perform a similar calculation for $\overline F_2^{(R)}$,
\begin{salign}
    \overline F_2^{(\sqrt{R})}(\calD)
    &= \sum_K \Expval_{\psi \in \calE} \bra \psi K \ket\psi \bra\psi K^\dag \ket\psi\\
    &= \sum_K \frac{\Tr(R^{1/2} K^\dag R^{1/2}) \Tr(R^{1/2} KR^{1/2}) + \Tr(R^{1/2} K^\dag R K R^{1/2})}{(\Tr R)^2 + \Tr R^2}\\
    &= \frac{\sum_K\abs{\Tr(R K)}^2 + \sum_K\Tr(K R K^\dag R)}{(\Tr R)^2 + \Tr R^2}\\
    &= \frac{d_R\sum_K\abs{\Tr(\rho_R K)}^2 + d_R\sum_K\Tr(K \rho_R K^\dag \rho_R)}{d_R + 1}\\
    &= \frac{d_R F_e^{(R)}(\calD) + d_R\Tr(\calD(\rho_R) \rho_R)}{d_R + 1}.\label{eq:f2-fe-r}
\end{salign}

When $R = P_d$ is the projector and $\calD$ is trace-preserving on the restricted $d$-dimensional subspace, both relations reduce to the finite dimensional relation. When $R$ is invertible, such as the case when $R = \e^{-\beta \hat n}$, we find
\begin{equation}
    \label{eq:f1-fe-r}
    \overline F_1^{(R)}(\calD) = \frac{d_R F_e^{(R)}(\calD) + 1}{d_R + 1}.
\end{equation}

\subsubsection{Loss channel}
\label{ap:loss-channel}

We now compute the various average fidelity quantities for the pure-loss channel $\calL^\kappa$ defined in \cref{sec:loss-channel} and shown in \cref{fig:loss}. From \cite[eq.~4.6]{ivan_operator-sum_2011}, the Kraus operators for $\calL^\kappa$ are
\begin{equation}
    K_i = \sum_{m=0}^\infty \sqrt{\binom{m+i}{i}}(1-\kappa^2)^{i/2}\kappa^m \ket m \bra{m+i}
\end{equation}
for $i \in \bbN_0$.

We begin with $\overline F_{\rm coh}^{(\bar n)}(\calL^\kappa)$. Let $\ket\alpha$ be the coherent state specified by $\alpha\in \bbC$. Then, as calculated in \cite{sharmaCharacterizingPerformanceContinuousvariable2020},
\begin{salign}
    \overline F_{\rm coh}^{(\bar n)} 
    &= \frac{1}{\pi \bar n}\int_\bbC \e^{-\abs{\alpha}^2/\bar n} \bra\alpha \calL^\kappa(\ket\alpha\bra\alpha) \ket\alpha \Dd{2}{\alpha}\\
    &= \frac{1}{1+\bar n(1-\kappa)^2}.
\end{salign}
Next, we consider the entanglement fidelity $F_e^{(R)}(\calL^\kappa)$. Let $\ket{\phi_R} = \frac{1}{\sqrt{\Tr R}}(R^{1/4}\otimes R^{1/4}) \sum_{n=0}^\infty \ket n \otimes \ket n$ and assume that $R$ is diagonal in the $\ket n$ basis. Then from \cref{eq:ent-fid-kraus-operators},
\begin{equation}
    F_e^{(R)}(\calL^\kappa)
    = \frac{1}{(\Tr R)^2}\sum_{i=0}^\infty \abs{\Tr(R K_i)}^2.
\end{equation}
When $R = R_\beta = \e^{- \beta \hat n}$, one easily finds this to be
$\frac{(\e^\beta - 1)^2}{(\e^\beta - \kappa)^2}$.
Recall that in \cref{sec:loss-channel} we required that $d_{R_\beta} \coloneqq (\Tr R_\beta)^2 / \Tr R_\beta^2 = 1+2\bar n$. Solving for $\beta$, we find that $\e^\beta = 1+1/\bar n$, and therefore
\begin{equation}
    F_e^{(R_\beta)}(\calL^\kappa) = \parentheses{ 1 + \bar n(1- \kappa) }^{-2}.
\end{equation}
On the other hand, when $R = P_d = \sum_{n=0}^{d-1} \ket n \bra n$, we find
\begin{equation}
    F_e^{(P_d)}(\calL^\kappa) = \frac{\parentheses{ 1-\kappa ^{d}}^2}{(1-\kappa )^2 d^2}.
\end{equation}

From \cref{eq:f1-fe-r}, $\overline F_1^{(R_\beta)}(\calL^\kappa)$ is the same as $F_e^{(R_\beta)}(\calL^\kappa)$ up to an offset. However, $\overline F_2^{(R_\beta)}(\calL^\kappa)$ is not as simple. Indeed, from \cref{eq:f2-fe-r}, we must compute
\begin{salign}
    \Tr [\calL^\kappa(\rho_{R_\beta}) \rho_{R_\beta} ]
    &= \frac{1}{(\Tr R_\beta)^2} \sum_i \Tr(K_i R_\beta K_i^\dag R_\beta)\\
    &= (1=\e^{-\beta})^2 \sum_i \sum_{a,b} \e^{-\beta (a+b)} \bra a K_i \ket b \bra b K_i^\dag  \ket a\\
    &= (1-\e^{-\beta})^2 \sum_{a\leq b} \e^{-\beta (a+b)} \binom{b}{b-a} (1-\kappa^2)^{b-a} \kappa^{2a}\\
    &= \frac{e^{\beta }-1}{e^{\beta }+\kappa ^2}.
\end{salign}
Therefore, from \cref{eq:f2-fe-r}, 
\begin{salign}
    \overline F_2^{(R_\beta)}(\calL^\kappa)
    &= \frac{d_{R_{2\beta}}}{d_{R_{2\beta}} + 1} \brackets{ d_{R_{2\beta}} F_e^{(R_{2\beta})}(\calL^\kappa)  + \frac{e^{\beta }-1}{e^{\beta }+\kappa ^2} }\\
    &= \frac{1}{d_{R_{2\beta}} + 1} \brackets{ \frac{(\e^{2\beta} - 1)^2}{(\e^{2\beta} - \kappa)^2}  + \frac{e^{2\beta }-1}{e^{2\beta }+\kappa ^2} }\\
    &= \frac{1}{\tanh\beta + 1} \brackets{ \frac{(\e^{2\beta} - 1)^2}{(\e^{2\beta} - \kappa)^2}  + \frac{e^{2\beta }-1}{e^{2\beta }+\kappa ^2} }.
\end{salign}
Requiring that $d_{R_\beta} = 1+2\bar n$ yields $\e^\beta = 1+1/\bar n$, giving
\begin{equation}
    \overline F_2^{(R_{\beta/2})}(\calL^\kappa)
    = \frac{(2 \bar n+1) \left((1-\kappa )^2 \bar n+2\right)}{2 ((1-\kappa ) \bar n+1)^2 \left(\left(\kappa ^2+1\right) \bar n+1\right)}.
\end{equation}

Finally, we compute $\overline F_{1,2}^{(P_d)}(\calL^\kappa) \equiv \overline F_1^{(P_d)}(\calL^\kappa) = \overline F_2^{(P_d)}(\calL^\kappa)$. From \cref{eq:f2-fe}, it only remains to compute $\Tr(\calL^\kappa(P_d/d) P_d)$, which is
\begin{salign}
    \Tr(\calL^\kappa(P_d/d) P_d)
    &= \frac{1}{d}\sum_{i=0}^\infty \sum_{a,b=0}^{d-1} \bra a K_i \ket b \bra b K_i^\dag  \ket a\\
    &= \frac{1}{d} \sum_{a\leq b=0}^{d-1} \binom{b}{b-a} (1-\kappa^2)^{b-a} \kappa^{2a}\\
    &= 1.
\end{salign}
Indeed, this is $1$ since $\calL^\kappa$ does not take a state that is defined in the subspace $P_d$ out of that subspace.
Using $d = 1+\bar n$, we find
\begin{equation}
    F_{1,2}^{(P_{\bar n + 1})}(\calL^\kappa) = \frac{\parentheses{ 1-\kappa ^{\bar n+1}}^2}{(1-\kappa )^2 (\bar n+1)(\bar n+2)} + \frac{1}{\bar n+2}.
\end{equation}

The plots of all of these fidelities as functions of $\kappa$ are shown in \cref{fig:loss}.

\section{Torus designs, trigonometric cubature, and mutually unbiased bases}
\label{ap:torus}

In this section, we prove the equivalence between our definition of a torus $t$-design (c.f.~\cref{def:torus-design}) and the definition given in Ref.~\cite{kuperberg_numerical_2004}. In Ref.~\cite{kuperberg_numerical_2004}, Kuperberg defines a general notion of torus cubature that generalizes the more established theory of trigonometric cubature \cite{cools_constructing_1997,cools_minimal_1996}. We will be interested in one of the cases of his definition; namely, our definition of a $T^{n+1}$ $t$-design is equivalent to his definition of a positive degree $t$ cubature rule on $T(\PSU(n+1))$. After showing this, we will compare a torus design to the more standard trigonometric cubature rules and find that a torus $t$-design lies somewhere between a degree $t$ and degree $2t$ positive trigonometric cubature rule. Finally, we prove a relationship between torus $2$-designs and complete sets of mutually unbiased bases.

\subsection{Equivalence to Kuperberg's definition}

We begin by describing Kuperberg's definition. Consider a group $\calT$ that is isomorphic to the torus $\calT \cong T^n = (S^1)^n$. 
Suppose $\rho\colon \calT \to \GL(V)$ is a free linear representation, with $V$ a real vector space $V \cong \bbR^N$. Since $\rho$ is free it follows that there is one or many faithful orbits $\calO$. Suppose $u \in V$ such that $\calO = \set{\rho(g) u \mid g \in \calT}$ is a faithful orbit. Since $\calO$ is faithful, $\calT$ can be identified with $\calO$ via $\calT \ni g \leftrightarrow \rho(g) u \in \calO$. With this identification, $\calT$ inherits an algebraic structure, since it is well-defined to consider addition such as $\rho(g)u + \rho(h)u \in V$ for every $g, h \in \calT$. With this structure, along with the unit normalized Haar measure on $\calT$ (since $\calT$ is compact), we can define cubature on $\calT$ as follows. A set $S \subset \calT$ and weight function $v\colon S \to \bbR_{> 0}$ is a (positive) cubature rule of degree $t$ on $\calT$ if
\begin{equation}
    \sum_{h \in S} v(h) f(\rho(h)u) = \int_\calT f(\rho(g)u) \dd{\mu(g)}
\end{equation}
for any polynomial $f\colon V \to \bbR$ of degree $t$ or less. Since $V\cong \bbR^N$, we can therefore view $f$ as a being a function of the entries of the vectors $\rho(g)u$. Kuperberg also states that this definition is independent of $u$ for generic choices of $u$ as long as the resulting $\calO$ is faithful \footnote{Personal communication with Greg Kuperberg.}.

We now apply this to $\calT = T(\PSU(n+1))$. $\PSU(n+1)$ is the projective special unitary group of $(n+1)\times (n+1)$ matrices defined by $\SU(n+1) / \U(1)$. Then $\calT = T(\PSU(n+1))$ is a maximal torus (maximal, compact, connected, abelian Lie subgroup) of $\PSU(n+1)$, which is the group of diagonal unitary matrices with determinant $1$ modulo the center of $\SU(n+1)$ (i.e.~modulo global phases). For a unitary $U \in \calT$, let $U_{ij}$ denote the entry in the $i^{\rm th}$ row and $j^{\rm th}$ column. The determinant condition implies that $U_{n+1,n+1}$ is uniquely determined by $U_{ii}$ for $i=1,\dots, n$. We therefore see that $\calT \cong T^n$. We can also take an alternative view of $\calT$; we can view $\calT$ as the group of diagonal unitary matrices with $U_{n+1,n+1} = 1$ modulo the center. This is the view we will take. Below we will consider the adjoint action of this group, and therefore we do not have to worry about modding out the center; ultimately, we will just end up integrating out global phases.

We let $N = 2(n+1)^2$ and identify $V$ with the vector space of $(n+1)\times (n+1)$ complex matrices $\bbC^{(n+1)\times (n+1)}$. We consider a linear action of $\calT$ defined by conjugation on $V$; in other words, $\rho(g)$ is defined by $A \mapsto g A g^\dag$. As mentioned, we can pick any $u\in V$ as a base point as long as the resulting orbit,
\begin{equation}
    \calO = \set{
    \begin{pmatrix} \e^{\i\phi_1}\\& \ddots\\&&\e^{\i\phi_n}\\&&&1 \end{pmatrix}
    u
    \begin{pmatrix} \e^{-\i\phi_1}\\& \ddots\\&&\e^{-\i\phi_n}\\&&&1 \end{pmatrix}
    \mid \phi_1,\dots,\phi_n \in [-\pi,\pi)
    },
\end{equation}
is faithful. We pick $u$ to be the matrix of all $1$'s, $u = \begin{pmatrix}
1&\dots&1\\
\vdots&\ddots&\vdots\\
1&\dots & 1
\end{pmatrix}$. One can then easily check that 
\begin{equation}
    \calO = \set{v \text{ where } v_{ij} = \e^{\i(\phi_i-\phi_j)} \mid \phi_1, \dots, \phi_n \in [-\pi,\pi), \phi_{n+1} = 0}.
\end{equation}
A degree $t$ positive cubature rule on $\calT$ is a set $S \subset \calT$ and weight function $v\colon S \to \bbR_{>0}$ that satisfies
\begin{equation}
    \sum_{h \in S} v(h) f(huh^\dag) = \int_{[-\pi,\pi)^n} f(gug^\dag) \frac{\dd{\phi_1}\dots \dd{\phi_n}}{(2\pi)^n}.
\end{equation}
$f$ is a polynomial of degree at most $t$ in the entries. By linearity, we can consider $f$ to be a monomial. From $\calO$, we consider monomials of degree $\leq t$ in the variables
\begin{equation}
    \set{\e^{\i(\phi_i-\phi_j)} \mid i,j=1,\dots,n+1} 
\end{equation}
where recall that $\phi_{n+1} = 0$. It follows that an equivalent definition of a degree $t$ positive cubature rule on $\calT$ is as follows. Let $S \subset [-\pi,\pi)^n$ and $v \colon S \to \bbR_{>0}$. Hence, for each $\phi \in S$, $\phi_i \in [-\pi,\pi)$, and we define $\phi_{n+1} = 0$. Then $(S,v)$ must satisfy
\begin{equation}
    \forall j_1,\dots, j_t,k_1,\dots, k_t \in \set{1,\dots, n+1} \colon ~\sum_{\theta \in S} v(\theta) \e^{\i(\theta_{j_1} + \dots + \theta_{j_t}-\theta_{k_1} - \dots - \theta_{k_t})} = \int_{[-\pi,\pi)^n} \e^{\i(\phi_{j_1} + \dots + \phi_{j_t}-\phi_{k_1} - \dots - \phi_{k_t})} \frac{\dd{\phi_1}\dots \dd{\phi_n}}{(2\pi)^n}.
\end{equation}
Notice that this takes care of \textit{all} monimials of degree $t$ or less. For example, consider the monomial $\e^{\i\phi_1}$. This is taken care of by setting $j_1=1$ and $j_2=\dots=j_t = k_1=\dots=k_t = n+1$. 

We easily see that the right hand side (i.e.~the integral) does not change if we integrate over $\phi_{n+1}$ instead of just fixing it to be $0$. Similarly, on the left hand side, for every $\theta \in S$, we can shift each $\theta_i$ by a constant $\theta_i \mapsto \theta_i + c$ without changing anything. Therefore, we can remove the definition that $\theta_{n+1} = 0$, and instead allow $\theta_{n+1}$ to be arbitrary. Thus, we arrive at an equivalent definition of a degree $t$ positive cubature rule on $\calT$ as follows. Let $S \subset [-\pi,\pi)^{n+1}$ and $v \colon S \to \bbR_{>0}$. Then $(S, v)$ must satisfy
\begin{equation}
    \sum_{\theta \in S}v(\theta) \prod_{i=1}^t \e^{\i(\theta_{j_i} - \theta_{k_i})} = \int_{T^m} \prod_{i=1}^t \e^{\i(\phi_{j_i} - \phi_{k_i})} \frac{\dd{\phi_1}\dots \dd{\phi_{n+1}}}{(2\pi)^{n+1}}.
\end{equation}
Notice that this is exactly our definition of an $(n+1)$-torus $t$-design per \cref{def:torus-design}. 

In conclusion, our definition of a $T^{n+1}$ $t$-design is equivalent to Kuperberg's definition of a degree $t$ positive cubature rule on $T(\PSU(n+1))$.

\subsection{Comparison to standard trigonometric cubature}

A degree $t$ positive trigonometric cubature rule $(S,v)$ on $T^n$ must satisfy
\begin{equation}
    \sum_{\theta \in S}v(\theta) \prod_{i=1}^n \e^{\i \alpha_i \theta_{i}} = \int_{T^n} \prod_{i=1}^n \e^{\i \alpha_i \phi_{i}} \dd{\phi}
\end{equation}
whenever $\sum_{i=1}^n \abs{\alpha_i} \leq t$. We see that our $T^n$ $t$-designs lie somewhere between a degree $t$ and degree $2t$ trigonometric cubature rule. To see the former, we show that a torus $t$-design must also be a degree $t$ trigonometric cubature rule. From the definition of torus designs,
\begin{equation}
    \sum_{\theta \in S}v(\theta) \prod_{i=1}^t \e^{\i(\theta_{a_i} - \theta_{b_i})} = \int_{T^n} \prod_{i=1}^t \e^{\i(\phi_{a_i} - \phi_{b_i})} \dd{\phi}.
\end{equation}
Suppose we consider a monimial $\prod_{i=1}^n \e^{\i \alpha_i \theta_{i}}$. If $\sum_i \abs{\alpha_i} \leq t$, then we can generate the monomial via a choice of $a_i$ and $b_i$. Indeed, recall that without loss of generality we can assume that $\theta_{n}=0$. Consider as an example $t=3$, $n=4$, and the task of generating the monoimal defined by $\alpha = (2,0,-1,0)$. Then we set $a_1 = 1$, $a_2 = 1$, $a_3=4$, $b_1=3$, and $b_2 = b_3 = 4$. Then $\prod_{i=1}^t \e^{\i(\theta_{a_i} - \theta_{b_i})} = \e^{\i(\theta_1+\theta_1 + \theta_4-\theta_3-\theta_4-\theta_4)}$, which is exactly $\prod_i\e^{\i\alpha_i \theta_i}$ since $\theta_4 = 0$. Hence, by a proper choice of $a_i$ and $b_i$, any monomial of degree $t$ or less can be generated by $\prod_{i=1}^t \e^{\i(\theta_{a_i} - \theta_{b_i})}$, and hence a torus $t$-design is also a degree $t$ trigonometric cubature rule.

On the contrary, if we allow $\sum_i \abs{\alpha_i} > t$, we find that there are some monomials that cannot be generated by a sufficient choice of $a_i$ and $b_i$. So even though a torus $t$-design involves monomials of degree up to $2t$, it is not in general a degree $2t$ trigonometric cubature rule. However, since torus $t$-designs only involve certain monomials up to degree $2t$, a trigonometric cubature rule of degree $\geq 2t$ is a torus $t$-design.

\subsection{Relation to MUBs}

We begin by recalling the definition of a complete set of mutually unbiased bases (MUBs) \cite{durtMutuallyUnbiasedBases2010}.

\medskip

\begin{definition}[Complete set of MUBs]
Suppose that $B_0, \dots, B_n$ are each orthonormal bases of $\bbC^n$. $B_i$ and $B_j$ are called \emph{mutually unbiased} if
\begin{equation}
    \forall \ket\psi \in B_i, \ket\phi \in B_j\colon~~ \abs{\braket{\psi\vert\phi}}^2 = 1/n.
\end{equation}
The collection $B_0, \dots, B_n$ is called a \emph{complete set of MUBs} if the bases are pairwise mutually unbiased. This can be equivalently stated in term of the phases $\theta^i_{j,k}$ involved in the bases (see below);
\begin{enumerate}
    \item (Orthonormality). $\forall i,j,k \in \set{0,\dots, n-1}\colon$  $\frac{1}{n}\sum_{l=0}^{n-1} \e^{\i(\theta^i_{j,l}-\theta^i_{k,l})} = \delta_{jk}$;
    \item (Mutually unbiasedness). $\forall i\neq j, k, m \in \set{0,\dots, n-1}\colon$  $\abs{\sum_{l=0}^{n-1} \e^{\i(\theta^i_{k,l}-\theta^j_{m,l})}}^2 = n$.
\end{enumerate}
\end{definition}

\medskip

We now show the relationship between complete sets of MUBs and torus $2$-designs. Recall the matrix $\Pi_2$ that, for any orthonormal basis $\set{\ket 0, \dots, \ket{n-1}}$, has matrix elements 
\begin{equation}
    \Pi_2(a,b;c,d) \coloneqq \bra{a}\otimes \bra{b} \Pi_2 \ket{c}\otimes \ket d = \frac{1}{2}(\delta_{ac}\delta_{bd} - \delta_{ad}\delta_{bc}).
\end{equation}
By simply doing the integration, one finds that $S$ is an equal weight (i.e.~$v(\theta) = 1/\abs{S}$) $T^n$ $2$-design if and only if
\begin{equation}
    \frac{1}{\abs{S}}\sum_{\theta \in S} \e^{\i(\theta_a+\theta_b-\theta_c-\theta_d)} = 2\Pi_2(a,b;c,d)-\delta_{ab}\delta_{ac}\delta_{ad}.
\end{equation}
Here we show a connection between equal weight torus $2$-designs and complete sets of MUBs.

\medskip

\begin{lemma}
The phases of a complete set of MUBs on $\bbC^n$ form an equal weighted $n$-torus $2$-design of size $n^2$.
\end{lemma}
\begin{proof}
Without loss of generality, we can assume that one of the bases is the computational basis. So assume that $B_0 = \{\ket 0, \dots, \ket {n-1} \}$. Then in order for $\lvert \braket{\psi \vert j} \rvert = 1/\sqrt n$ for each $j \in \{0,\dots, n-1 \}$, $\ket\psi \in B_i$, and $i \in \{1,\dots, n\}$, it must be that each other basis $B_i$ must only involve uniform superposition states over the computational basis. With this in mind, define $\ket{\psi_j^i}$ so that
\begin{equation}
B_i = \{\ket{\psi_j^i} \mid j \in \{1,\dots, n \} \}.
\end{equation}
Define $\theta^i_{j,k}$ so that
\begin{equation}
\ket{\psi_j^i} = \frac{1}{\sqrt n}\sum_{k=0}^{n-1} \mathrm{e}^{\mathrm i \theta^i_{j,k}} \ket k.
\end{equation}
From \cite{klappenecker2005mutually}, we know that the complete set of MUBs forms a complex-projective $2$-design. Therefore, $D = \{\ket 0,\dots,\ket{n-1} \} \cup \{\ket{\psi_j^i} \mid i,j \in \{ 1,\dots, n\} \}$ is a complex-projective $2$-design. We therefore find that
\begin{equation}
\frac{1}{n(n+1)}\left(\sum_{k=0}^{n-1} (\ket{k}\bra k)^{\otimes 2} + \sum_{i,j=1}^n (\ket{\psi^i_j}\bra{\psi^i_j})^{\otimes 2} \right) = \frac{2}{n(n+1)}\Pi_2.
\end{equation}
Let $a, b,c,d \in \{0,\dots, n-1 \}$. Applying $\bra a \otimes \bra b$ on the left hand side and $\ket c \otimes \ket d$ on the right hand side, we find
\begin{equation}
\delta_{ab}\delta_{ac}\delta_{ad} + \frac{1}{n^2}\sum_{i,j=1}^n \mathrm e^{\mathrm i(\theta^i_{j,a}+\theta^i_{j,b}-\theta^i_{j,c}-\theta^i_{j,d})} = 2\Pi_2(a,b;c,d).
\end{equation}
Per the definition of an $n$-torus $2$-design from above, we see that the angles $\theta^i_{j,k}$ form an $n$-torus $2$-design with size $n^2$.
\end{proof}

\medskip

Furthermore, Theorem~3.3 from Ref.~\cite{royWeightedComplexProjective2007} states that if $B_0, \dots, B_m$ are each orthonormal bases of $\bbC^n$ and $\bigcup_i B_i$ is an unweighted complex-projective $2$-design, then $m=n$ only if the bases are mutually unbiased. The following lemma therefore follows.

\medskip

\begin{lemma}
If an equal weighted $n$-torus $2$-design exists such that the phases in the design define $n$ orthonormal bases, then there exists a complete set of MUBs in $\bbC^n$.
\end{lemma}

\medskip

Therefore, we have an if and only if.

\medskip

\begin{corollary}
    There exists a complete set of MUBs in $\bbC^n$ if and only if there exists an equal weighted $n$-torus $2$-design such that the phases in the design define $n$ orthonormal bases. Concretely, there exists a complete set of MUBs in $\bbC^n$ if and only there exists angles $\theta^i_{j,k}$ such that
    \begin{enumerate}
        \item $\forall i,j,k \in \set{0,\dots, n-1}\colon$  $\frac{1}{n}\sum_{l=0}^{n-1} \e^{\i(\theta^i_{j,l}-\theta^i_{k,l})} = \delta_{jk}$;
        \item $\forall a,b,c,d\in \set{0,\dots, n-1}\colon$ $\frac{1}{n^2}\sum_{i,j=0}^{n-1} \e^{\i(\theta^i_{j,a}+\theta^i_{j,b}-\theta^i_{j,c}-\theta^i_{j,d})} = \begin{cases}
        1&\text{if } (a=c \text{ and } b=d) \text{  or  } (a=d \text{ and } b=c)\\
        0&\text{otherwise}
        \end{cases}$.
    \end{enumerate}
\end{corollary}

\medskip

In summary, the definition of a complete set of MUBs has two conditions: orthonormality and mutually unbiasedness. We have shown that the mutually unbiased condition can be replaced with the condition that the phases must form a torus $2$-design of size exactly $n^2$.

\twocolumngrid
\bibliography{references}

\begin{thebibliography}{137}%
\makeatletter
\providecommand \@ifxundefined [1]{%
 \@ifx{#1\undefined}
}%
\providecommand \@ifnum [1]{%
 \ifnum #1\expandafter \@firstoftwo
 \else \expandafter \@secondoftwo
 \fi
}%
\providecommand \@ifx [1]{%
 \ifx #1\expandafter \@firstoftwo
 \else \expandafter \@secondoftwo
 \fi
}%
\providecommand \natexlab [1]{#1}%
\providecommand \enquote  [1]{``#1''}%
\providecommand \bibnamefont  [1]{#1}%
\providecommand \bibfnamefont [1]{#1}%
\providecommand \citenamefont [1]{#1}%
\providecommand \href@noop [0]{\@secondoftwo}%
\providecommand \href [0]{\begingroup \@sanitize@url \@href}%
\providecommand \@href[1]{\@@startlink{#1}\@@href}%
\providecommand \@@href[1]{\endgroup#1\@@endlink}%
\providecommand \@sanitize@url [0]{\catcode `\\12\catcode `\$12\catcode
  `\&12\catcode `\#12\catcode `\^12\catcode `\_12\catcode `\%12\relax}%
\providecommand \@@startlink[1]{}%
\providecommand \@@endlink[0]{}%
\providecommand \url  [0]{\begingroup\@sanitize@url \@url }%
\providecommand \@url [1]{\endgroup\@href {#1}{\urlprefix }}%
\providecommand \urlprefix  [0]{URL }%
\providecommand \Eprint [0]{\href }%
\providecommand \doibase [0]{https://doi.org/}%
\providecommand \selectlanguage [0]{\@gobble}%
\providecommand \bibinfo  [0]{\@secondoftwo}%
\providecommand \bibfield  [0]{\@secondoftwo}%
\providecommand \translation [1]{[#1]}%
\providecommand \BibitemOpen [0]{}%
\providecommand \bibitemStop [0]{}%
\providecommand \bibitemNoStop [0]{.\EOS\space}%
\providecommand \EOS [0]{\spacefactor3000\relax}%
\providecommand \BibitemShut  [1]{\csname bibitem#1\endcsname}%
\let\auto@bib@innerbib\@empty
\bibitem [{\citenamefont {Gauss}(1866)}]{Gauss1866}%
  \BibitemOpen
  \bibfield  {author} {\bibinfo {author} {\bibfnamefont {C.~F.}\ \bibnamefont
  {Gauss}},\ }\bibfield  {title} {\bibinfo {title} {{Methodus nova integralium
  valores per approximationem inveniendi}},\ }in\ \href
  {https://doi.org/10.1017/CBO9781139058247.008} {\emph {\bibinfo {booktitle}
  {Werke}}}\ (\bibinfo  {publisher} {Cambridge University Press},\ \bibinfo
  {year} {1866})\ pp.\ \bibinfo {pages} {165--196}\BibitemShut {NoStop}%
\bibitem [{\citenamefont {Delsarte}\ \emph {et~al.}(1977)\citenamefont
  {Delsarte}, \citenamefont {Goethals},\ and\ \citenamefont
  {Seidel}}]{Delsarte1977}%
  \BibitemOpen
  \bibfield  {author} {\bibinfo {author} {\bibfnamefont {P.}~\bibnamefont
  {Delsarte}}, \bibinfo {author} {\bibfnamefont {J.~M.}\ \bibnamefont
  {Goethals}},\ and\ \bibinfo {author} {\bibfnamefont {J.~J.}\ \bibnamefont
  {Seidel}},\ }\bibfield  {title} {\bibinfo {title} {{Spherical codes and
  designs}},\ }\href {https://doi.org/10.1007/BF03187604} {\bibfield  {journal}
  {\bibinfo  {journal} {Geometriae Dedicata}\ }\textbf {\bibinfo {volume}
  {6}},\ \bibinfo {pages} {363} (\bibinfo {year} {1977})}\BibitemShut {NoStop}%
\bibitem [{\citenamefont {Hardin}\ and\ \citenamefont
  {Sloane}(1996)}]{Hardin1996}%
  \BibitemOpen
  \bibfield  {author} {\bibinfo {author} {\bibfnamefont {R.~H.}\ \bibnamefont
  {Hardin}}\ and\ \bibinfo {author} {\bibfnamefont {N.~J.~A.}\ \bibnamefont
  {Sloane}},\ }\bibfield  {title} {\bibinfo {title} {{McLaren's improved snub
  cube and other new spherical designs in three dimensions}},\ }\href
  {https://doi.org/10.1007/BF02711518} {\bibfield  {journal} {\bibinfo
  {journal} {Discrete \& Computational Geometry}\ }\textbf {\bibinfo {volume}
  {15}},\ \bibinfo {pages} {429} (\bibinfo {year} {1996})}\BibitemShut
  {NoStop}%
\bibitem [{\citenamefont {Stroud}(1971)}]{stroud_approximate_1971}%
  \BibitemOpen
  \bibfield  {author} {\bibinfo {author} {\bibfnamefont {A.~H.}\ \bibnamefont
  {Stroud}},\ }\href@noop {} {\emph {\bibinfo {title} {Approximate Calculation
  of Multiple Integrals}}}\ (\bibinfo  {publisher} {Prentice-Hall},\ \bibinfo
  {year} {1971})\BibitemShut {NoStop}%
\bibitem [{\citenamefont {Cools}(1997)}]{cools_constructing_1997}%
  \BibitemOpen
  \bibfield  {author} {\bibinfo {author} {\bibfnamefont {R.}~\bibnamefont
  {Cools}},\ }\bibfield  {title} {\bibinfo {title} {Constructing cubature
  formulae: the science behind the art},\ }\href
  {https://doi.org/10.1017/S0962492900002701} {\bibfield  {journal} {\bibinfo
  {journal} {Acta Numerica}\ }\textbf {\bibinfo {volume} {6}},\ \bibinfo
  {pages} {1} (\bibinfo {year} {1997})}\BibitemShut {NoStop}%
\bibitem [{\citenamefont {Hammer}\ and\ \citenamefont
  {Stroud}(1956)}]{hammer_numerical_1956}%
  \BibitemOpen
  \bibfield  {author} {\bibinfo {author} {\bibfnamefont {P.~C.}\ \bibnamefont
  {Hammer}}\ and\ \bibinfo {author} {\bibfnamefont {A.~H.}\ \bibnamefont
  {Stroud}},\ }\bibfield  {title} {\bibinfo {title} {Numerical integration over
  simplexes},\ }\href@noop {} {\bibfield  {journal} {\bibinfo  {journal}
  {Mathematical tables and other aids to computation}\ }\textbf {\bibinfo
  {volume} {10}},\ \bibinfo {pages} {137} (\bibinfo {year} {1956})}\BibitemShut
  {NoStop}%
\bibitem [{\citenamefont {Baladram}(2018)}]{baladram_on_2018}%
  \BibitemOpen
  \bibfield  {author} {\bibinfo {author} {\bibfnamefont {M.~S.}\ \bibnamefont
  {Baladram}},\ }\bibfield  {title} {\bibinfo {title} {On explicit construction
  of simplex t-designs},\ }\href@noop {} {\bibfield  {journal} {\bibinfo
  {journal} {Interdisciplinary Information Sciences}\ }\textbf {\bibinfo
  {volume} {24}},\ \bibinfo {pages} {181} (\bibinfo {year} {2018})}\BibitemShut
  {NoStop}%
\bibitem [{\citenamefont
  {Kuperberg}(2004{\natexlab{a}})}]{kuperberg_numerical_2004}%
  \BibitemOpen
  \bibfield  {author} {\bibinfo {author} {\bibfnamefont {G.}~\bibnamefont
  {Kuperberg}},\ }\bibfield  {title} {\bibinfo {title} {Numerical cubature from
  {Archimedes}' hat-box theorem},\ }\href {http://arxiv.org/abs/math/0405366}
  {\bibfield  {journal} {\bibinfo  {journal} {arXiv:math/0405366}\ } (\bibinfo
  {year} {2004}{\natexlab{a}})}\BibitemShut {NoStop}%
\bibitem [{\citenamefont
  {Kuperberg}(2004{\natexlab{b}})}]{kuperberg_numerical_ec_2004}%
  \BibitemOpen
  \bibfield  {author} {\bibinfo {author} {\bibfnamefont {G.}~\bibnamefont
  {Kuperberg}},\ }\bibfield  {title} {\bibinfo {title} {Numerical cubature
  using error-correcting codes},\ }\href {http://arxiv.org/abs/math/0402047}
  {\bibfield  {journal} {\bibinfo  {journal} {arXiv:math/0402047}\ } (\bibinfo
  {year} {2004}{\natexlab{b}})}\BibitemShut {NoStop}%
\bibitem [{\citenamefont {Victoir}(2004)}]{victoir}%
  \BibitemOpen
  \bibfield  {author} {\bibinfo {author} {\bibfnamefont {N.}~\bibnamefont
  {Victoir}},\ }\bibfield  {title} {\bibinfo {title} {Asymmetric cubature
  formulae with few points in high dimension for symmetric measures},\ }\href
  {https://doi.org/10.1137/S0036142902407952} {\bibfield  {journal} {\bibinfo
  {journal} {SIAM Journal on Numerical Analysis}\ }\textbf {\bibinfo {volume}
  {42}},\ \bibinfo {pages} {209} (\bibinfo {year} {2004})},\ \Eprint
  {https://arxiv.org/abs/https://doi.org/10.1137/S0036142902407952}
  {https://doi.org/10.1137/S0036142902407952} \BibitemShut {NoStop}%
\bibitem [{\citenamefont {Seymour}\ and\ \citenamefont
  {Zaslavsky}(1984)}]{seymour1984averaging}%
  \BibitemOpen
  \bibfield  {author} {\bibinfo {author} {\bibfnamefont {P.~D.}\ \bibnamefont
  {Seymour}}\ and\ \bibinfo {author} {\bibfnamefont {T.}~\bibnamefont
  {Zaslavsky}},\ }\bibfield  {title} {\bibinfo {title} {Averaging sets: a
  generalization of mean values and spherical designs},\ }\href@noop {}
  {\bibfield  {journal} {\bibinfo  {journal} {Advances in Mathematics}\
  }\textbf {\bibinfo {volume} {52}},\ \bibinfo {pages} {213} (\bibinfo {year}
  {1984})}\BibitemShut {NoStop}%
\bibitem [{\citenamefont {Hoggar}(1982)}]{hoggarTDesignsProjectiveSpaces1982}%
  \BibitemOpen
  \bibfield  {author} {\bibinfo {author} {\bibfnamefont {S.~G.}\ \bibnamefont
  {Hoggar}},\ }\bibfield  {title} {\bibinfo {title} {T-{{Designs}} in
  {{Projective Spaces}}},\ }\href
  {https://doi.org/10.1016/S0195-6698(82)80035-8} {\bibfield  {journal}
  {\bibinfo  {journal} {European Journal of Combinatorics}\ }\textbf {\bibinfo
  {volume} {3}},\ \bibinfo {pages} {233} (\bibinfo {year} {1982})}\BibitemShut
  {NoStop}%
\bibitem [{\citenamefont {Hoggar}(1984)}]{hoggarParametersTDesignsFPd1984}%
  \BibitemOpen
  \bibfield  {author} {\bibinfo {author} {\bibfnamefont {S.~G.}\ \bibnamefont
  {Hoggar}},\ }\bibfield  {title} {\bibinfo {title} {Parameters of
  t-{{Designs}} in {{FPd}}-1},\ }\href
  {https://doi.org/10.1016/S0195-6698(84)80015-3} {\bibfield  {journal}
  {\bibinfo  {journal} {European Journal of Combinatorics}\ }\textbf {\bibinfo
  {volume} {5}},\ \bibinfo {pages} {29} (\bibinfo {year} {1984})}\BibitemShut
  {NoStop}%
\bibitem [{\citenamefont {Bannai}\ and\ \citenamefont
  {Hoggar}(1985)}]{bannaiTightDesignsCompact1985}%
  \BibitemOpen
  \bibfield  {author} {\bibinfo {author} {\bibfnamefont {E.}~\bibnamefont
  {Bannai}}\ and\ \bibinfo {author} {\bibfnamefont {S.~G.}\ \bibnamefont
  {Hoggar}},\ }\bibfield  {title} {\bibinfo {title} {On tight \$t\$-designs in
  compact symmetric spaces of rank one},\ }\bibfield  {journal} {\bibinfo
  {journal} {Proceedings of the Japan Academy, Series A, Mathematical
  Sciences}\ }\textbf {\bibinfo {volume} {61}},\ \href
  {https://doi.org/10.3792/pjaa.61.78} {10.3792/pjaa.61.78} (\bibinfo {year}
  {1985})\BibitemShut {NoStop}%
\bibitem [{\citenamefont {Wootters}\ and\ \citenamefont
  {Fields}(1989{\natexlab{a}})}]{wootters1989optimal}%
  \BibitemOpen
  \bibfield  {author} {\bibinfo {author} {\bibfnamefont {W.~K.}\ \bibnamefont
  {Wootters}}\ and\ \bibinfo {author} {\bibfnamefont {B.~D.}\ \bibnamefont
  {Fields}},\ }\bibfield  {title} {\bibinfo {title} {Optimal
  state-determination by mutually unbiased measurements},\ }\href@noop {}
  {\bibfield  {journal} {\bibinfo  {journal} {Annals of Physics}\ }\textbf
  {\bibinfo {volume} {191}},\ \bibinfo {pages} {363} (\bibinfo {year}
  {1989}{\natexlab{a}})}\BibitemShut {NoStop}%
\bibitem [{\citenamefont {Renes}\ \emph {et~al.}(2004)\citenamefont {Renes},
  \citenamefont {{Blume-Kohout}}, \citenamefont {Scott},\ and\ \citenamefont
  {Caves}}]{renesSymmetricInformationallyComplete2004}%
  \BibitemOpen
  \bibfield  {author} {\bibinfo {author} {\bibfnamefont {J.~M.}\ \bibnamefont
  {Renes}}, \bibinfo {author} {\bibfnamefont {R.}~\bibnamefont
  {{Blume-Kohout}}}, \bibinfo {author} {\bibfnamefont {A.~J.}\ \bibnamefont
  {Scott}},\ and\ \bibinfo {author} {\bibfnamefont {C.~M.}\ \bibnamefont
  {Caves}},\ }\bibfield  {title} {\bibinfo {title} {Symmetric informationally
  complete quantum measurements},\ }\href {https://doi.org/10.1063/1.1737053}
  {\bibfield  {journal} {\bibinfo  {journal} {Journal of Mathematical Physics}\
  }\textbf {\bibinfo {volume} {45}},\ \bibinfo {pages} {2171} (\bibinfo {year}
  {2004})}\BibitemShut {NoStop}%
\bibitem [{\citenamefont {Klappenecker}\ and\ \citenamefont
  {Rotteler}(2005)}]{klappenecker2005mutually}%
  \BibitemOpen
  \bibfield  {author} {\bibinfo {author} {\bibfnamefont {A.}~\bibnamefont
  {Klappenecker}}\ and\ \bibinfo {author} {\bibfnamefont {M.}~\bibnamefont
  {Rotteler}},\ }\bibfield  {title} {\bibinfo {title} {Mutually unbiased bases
  are complex projective 2-designs},\ }in\ \href@noop {} {\emph {\bibinfo
  {booktitle} {Proceedings. International Symposium on Information Theory,
  2005. ISIT 2005.}}}\ (\bibinfo {organization} {IEEE},\ \bibinfo {year}
  {2005})\ pp.\ \bibinfo {pages} {1740--1744}\BibitemShut {NoStop}%
\bibitem [{\citenamefont {Dankert}(2005)}]{dankert2005efficient}%
  \BibitemOpen
  \bibfield  {author} {\bibinfo {author} {\bibfnamefont {C.}~\bibnamefont
  {Dankert}},\ }\href {https://doi.org/10.48550/arXiv.quant-ph/0512217}
  {\bibinfo {title} {Efficient simulation of random quantum states and
  operators}} (\bibinfo {year} {2005}),\ \Eprint
  {https://arxiv.org/abs/0512217} {0512217 [quant-ph]} \BibitemShut {NoStop}%
\bibitem [{\citenamefont {Scott}(2006)}]{scott_tight_2006}%
  \BibitemOpen
  \bibfield  {author} {\bibinfo {author} {\bibfnamefont {A.~J.}\ \bibnamefont
  {Scott}},\ }\bibfield  {title} {\bibinfo {title} {Tight informationally
  complete quantum measurements},\ }\href
  {https://doi.org/10.1088/0305-4470/39/43/009} {\bibfield  {journal} {\bibinfo
   {journal} {Journal of Physics A: Mathematical and General}\ }\textbf
  {\bibinfo {volume} {39}},\ \bibinfo {pages} {13507} (\bibinfo {year}
  {2006})}\BibitemShut {NoStop}%
\bibitem [{\citenamefont {Ambainis}\ and\ \citenamefont
  {Emerson}(2007)}]{ambainisQuantumTdesignsTwise2007}%
  \BibitemOpen
  \bibfield  {author} {\bibinfo {author} {\bibfnamefont {A.}~\bibnamefont
  {Ambainis}}\ and\ \bibinfo {author} {\bibfnamefont {J.}~\bibnamefont
  {Emerson}},\ }\bibfield  {title} {\bibinfo {title} {Quantum t-designs: T-wise
  independence in the quantum world},\ }in\ \href@noop {} {\emph {\bibinfo
  {booktitle} {Twenty-{{Second Annual IEEE Conference}} on {{Computational
  Complexity}} ({{CCC}}'07)}}}\ (\bibinfo  {publisher} {{IEEE}},\ \bibinfo
  {year} {2007})\ pp.\ \bibinfo {pages} {129--140}\BibitemShut {NoStop}%
\bibitem [{\citenamefont {Roberts}\ and\ \citenamefont
  {Yoshida}(2017)}]{roberts_chaos_2017}%
  \BibitemOpen
  \bibfield  {author} {\bibinfo {author} {\bibfnamefont {D.~A.}\ \bibnamefont
  {Roberts}}\ and\ \bibinfo {author} {\bibfnamefont {B.}~\bibnamefont
  {Yoshida}},\ }\bibfield  {title} {\bibinfo {title} {Chaos and complexity by
  design},\ }\href {https://doi.org/10.1007/JHEP04(2017)121} {\bibfield
  {journal} {\bibinfo  {journal} {Journal of High Energy Physics}\ }\textbf
  {\bibinfo {volume} {2017}},\ \bibinfo {pages} {121} (\bibinfo {year}
  {2017})}\BibitemShut {NoStop}%
\bibitem [{\citenamefont {Kueng}\ and\ \citenamefont
  {Gross}(2015)}]{Kueng2015}%
  \BibitemOpen
  \bibfield  {author} {\bibinfo {author} {\bibfnamefont {R.}~\bibnamefont
  {Kueng}}\ and\ \bibinfo {author} {\bibfnamefont {D.}~\bibnamefont {Gross}},\
  }\bibfield  {title} {\bibinfo {title} {Qubit stabilizer states are complex
  projective 3-designs},\ }\href@noop {} {\bibfield  {journal} {\bibinfo
  {journal} {arXiv preprint arXiv:1510.02767}\ } (\bibinfo {year} {2015})},\
  \Eprint {https://arxiv.org/abs/1510.02767} {arXiv:1510.02767} \BibitemShut
  {NoStop}%
\bibitem [{\citenamefont {Dankert}\ \emph {et~al.}(2009)\citenamefont
  {Dankert}, \citenamefont {Cleve}, \citenamefont {Emerson},\ and\
  \citenamefont {Livine}}]{dankertExactApproximateUnitary2009}%
  \BibitemOpen
  \bibfield  {author} {\bibinfo {author} {\bibfnamefont {C.}~\bibnamefont
  {Dankert}}, \bibinfo {author} {\bibfnamefont {R.}~\bibnamefont {Cleve}},
  \bibinfo {author} {\bibfnamefont {J.}~\bibnamefont {Emerson}},\ and\ \bibinfo
  {author} {\bibfnamefont {E.}~\bibnamefont {Livine}},\ }\bibfield  {title}
  {\bibinfo {title} {Exact and approximate unitary 2-designs and their
  application to fidelity estimation},\ }\href
  {https://doi.org/10.1103/PhysRevA.80.012304} {\bibfield  {journal} {\bibinfo
  {journal} {Physical Review A}\ }\textbf {\bibinfo {volume} {80}},\ \bibinfo
  {pages} {012304} (\bibinfo {year} {2009})}\BibitemShut {NoStop}%
\bibitem [{\citenamefont {van Enk}\ and\ \citenamefont
  {Beenakker}(2012)}]{PhysRevLett.108.110503}%
  \BibitemOpen
  \bibfield  {author} {\bibinfo {author} {\bibfnamefont {S.~J.}\ \bibnamefont
  {van Enk}}\ and\ \bibinfo {author} {\bibfnamefont {C.~W.~J.}\ \bibnamefont
  {Beenakker}},\ }\bibfield  {title} {\bibinfo {title} {Measuring
  $\mathrm{Tr}{\ensuremath{\rho}}^{n}$ on single copies of $\ensuremath{\rho}$
  using random measurements},\ }\href
  {https://doi.org/10.1103/PhysRevLett.108.110503} {\bibfield  {journal}
  {\bibinfo  {journal} {Phys. Rev. Lett.}\ }\textbf {\bibinfo {volume} {108}},\
  \bibinfo {pages} {110503} (\bibinfo {year} {2012})}\BibitemShut {NoStop}%
\bibitem [{\citenamefont
  {Aaronson}(2018)}]{aaronsonShadowTomographyQuantum2018}%
  \BibitemOpen
  \bibfield  {author} {\bibinfo {author} {\bibfnamefont {S.}~\bibnamefont
  {Aaronson}},\ }\bibfield  {title} {\bibinfo {title} {{Shadow tomography of
  quantum states}},\ }in\ \href {https://doi.org/10.1145/3188745.3188802}
  {\emph {\bibinfo {booktitle} {Proc. 50th Annu. ACM SIGACT Symp. Theory
  Comput.}}}\ (\bibinfo  {publisher} {ACM},\ \bibinfo {address} {New York, NY,
  USA},\ \bibinfo {year} {2018})\ pp.\ \bibinfo {pages} {325--338},\ \Eprint
  {https://arxiv.org/abs/1711.01053} {arXiv:1711.01053} \BibitemShut {NoStop}%
\bibitem [{\citenamefont {Huang}\ \emph {et~al.}(2020)\citenamefont {Huang},
  \citenamefont {Kueng},\ and\ \citenamefont
  {Preskill}}]{huangPredictingManyProperties2020}%
  \BibitemOpen
  \bibfield  {author} {\bibinfo {author} {\bibfnamefont {H.-Y.}\ \bibnamefont
  {Huang}}, \bibinfo {author} {\bibfnamefont {R.}~\bibnamefont {Kueng}},\ and\
  \bibinfo {author} {\bibfnamefont {J.}~\bibnamefont {Preskill}},\ }\bibfield
  {title} {\bibinfo {title} {Predicting many properties of a quantum system
  from very few measurements},\ }\href
  {https://doi.org/10.1038/s41567-020-0932-7} {\bibfield  {journal} {\bibinfo
  {journal} {Nature Physics}\ }\textbf {\bibinfo {volume} {16}},\ \bibinfo
  {pages} {1050} (\bibinfo {year} {2020})}\BibitemShut {NoStop}%
\bibitem [{\citenamefont {Huang}\ \emph {et~al.}(2022)\citenamefont {Huang},
  \citenamefont {Kueng}, \citenamefont {Torlai}, \citenamefont {Albert},\ and\
  \citenamefont {Preskill}}]{huangProvablyEfficientMachine2022}%
  \BibitemOpen
  \bibfield  {author} {\bibinfo {author} {\bibfnamefont {H.-Y.}\ \bibnamefont
  {Huang}}, \bibinfo {author} {\bibfnamefont {R.}~\bibnamefont {Kueng}},
  \bibinfo {author} {\bibfnamefont {G.}~\bibnamefont {Torlai}}, \bibinfo
  {author} {\bibfnamefont {V.~V.}\ \bibnamefont {Albert}},\ and\ \bibinfo
  {author} {\bibfnamefont {J.}~\bibnamefont {Preskill}},\ }\href
  {http://arxiv.org/abs/2106.12627} {\bibinfo {title} {Provably efficient
  machine learning for quantum many-body problems}} (\bibinfo {year} {2022}),\
  \Eprint {https://arxiv.org/abs/2106.12627} {arXiv:2106.12627 [quant-ph]}
  \BibitemShut {NoStop}%
\bibitem [{\citenamefont {Acharya}\ \emph
  {et~al.}(2021{\natexlab{a}})\citenamefont {Acharya}, \citenamefont {Saha},\
  and\ \citenamefont {Sengupta}}]{Acharya2021}%
  \BibitemOpen
  \bibfield  {author} {\bibinfo {author} {\bibfnamefont {A.}~\bibnamefont
  {Acharya}}, \bibinfo {author} {\bibfnamefont {S.}~\bibnamefont {Saha}},\ and\
  \bibinfo {author} {\bibfnamefont {A.~M.}\ \bibnamefont {Sengupta}},\
  }\bibfield  {title} {\bibinfo {title} {{Informationally complete POVM-based
  shadow tomography}},\ }\href {http://arxiv.org/abs/2105.05992} {\bibfield
  {journal} {\bibinfo  {journal} {{arXiv}}\ } (\bibinfo {year}
  {2021}{\natexlab{a}})},\ \Eprint {https://arxiv.org/abs/2105.05992}
  {arXiv:2105.05992} \BibitemShut {NoStop}%
\bibitem [{\citenamefont {Kueng}\ \emph {et~al.}(2016)\citenamefont {Kueng},
  \citenamefont {Zhu},\ and\ \citenamefont {Gross}}]{kueng2016distinguishing}%
  \BibitemOpen
  \bibfield  {author} {\bibinfo {author} {\bibfnamefont {R.}~\bibnamefont
  {Kueng}}, \bibinfo {author} {\bibfnamefont {H.}~\bibnamefont {Zhu}},\ and\
  \bibinfo {author} {\bibfnamefont {D.}~\bibnamefont {Gross}},\ }\bibfield
  {title} {\bibinfo {title} {Distinguishing quantum states using clifford
  orbits},\ }\href@noop {} {\bibfield  {journal} {\bibinfo  {journal} {arXiv
  preprint arXiv:1609.08595}\ } (\bibinfo {year} {2016})}\BibitemShut {NoStop}%
\bibitem [{\citenamefont {Emerson}\ \emph {et~al.}(2005)\citenamefont
  {Emerson}, \citenamefont {Alicki},\ and\ \citenamefont
  {{\.Z}yczkowski}}]{emersonScalableNoiseEstimation2005}%
  \BibitemOpen
  \bibfield  {author} {\bibinfo {author} {\bibfnamefont {J.}~\bibnamefont
  {Emerson}}, \bibinfo {author} {\bibfnamefont {R.}~\bibnamefont {Alicki}},\
  and\ \bibinfo {author} {\bibfnamefont {K.}~\bibnamefont {{\.Z}yczkowski}},\
  }\bibfield  {title} {\bibinfo {title} {Scalable noise estimation with random
  unitary operators},\ }\href {https://doi.org/10.1088/1464-4266/7/10/021}
  {\bibfield  {journal} {\bibinfo  {journal} {Journal of Optics B: Quantum and
  Semiclassical Optics}\ }\textbf {\bibinfo {volume} {7}},\ \bibinfo {pages}
  {S347} (\bibinfo {year} {2005})}\BibitemShut {NoStop}%
\bibitem [{\citenamefont {Knill}\ \emph {et~al.}(2008)\citenamefont {Knill},
  \citenamefont {Leibfried}, \citenamefont {Reichle}, \citenamefont {Britton},
  \citenamefont {Blakestad}, \citenamefont {Jost}, \citenamefont {Langer},
  \citenamefont {Ozeri}, \citenamefont {Seidelin},\ and\ \citenamefont
  {Wineland}}]{knillRandomizedBenchmarkingQuantum2008}%
  \BibitemOpen
  \bibfield  {author} {\bibinfo {author} {\bibfnamefont {E.}~\bibnamefont
  {Knill}}, \bibinfo {author} {\bibfnamefont {D.}~\bibnamefont {Leibfried}},
  \bibinfo {author} {\bibfnamefont {R.}~\bibnamefont {Reichle}}, \bibinfo
  {author} {\bibfnamefont {J.}~\bibnamefont {Britton}}, \bibinfo {author}
  {\bibfnamefont {R.~B.}\ \bibnamefont {Blakestad}}, \bibinfo {author}
  {\bibfnamefont {J.~D.}\ \bibnamefont {Jost}}, \bibinfo {author}
  {\bibfnamefont {C.}~\bibnamefont {Langer}}, \bibinfo {author} {\bibfnamefont
  {R.}~\bibnamefont {Ozeri}}, \bibinfo {author} {\bibfnamefont
  {S.}~\bibnamefont {Seidelin}},\ and\ \bibinfo {author} {\bibfnamefont
  {D.~J.}\ \bibnamefont {Wineland}},\ }\bibfield  {title} {\bibinfo {title}
  {Randomized benchmarking of quantum gates},\ }\href
  {https://doi.org/10.1103/PhysRevA.77.012307} {\bibfield  {journal} {\bibinfo
  {journal} {Physical Review A}\ }\textbf {\bibinfo {volume} {77}},\ \bibinfo
  {pages} {012307} (\bibinfo {year} {2008})}\BibitemShut {NoStop}%
\bibitem [{\citenamefont {Magesan}\ \emph
  {et~al.}(2011{\natexlab{a}})\citenamefont {Magesan}, \citenamefont
  {Gambetta},\ and\ \citenamefont {Emerson}}]{scalablemagesan}%
  \BibitemOpen
  \bibfield  {author} {\bibinfo {author} {\bibfnamefont {E.}~\bibnamefont
  {Magesan}}, \bibinfo {author} {\bibfnamefont {J.~M.}\ \bibnamefont
  {Gambetta}},\ and\ \bibinfo {author} {\bibfnamefont {J.}~\bibnamefont
  {Emerson}},\ }\bibfield  {title} {\bibinfo {title} {Scalable and robust
  randomized benchmarking of quantum processes},\ }\href
  {https://doi.org/10.1103/PhysRevLett.106.180504} {\bibfield  {journal}
  {\bibinfo  {journal} {Phys. Rev. Lett.}\ }\textbf {\bibinfo {volume} {106}},\
  \bibinfo {pages} {180504} (\bibinfo {year} {2011}{\natexlab{a}})}\BibitemShut
  {NoStop}%
\bibitem [{\citenamefont {Cross}\ \emph {et~al.}(2016)\citenamefont {Cross},
  \citenamefont {Magesan}, \citenamefont {Bishop}, \citenamefont {Smolin},\
  and\ \citenamefont {Gambetta}}]{cross2016scalable}%
  \BibitemOpen
  \bibfield  {author} {\bibinfo {author} {\bibfnamefont {A.~W.}\ \bibnamefont
  {Cross}}, \bibinfo {author} {\bibfnamefont {E.}~\bibnamefont {Magesan}},
  \bibinfo {author} {\bibfnamefont {L.~S.}\ \bibnamefont {Bishop}}, \bibinfo
  {author} {\bibfnamefont {J.~A.}\ \bibnamefont {Smolin}},\ and\ \bibinfo
  {author} {\bibfnamefont {J.~M.}\ \bibnamefont {Gambetta}},\ }\bibfield
  {title} {\bibinfo {title} {Scalable randomised benchmarking of non-clifford
  gates},\ }\href@noop {} {\bibfield  {journal} {\bibinfo  {journal} {npj
  Quantum Information}\ }\textbf {\bibinfo {volume} {2}},\ \bibinfo {pages} {1}
  (\bibinfo {year} {2016})}\BibitemShut {NoStop}%
\bibitem [{\citenamefont
  {Nielsen}(1996)}]{nielsenEntanglementFidelityQuantum1996}%
  \BibitemOpen
  \bibfield  {author} {\bibinfo {author} {\bibfnamefont {M.~A.}\ \bibnamefont
  {Nielsen}},\ }\bibfield  {title} {\bibinfo {title} {The entanglement fidelity
  and quantum error correction},\ }\bibfield  {journal} {\bibinfo  {journal}
  {{arXiv}}\ }\href {https://doi.org/10.48550/arxiv.quant-ph/9606012}
  {10.48550/arxiv.quant-ph/9606012} (\bibinfo {year} {1996})\BibitemShut
  {NoStop}%
\bibitem [{\citenamefont {Horodecki}\ \emph {et~al.}(1999)\citenamefont
  {Horodecki}, \citenamefont {Horodecki},\ and\ \citenamefont
  {Horodecki}}]{horodeckiGeneralTeleportationChannel1999}%
  \BibitemOpen
  \bibfield  {author} {\bibinfo {author} {\bibfnamefont {M.}~\bibnamefont
  {Horodecki}}, \bibinfo {author} {\bibfnamefont {P.}~\bibnamefont
  {Horodecki}},\ and\ \bibinfo {author} {\bibfnamefont {R.}~\bibnamefont
  {Horodecki}},\ }\bibfield  {title} {\bibinfo {title} {General teleportation
  channel, singlet fraction, and quasidistillation},\ }\href
  {https://doi.org/10.1103/PhysRevA.60.1888} {\bibfield  {journal} {\bibinfo
  {journal} {Physical Review A}\ }\textbf {\bibinfo {volume} {60}},\ \bibinfo
  {pages} {1888} (\bibinfo {year} {1999})}\BibitemShut {NoStop}%
\bibitem [{\citenamefont {Nielsen}(2002)}]{nielsenSimpleFormulaAverage2002}%
  \BibitemOpen
  \bibfield  {author} {\bibinfo {author} {\bibfnamefont {M.~A.}\ \bibnamefont
  {Nielsen}},\ }\bibfield  {title} {\bibinfo {title} {A simple formula for the
  average gate fidelity of a quantum dynamical operation},\ }\href
  {https://doi.org/10.1016/S0375-9601(02)01272-0} {\bibfield  {journal}
  {\bibinfo  {journal} {Physics Letters A}\ }\textbf {\bibinfo {volume}
  {303}},\ \bibinfo {pages} {249} (\bibinfo {year} {2002})}\BibitemShut
  {NoStop}%
\bibitem [{\citenamefont {Magesan}\ \emph
  {et~al.}(2011{\natexlab{b}})\citenamefont {Magesan}, \citenamefont
  {{Blume-Kohout}},\ and\ \citenamefont
  {Emerson}}]{magesanGateFidelityFluctuations2011}%
  \BibitemOpen
  \bibfield  {author} {\bibinfo {author} {\bibfnamefont {E.}~\bibnamefont
  {Magesan}}, \bibinfo {author} {\bibfnamefont {R.}~\bibnamefont
  {{Blume-Kohout}}},\ and\ \bibinfo {author} {\bibfnamefont {J.}~\bibnamefont
  {Emerson}},\ }\bibfield  {title} {\bibinfo {title} {Gate fidelity
  fluctuations and quantum process invariants},\ }\href
  {https://doi.org/10.1103/PhysRevA.84.012309} {\bibfield  {journal} {\bibinfo
  {journal} {Physical Review A}\ }\textbf {\bibinfo {volume} {84}},\ \bibinfo
  {pages} {012309} (\bibinfo {year} {2011}{\natexlab{b}})}\BibitemShut
  {NoStop}%
\bibitem [{\citenamefont {Lu}\ \emph {et~al.}(2015)\citenamefont {Lu},
  \citenamefont {Li}, \citenamefont {Trottier}, \citenamefont {Li},
  \citenamefont {Brodutch}, \citenamefont {Krismanich}, \citenamefont
  {Ghavami}, \citenamefont {Dmitrienko}, \citenamefont {Long}, \citenamefont
  {Baugh},\ and\ \citenamefont
  {Laflamme}}]{luExperimentalEstimationAverage2015}%
  \BibitemOpen
  \bibfield  {author} {\bibinfo {author} {\bibfnamefont {D.}~\bibnamefont
  {Lu}}, \bibinfo {author} {\bibfnamefont {H.}~\bibnamefont {Li}}, \bibinfo
  {author} {\bibfnamefont {D.-A.}\ \bibnamefont {Trottier}}, \bibinfo {author}
  {\bibfnamefont {J.}~\bibnamefont {Li}}, \bibinfo {author} {\bibfnamefont
  {A.}~\bibnamefont {Brodutch}}, \bibinfo {author} {\bibfnamefont {A.~P.}\
  \bibnamefont {Krismanich}}, \bibinfo {author} {\bibfnamefont
  {A.}~\bibnamefont {Ghavami}}, \bibinfo {author} {\bibfnamefont {G.~I.}\
  \bibnamefont {Dmitrienko}}, \bibinfo {author} {\bibfnamefont
  {G.}~\bibnamefont {Long}}, \bibinfo {author} {\bibfnamefont {J.}~\bibnamefont
  {Baugh}},\ and\ \bibinfo {author} {\bibfnamefont {R.}~\bibnamefont
  {Laflamme}},\ }\bibfield  {title} {\bibinfo {title} {Experimental
  {{Estimation}} of {{Average Fidelity}} of a {{Clifford Gate}} on a 7-{{Qubit
  Quantum Processor}}},\ }\href
  {https://doi.org/10.1103/PhysRevLett.114.140505} {\bibfield  {journal}
  {\bibinfo  {journal} {Physical Review Letters}\ }\textbf {\bibinfo {volume}
  {114}},\ \bibinfo {pages} {140505} (\bibinfo {year} {2015})}\BibitemShut
  {NoStop}%
\bibitem [{\citenamefont {Bravyi}\ \emph {et~al.}(2021)\citenamefont {Bravyi},
  \citenamefont {Chowdhury}, \citenamefont {Gosset},\ and\ \citenamefont
  {Wocjan}}]{Bravyi2021}%
  \BibitemOpen
  \bibfield  {author} {\bibinfo {author} {\bibfnamefont {S.}~\bibnamefont
  {Bravyi}}, \bibinfo {author} {\bibfnamefont {A.}~\bibnamefont {Chowdhury}},
  \bibinfo {author} {\bibfnamefont {D.}~\bibnamefont {Gosset}},\ and\ \bibinfo
  {author} {\bibfnamefont {P.}~\bibnamefont {Wocjan}},\ }\bibfield  {title}
  {\bibinfo {title} {{On the complexity of quantum partition functions}},\
  }\href {http://arxiv.org/abs/2110.15466} {\bibfield  {journal} {\bibinfo
  {journal} {{arXiv}}\ } (\bibinfo {year} {2021})},\ \Eprint
  {https://arxiv.org/abs/2110.15466} {arXiv:2110.15466} \BibitemShut {NoStop}%
\bibitem [{\citenamefont {Ambainis}\ and\ \citenamefont
  {Smith}(2004)}]{ambainis2004small}%
  \BibitemOpen
  \bibfield  {author} {\bibinfo {author} {\bibfnamefont {A.}~\bibnamefont
  {Ambainis}}\ and\ \bibinfo {author} {\bibfnamefont {A.}~\bibnamefont
  {Smith}},\ }\bibfield  {title} {\bibinfo {title} {Small pseudo-random
  families of matrices: Derandomizing approximate quantum encryption},\ }in\
  \href@noop {} {\emph {\bibinfo {booktitle} {Approximation, Randomization, and
  Combinatorial Optimization. Algorithms and Techniques}}}\ (\bibinfo
  {publisher} {Springer},\ \bibinfo {year} {2004})\ pp.\ \bibinfo {pages}
  {249--260}\BibitemShut {NoStop}%
\bibitem [{\citenamefont {Hayden}\ \emph {et~al.}(2004)\citenamefont {Hayden},
  \citenamefont {Leung}, \citenamefont {Shor},\ and\ \citenamefont
  {Winter}}]{hayden2004randomizing}%
  \BibitemOpen
  \bibfield  {author} {\bibinfo {author} {\bibfnamefont {P.}~\bibnamefont
  {Hayden}}, \bibinfo {author} {\bibfnamefont {D.}~\bibnamefont {Leung}},
  \bibinfo {author} {\bibfnamefont {P.~W.}\ \bibnamefont {Shor}},\ and\
  \bibinfo {author} {\bibfnamefont {A.}~\bibnamefont {Winter}},\ }\bibfield
  {title} {\bibinfo {title} {Randomizing quantum states: Constructions and
  applications},\ }\href@noop {} {\bibfield  {journal} {\bibinfo  {journal}
  {Communications in Mathematical Physics}\ }\textbf {\bibinfo {volume}
  {250}},\ \bibinfo {pages} {371} (\bibinfo {year} {2004})}\BibitemShut
  {NoStop}%
\bibitem [{\citenamefont {Oszmaniec}\ \emph {et~al.}(2016)\citenamefont
  {Oszmaniec}, \citenamefont {Augusiak}, \citenamefont {Gogolin}, \citenamefont
  {Ko\l{}ody\ifmmode~\acute{n}\else \'{n}\fi{}ski}, \citenamefont {Ac\'{\i}n},\
  and\ \citenamefont {Lewenstein}}]{OAG16}%
  \BibitemOpen
  \bibfield  {author} {\bibinfo {author} {\bibfnamefont {M.}~\bibnamefont
  {Oszmaniec}}, \bibinfo {author} {\bibfnamefont {R.}~\bibnamefont {Augusiak}},
  \bibinfo {author} {\bibfnamefont {C.}~\bibnamefont {Gogolin}}, \bibinfo
  {author} {\bibfnamefont {J.}~\bibnamefont {Ko\l{}ody\ifmmode~\acute{n}\else
  \'{n}\fi{}ski}}, \bibinfo {author} {\bibfnamefont {A.}~\bibnamefont
  {Ac\'{\i}n}},\ and\ \bibinfo {author} {\bibfnamefont {M.}~\bibnamefont
  {Lewenstein}},\ }\bibfield  {title} {\bibinfo {title} {Random bosonic states
  for robust quantum metrology},\ }\href
  {https://doi.org/10.1103/PhysRevX.6.041044} {\bibfield  {journal} {\bibinfo
  {journal} {Phys. Rev. X}\ }\textbf {\bibinfo {volume} {6}},\ \bibinfo {pages}
  {041044} (\bibinfo {year} {2016})}\BibitemShut {NoStop}%
\bibitem [{\citenamefont {Kimmel}\ and\ \citenamefont
  {Liu}(2017)}]{kimmel2017phase}%
  \BibitemOpen
  \bibfield  {author} {\bibinfo {author} {\bibfnamefont {S.}~\bibnamefont
  {Kimmel}}\ and\ \bibinfo {author} {\bibfnamefont {Y.-K.}\ \bibnamefont
  {Liu}},\ }\bibfield  {title} {\bibinfo {title} {Phase retrieval using unitary
  2-designs},\ }in\ \href@noop {} {\emph {\bibinfo {booktitle} {2017
  International Conference on Sampling Theory and Applications (SampTA)}}}\
  (\bibinfo {organization} {IEEE},\ \bibinfo {year} {2017})\ pp.\ \bibinfo
  {pages} {345--349}\BibitemShut {NoStop}%
\bibitem [{\citenamefont {Mi}\ \emph {et~al.}(2021)\citenamefont {Mi},
  \citenamefont {Roushan}, \citenamefont {Quintana}, \citenamefont {Mandra},
  \citenamefont {Marshall}, \citenamefont {Neill}, \citenamefont {Arute},
  \citenamefont {Arya}, \citenamefont {Atalaya}, \citenamefont {Babbush},
  \citenamefont {Bardin}, \citenamefont {Barends}, \citenamefont {Bengtsson},
  \citenamefont {Boixo}, \citenamefont {Bourassa}, \citenamefont {Broughton},
  \citenamefont {Buckley}, \citenamefont {Buell}, \citenamefont {Burkett},
  \citenamefont {Bushnell}, \citenamefont {Chen}, \citenamefont {Chiaro},
  \citenamefont {Collins}, \citenamefont {Courtney}, \citenamefont {Demura},
  \citenamefont {Derk}, \citenamefont {Dunsworth}, \citenamefont {Eppens},
  \citenamefont {Erickson}, \citenamefont {Farhi}, \citenamefont {Fowler},
  \citenamefont {Foxen}, \citenamefont {Gidney}, \citenamefont {Giustina},
  \citenamefont {Gross}, \citenamefont {Harrigan}, \citenamefont {Harrington},
  \citenamefont {Hilton}, \citenamefont {Ho}, \citenamefont {Hong},
  \citenamefont {Huang}, \citenamefont {Huggins}, \citenamefont {Ioffe},
  \citenamefont {Isakov}, \citenamefont {Jeffrey}, \citenamefont {Jiang},
  \citenamefont {Jones}, \citenamefont {Kafri}, \citenamefont {Kelly},
  \citenamefont {Kim}, \citenamefont {Kitaev}, \citenamefont {Klimov},
  \citenamefont {Korotkov}, \citenamefont {Kostritsa}, \citenamefont
  {Landhuis}, \citenamefont {Laptev}, \citenamefont {Lucero}, \citenamefont
  {Martin}, \citenamefont {McClean}, \citenamefont {McCourt}, \citenamefont
  {McEwen}, \citenamefont {Megrant}, \citenamefont {Miao}, \citenamefont
  {Mohseni}, \citenamefont {Mruczkiewicz}, \citenamefont {Mutus}, \citenamefont
  {Naaman}, \citenamefont {Neeley}, \citenamefont {Newman}, \citenamefont
  {Niu}, \citenamefont {O'Brien}, \citenamefont {Opremcak}, \citenamefont
  {Ostby}, \citenamefont {Pato}, \citenamefont {Petukhov}, \citenamefont
  {Redd}, \citenamefont {Rubin}, \citenamefont {Sank}, \citenamefont
  {Satzinger}, \citenamefont {Shvarts}, \citenamefont {Strain}, \citenamefont
  {Szalay}, \citenamefont {Trevithick}, \citenamefont {Villalonga},
  \citenamefont {White}, \citenamefont {Yao}, \citenamefont {Yeh},
  \citenamefont {Zalcman}, \citenamefont {Neven}, \citenamefont {Aleiner},
  \citenamefont {Kechedzhi}, \citenamefont {Smelyanskiy},\ and\ \citenamefont
  {Chen}}]{miInformationScramblingComputationally2021}%
  \BibitemOpen
  \bibfield  {author} {\bibinfo {author} {\bibfnamefont {X.}~\bibnamefont
  {Mi}}, \bibinfo {author} {\bibfnamefont {P.}~\bibnamefont {Roushan}},
  \bibinfo {author} {\bibfnamefont {C.}~\bibnamefont {Quintana}}, \bibinfo
  {author} {\bibfnamefont {S.}~\bibnamefont {Mandra}}, \bibinfo {author}
  {\bibfnamefont {J.}~\bibnamefont {Marshall}}, \bibinfo {author}
  {\bibfnamefont {C.}~\bibnamefont {Neill}}, \bibinfo {author} {\bibfnamefont
  {F.}~\bibnamefont {Arute}}, \bibinfo {author} {\bibfnamefont
  {K.}~\bibnamefont {Arya}}, \bibinfo {author} {\bibfnamefont {J.}~\bibnamefont
  {Atalaya}}, \bibinfo {author} {\bibfnamefont {R.}~\bibnamefont {Babbush}},
  \bibinfo {author} {\bibfnamefont {J.~C.}\ \bibnamefont {Bardin}}, \bibinfo
  {author} {\bibfnamefont {R.}~\bibnamefont {Barends}}, \bibinfo {author}
  {\bibfnamefont {A.}~\bibnamefont {Bengtsson}}, \bibinfo {author}
  {\bibfnamefont {S.}~\bibnamefont {Boixo}}, \bibinfo {author} {\bibfnamefont
  {A.}~\bibnamefont {Bourassa}}, \bibinfo {author} {\bibfnamefont
  {M.}~\bibnamefont {Broughton}}, \bibinfo {author} {\bibfnamefont {B.~B.}\
  \bibnamefont {Buckley}}, \bibinfo {author} {\bibfnamefont {D.~A.}\
  \bibnamefont {Buell}}, \bibinfo {author} {\bibfnamefont {B.}~\bibnamefont
  {Burkett}}, \bibinfo {author} {\bibfnamefont {N.}~\bibnamefont {Bushnell}},
  \bibinfo {author} {\bibfnamefont {Z.}~\bibnamefont {Chen}}, \bibinfo {author}
  {\bibfnamefont {B.}~\bibnamefont {Chiaro}}, \bibinfo {author} {\bibfnamefont
  {R.}~\bibnamefont {Collins}}, \bibinfo {author} {\bibfnamefont
  {W.}~\bibnamefont {Courtney}}, \bibinfo {author} {\bibfnamefont
  {S.}~\bibnamefont {Demura}}, \bibinfo {author} {\bibfnamefont {A.~R.}\
  \bibnamefont {Derk}}, \bibinfo {author} {\bibfnamefont {A.}~\bibnamefont
  {Dunsworth}}, \bibinfo {author} {\bibfnamefont {D.}~\bibnamefont {Eppens}},
  \bibinfo {author} {\bibfnamefont {C.}~\bibnamefont {Erickson}}, \bibinfo
  {author} {\bibfnamefont {E.}~\bibnamefont {Farhi}}, \bibinfo {author}
  {\bibfnamefont {A.~G.}\ \bibnamefont {Fowler}}, \bibinfo {author}
  {\bibfnamefont {B.}~\bibnamefont {Foxen}}, \bibinfo {author} {\bibfnamefont
  {C.}~\bibnamefont {Gidney}}, \bibinfo {author} {\bibfnamefont
  {M.}~\bibnamefont {Giustina}}, \bibinfo {author} {\bibfnamefont {J.~A.}\
  \bibnamefont {Gross}}, \bibinfo {author} {\bibfnamefont {M.~P.}\ \bibnamefont
  {Harrigan}}, \bibinfo {author} {\bibfnamefont {S.~D.}\ \bibnamefont
  {Harrington}}, \bibinfo {author} {\bibfnamefont {J.}~\bibnamefont {Hilton}},
  \bibinfo {author} {\bibfnamefont {A.}~\bibnamefont {Ho}}, \bibinfo {author}
  {\bibfnamefont {S.}~\bibnamefont {Hong}}, \bibinfo {author} {\bibfnamefont
  {T.}~\bibnamefont {Huang}}, \bibinfo {author} {\bibfnamefont {W.~J.}\
  \bibnamefont {Huggins}}, \bibinfo {author} {\bibfnamefont {L.~B.}\
  \bibnamefont {Ioffe}}, \bibinfo {author} {\bibfnamefont {S.~V.}\ \bibnamefont
  {Isakov}}, \bibinfo {author} {\bibfnamefont {E.}~\bibnamefont {Jeffrey}},
  \bibinfo {author} {\bibfnamefont {Z.}~\bibnamefont {Jiang}}, \bibinfo
  {author} {\bibfnamefont {C.}~\bibnamefont {Jones}}, \bibinfo {author}
  {\bibfnamefont {D.}~\bibnamefont {Kafri}}, \bibinfo {author} {\bibfnamefont
  {J.}~\bibnamefont {Kelly}}, \bibinfo {author} {\bibfnamefont
  {S.}~\bibnamefont {Kim}}, \bibinfo {author} {\bibfnamefont {A.}~\bibnamefont
  {Kitaev}}, \bibinfo {author} {\bibfnamefont {P.~V.}\ \bibnamefont {Klimov}},
  \bibinfo {author} {\bibfnamefont {A.~N.}\ \bibnamefont {Korotkov}}, \bibinfo
  {author} {\bibfnamefont {F.}~\bibnamefont {Kostritsa}}, \bibinfo {author}
  {\bibfnamefont {D.}~\bibnamefont {Landhuis}}, \bibinfo {author}
  {\bibfnamefont {P.}~\bibnamefont {Laptev}}, \bibinfo {author} {\bibfnamefont
  {E.}~\bibnamefont {Lucero}}, \bibinfo {author} {\bibfnamefont
  {O.}~\bibnamefont {Martin}}, \bibinfo {author} {\bibfnamefont {J.~R.}\
  \bibnamefont {McClean}}, \bibinfo {author} {\bibfnamefont {T.}~\bibnamefont
  {McCourt}}, \bibinfo {author} {\bibfnamefont {M.}~\bibnamefont {McEwen}},
  \bibinfo {author} {\bibfnamefont {A.}~\bibnamefont {Megrant}}, \bibinfo
  {author} {\bibfnamefont {K.~C.}\ \bibnamefont {Miao}}, \bibinfo {author}
  {\bibfnamefont {M.}~\bibnamefont {Mohseni}}, \bibinfo {author} {\bibfnamefont
  {W.}~\bibnamefont {Mruczkiewicz}}, \bibinfo {author} {\bibfnamefont
  {J.}~\bibnamefont {Mutus}}, \bibinfo {author} {\bibfnamefont
  {O.}~\bibnamefont {Naaman}}, \bibinfo {author} {\bibfnamefont
  {M.}~\bibnamefont {Neeley}}, \bibinfo {author} {\bibfnamefont
  {M.}~\bibnamefont {Newman}}, \bibinfo {author} {\bibfnamefont {M.~Y.}\
  \bibnamefont {Niu}}, \bibinfo {author} {\bibfnamefont {T.~E.}\ \bibnamefont
  {O'Brien}}, \bibinfo {author} {\bibfnamefont {A.}~\bibnamefont {Opremcak}},
  \bibinfo {author} {\bibfnamefont {E.}~\bibnamefont {Ostby}}, \bibinfo
  {author} {\bibfnamefont {B.}~\bibnamefont {Pato}}, \bibinfo {author}
  {\bibfnamefont {A.}~\bibnamefont {Petukhov}}, \bibinfo {author}
  {\bibfnamefont {N.}~\bibnamefont {Redd}}, \bibinfo {author} {\bibfnamefont
  {N.~C.}\ \bibnamefont {Rubin}}, \bibinfo {author} {\bibfnamefont
  {D.}~\bibnamefont {Sank}}, \bibinfo {author} {\bibfnamefont {K.~J.}\
  \bibnamefont {Satzinger}}, \bibinfo {author} {\bibfnamefont {V.}~\bibnamefont
  {Shvarts}}, \bibinfo {author} {\bibfnamefont {D.}~\bibnamefont {Strain}},
  \bibinfo {author} {\bibfnamefont {M.}~\bibnamefont {Szalay}}, \bibinfo
  {author} {\bibfnamefont {M.~D.}\ \bibnamefont {Trevithick}}, \bibinfo
  {author} {\bibfnamefont {B.}~\bibnamefont {Villalonga}}, \bibinfo {author}
  {\bibfnamefont {T.}~\bibnamefont {White}}, \bibinfo {author} {\bibfnamefont
  {Z.~J.}\ \bibnamefont {Yao}}, \bibinfo {author} {\bibfnamefont
  {P.}~\bibnamefont {Yeh}}, \bibinfo {author} {\bibfnamefont {A.}~\bibnamefont
  {Zalcman}}, \bibinfo {author} {\bibfnamefont {H.}~\bibnamefont {Neven}},
  \bibinfo {author} {\bibfnamefont {I.}~\bibnamefont {Aleiner}}, \bibinfo
  {author} {\bibfnamefont {K.}~\bibnamefont {Kechedzhi}}, \bibinfo {author}
  {\bibfnamefont {V.}~\bibnamefont {Smelyanskiy}},\ and\ \bibinfo {author}
  {\bibfnamefont {Y.}~\bibnamefont {Chen}},\ }\bibfield  {title} {\bibinfo
  {title} {Information {{Scrambling}} in {{Computationally Complex Quantum
  Circuits}}},\ }\href {https://doi.org/10.1126/science.abg5029} {\bibfield
  {journal} {\bibinfo  {journal} {Science}\ }\textbf {\bibinfo {volume}
  {374}},\ \bibinfo {pages} {1479} (\bibinfo {year} {2021})},\ \Eprint
  {https://arxiv.org/abs/2101.08870} {arXiv:2101.08870 [cond-mat,
  physics:hep-th, physics:quant-ph]} \BibitemShut {NoStop}%
\bibitem [{\citenamefont {Sekino}\ and\ \citenamefont
  {Susskind}(2008)}]{sekino2008fast}%
  \BibitemOpen
  \bibfield  {author} {\bibinfo {author} {\bibfnamefont {Y.}~\bibnamefont
  {Sekino}}\ and\ \bibinfo {author} {\bibfnamefont {L.}~\bibnamefont
  {Susskind}},\ }\bibfield  {title} {\bibinfo {title} {Fast scramblers},\
  }\href@noop {} {\bibfield  {journal} {\bibinfo  {journal} {Journal of High
  Energy Physics}\ }\textbf {\bibinfo {volume} {2008}},\ \bibinfo {pages} {065}
  (\bibinfo {year} {2008})}\BibitemShut {NoStop}%
\bibitem [{\citenamefont {Hayden}\ and\ \citenamefont
  {Preskill}(2007)}]{hayden2007black}%
  \BibitemOpen
  \bibfield  {author} {\bibinfo {author} {\bibfnamefont {P.}~\bibnamefont
  {Hayden}}\ and\ \bibinfo {author} {\bibfnamefont {J.}~\bibnamefont
  {Preskill}},\ }\bibfield  {title} {\bibinfo {title} {Black holes as mirrors:
  quantum information in random subsystems},\ }\href@noop {} {\bibfield
  {journal} {\bibinfo  {journal} {Journal of high energy physics}\ }\textbf
  {\bibinfo {volume} {2007}},\ \bibinfo {pages} {120} (\bibinfo {year}
  {2007})}\BibitemShut {NoStop}%
\bibitem [{\citenamefont {Conway}\ and\ \citenamefont
  {Sloane}(1999)}]{conwaySpherePackingsLattices1999}%
  \BibitemOpen
  \bibfield  {author} {\bibinfo {author} {\bibfnamefont {J.~H.}\ \bibnamefont
  {Conway}}\ and\ \bibinfo {author} {\bibfnamefont {N.~J.~A.}\ \bibnamefont
  {Sloane}},\ }\href
  {https://archive.org/details/spherepackingsla0000conw_b8u0} {\emph {\bibinfo
  {title} {Sphere Packings, Lattices, and Groups}}}\ (\bibinfo  {publisher}
  {{Springer}},\ \bibinfo {address} {{New York}},\ \bibinfo {year}
  {1999})\BibitemShut {NoStop}%
\bibitem [{\citenamefont {Wu}\ and\ \citenamefont {Verdu}(2010)}]{Wu2010}%
  \BibitemOpen
  \bibfield  {author} {\bibinfo {author} {\bibfnamefont {Y.}~\bibnamefont
  {Wu}}\ and\ \bibinfo {author} {\bibfnamefont {S.}~\bibnamefont {Verdu}},\
  }\bibfield  {title} {\bibinfo {title} {{The impact of constellation
  cardinality on Gaussian channel capacity}},\ }in\ \href
  {https://doi.org/10.1109/ALLERTON.2010.5706965} {\emph {\bibinfo {booktitle}
  {2010 48th Annu. Allert. Conf. Commun. Control. Comput.}}}\ (\bibinfo
  {publisher} {IEEE},\ \bibinfo {year} {2010})\ pp.\ \bibinfo {pages}
  {620--628}\BibitemShut {NoStop}%
\bibitem [{\citenamefont {Lacerda}\ \emph {et~al.}(2016)\citenamefont
  {Lacerda}, \citenamefont {Renes},\ and\ \citenamefont
  {Scholz}}]{Lacerda2016}%
  \BibitemOpen
  \bibfield  {author} {\bibinfo {author} {\bibfnamefont {F.}~\bibnamefont
  {Lacerda}}, \bibinfo {author} {\bibfnamefont {J.~M.}\ \bibnamefont {Renes}},\
  and\ \bibinfo {author} {\bibfnamefont {V.~B.}\ \bibnamefont {Scholz}},\
  }\bibfield  {title} {\bibinfo {title} {{Coherent state constellations for
  Bosonic Gaussian channels}},\ }\href
  {https://doi.org/10.1109/ISIT.2016.7541749} {\bibfield  {journal} {\bibinfo
  {journal} {2016 IEEE Int. Symp. Inf. Theory}\ ,\ \bibinfo {pages} {2499}}
  (\bibinfo {year} {2016})}\BibitemShut {NoStop}%
\bibitem [{\citenamefont {Lacerda}\ \emph {et~al.}(2017)\citenamefont
  {Lacerda}, \citenamefont {Renes},\ and\ \citenamefont
  {Scholz}}]{Lacerda2017}%
  \BibitemOpen
  \bibfield  {author} {\bibinfo {author} {\bibfnamefont {F.}~\bibnamefont
  {Lacerda}}, \bibinfo {author} {\bibfnamefont {J.~M.}\ \bibnamefont {Renes}},\
  and\ \bibinfo {author} {\bibfnamefont {V.~B.}\ \bibnamefont {Scholz}},\
  }\bibfield  {title} {\bibinfo {title} {{Coherent-state constellations and
  polar codes for thermal Gaussian channels}},\ }\href
  {https://doi.org/10.1103/PhysRevA.95.062343} {\bibfield  {journal} {\bibinfo
  {journal} {Phys. Rev. A}\ }\textbf {\bibinfo {volume} {95}},\ \bibinfo
  {pages} {062343} (\bibinfo {year} {2017})},\ \Eprint
  {https://arxiv.org/abs/1603.05970} {arXiv:1603.05970} \BibitemShut {NoStop}%
\bibitem [{\citenamefont {Blume-Kohout}\ and\ \citenamefont
  {Turner}(2014)}]{blume-kohout_curious_2014}%
  \BibitemOpen
  \bibfield  {author} {\bibinfo {author} {\bibfnamefont {R.}~\bibnamefont
  {Blume-Kohout}}\ and\ \bibinfo {author} {\bibfnamefont {P.~S.}\ \bibnamefont
  {Turner}},\ }\bibfield  {title} {\bibinfo {title} {The curious nonexistence
  of {G}aussian 2-designs},\ }\href@noop {} {\bibfield  {journal} {\bibinfo
  {journal} {Communications in Mathematical Physics}\ }\textbf {\bibinfo
  {volume} {326}},\ \bibinfo {pages} {755} (\bibinfo {year}
  {2014})}\BibitemShut {NoStop}%
\bibitem [{\citenamefont {Weedbrook}\ \emph {et~al.}(2012)\citenamefont
  {Weedbrook}, \citenamefont {Pirandola}, \citenamefont
  {Garc{\'{i}}a-Patr{\'{o}}n}, \citenamefont {Cerf}, \citenamefont {Ralph},
  \citenamefont {Shapiro},\ and\ \citenamefont {Lloyd}}]{Weedbrook2012}%
  \BibitemOpen
  \bibfield  {author} {\bibinfo {author} {\bibfnamefont {C.}~\bibnamefont
  {Weedbrook}}, \bibinfo {author} {\bibfnamefont {S.}~\bibnamefont
  {Pirandola}}, \bibinfo {author} {\bibfnamefont {R.}~\bibnamefont
  {Garc{\'{i}}a-Patr{\'{o}}n}}, \bibinfo {author} {\bibfnamefont {N.~J.}\
  \bibnamefont {Cerf}}, \bibinfo {author} {\bibfnamefont {T.~C.}\ \bibnamefont
  {Ralph}}, \bibinfo {author} {\bibfnamefont {J.~H.}\ \bibnamefont {Shapiro}},\
  and\ \bibinfo {author} {\bibfnamefont {S.}~\bibnamefont {Lloyd}},\ }\bibfield
   {title} {\bibinfo {title} {{Gaussian quantum information}},\ }\href
  {https://doi.org/10.1103/RevModPhys.84.621} {\bibfield  {journal} {\bibinfo
  {journal} {Rev. Mod. Phys.}\ }\textbf {\bibinfo {volume} {84}},\ \bibinfo
  {pages} {621} (\bibinfo {year} {2012})}\BibitemShut {NoStop}%
\bibitem [{\citenamefont {Webb}(2016)}]{webbCliffordGroupForms2016}%
  \BibitemOpen
  \bibfield  {author} {\bibinfo {author} {\bibfnamefont {Z.}~\bibnamefont
  {Webb}},\ }\href {http://arxiv.org/abs/1510.02769} {\bibinfo {title} {The
  {{Clifford}} group forms a unitary 3-design}} (\bibinfo {year} {2016}),\
  \Eprint {https://arxiv.org/abs/1510.02769} {arXiv:1510.02769 [quant-ph]}
  \BibitemShut {NoStop}%
\bibitem [{\citenamefont {Zhu}\ \emph {et~al.}(2016)\citenamefont {Zhu},
  \citenamefont {Kueng}, \citenamefont {Grassl},\ and\ \citenamefont
  {Gross}}]{zhuCliffordGroupFails2016}%
  \BibitemOpen
  \bibfield  {author} {\bibinfo {author} {\bibfnamefont {H.}~\bibnamefont
  {Zhu}}, \bibinfo {author} {\bibfnamefont {R.}~\bibnamefont {Kueng}}, \bibinfo
  {author} {\bibfnamefont {M.}~\bibnamefont {Grassl}},\ and\ \bibinfo {author}
  {\bibfnamefont {D.}~\bibnamefont {Gross}},\ }\href
  {http://arxiv.org/abs/1609.08172} {\bibinfo {title} {The {{Clifford}} group
  fails gracefully to be a unitary 4-design}} (\bibinfo {year} {2016}),\
  \Eprint {https://arxiv.org/abs/1609.08172} {arXiv:1609.08172 [quant-ph]}
  \BibitemShut {NoStop}%
\bibitem [{\citenamefont {Zhu}(2017)}]{zhuMultiqubitCliffordGroups2017}%
  \BibitemOpen
  \bibfield  {author} {\bibinfo {author} {\bibfnamefont {H.}~\bibnamefont
  {Zhu}},\ }\bibfield  {title} {\bibinfo {title} {Multiqubit {{Clifford}}
  groups are unitary 3-designs},\ }\href
  {https://doi.org/10.1103/PhysRevA.96.062336} {\bibfield  {journal} {\bibinfo
  {journal} {Physical Review A}\ }\textbf {\bibinfo {volume} {96}},\ \bibinfo
  {pages} {062336} (\bibinfo {year} {2017})}\BibitemShut {NoStop}%
\bibitem [{\citenamefont {Graydon}\ \emph {et~al.}(2021)\citenamefont
  {Graydon}, \citenamefont {{Skanes-Norman}},\ and\ \citenamefont
  {Wallman}}]{graydonCliffordGroupsAre2021}%
  \BibitemOpen
  \bibfield  {author} {\bibinfo {author} {\bibfnamefont {M.~A.}\ \bibnamefont
  {Graydon}}, \bibinfo {author} {\bibfnamefont {J.}~\bibnamefont
  {{Skanes-Norman}}},\ and\ \bibinfo {author} {\bibfnamefont {J.~J.}\
  \bibnamefont {Wallman}},\ }\href {http://arxiv.org/abs/2108.04200} {\bibinfo
  {title} {Clifford groups are not always 2-designs}} (\bibinfo {year}
  {2021}),\ \Eprint {https://arxiv.org/abs/2108.04200} {arXiv:2108.04200
  [quant-ph]} \BibitemShut {NoStop}%
\bibitem [{\citenamefont {Zhuang}\ \emph {et~al.}(2019)\citenamefont {Zhuang},
  \citenamefont {Schuster}, \citenamefont {Yoshida},\ and\ \citenamefont
  {Yao}}]{zhuang_scrambling_2019}%
  \BibitemOpen
  \bibfield  {author} {\bibinfo {author} {\bibfnamefont {Q.}~\bibnamefont
  {Zhuang}}, \bibinfo {author} {\bibfnamefont {T.}~\bibnamefont {Schuster}},
  \bibinfo {author} {\bibfnamefont {B.}~\bibnamefont {Yoshida}},\ and\ \bibinfo
  {author} {\bibfnamefont {N.~Y.}\ \bibnamefont {Yao}},\ }\bibfield  {title}
  {\bibinfo {title} {Scrambling and complexity in phase space},\ }\href@noop {}
  {\bibfield  {journal} {\bibinfo  {journal} {Physical Review A}\ }\textbf
  {\bibinfo {volume} {99}},\ \bibinfo {pages} {062334} (\bibinfo {year}
  {2019})}\BibitemShut {NoStop}%
\bibitem [{\citenamefont {Susskind}\ and\ \citenamefont
  {Glogower}(1964)}]{susskindQuantumMechanicalPhase1964}%
  \BibitemOpen
  \bibfield  {author} {\bibinfo {author} {\bibfnamefont {L.}~\bibnamefont
  {Susskind}}\ and\ \bibinfo {author} {\bibfnamefont {J.}~\bibnamefont
  {Glogower}},\ }\bibfield  {title} {\bibinfo {title} {Quantum mechanical phase
  and time operator},\ }\href
  {https://doi.org/10.1103/PhysicsPhysiqueFizika.1.49} {\bibfield  {journal}
  {\bibinfo  {journal} {Physics Physique Fizika}\ }\textbf {\bibinfo {volume}
  {1}},\ \bibinfo {pages} {49} (\bibinfo {year} {1964})}\BibitemShut {NoStop}%
\bibitem [{\citenamefont {Carruthers}\ and\ \citenamefont
  {Nieto}(1968)}]{carruthers_phase_1968}%
  \BibitemOpen
  \bibfield  {author} {\bibinfo {author} {\bibfnamefont {P.}~\bibnamefont
  {Carruthers}}\ and\ \bibinfo {author} {\bibfnamefont {M.~M.}\ \bibnamefont
  {Nieto}},\ }\bibfield  {title} {\bibinfo {title} {Phase and {Angle}
  {Variables} in {Quantum} {Mechanics}},\ }\href
  {https://doi.org/10.1103/RevModPhys.40.411} {\bibfield  {journal} {\bibinfo
  {journal} {Reviews of Modern Physics}\ }\textbf {\bibinfo {volume} {40}},\
  \bibinfo {pages} {411} (\bibinfo {year} {1968})}\BibitemShut {NoStop}%
\bibitem [{\citenamefont
  {Helstrom}(1969)}]{helstromQuantumDetectionEstimation1969}%
  \BibitemOpen
  \bibfield  {author} {\bibinfo {author} {\bibfnamefont {C.~W.}\ \bibnamefont
  {Helstrom}},\ }\bibfield  {title} {\bibinfo {title} {Quantum detection and
  estimation theory},\ }\href {https://doi.org/10.1007/BF01007479} {\bibfield
  {journal} {\bibinfo  {journal} {Journal of Statistical Physics}\ }\textbf
  {\bibinfo {volume} {1}},\ \bibinfo {pages} {231} (\bibinfo {year}
  {1969})}\BibitemShut {NoStop}%
\bibitem [{\citenamefont
  {Holevo}(1984)}]{holevoCovariantMeasurementsImprimitivity1984}%
  \BibitemOpen
  \bibfield  {author} {\bibinfo {author} {\bibfnamefont {A.~S.}\ \bibnamefont
  {Holevo}},\ }\bibfield  {title} {\bibinfo {title} {Covariant measurements and
  imprimitivity systems},\ }in\ \href {https://doi.org/10.1007/BFb0071720}
  {\emph {\bibinfo {booktitle} {Quantum {{Probability}} and {{Applications}} to
  the {{Quantum Theory}} of {{Irreversible Processes}}}}},\ Vol.\ \bibinfo
  {volume} {1055},\ \bibinfo {editor} {edited by\ \bibinfo {editor}
  {\bibfnamefont {L.}~\bibnamefont {Accardi}}, \bibinfo {editor} {\bibfnamefont
  {A.}~\bibnamefont {Frigerio}},\ and\ \bibinfo {editor} {\bibfnamefont
  {V.}~\bibnamefont {Gorini}}}\ (\bibinfo  {publisher} {{Springer Berlin
  Heidelberg}},\ \bibinfo {address} {{Berlin, Heidelberg}},\ \bibinfo {year}
  {1984})\ pp.\ \bibinfo {pages} {153--172}\BibitemShut {NoStop}%
\bibitem [{\citenamefont {Bergou}\ and\ \citenamefont
  {Englert}(1991)}]{Bergou1991}%
  \BibitemOpen
  \bibfield  {author} {\bibinfo {author} {\bibfnamefont {J.}~\bibnamefont
  {Bergou}}\ and\ \bibinfo {author} {\bibfnamefont {B.-G.}\ \bibnamefont
  {Englert}},\ }\bibfield  {title} {\bibinfo {title} {{Operators of the phase.
  Fundamentals}},\ }\href {https://doi.org/10.1016/0003-4916(91)90037-9}
  {\bibfield  {journal} {\bibinfo  {journal} {Annals of Physics}\ }\textbf
  {\bibinfo {volume} {209}},\ \bibinfo {pages} {479} (\bibinfo {year}
  {1991})}\BibitemShut {NoStop}%
\bibitem [{\citenamefont {Mathews}\ and\ \citenamefont
  {Eswaran}(1974)}]{mathewsSimultaneousUncertaintiesCosine1974}%
  \BibitemOpen
  \bibfield  {author} {\bibinfo {author} {\bibfnamefont {P.~M.}\ \bibnamefont
  {Mathews}}\ and\ \bibinfo {author} {\bibfnamefont {K.}~\bibnamefont
  {Eswaran}},\ }\bibfield  {title} {\bibinfo {title} {Simultaneous
  uncertainties of the cosine and sine operators},\ }\href
  {https://doi.org/10.1007/BF02749759} {\bibfield  {journal} {\bibinfo
  {journal} {Il Nuovo Cimento B (1971-1996)}\ }\textbf {\bibinfo {volume}
  {19}},\ \bibinfo {pages} {99} (\bibinfo {year} {1974})}\BibitemShut {NoStop}%
\bibitem [{\citenamefont {Shapiro}\ and\ \citenamefont
  {Shepard}(1991)}]{shapiroQuantumPhaseMeasurement1991}%
  \BibitemOpen
  \bibfield  {author} {\bibinfo {author} {\bibfnamefont {J.~H.}\ \bibnamefont
  {Shapiro}}\ and\ \bibinfo {author} {\bibfnamefont {S.~R.}\ \bibnamefont
  {Shepard}},\ }\bibfield  {title} {\bibinfo {title} {Quantum phase
  measurement: {{A}} system-theory perspective},\ }\href
  {https://doi.org/10.1103/PhysRevA.43.3795} {\bibfield  {journal} {\bibinfo
  {journal} {Physical Review A}\ }\textbf {\bibinfo {volume} {43}},\ \bibinfo
  {pages} {3795} (\bibinfo {year} {1991})}\BibitemShut {NoStop}%
\bibitem [{\citenamefont {Holevo}(2011)}]{Holevo2011}%
  \BibitemOpen
  \bibfield  {author} {\bibinfo {author} {\bibfnamefont {A.}~\bibnamefont
  {Holevo}},\ }\href {https://doi.org/10.1007/978-88-7642-378-9} {\emph
  {\bibinfo {title} {{Probabilistic and Statistical Aspects of Quantum
  Theory}}}}\ (\bibinfo  {publisher} {Edizioni della Normale},\ \bibinfo
  {address} {Pisa},\ \bibinfo {year} {2011})\BibitemShut {NoStop}%
\bibitem [{\citenamefont {Girvin}(2014)}]{girvin2014circuit}%
  \BibitemOpen
  \bibfield  {author} {\bibinfo {author} {\bibfnamefont {S.~M.}\ \bibnamefont
  {Girvin}},\ }\bibfield  {title} {\bibinfo {title} {Circuit qed:
  superconducting qubits coupled to microwave photons},\ }\href@noop {}
  {\bibfield  {journal} {\bibinfo  {journal} {Quantum machines: measurement and
  control of engineered quantum systems}\ ,\ \bibinfo {pages} {113}} (\bibinfo
  {year} {2014})}\BibitemShut {NoStop}%
\bibitem [{\citenamefont {Gandhari}\ \emph {et~al.}(2022)\citenamefont
  {Gandhari}, \citenamefont {Albert}, \citenamefont {Gerrits}, \citenamefont
  {Taylor},\ and\ \citenamefont {Gullans}}]{Gandhari2022}%
  \BibitemOpen
  \bibfield  {author} {\bibinfo {author} {\bibfnamefont {S.}~\bibnamefont
  {Gandhari}}, \bibinfo {author} {\bibfnamefont {V.~V.}\ \bibnamefont
  {Albert}}, \bibinfo {author} {\bibfnamefont {T.}~\bibnamefont {Gerrits}},
  \bibinfo {author} {\bibfnamefont {J.~M.}\ \bibnamefont {Taylor}},\ and\
  \bibinfo {author} {\bibfnamefont {M.~J.}\ \bibnamefont {Gullans}},\ }\href
  {https://doi.org/10.48550/arXiv.2211.05149} {\bibinfo {title}
  {Continuous-{{Variable Shadow Tomography}}}} (\bibinfo {year} {2022}),\
  \Eprint {https://arxiv.org/abs/2211.05149} {arXiv:2211.05149 [quant-ph]}
  \BibitemShut {NoStop}%
\bibitem [{\citenamefont {Wiseman}(1995)}]{Wiseman95}%
  \BibitemOpen
  \bibfield  {author} {\bibinfo {author} {\bibfnamefont {H.~M.}\ \bibnamefont
  {Wiseman}},\ }\bibfield  {title} {\bibinfo {title} {Adaptive phase
  measurements of optical modes: Going beyond the marginal $q$ distribution},\
  }\href {https://doi.org/10.1103/PhysRevLett.75.4587} {\bibfield  {journal}
  {\bibinfo  {journal} {Phys. Rev. Lett.}\ }\textbf {\bibinfo {volume} {75}},\
  \bibinfo {pages} {4587} (\bibinfo {year} {1995})}\BibitemShut {NoStop}%
\bibitem [{\citenamefont {Martin}\ \emph {et~al.}(2020)\citenamefont {Martin},
  \citenamefont {Livingston}, \citenamefont {Hacohen-Gourgy}, \citenamefont
  {Wiseman},\ and\ \citenamefont {Siddiqi}}]{Martin2020}%
  \BibitemOpen
  \bibfield  {author} {\bibinfo {author} {\bibfnamefont {L.~S.}\ \bibnamefont
  {Martin}}, \bibinfo {author} {\bibfnamefont {W.~P.}\ \bibnamefont
  {Livingston}}, \bibinfo {author} {\bibfnamefont {S.}~\bibnamefont
  {Hacohen-Gourgy}}, \bibinfo {author} {\bibfnamefont {H.~M.}\ \bibnamefont
  {Wiseman}},\ and\ \bibinfo {author} {\bibfnamefont {I.}~\bibnamefont
  {Siddiqi}},\ }\bibfield  {title} {\bibinfo {title} {{Implementation of a
  canonical phase measurement with quantum feedback}},\ }\href
  {https://doi.org/10.1038/s41567-020-0939-0} {\bibfield  {journal} {\bibinfo
  {journal} {Nature Physics}\ }\textbf {\bibinfo {volume} {16}},\ \bibinfo
  {pages} {1046} (\bibinfo {year} {2020})}\BibitemShut {NoStop}%
\bibitem [{\citenamefont {Schuster}\ \emph {et~al.}(2007)\citenamefont
  {Schuster}, \citenamefont {Houck}, \citenamefont {Schreier}, \citenamefont
  {Wallraff}, \citenamefont {Gambetta}, \citenamefont {Blais}, \citenamefont
  {Frunzio}, \citenamefont {Majer}, \citenamefont {Johnson}, \citenamefont
  {Devoret}, \citenamefont {Girvin},\ and\ \citenamefont
  {Schoelkopf}}]{Schuster2007}%
  \BibitemOpen
  \bibfield  {author} {\bibinfo {author} {\bibfnamefont {D.~I.}\ \bibnamefont
  {Schuster}}, \bibinfo {author} {\bibfnamefont {A.~A.}\ \bibnamefont {Houck}},
  \bibinfo {author} {\bibfnamefont {J.~A.}\ \bibnamefont {Schreier}}, \bibinfo
  {author} {\bibfnamefont {A.}~\bibnamefont {Wallraff}}, \bibinfo {author}
  {\bibfnamefont {J.~M.}\ \bibnamefont {Gambetta}}, \bibinfo {author}
  {\bibfnamefont {A.}~\bibnamefont {Blais}}, \bibinfo {author} {\bibfnamefont
  {L.}~\bibnamefont {Frunzio}}, \bibinfo {author} {\bibfnamefont
  {J.}~\bibnamefont {Majer}}, \bibinfo {author} {\bibfnamefont
  {B.}~\bibnamefont {Johnson}}, \bibinfo {author} {\bibfnamefont {M.~H.}\
  \bibnamefont {Devoret}}, \bibinfo {author} {\bibfnamefont {S.~M.}\
  \bibnamefont {Girvin}},\ and\ \bibinfo {author} {\bibfnamefont {R.~J.}\
  \bibnamefont {Schoelkopf}},\ }\bibfield  {title} {\bibinfo {title}
  {{Resolving photon number states in a superconducting circuit}},\ }\href
  {http://dx.doi.org/10.1038/nature05461} {\bibfield  {journal} {\bibinfo
  {journal} {Nature}\ }\textbf {\bibinfo {volume} {445}},\ \bibinfo {pages}
  {515} (\bibinfo {year} {2007})}\BibitemShut {NoStop}%
\bibitem [{\citenamefont {Holland}\ \emph {et~al.}(2015)\citenamefont
  {Holland}, \citenamefont {Vlastakis}, \citenamefont {Heeres}, \citenamefont
  {Reagor}, \citenamefont {Vool}, \citenamefont {Leghtas}, \citenamefont
  {Frunzio}, \citenamefont {Kirchmair}, \citenamefont {Devoret}, \citenamefont
  {Mirrahimi},\ and\ \citenamefont {Schoelkopf}}]{Holland2015}%
  \BibitemOpen
  \bibfield  {author} {\bibinfo {author} {\bibfnamefont {E.~T.}\ \bibnamefont
  {Holland}}, \bibinfo {author} {\bibfnamefont {B.}~\bibnamefont {Vlastakis}},
  \bibinfo {author} {\bibfnamefont {R.~W.}\ \bibnamefont {Heeres}}, \bibinfo
  {author} {\bibfnamefont {M.~J.}\ \bibnamefont {Reagor}}, \bibinfo {author}
  {\bibfnamefont {U.}~\bibnamefont {Vool}}, \bibinfo {author} {\bibfnamefont
  {Z.}~\bibnamefont {Leghtas}}, \bibinfo {author} {\bibfnamefont
  {L.}~\bibnamefont {Frunzio}}, \bibinfo {author} {\bibfnamefont
  {G.}~\bibnamefont {Kirchmair}}, \bibinfo {author} {\bibfnamefont {M.~H.}\
  \bibnamefont {Devoret}}, \bibinfo {author} {\bibfnamefont {M.}~\bibnamefont
  {Mirrahimi}},\ and\ \bibinfo {author} {\bibfnamefont {R.~J.}\ \bibnamefont
  {Schoelkopf}},\ }\bibfield  {title} {\bibinfo {title}
  {{Single-Photon-Resolved Cross-Kerr Interaction for Autonomous Stabilization
  of Photon-Number States}},\ }\href
  {https://doi.org/10.1103/PhysRevLett.115.180501} {\bibfield  {journal}
  {\bibinfo  {journal} {Phys. Rev. Lett.}\ }\textbf {\bibinfo {volume} {115}},\
  \bibinfo {pages} {180501} (\bibinfo {year} {2015})}\BibitemShut {NoStop}%
\bibitem [{\citenamefont {Elliott}\ \emph {et~al.}(2018)\citenamefont
  {Elliott}, \citenamefont {Joo},\ and\ \citenamefont
  {Ginossar}}]{Elliott2018}%
  \BibitemOpen
  \bibfield  {author} {\bibinfo {author} {\bibfnamefont {M.}~\bibnamefont
  {Elliott}}, \bibinfo {author} {\bibfnamefont {J.}~\bibnamefont {Joo}},\ and\
  \bibinfo {author} {\bibfnamefont {E.}~\bibnamefont {Ginossar}},\ }\bibfield
  {title} {\bibinfo {title} {{Designing Kerr interactions using multiple
  superconducting qubit types in a single circuit}},\ }\href
  {https://doi.org/10.1088/1367-2630/aa9243} {\bibfield  {journal} {\bibinfo
  {journal} {New Journal of Physics}\ }\textbf {\bibinfo {volume} {20}},\
  \bibinfo {pages} {023037} (\bibinfo {year} {2018})}\BibitemShut {NoStop}%
\bibitem [{\citenamefont {Zhang}\ \emph {et~al.}(2022)\citenamefont {Zhang},
  \citenamefont {Curtis}, \citenamefont {Wang}, \citenamefont {Schoelkopf},\
  and\ \citenamefont {Girvin}}]{Zhang2022}%
  \BibitemOpen
  \bibfield  {author} {\bibinfo {author} {\bibfnamefont {Y.}~\bibnamefont
  {Zhang}}, \bibinfo {author} {\bibfnamefont {J.~C.}\ \bibnamefont {Curtis}},
  \bibinfo {author} {\bibfnamefont {C.~S.}\ \bibnamefont {Wang}}, \bibinfo
  {author} {\bibfnamefont {R.~J.}\ \bibnamefont {Schoelkopf}},\ and\ \bibinfo
  {author} {\bibfnamefont {S.~M.}\ \bibnamefont {Girvin}},\ }\bibfield  {title}
  {\bibinfo {title} {{Drive-induced nonlinearities of cavity modes coupled to a
  transmon ancilla}},\ }\href {https://doi.org/10.1103/PhysRevA.105.022423}
  {\bibfield  {journal} {\bibinfo  {journal} {Physical Review A}\ }\textbf
  {\bibinfo {volume} {105}},\ \bibinfo {pages} {022423} (\bibinfo {year}
  {2022})}\BibitemShut {NoStop}%
\bibitem [{\citenamefont {Scott}(2008)}]{scottOptimizingQuantumProcess2008}%
  \BibitemOpen
  \bibfield  {author} {\bibinfo {author} {\bibfnamefont {A.~J.}\ \bibnamefont
  {Scott}},\ }\bibfield  {title} {\bibinfo {title} {Optimizing quantum process
  tomography with unitary 2-designs},\ }\href
  {https://doi.org/10.1088/1751-8113/41/5/055308} {\bibfield  {journal}
  {\bibinfo  {journal} {Journal of Physics A: Mathematical and Theoretical}\
  }\textbf {\bibinfo {volume} {41}},\ \bibinfo {pages} {055308} (\bibinfo
  {year} {2008})}\BibitemShut {NoStop}%
\bibitem [{\citenamefont {Gottesman}\ \emph {et~al.}(2001)\citenamefont
  {Gottesman}, \citenamefont {{Yu. Kitaev}},\ and\ \citenamefont
  {Preskill}}]{Gottesman2001}%
  \BibitemOpen
  \bibfield  {author} {\bibinfo {author} {\bibfnamefont {D.}~\bibnamefont
  {Gottesman}}, \bibinfo {author} {\bibfnamefont {A.}~\bibnamefont {{Yu.
  Kitaev}}},\ and\ \bibinfo {author} {\bibfnamefont {J.}~\bibnamefont
  {Preskill}},\ }\bibfield  {title} {\bibinfo {title} {{Encoding a qubit in an
  oscillator}},\ }\href {https://doi.org/10.1103/PhysRevA.64.012310} {\bibfield
   {journal} {\bibinfo  {journal} {Phys. Rev. A}\ }\textbf {\bibinfo {volume}
  {64}},\ \bibinfo {pages} {012310} (\bibinfo {year} {2001})}\BibitemShut
  {NoStop}%
\bibitem [{\citenamefont {Albert}\ \emph {et~al.}(2020)\citenamefont {Albert},
  \citenamefont {Covey},\ and\ \citenamefont {Preskill}}]{mol}%
  \BibitemOpen
  \bibfield  {author} {\bibinfo {author} {\bibfnamefont {V.~V.}\ \bibnamefont
  {Albert}}, \bibinfo {author} {\bibfnamefont {J.~P.}\ \bibnamefont {Covey}},\
  and\ \bibinfo {author} {\bibfnamefont {J.}~\bibnamefont {Preskill}},\
  }\bibfield  {title} {\bibinfo {title} {{Robust Encoding of a Qubit in a
  Molecule}},\ }\href {https://doi.org/10.1103/PhysRevX.10.031050} {\bibfield
  {journal} {\bibinfo  {journal} {Physical Review X}\ }\textbf {\bibinfo
  {volume} {10}},\ \bibinfo {pages} {031050} (\bibinfo {year} {2020})},\
  \Eprint {https://arxiv.org/abs/1911.00099} {arXiv:1911.00099} \BibitemShut
  {NoStop}%
\bibitem [{\citenamefont {Menicucci}(2014)}]{Menicucci2014}%
  \BibitemOpen
  \bibfield  {author} {\bibinfo {author} {\bibfnamefont {N.~C.}\ \bibnamefont
  {Menicucci}},\ }\bibfield  {title} {\bibinfo {title} {{Fault-Tolerant
  Measurement-Based Quantum Computing with Continuous-Variable Cluster
  States}},\ }\href {https://doi.org/10.1103/PhysRevLett.112.120504} {\bibfield
   {journal} {\bibinfo  {journal} {Phys. Rev. Lett.}\ }\textbf {\bibinfo
  {volume} {112}},\ \bibinfo {pages} {120504} (\bibinfo {year}
  {2014})}\BibitemShut {NoStop}%
\bibitem [{\citenamefont {Winter}(2016)}]{winter2016tight}%
  \BibitemOpen
  \bibfield  {author} {\bibinfo {author} {\bibfnamefont {A.}~\bibnamefont
  {Winter}},\ }\bibfield  {title} {\bibinfo {title} {Tight uniform continuity
  bounds for quantum entropies: conditional entropy, relative entropy distance
  and energy constraints},\ }\href@noop {} {\bibfield  {journal} {\bibinfo
  {journal} {Communications in Mathematical Physics}\ }\textbf {\bibinfo
  {volume} {347}},\ \bibinfo {pages} {291} (\bibinfo {year}
  {2016})}\BibitemShut {NoStop}%
\bibitem [{\citenamefont {Sharma}\ \emph {et~al.}(2018)\citenamefont {Sharma},
  \citenamefont {Wilde}, \citenamefont {Adhikari},\ and\ \citenamefont
  {Takeoka}}]{sharma2018bounding}%
  \BibitemOpen
  \bibfield  {author} {\bibinfo {author} {\bibfnamefont {K.}~\bibnamefont
  {Sharma}}, \bibinfo {author} {\bibfnamefont {M.~M.}\ \bibnamefont {Wilde}},
  \bibinfo {author} {\bibfnamefont {S.}~\bibnamefont {Adhikari}},\ and\
  \bibinfo {author} {\bibfnamefont {M.}~\bibnamefont {Takeoka}},\ }\bibfield
  {title} {\bibinfo {title} {Bounding the energy-constrained quantum and
  private capacities of phase-insensitive bosonic gaussian channels},\
  }\href@noop {} {\bibfield  {journal} {\bibinfo  {journal} {New Journal of
  Physics}\ }\textbf {\bibinfo {volume} {20}},\ \bibinfo {pages} {063025}
  (\bibinfo {year} {2018})}\BibitemShut {NoStop}%
\bibitem [{\citenamefont {Winter}(2017)}]{winter2017energy}%
  \BibitemOpen
  \bibfield  {author} {\bibinfo {author} {\bibfnamefont {A.}~\bibnamefont
  {Winter}},\ }\bibfield  {title} {\bibinfo {title} {Energy-constrained diamond
  norm with applications to the uniform continuity of continuous variable
  channel capacities},\ }\href@noop {} {\bibfield  {journal} {\bibinfo
  {journal} {arXiv preprint arXiv:1712.10267}\ } (\bibinfo {year}
  {2017})}\BibitemShut {NoStop}%
\bibitem [{\citenamefont {Shirokov}(2018)}]{shirokov2018energy}%
  \BibitemOpen
  \bibfield  {author} {\bibinfo {author} {\bibfnamefont {M.~E.}\ \bibnamefont
  {Shirokov}},\ }\bibfield  {title} {\bibinfo {title} {On the
  energy-constrained diamond norm and its application in quantum information
  theory},\ }\href@noop {} {\bibfield  {journal} {\bibinfo  {journal} {Problems
  of Information Transmission}\ }\textbf {\bibinfo {volume} {54}},\ \bibinfo
  {pages} {20} (\bibinfo {year} {2018})}\BibitemShut {NoStop}%
\bibitem [{\citenamefont {Culf}\ \emph {et~al.}(2022)\citenamefont {Culf},
  \citenamefont {Vidick},\ and\ \citenamefont {Albert}}]{Culf}%
  \BibitemOpen
  \bibfield  {author} {\bibinfo {author} {\bibfnamefont {E.}~\bibnamefont
  {Culf}}, \bibinfo {author} {\bibfnamefont {T.}~\bibnamefont {Vidick}},\ and\
  \bibinfo {author} {\bibfnamefont {V.~V.}\ \bibnamefont {Albert}},\ }\bibfield
   {title} {\bibinfo {title} {{Group coset monogamy games and an application to
  device-independent continuous-variable QKD}}} (\bibinfo {year} {2022}),\
  \bibinfo {note} {to appear}\BibitemShut {NoStop}%
\bibitem [{\citenamefont {Braunstein}\ \emph {et~al.}(2000)\citenamefont
  {Braunstein}, \citenamefont {Fuchs},\ and\ \citenamefont
  {Kimble}}]{braunstein2000criteria}%
  \BibitemOpen
  \bibfield  {author} {\bibinfo {author} {\bibfnamefont {S.~L.}\ \bibnamefont
  {Braunstein}}, \bibinfo {author} {\bibfnamefont {C.~A.}\ \bibnamefont
  {Fuchs}},\ and\ \bibinfo {author} {\bibfnamefont {H.~J.}\ \bibnamefont
  {Kimble}},\ }\bibfield  {title} {\bibinfo {title} {Criteria for
  continuous-variable quantum teleportation},\ }\href@noop {} {\bibfield
  {journal} {\bibinfo  {journal} {Journal of Modern Optics}\ }\textbf {\bibinfo
  {volume} {47}},\ \bibinfo {pages} {267} (\bibinfo {year} {2000})}\BibitemShut
  {NoStop}%
\bibitem [{\citenamefont {Furusawa}\ \emph {et~al.}(1998)\citenamefont
  {Furusawa}, \citenamefont {S{\o}rensen}, \citenamefont {Braunstein},
  \citenamefont {Fuchs}, \citenamefont {Kimble},\ and\ \citenamefont
  {Polzik}}]{furusawa1998unconditional}%
  \BibitemOpen
  \bibfield  {author} {\bibinfo {author} {\bibfnamefont {A.}~\bibnamefont
  {Furusawa}}, \bibinfo {author} {\bibfnamefont {J.~L.}\ \bibnamefont
  {S{\o}rensen}}, \bibinfo {author} {\bibfnamefont {S.~L.}\ \bibnamefont
  {Braunstein}}, \bibinfo {author} {\bibfnamefont {C.~A.}\ \bibnamefont
  {Fuchs}}, \bibinfo {author} {\bibfnamefont {H.~J.}\ \bibnamefont {Kimble}},\
  and\ \bibinfo {author} {\bibfnamefont {E.~S.}\ \bibnamefont {Polzik}},\
  }\bibfield  {title} {\bibinfo {title} {Unconditional quantum teleportation},\
  }\href@noop {} {\bibfield  {journal} {\bibinfo  {journal} {science}\ }\textbf
  {\bibinfo {volume} {282}},\ \bibinfo {pages} {706} (\bibinfo {year}
  {1998})}\BibitemShut {NoStop}%
\bibitem [{\citenamefont {Namiki}\ \emph {et~al.}(2008)\citenamefont {Namiki},
  \citenamefont {Koashi},\ and\ \citenamefont {Imoto}}]{namiki2008fidelity}%
  \BibitemOpen
  \bibfield  {author} {\bibinfo {author} {\bibfnamefont {R.}~\bibnamefont
  {Namiki}}, \bibinfo {author} {\bibfnamefont {M.}~\bibnamefont {Koashi}},\
  and\ \bibinfo {author} {\bibfnamefont {N.}~\bibnamefont {Imoto}},\ }\bibfield
   {title} {\bibinfo {title} {Fidelity criterion for quantum-domain
  transmission and storage of coherent states beyond the unit-gain
  constraint},\ }\href@noop {} {\bibfield  {journal} {\bibinfo  {journal}
  {Physical review letters}\ }\textbf {\bibinfo {volume} {101}},\ \bibinfo
  {pages} {100502} (\bibinfo {year} {2008})}\BibitemShut {NoStop}%
\bibitem [{\citenamefont {Lvovsky}\ and\ \citenamefont
  {Raymer}(2009{\natexlab{a}})}]{lvovsky2009continuous}%
  \BibitemOpen
  \bibfield  {author} {\bibinfo {author} {\bibfnamefont {A.~I.}\ \bibnamefont
  {Lvovsky}}\ and\ \bibinfo {author} {\bibfnamefont {M.~G.}\ \bibnamefont
  {Raymer}},\ }\bibfield  {title} {\bibinfo {title} {Continuous-variable
  optical quantum-state tomography},\ }\href
  {https://doi.org/10.1103/RevModPhys.81.299} {\bibfield  {journal} {\bibinfo
  {journal} {Rev. Mod. Phys.}\ }\textbf {\bibinfo {volume} {81}},\ \bibinfo
  {pages} {299} (\bibinfo {year} {2009}{\natexlab{a}})}\BibitemShut {NoStop}%
\bibitem [{\citenamefont {Chiribella}\ and\ \citenamefont
  {Xie}(2013)}]{chiribella2013optimal}%
  \BibitemOpen
  \bibfield  {author} {\bibinfo {author} {\bibfnamefont {G.}~\bibnamefont
  {Chiribella}}\ and\ \bibinfo {author} {\bibfnamefont {J.}~\bibnamefont
  {Xie}},\ }\bibfield  {title} {\bibinfo {title} {Optimal design and quantum
  benchmarks for coherent state amplifiers},\ }\href@noop {} {\bibfield
  {journal} {\bibinfo  {journal} {Physical review letters}\ }\textbf {\bibinfo
  {volume} {110}},\ \bibinfo {pages} {213602} (\bibinfo {year}
  {2013})}\BibitemShut {NoStop}%
\bibitem [{\citenamefont {Bai}\ and\ \citenamefont
  {Chiribella}(2018)}]{bai2018test}%
  \BibitemOpen
  \bibfield  {author} {\bibinfo {author} {\bibfnamefont {G.}~\bibnamefont
  {Bai}}\ and\ \bibinfo {author} {\bibfnamefont {G.}~\bibnamefont
  {Chiribella}},\ }\bibfield  {title} {\bibinfo {title} {Test one to test many:
  a unified approach to quantum benchmarks},\ }\href@noop {} {\bibfield
  {journal} {\bibinfo  {journal} {Physical Review Letters}\ }\textbf {\bibinfo
  {volume} {120}},\ \bibinfo {pages} {150502} (\bibinfo {year}
  {2018})}\BibitemShut {NoStop}%
\bibitem [{\citenamefont {Wu}\ and\ \citenamefont
  {Sanders}(2019)}]{wu2019efficient}%
  \BibitemOpen
  \bibfield  {author} {\bibinfo {author} {\bibfnamefont {Y.-D.}\ \bibnamefont
  {Wu}}\ and\ \bibinfo {author} {\bibfnamefont {B.~C.}\ \bibnamefont
  {Sanders}},\ }\bibfield  {title} {\bibinfo {title} {Efficient verification of
  bosonic quantum channels via benchmarking},\ }\href@noop {} {\bibfield
  {journal} {\bibinfo  {journal} {New Journal of Physics}\ }\textbf {\bibinfo
  {volume} {21}},\ \bibinfo {pages} {073026} (\bibinfo {year}
  {2019})}\BibitemShut {NoStop}%
\bibitem [{\citenamefont {Farias}\ and\ \citenamefont
  {Aolita}(2021)}]{farias2021certification}%
  \BibitemOpen
  \bibfield  {author} {\bibinfo {author} {\bibfnamefont {R.~M.}\ \bibnamefont
  {Farias}}\ and\ \bibinfo {author} {\bibfnamefont {L.}~\bibnamefont
  {Aolita}},\ }\bibfield  {title} {\bibinfo {title} {Certification of
  continuous-variable gates using average channel-fidelity witnesses},\
  }\href@noop {} {\bibfield  {journal} {\bibinfo  {journal} {Quantum Science
  and Technology}\ }\textbf {\bibinfo {volume} {6}},\ \bibinfo {pages} {035014}
  (\bibinfo {year} {2021})}\BibitemShut {NoStop}%
\bibitem [{\citenamefont {Sharma}\ \emph {et~al.}(2022)\citenamefont {Sharma},
  \citenamefont {Sanders},\ and\ \citenamefont {Wilde}}]{sharma2022optimal}%
  \BibitemOpen
  \bibfield  {author} {\bibinfo {author} {\bibfnamefont {K.}~\bibnamefont
  {Sharma}}, \bibinfo {author} {\bibfnamefont {B.~C.}\ \bibnamefont
  {Sanders}},\ and\ \bibinfo {author} {\bibfnamefont {M.~M.}\ \bibnamefont
  {Wilde}},\ }\bibfield  {title} {\bibinfo {title} {Optimal tests for
  continuous-variable quantum teleportation and photodetectors},\ }\href@noop
  {} {\bibfield  {journal} {\bibinfo  {journal} {Physical Review Research}\
  }\textbf {\bibinfo {volume} {4}},\ \bibinfo {pages} {023066} (\bibinfo {year}
  {2022})}\BibitemShut {NoStop}%
\bibitem [{\citenamefont {Sharma}\ and\ \citenamefont
  {Wilde}(2020)}]{sharmaCharacterizingPerformanceContinuousvariable2020}%
  \BibitemOpen
  \bibfield  {author} {\bibinfo {author} {\bibfnamefont {K.}~\bibnamefont
  {Sharma}}\ and\ \bibinfo {author} {\bibfnamefont {M.~M.}\ \bibnamefont
  {Wilde}},\ }\bibfield  {title} {\bibinfo {title} {Characterizing the
  performance of continuous-variable {{Gaussian}} quantum gates},\ }\href
  {https://doi.org/10.1103/PhysRevResearch.2.013126} {\bibfield  {journal}
  {\bibinfo  {journal} {Physical Review Research}\ }\textbf {\bibinfo {volume}
  {2}},\ \bibinfo {pages} {013126} (\bibinfo {year} {2020})}\BibitemShut
  {NoStop}%
\bibitem [{\citenamefont {Becker}\ and\ \citenamefont
  {Datta}(2020)}]{becker2020convergence}%
  \BibitemOpen
  \bibfield  {author} {\bibinfo {author} {\bibfnamefont {S.}~\bibnamefont
  {Becker}}\ and\ \bibinfo {author} {\bibfnamefont {N.}~\bibnamefont {Datta}},\
  }\bibfield  {title} {\bibinfo {title} {Convergence rates for quantum
  evolution and entropic continuity bounds in infinite dimensions},\
  }\href@noop {} {\bibfield  {journal} {\bibinfo  {journal} {Communications in
  Mathematical Physics}\ }\textbf {\bibinfo {volume} {374}},\ \bibinfo {pages}
  {823} (\bibinfo {year} {2020})}\BibitemShut {NoStop}%
\bibitem [{\citenamefont {Becker}\ \emph {et~al.}(2021)\citenamefont {Becker},
  \citenamefont {Datta}, \citenamefont {Lami},\ and\ \citenamefont
  {Rouz\'e}}]{becker2021energy}%
  \BibitemOpen
  \bibfield  {author} {\bibinfo {author} {\bibfnamefont {S.}~\bibnamefont
  {Becker}}, \bibinfo {author} {\bibfnamefont {N.}~\bibnamefont {Datta}},
  \bibinfo {author} {\bibfnamefont {L.}~\bibnamefont {Lami}},\ and\ \bibinfo
  {author} {\bibfnamefont {C.}~\bibnamefont {Rouz\'e}},\ }\bibfield  {title}
  {\bibinfo {title} {Energy-constrained discrimination of unitaries, quantum
  speed limits, and a gaussian solovay-kitaev theorem},\ }\href
  {https://doi.org/10.1103/PhysRevLett.126.190504} {\bibfield  {journal}
  {\bibinfo  {journal} {Phys. Rev. Lett.}\ }\textbf {\bibinfo {volume} {126}},\
  \bibinfo {pages} {190504} (\bibinfo {year} {2021})}\BibitemShut {NoStop}%
\bibitem [{\citenamefont {Lami}(2021)}]{lami2021quantum}%
  \BibitemOpen
  \bibfield  {author} {\bibinfo {author} {\bibfnamefont {L.}~\bibnamefont
  {Lami}},\ }\bibfield  {title} {\bibinfo {title} {Quantum data hiding with
  continuous-variable systems},\ }\href@noop {} {\bibfield  {journal} {\bibinfo
   {journal} {Physical Review A}\ }\textbf {\bibinfo {volume} {104}},\ \bibinfo
  {pages} {052428} (\bibinfo {year} {2021})}\BibitemShut {NoStop}%
\bibitem [{\citenamefont {Mishra}\ \emph {et~al.}(2022)\citenamefont {Mishra},
  \citenamefont {Oskouei},\ and\ \citenamefont {Wilde}}]{mishra2022optimal}%
  \BibitemOpen
  \bibfield  {author} {\bibinfo {author} {\bibfnamefont {H.~K.}\ \bibnamefont
  {Mishra}}, \bibinfo {author} {\bibfnamefont {S.~K.}\ \bibnamefont
  {Oskouei}},\ and\ \bibinfo {author} {\bibfnamefont {M.~M.}\ \bibnamefont
  {Wilde}},\ }\bibfield  {title} {\bibinfo {title} {Optimal input states for
  quantifying the performance of continuous-variable unidirectional and
  bidirectional teleportation},\ }\href@noop {} {\bibfield  {journal} {\bibinfo
   {journal} {arXiv preprint arXiv:2210.05007}\ } (\bibinfo {year}
  {2022})}\BibitemShut {NoStop}%
\bibitem [{\citenamefont {Holevo}(2012)}]{H12}%
  \BibitemOpen
  \bibfield  {author} {\bibinfo {author} {\bibfnamefont {A.~S.}\ \bibnamefont
  {Holevo}},\ }\href@noop {} {\emph {\bibinfo {title} {Quantum Systems,
  Channels, Information}}},\ de Gruyter Studies in Mathematical Physics (Book
  16)\ (\bibinfo  {publisher} {de Gruyter},\ \bibinfo {year} {2012})\ p.\
  \bibinfo {pages} {349}\BibitemShut {NoStop}%
\bibitem [{\citenamefont {Serafini}(2017)}]{AS17}%
  \BibitemOpen
  \bibfield  {author} {\bibinfo {author} {\bibfnamefont {A.}~\bibnamefont
  {Serafini}},\ }\href@noop {} {\emph {\bibinfo {title} {Quantum Continuous
  Variables: A Primer of Theoretical Methods}}}\ (\bibinfo  {publisher} {CRC
  Press},\ \bibinfo {year} {2017})\BibitemShut {NoStop}%
\bibitem [{\citenamefont {Durt}\ \emph {et~al.}(2010)\citenamefont {Durt},
  \citenamefont {Englert}, \citenamefont {Bengtsson},\ and\ \citenamefont
  {{\.Z}yczkowski}}]{durtMutuallyUnbiasedBases2010}%
  \BibitemOpen
  \bibfield  {author} {\bibinfo {author} {\bibfnamefont {T.}~\bibnamefont
  {Durt}}, \bibinfo {author} {\bibfnamefont {B.-G.}\ \bibnamefont {Englert}},
  \bibinfo {author} {\bibfnamefont {I.}~\bibnamefont {Bengtsson}},\ and\
  \bibinfo {author} {\bibfnamefont {K.}~\bibnamefont {{\.Z}yczkowski}},\
  }\bibfield  {title} {\bibinfo {title} {On mutually unbiased bases},\ }\href
  {https://doi.org/10.1142/S0219749910006502} {\bibfield  {journal} {\bibinfo
  {journal} {International Journal of Quantum Information}\ }\textbf {\bibinfo
  {volume} {08}},\ \bibinfo {pages} {535} (\bibinfo {year} {2010})}\BibitemShut
  {NoStop}%
\bibitem [{\citenamefont {Bengtsson}\ and\ \citenamefont
  {Życzkowski}(2008)}]{bengtsson_geometry_2008}%
  \BibitemOpen
  \bibfield  {author} {\bibinfo {author} {\bibfnamefont {I.}~\bibnamefont
  {Bengtsson}}\ and\ \bibinfo {author} {\bibfnamefont {K.}~\bibnamefont
  {Życzkowski}},\ }\href@noop {} {\emph {\bibinfo {title} {Geometry of quantum
  states: an introduction to quantum entanglement}}},\ \bibinfo {edition}
  {reprinted with corr}\ ed.\ (\bibinfo  {publisher} {Cambridge University
  Press},\ \bibinfo {address} {Cambridge},\ \bibinfo {year} {2008})\BibitemShut
  {NoStop}%
\bibitem [{\citenamefont {Nakahara}(2018)}]{nakahara_geometry_2018}%
  \BibitemOpen
  \bibfield  {author} {\bibinfo {author} {\bibfnamefont {M.}~\bibnamefont
  {Nakahara}},\ }\href@noop {} {\emph {\bibinfo {title} {Geometry, topology and
  physics}}}\ (\bibinfo  {publisher} {CRC press},\ \bibinfo {year}
  {2018})\BibitemShut {NoStop}%
\bibitem [{\citenamefont {Harrow}(2013)}]{harrow_church_2013}%
  \BibitemOpen
  \bibfield  {author} {\bibinfo {author} {\bibfnamefont {A.~W.}\ \bibnamefont
  {Harrow}},\ }\bibfield  {title} {\bibinfo {title} {The {Church} of the
  {Symmetric} {Subspace}},\ }\href {http://arxiv.org/abs/1308.6595} {\bibfield
  {journal} {\bibinfo  {journal} {arXiv:1308.6595 [quant-ph]}\ } (\bibinfo
  {year} {2013})}\BibitemShut {NoStop}%
\bibitem [{\citenamefont {Czartowski}\ \emph {et~al.}(2020)\citenamefont
  {Czartowski}, \citenamefont {Goyeneche}, \citenamefont {Grassl},\ and\
  \citenamefont {Zyczkowski}}]{czartowski_isoentangled_2020}%
  \BibitemOpen
  \bibfield  {author} {\bibinfo {author} {\bibfnamefont {J.}~\bibnamefont
  {Czartowski}}, \bibinfo {author} {\bibfnamefont {D.}~\bibnamefont
  {Goyeneche}}, \bibinfo {author} {\bibfnamefont {M.}~\bibnamefont {Grassl}},\
  and\ \bibinfo {author} {\bibfnamefont {K.}~\bibnamefont {Zyczkowski}},\
  }\bibfield  {title} {\bibinfo {title} {Isoentangled {Mutually} {Unbiased}
  {Bases}, {Symmetric} {Quantum} {Measurements}, and {Mixed}-{State}
  {Designs}},\ }\href {https://doi.org/10.1103/PhysRevLett.124.090503}
  {\bibfield  {journal} {\bibinfo  {journal} {Physical Review Letters}\
  }\textbf {\bibinfo {volume} {124}},\ \bibinfo {pages} {090503} (\bibinfo
  {year} {2020})}\BibitemShut {NoStop}%
\bibitem [{\citenamefont {Hunt}\ \emph {et~al.}(1992)\citenamefont {Hunt},
  \citenamefont {Sauer},\ and\ \citenamefont
  {Yorke}}]{huntPrevalenceTranslationinvariantAlmost1992}%
  \BibitemOpen
  \bibfield  {author} {\bibinfo {author} {\bibfnamefont {B.~R.}\ \bibnamefont
  {Hunt}}, \bibinfo {author} {\bibfnamefont {T.}~\bibnamefont {Sauer}},\ and\
  \bibinfo {author} {\bibfnamefont {J.~A.}\ \bibnamefont {Yorke}},\ }\bibfield
  {title} {\bibinfo {title} {Prevalence: A translation-invariant ``almost
  every'' on infinite-dimensional spaces},\ }\href
  {https://doi.org/10.1090/S0273-0979-1992-00328-2} {\bibfield  {journal}
  {\bibinfo  {journal} {Bulletin of the American Mathematical Society}\
  }\textbf {\bibinfo {volume} {27}},\ \bibinfo {pages} {217} (\bibinfo {year}
  {1992})}\BibitemShut {NoStop}%
\bibitem [{\citenamefont {Grigorchuk}\ and\ \citenamefont {{de la
  Harpe}}(2015)}]{grigorchukAmenabilityErgodicProperties2015}%
  \BibitemOpen
  \bibfield  {author} {\bibinfo {author} {\bibfnamefont {R.}~\bibnamefont
  {Grigorchuk}}\ and\ \bibinfo {author} {\bibfnamefont {P.}~\bibnamefont {{de
  la Harpe}}},\ }\href {http://arxiv.org/abs/1404.7030} {\bibinfo {title}
  {Amenability and ergodic properties of topological groups: From
  {{Bogolyubov}} onwards}} (\bibinfo {year} {2015}),\ \Eprint
  {https://arxiv.org/abs/1404.7030} {arXiv:1404.7030 [math]} \BibitemShut
  {NoStop}%
\bibitem [{\citenamefont {Becnel}\ and\ \citenamefont
  {Sengupta}(2015)}]{becnel_schwartz_2015}%
  \BibitemOpen
  \bibfield  {author} {\bibinfo {author} {\bibfnamefont {J.}~\bibnamefont
  {Becnel}}\ and\ \bibinfo {author} {\bibfnamefont {A.}~\bibnamefont
  {Sengupta}},\ }\bibfield  {title} {\bibinfo {title} {The {Schwartz} {Space}:
  {Tools} for {Quantum} {Mechanics} and {Infinite} {Dimensional} {Analysis}},\
  }\href {https://doi.org/10.3390/math3020527} {\bibfield  {journal} {\bibinfo
  {journal} {Mathematics}\ }\textbf {\bibinfo {volume} {3}},\ \bibinfo {pages}
  {527} (\bibinfo {year} {2015})}\BibitemShut {NoStop}%
\bibitem [{\citenamefont {Rudin}(1991)}]{rudin_functional_1991}%
  \BibitemOpen
  \bibfield  {author} {\bibinfo {author} {\bibfnamefont {W.}~\bibnamefont
  {Rudin}},\ }\href@noop {} {\emph {\bibinfo {title} {Functional analysis}}},\
  \bibinfo {edition} {2nd}\ ed.,\ International series in pure and applied
  mathematics\ (\bibinfo  {publisher} {McGraw-Hill},\ \bibinfo {address} {New
  York},\ \bibinfo {year} {1991})\BibitemShut {NoStop}%
\bibitem [{\citenamefont {Gieres}(2000)}]{gieres_mathematical_2000}%
  \BibitemOpen
  \bibfield  {author} {\bibinfo {author} {\bibfnamefont {F.}~\bibnamefont
  {Gieres}},\ }\bibfield  {title} {\bibinfo {title} {Mathematical surprises and
  {Dirac}'s formalism in quantum mechanics},\ }\href
  {https://doi.org/10.1088/0034-4885/63/12/201} {\bibfield  {journal} {\bibinfo
   {journal} {Reports on Progress in Physics}\ }\textbf {\bibinfo {volume}
  {63}},\ \bibinfo {pages} {1893} (\bibinfo {year} {2000})}\BibitemShut
  {NoStop}%
\bibitem [{\citenamefont {Madrid}(2005)}]{madrid_role_2005}%
  \BibitemOpen
  \bibfield  {author} {\bibinfo {author} {\bibfnamefont {R.~d.~l.}\
  \bibnamefont {Madrid}},\ }\bibfield  {title} {\bibinfo {title} {The role of
  the rigged {Hilbert} space in quantum mechanics},\ }\href
  {https://doi.org/10.1088/0143-0807/26/2/008} {\bibfield  {journal} {\bibinfo
  {journal} {European Journal of Physics}\ }\textbf {\bibinfo {volume} {26}},\
  \bibinfo {pages} {287} (\bibinfo {year} {2005})}\BibitemShut {NoStop}%
\bibitem [{\citenamefont {Dodonov}(2002)}]{dodonov_nonclassical_2002}%
  \BibitemOpen
  \bibfield  {author} {\bibinfo {author} {\bibfnamefont {V.~V.}\ \bibnamefont
  {Dodonov}},\ }\bibfield  {title} {\bibinfo {title} {`{Nonclassical}' states
  in quantum optics: a `squeezed' review of the first 75 years},\ }\href
  {https://doi.org/10.1088/1464-4266/4/1/201} {\bibfield  {journal} {\bibinfo
  {journal} {Journal of Optics B: Quantum and Semiclassical Optics}\ }\textbf
  {\bibinfo {volume} {4}},\ \bibinfo {pages} {R1} (\bibinfo {year}
  {2002})}\BibitemShut {NoStop}%
\bibitem [{\citenamefont {Acharya}\ \emph
  {et~al.}(2021{\natexlab{b}})\citenamefont {Acharya}, \citenamefont {Saha},\
  and\ \citenamefont {Sengupta}}]{acharyaInformationallyCompletePOVMbased2021}%
  \BibitemOpen
  \bibfield  {author} {\bibinfo {author} {\bibfnamefont {A.}~\bibnamefont
  {Acharya}}, \bibinfo {author} {\bibfnamefont {S.}~\bibnamefont {Saha}},\ and\
  \bibinfo {author} {\bibfnamefont {A.~M.}\ \bibnamefont {Sengupta}},\ }\href
  {https://doi.org/10.48550/arXiv.2105.05992} {\bibinfo {title}
  {Informationally complete {{POVM-based}} shadow tomography}} (\bibinfo {year}
  {2021}{\natexlab{b}}),\ \Eprint {https://arxiv.org/abs/2105.05992}
  {arXiv:2105.05992 [quant-ph]} \BibitemShut {NoStop}%
\bibitem [{\citenamefont {Lvovsky}\ and\ \citenamefont
  {Raymer}(2009{\natexlab{b}})}]{Raymer_Lvovsky}%
  \BibitemOpen
  \bibfield  {author} {\bibinfo {author} {\bibfnamefont {A.~I.}\ \bibnamefont
  {Lvovsky}}\ and\ \bibinfo {author} {\bibfnamefont {M.~G.}\ \bibnamefont
  {Raymer}},\ }\bibfield  {title} {\bibinfo {title} {Continuous-variable
  optical quantum-state tomography},\ }\href
  {https://doi.org/10.1103/RevModPhys.81.299} {\bibfield  {journal} {\bibinfo
  {journal} {Rev. Mod. Phys.}\ }\textbf {\bibinfo {volume} {81}},\ \bibinfo
  {pages} {299} (\bibinfo {year} {2009}{\natexlab{b}})}\BibitemShut {NoStop}%
\bibitem [{\citenamefont {Hou}\ and\ \citenamefont
  {Qi}(2010)}]{houConstructingEntanglementWitnesses2010}%
  \BibitemOpen
  \bibfield  {author} {\bibinfo {author} {\bibfnamefont {J.}~\bibnamefont
  {Hou}}\ and\ \bibinfo {author} {\bibfnamefont {X.}~\bibnamefont {Qi}},\
  }\bibfield  {title} {\bibinfo {title} {Constructing entanglement witnesses
  for infinite-dimensional systems},\ }\href
  {https://doi.org/10.1103/PhysRevA.81.062351} {\bibfield  {journal} {\bibinfo
  {journal} {Physical Review A}\ }\textbf {\bibinfo {volume} {81}},\ \bibinfo
  {pages} {062351} (\bibinfo {year} {2010})}\BibitemShut {NoStop}%
\bibitem [{\citenamefont {Albert}\ \emph {et~al.}(2018)\citenamefont {Albert},
  \citenamefont {Noh}, \citenamefont {Duivenvoorden}, \citenamefont {Young},
  \citenamefont {Brierley}, \citenamefont {Reinhold}, \citenamefont {Vuillot},
  \citenamefont {Li}, \citenamefont {Shen}, \citenamefont {Girvin} \emph
  {et~al.}}]{albertPerformanceStructureSinglemode2018}%
  \BibitemOpen
  \bibfield  {author} {\bibinfo {author} {\bibfnamefont {V.~V.}\ \bibnamefont
  {Albert}}, \bibinfo {author} {\bibfnamefont {K.}~\bibnamefont {Noh}},
  \bibinfo {author} {\bibfnamefont {K.}~\bibnamefont {Duivenvoorden}}, \bibinfo
  {author} {\bibfnamefont {D.~J.}\ \bibnamefont {Young}}, \bibinfo {author}
  {\bibfnamefont {{\relax RT}.}~\bibnamefont {Brierley}}, \bibinfo {author}
  {\bibfnamefont {P.}~\bibnamefont {Reinhold}}, \bibinfo {author}
  {\bibfnamefont {C.}~\bibnamefont {Vuillot}}, \bibinfo {author} {\bibfnamefont
  {L.}~\bibnamefont {Li}}, \bibinfo {author} {\bibfnamefont {C.}~\bibnamefont
  {Shen}}, \bibinfo {author} {\bibfnamefont {{\relax SM}.}~\bibnamefont
  {Girvin}}, \emph {et~al.},\ }\bibfield  {title} {\bibinfo {title}
  {Performance and structure of single-mode bosonic codes},\ }\href@noop {}
  {\bibfield  {journal} {\bibinfo  {journal} {Physical Review A}\ }\textbf
  {\bibinfo {volume} {97}},\ \bibinfo {pages} {032346} (\bibinfo {year}
  {2018})}\BibitemShut {NoStop}%
\bibitem [{\citenamefont {Ivan}\ \emph {et~al.}(2011)\citenamefont {Ivan},
  \citenamefont {Sabapathy},\ and\ \citenamefont
  {Simon}}]{ivan_operator-sum_2011}%
  \BibitemOpen
  \bibfield  {author} {\bibinfo {author} {\bibfnamefont {J.~S.}\ \bibnamefont
  {Ivan}}, \bibinfo {author} {\bibfnamefont {K.~K.}\ \bibnamefont
  {Sabapathy}},\ and\ \bibinfo {author} {\bibfnamefont {R.}~\bibnamefont
  {Simon}},\ }\bibfield  {title} {\bibinfo {title} {Operator-sum representation
  for bosonic {Gaussian} channels},\ }\href
  {https://doi.org/10.1103/PhysRevA.84.042311} {\bibfield  {journal} {\bibinfo
  {journal} {Physical Review A}\ }\textbf {\bibinfo {volume} {84}},\ \bibinfo
  {pages} {042311} (\bibinfo {year} {2011})}\BibitemShut {NoStop}%
\bibitem [{\citenamefont {Paris}(1996)}]{paris1996displacement}%
  \BibitemOpen
  \bibfield  {author} {\bibinfo {author} {\bibfnamefont {M.~G.}\ \bibnamefont
  {Paris}},\ }\bibfield  {title} {\bibinfo {title} {Displacement operator by
  beam splitter},\ }\href {https://doi.org/10.1016/0375-9601(96)00339-8}
  {\bibfield  {journal} {\bibinfo  {journal} {Physics Letters A}\ }\textbf
  {\bibinfo {volume} {217}},\ \bibinfo {pages} {78} (\bibinfo {year}
  {1996})}\BibitemShut {NoStop}%
\bibitem [{\citenamefont {Nair}(2018)}]{nair2018}%
  \BibitemOpen
  \bibfield  {author} {\bibinfo {author} {\bibfnamefont {R.}~\bibnamefont
  {Nair}},\ }\bibfield  {title} {\bibinfo {title} {Quantum-limited loss
  sensing: Multiparameter estimation and bures distance between loss
  channels},\ }\href {https://doi.org/10.1103/PhysRevLett.121.230801}
  {\bibfield  {journal} {\bibinfo  {journal} {Phys. Rev. Lett.}\ }\textbf
  {\bibinfo {volume} {121}},\ \bibinfo {pages} {230801} (\bibinfo {year}
  {2018})}\BibitemShut {NoStop}%
\bibitem [{\citenamefont {Nielsen}\ and\ \citenamefont
  {Chuang}(2010)}]{nielsen_quantum_2010}%
  \BibitemOpen
  \bibfield  {author} {\bibinfo {author} {\bibfnamefont {M.~A.}\ \bibnamefont
  {Nielsen}}\ and\ \bibinfo {author} {\bibfnamefont {I.~L.}\ \bibnamefont
  {Chuang}},\ }\href@noop {} {\emph {\bibinfo {title} {Quantum computation and
  quantum information}}},\ \bibinfo {edition} {10th}\ ed.\ (\bibinfo
  {publisher} {Cambridge University Press},\ \bibinfo {address} {Cambridge ;
  New York},\ \bibinfo {year} {2010})\BibitemShut {NoStop}%
\bibitem [{\citenamefont {Cameron}(1951)}]{cameronSimpsonRuleNumerical1951}%
  \BibitemOpen
  \bibfield  {author} {\bibinfo {author} {\bibfnamefont {R.~H.}\ \bibnamefont
  {Cameron}},\ }\bibfield  {title} {\bibinfo {title} {A ``{{Simpson}}'s rule''
  for the numerical evaluation of {{Wiener}}'s integrals in function space},\
  }\bibfield  {journal} {\bibinfo  {journal} {Duke Mathematical Journal}\
  }\textbf {\bibinfo {volume} {18}},\ \href
  {https://doi.org/10.1215/S0012-7094-51-01810-8}
  {10.1215/S0012-7094-51-01810-8} (\bibinfo {year} {1951})\BibitemShut
  {NoStop}%
\bibitem [{\citenamefont {Gel'fand}\ and\ \citenamefont
  {Yaglom}(1960)}]{gelfandIntegrationFunctionalSpaces1960}%
  \BibitemOpen
  \bibfield  {author} {\bibinfo {author} {\bibfnamefont {I.~M.}\ \bibnamefont
  {Gel'fand}}\ and\ \bibinfo {author} {\bibfnamefont {A.~M.}\ \bibnamefont
  {Yaglom}},\ }\bibfield  {title} {\bibinfo {title} {Integration in
  {{Functional Spaces}} and its {{Applications}} in {{Quantum Physics}}},\
  }\href {https://doi.org/10.1063/1.1703636} {\bibfield  {journal} {\bibinfo
  {journal} {Journal of Mathematical Physics}\ }\textbf {\bibinfo {volume}
  {1}},\ \bibinfo {pages} {48} (\bibinfo {year} {1960})}\BibitemShut {NoStop}%
\bibitem [{\citenamefont
  {Brush}(1961)}]{brushFunctionalIntegralsStatistical1961}%
  \BibitemOpen
  \bibfield  {author} {\bibinfo {author} {\bibfnamefont {S.~G.}\ \bibnamefont
  {Brush}},\ }\bibfield  {title} {\bibinfo {title} {Functional {{Integrals}}
  and {{Statistical Physics}}},\ }\href
  {https://doi.org/10.1103/RevModPhys.33.79} {\bibfield  {journal} {\bibinfo
  {journal} {Reviews of Modern Physics}\ }\textbf {\bibinfo {volume} {33}},\
  \bibinfo {pages} {79} (\bibinfo {year} {1961})}\BibitemShut {NoStop}%
\bibitem [{\citenamefont {Konheim}\ and\ \citenamefont
  {Miranker}(1967)}]{konheimNumericalEvaluationWiener1967}%
  \BibitemOpen
  \bibfield  {author} {\bibinfo {author} {\bibfnamefont {A.~G.}\ \bibnamefont
  {Konheim}}\ and\ \bibinfo {author} {\bibfnamefont {W.~L.}\ \bibnamefont
  {Miranker}},\ }\bibfield  {title} {\bibinfo {title} {Numerical evaluation of
  {{Wiener}} integrals},\ }\href
  {https://doi.org/10.1090/S0025-5718-1967-0221753-0} {\bibfield  {journal}
  {\bibinfo  {journal} {Mathematics of Computation}\ }\textbf {\bibinfo
  {volume} {21}},\ \bibinfo {pages} {49} (\bibinfo {year} {1967})}\BibitemShut
  {NoStop}%
\bibitem [{\citenamefont {Royden}\ and\ \citenamefont
  {Fitzpatrick}(1988)}]{royden_real_2010}%
  \BibitemOpen
  \bibfield  {author} {\bibinfo {author} {\bibfnamefont {H.~L.}\ \bibnamefont
  {Royden}}\ and\ \bibinfo {author} {\bibfnamefont {P.}~\bibnamefont
  {Fitzpatrick}},\ }\href@noop {} {\emph {\bibinfo {title} {Real analysis}}},\
  Vol.~\bibinfo {volume} {32}\ (\bibinfo  {publisher} {Pearson},\ \bibinfo
  {year} {1988})\BibitemShut {NoStop}%
\bibitem [{\citenamefont {Blanchard}\ and\ \citenamefont
  {Brüning}(2015)}]{blanchard_mathematical_2015}%
  \BibitemOpen
  \bibfield  {author} {\bibinfo {author} {\bibfnamefont {P.}~\bibnamefont
  {Blanchard}}\ and\ \bibinfo {author} {\bibfnamefont {E.}~\bibnamefont
  {Brüning}},\ }\href {https://doi.org/10.1007/978-3-319-14045-2} {\emph
  {\bibinfo {title} {Mathematical methods in physics: distributions, {Hilbert}
  space operators, variational methods, and applications in quantum
  physics}}},\ \bibinfo {edition} {2nd}\ ed.,\ \bibinfo {series} {Progress in
  mathematical physics}\ No.\ \bibinfo {number} {Vol. 69}\ (\bibinfo
  {publisher} {Birkhäuser},\ \bibinfo {address} {Cham Heidelberg},\ \bibinfo
  {year} {2015})\BibitemShut {NoStop}%
\bibitem [{\citenamefont {Grossmann}(2019)}]{brighter-side-of-mathematics}%
  \BibitemOpen
  \bibfield  {author} {\bibinfo {author} {\bibfnamefont {J.~P.}\ \bibnamefont
  {Grossmann}},\ }\href
  {https://youtube.com/playlist?list=PLBh2i93oe2qvMVqAzsX1Kuv6-4fjazZ8j}
  {\bibinfo {title} {{M}easure {T}heory {Y}ou{T}ube playlist}} (\bibinfo {year}
  {2019})\BibitemShut {NoStop}%
\bibitem [{\citenamefont {Sternberg}(2003)}]{sternbergGroupTheoryPhysics2003}%
  \BibitemOpen
  \bibfield  {author} {\bibinfo {author} {\bibfnamefont {S.}~\bibnamefont
  {Sternberg}},\ }\href@noop {} {\emph {\bibinfo {title} {Group Theory and
  Physics}}},\ \bibinfo {edition} {transferred to digital printing}\ ed.\
  (\bibinfo  {publisher} {{Cambridge Univ. Press}},\ \bibinfo {address}
  {{Cambridge}},\ \bibinfo {year} {2003})\BibitemShut {NoStop}%
\bibitem [{\citenamefont {van Dam}\ and\ \citenamefont
  {Russell}(2011)}]{van_dam_mutually_2011}%
  \BibitemOpen
  \bibfield  {author} {\bibinfo {author} {\bibfnamefont {W.}~\bibnamefont {van
  Dam}}\ and\ \bibinfo {author} {\bibfnamefont {A.}~\bibnamefont {Russell}},\
  }\bibfield  {title} {\bibinfo {title} {Mutually unbiased bases for quantum
  states defined over p-adic numbers},\ }\href {http://arxiv.org/abs/1109.0060}
  {\bibfield  {journal} {\bibinfo  {journal} {arXiv:1109.0060 [quant-ph]}\ }
  (\bibinfo {year} {2011})}\BibitemShut {NoStop}%
\bibitem [{\citenamefont {Wootters}\ and\ \citenamefont
  {Fields}(1989{\natexlab{b}})}]{wootters_optimal_1989}%
  \BibitemOpen
  \bibfield  {author} {\bibinfo {author} {\bibfnamefont {W.~K.}\ \bibnamefont
  {Wootters}}\ and\ \bibinfo {author} {\bibfnamefont {B.~D.}\ \bibnamefont
  {Fields}},\ }\bibfield  {title} {\bibinfo {title} {Optimal
  state-determination by mutually unbiased measurements},\ }\href@noop {}
  {\bibfield  {journal} {\bibinfo  {journal} {Annals of Physics}\ }\textbf
  {\bibinfo {volume} {191}},\ \bibinfo {pages} {363} (\bibinfo {year}
  {1989}{\natexlab{b}})}\BibitemShut {NoStop}%
\bibitem [{\citenamefont {Grundmann}\ and\ \citenamefont
  {Möller}(1978)}]{grundmann_invariant_1978}%
  \BibitemOpen
  \bibfield  {author} {\bibinfo {author} {\bibfnamefont {A.}~\bibnamefont
  {Grundmann}}\ and\ \bibinfo {author} {\bibfnamefont {H.-M.}\ \bibnamefont
  {Möller}},\ }\bibfield  {title} {\bibinfo {title} {Invariant integration
  formulas for the n-simplex by combinatorial methods},\ }\href@noop {}
  {\bibfield  {journal} {\bibinfo  {journal} {SIAM Journal on Numerical
  Analysis}\ }\textbf {\bibinfo {volume} {15}},\ \bibinfo {pages} {282}
  (\bibinfo {year} {1978})}\BibitemShut {NoStop}%
\bibitem [{\citenamefont {Benedetto}\ and\ \citenamefont
  {Czaja}(2009)}]{benedettoIntegrationModernAnalysis2009}%
  \BibitemOpen
  \bibfield  {author} {\bibinfo {author} {\bibfnamefont {J.}~\bibnamefont
  {Benedetto}}\ and\ \bibinfo {author} {\bibfnamefont {W.}~\bibnamefont
  {Czaja}},\ }\href@noop {} {\emph {\bibinfo {title} {Integration and Modern
  Analysis}}},\ Birkh\"auser Advanced Texts\ (\bibinfo  {publisher}
  {{Birkh\"auser}},\ \bibinfo {address} {{Boston}},\ \bibinfo {year}
  {2009})\BibitemShut {NoStop}%
\bibitem [{\citenamefont {Simon}(1971)}]{simonDistributionsTheirHermite1971}%
  \BibitemOpen
  \bibfield  {author} {\bibinfo {author} {\bibfnamefont {B.}~\bibnamefont
  {Simon}},\ }\bibfield  {title} {\bibinfo {title} {Distributions and {{Their
  Hermite Expansions}}},\ }\href {https://doi.org/10.1063/1.1665472} {\bibfield
   {journal} {\bibinfo  {journal} {Journal of Mathematical Physics}\ }\textbf
  {\bibinfo {volume} {12}},\ \bibinfo {pages} {140} (\bibinfo {year}
  {1971})}\BibitemShut {NoStop}%
\bibitem [{\citenamefont {Lundari}\ \emph {et~al.}(2015)\citenamefont
  {Lundari}, \citenamefont {Miranda},\ and\ \citenamefont
  {Pallara}}]{lunardiInfiniteDimAnalysis}%
  \BibitemOpen
  \bibfield  {author} {\bibinfo {author} {\bibfnamefont {A.}~\bibnamefont
  {Lundari}}, \bibinfo {author} {\bibfnamefont {M.}~\bibnamefont {Miranda}},\
  and\ \bibinfo {author} {\bibfnamefont {D.}~\bibnamefont {Pallara}},\
  }\bibfield  {title} {\bibinfo {title} {Infinite dimensional analysis},\
  }\href
  {https://www.mathematik.tu-darmstadt.de/media/analysis/lehrmaterial_anapde/hallerd/Lectures.pdf}
  {\bibfield  {journal} {\bibinfo  {journal}
  {\url{https://www.mathematik.tu-darmstadt.de/media/analysis/lehrmaterial_anapde/hallerd/Lectures.pdf}}\
  } (\bibinfo {year} {2015})}\BibitemShut {NoStop}%
\bibitem [{\citenamefont {Cohn}(2013)}]{cohnMeasureTheory2013}%
  \BibitemOpen
  \bibfield  {author} {\bibinfo {author} {\bibfnamefont {D.~L.}\ \bibnamefont
  {Cohn}},\ }\href@noop {} {\emph {\bibinfo {title} {Measure Theory}}},\
  \bibinfo {edition} {second edition}\ ed.,\ Birkh\"auser Advanced Texts\
  (\bibinfo  {publisher} {{Birkh\"auser}},\ \bibinfo {address} {{Boston}},\
  \bibinfo {year} {2013})\BibitemShut {NoStop}%
\bibitem [{\citenamefont {Varshalovich}\ \emph {et~al.}(1988)\citenamefont
  {Varshalovich}, \citenamefont {Moskalev},\ and\ \citenamefont
  {Khersonskii}}]{VMH}%
  \BibitemOpen
  \bibfield  {author} {\bibinfo {author} {\bibfnamefont {D.~A.}\ \bibnamefont
  {Varshalovich}}, \bibinfo {author} {\bibfnamefont {A.~N.}\ \bibnamefont
  {Moskalev}},\ and\ \bibinfo {author} {\bibfnamefont {V.~K.}\ \bibnamefont
  {Khersonskii}},\ }\href {https://doi.org/10.1142/0270} {\emph {\bibinfo
  {title} {{Quantum Theory of Angular Momentum}}}}\ (\bibinfo  {publisher}
  {World Scientific},\ \bibinfo {year} {1988})\BibitemShut {NoStop}%
\bibitem [{\citenamefont {Temme}\ \emph {et~al.}(2017)\citenamefont {Temme},
  \citenamefont {Bravyi},\ and\ \citenamefont
  {Gambetta}}]{temmeErrorMitigationShortDepth2017}%
  \BibitemOpen
  \bibfield  {author} {\bibinfo {author} {\bibfnamefont {K.}~\bibnamefont
  {Temme}}, \bibinfo {author} {\bibfnamefont {S.}~\bibnamefont {Bravyi}},\ and\
  \bibinfo {author} {\bibfnamefont {J.~M.}\ \bibnamefont {Gambetta}},\
  }\bibfield  {title} {\bibinfo {title} {Error {{Mitigation}} for {{Short-Depth
  Quantum Circuits}}},\ }\href {https://doi.org/10.1103/PhysRevLett.119.180509}
  {\bibfield  {journal} {\bibinfo  {journal} {Physical Review Letters}\
  }\textbf {\bibinfo {volume} {119}},\ \bibinfo {pages} {180509} (\bibinfo
  {year} {2017})}\BibitemShut {NoStop}%
\bibitem [{\citenamefont {Cools}\ and\ \citenamefont
  {Sloan}(1996)}]{cools_minimal_1996}%
  \BibitemOpen
  \bibfield  {author} {\bibinfo {author} {\bibfnamefont {R.}~\bibnamefont
  {Cools}}\ and\ \bibinfo {author} {\bibfnamefont {I.~H.}\ \bibnamefont
  {Sloan}},\ }\bibfield  {title} {\bibinfo {title} {Minimal cubature formulae
  of trigonometric degree},\ }\href
  {https://doi.org/10.1090/S0025-5718-96-00767-3} {\bibfield  {journal}
  {\bibinfo  {journal} {Mathematics of Computation}\ }\textbf {\bibinfo
  {volume} {65}},\ \bibinfo {pages} {1583} (\bibinfo {year}
  {1996})}\BibitemShut {NoStop}%
\bibitem [{\citenamefont {Roy}\ and\ \citenamefont
  {Scott}(2007)}]{royWeightedComplexProjective2007}%
  \BibitemOpen
  \bibfield  {author} {\bibinfo {author} {\bibfnamefont {A.}~\bibnamefont
  {Roy}}\ and\ \bibinfo {author} {\bibfnamefont {A.~J.}\ \bibnamefont
  {Scott}},\ }\bibfield  {title} {\bibinfo {title} {Weighted complex projective
  2-designs from bases: Optimal state determination by orthogonal
  measurements},\ }\href {https://doi.org/10.1063/1.2748617} {\bibfield
  {journal} {\bibinfo  {journal} {Journal of Mathematical Physics}\ }\textbf
  {\bibinfo {volume} {48}},\ \bibinfo {pages} {072110} (\bibinfo {year}
  {2007})},\ \Eprint {https://arxiv.org/abs/quant-ph/0703025}
  {arXiv:quant-ph/0703025} \BibitemShut {NoStop}%
\end{thebibliography}%

\end{document}